\documentclass[12pt]{article}
 
\usepackage{epsfig}
\usepackage{psfig}

\def\vcB{\vec{{\cal B}}}
\def\vcE{\vec{{\cal E}}}
\def\vcJ{\vec{{\cal J}}}
\def\vcH{\vec{{\cal H}}}
\def\vcY{\vec{{\cal Y}}}

\begin{document}   
\begin{titlepage}
\begin{flushright}
CERN-TH/2003-307
\end{flushright}
\vspace*{1cm}

\begin{center}
{\LARGE {\bf The Magnetized Universe}}
\vskip2.cm
{\large Massimo Giovannini\footnote{e-mail address: massimo.giovannini@cern.ch}}
\vskip1.cm
{\it CERN, Theoretical Physics Division, CH-1211 Geneva 23, Switzerland}
\end{center}
\begin{abstract}
Cosmology, high-energy physics and 
astrophysics are  converging on the study of large-scale magnetic fields.
While the experimental evidence for the existence of large-scale 
magnetization in galaxies, clusters and superclusters 
is rather compelling, the origin of the phenomenon 
remains puzzling especially in light of the most recent observations. 
The purpose of the present review is to describe the physical motivations and some of 
the open theoretical 
problems related to the existence of large-scale magnetic fields.
\end{abstract}
\end{titlepage}
\newpage

\pagenumbering{arabic}

\tableofcontents

\newpage

\section{A triple point}
\renewcommand{\theequation}{1.\arabic{equation}}
\setcounter{equation}{0}
Why do we observe magnetic fields in the sky? 
Why do we live in a magnetized Universe?
A variety of observations imply that stars, planets, galaxies,
clusters of galaxies are all magnetized. The typical 
magnetic field strengths\footnote{In this review magnetic fields 
will be expressed in gauss. In the SI units $ 1 {\rm T} = 10^{4} {\rm G}$.
For practical reasons, in cosmic ray physics and in cosmology it is also useful to 
express the magnetic field in ${\rm GeV}^2$ (in units $\hbar=c=1$).
Recalling that the Bohr magneton is about $ 5.7\times 10^{-11} 
{\rm MeV}/{\rm T}$ the conversion factor will then be $1 {\rm G} = 1.95 \times 
10^{-20} {\rm GeV}^2 $.} range from few $\mu$ G 
(in the case of galaxies and galaxy clusters) to few G (in the case of 
planets, like the earth)  up to 
$10^{12}$ G (in the case of neutron stars).
Physical phenomena are characterized depending 
upon the typical time and length scales where they take place. Magnetic 
fields of stars and planets are related to  length-scales
 which are much smaller than the diameter of the Milky Way (of the 
order of $30$ Kpc) or of the local supercluster.
In this sense, a magnetic field of the order of $\mu$ G is 
minute on the terrestrial scale but it is sizable over the scale of the 
supercluster. 

In the present review only large-scale magnetic fields will be considered, i.e.  
magnetic fields whose typical scale exceeds  the AU ( $1 {\rm AU} =1.49\times 10^{13} {\rm cm}$). 
Large-scale magnetic fields must be understood 
and treated as an essential part of the largest structures 
observed in the sky. Legitimate 
questions arising in the study of magnetized structures concern  
their formation and evolution. 
The answers to these questions are still not completely settled. 
Specific observations will eventually allow to discriminate 
between the different competing explanations. 

Magnetic fields of distant spiral galaxies are in the $\mu$ G range.
There is also compelling evidence of large-scale magnetic fields
which are not associated with individual galaxies. This empirical coincidence 
reminds a bit of one of the motivations of the standard hot big-bang model, namely 
the observation that the light elements are equally abundant 
in rather different parts of our Universe.  
The approximate equality of the abundances implies
that, unlike the heavier elements, the light elements 
have primordial origin. The four light isotopes 
${\rm D}$, $^{3}{\rm He}$, $^{4}{\rm He}$ and $^{7} {\rm Li}$ 
are mainly produced at a specific stage of the hot big bang model 
named nucleosynthesis occurring below the a typical temperature of $0.8$ 
MeV when neutrinos decouple from the plasma and the neutron abundance 
evolves via free neutron decay. The abundances calculated in the 
simplest big-bang nucleosythesis model agree fairly well with the 
astronomical observations.
In similar terms it is plausible 
to argue that large-scale magnetic fields have similar strengths at 
large scales because the initial conditions for their evolutions 
were the same, for instance at the time of the gravitational collapse 
of the protogalaxy. The way the initial conditions for the evolution
of large-scale magnetic fields are set  is generically 
named magnetogenesis.
 
There is another comparison which might be useful. Back in the seventies
the so-called Harrison-Zeldovich spectrum was postulated. Later, with the developments 
of inflationary cosmology the origin of a flat spectrum of curvature and density 
profiles has been justified on the basis of a period of quasi-de Sitter expansion named inflation.
It is plausible that in some inflationary models not only the fluctuations 
of the geometry are amplified but also the fluctuations of the gauge fields. This 
happens if, for instance, gauge couplings are effectively dynamical. As the Harrison-Zeldovich 
spectrum can be used as initial condition for the subsequent Newtonian evolution, the primordial 
spectrum of the gauge fields can be used as initial condition for the subsequent
magnetohydrodynamical (MHD) evolution which may lead, eventually, to the observed 
large-scale magnetic fields. 

Cosmologists and theoretical physicists  would like to understand 
large-scale magnetization in terms of symmetries which are  broken.
There are other two different motivations leading, independently, to the study of 
large-scale magnetic fields. For instance it was observed long ago by Fermi that 
if cosmic rays are in equilibrium with the galaxy, their pressure density 
is comparable with the one of the magnetic field of the galaxy.
Large-scale magnetic fields are also 
extremely relevant for astrophysics. Magnetic fields in galaxies 
are sufficiently intense to affect the dynamics of interstellar gas both 
on the galactic scale and also on smaller scales characteristic 
of star formation processes. This is the third motivation leading 
to the study of large scale-magnetic fields: the astrophysical 
motivation which tries to combine our knowledge of the Universe in powerful
dynamical principles based on the microscopic laws
of nature. 

The cross-disciplinary character of the physical phenomena addressed in this 
review is apparent from the table of content.
In Section 2 a brief historical account of this fifty years old problem 
will be given. Then, to clarify the  nature  of the 
various theoretical constructions, the empirical evidence of large scale 
magnetic fields will be discussed in Section 3. Section 4 contains some 
background material on the evolution of magnetic fields in globally 
neutral plasmas in flat space.  Section 5 and 6 address the problem of the 
evolution  of large scale magnetic fields in curved backgrounds and of their origin.
In Section 7 the possible effects of magnetic fields on the 
thermodynamical history of the Universe are scrutinized. 
Section 8 will be concerned with the effects of large scale magnetic fields 
on the Cosmic Microwave Background radiation (CMB) and on the 
relic background of gravitational waves. In Section 9 the r\^ole 
of gravitating magnetic fields in cosmological solutions will be 
swiftly pointed out. 

The perspective of present  review  is theoretical.  In this 
sense various (very important) experimental results 
(for instance concerning optical polarization)
will receive only a swift attention. 

\section{From Alfv\'en  to ROSAT}
\renewcommand{\theequation}{2.\arabic{equation}}
\setcounter{equation}{0}

In January 1949 large-scale magnetic fields had no empirical 
evidence. The theoretical situation can be summarized as 
follows: 
\begin{itemize}
\item{} the seminal contributions of Alfv\'en \cite{alv1} convinced the community that 
magnetic fields can have a very large life-time in a highly conducting plasma;
\item{} using the new discoveries  of Alfv\'en,  Fermi \cite{fermi} postulated 
the existence of a large-scale magnetic field permeating 
the galaxy with approximate intensity of $\mu$ G and, hence, in equilibrium with the cosmic rays;
\item{} rather surprisingly Alfv\'en \cite{alv2} did not appreciate the implications 
of the Fermi idea and was led to conclude (incorrectly) that cosmic
rays are in equilibrium with stars disregarding completely the possibility 
of a galactic magnetic field;
\item{} in 1949 Hiltner \cite{hiltner} and, independently, 
Hall \cite{hall} observed polarization of starlight which was later on interpreted 
by Davis and Greenstein \cite{davis} as an effect of galactic magnetic field 
aligning the dust grains 
\footnote{It should be noticed that the observations of Hiltner \cite{hiltner} and Hall \cite{hall}
took place from November 1948 to January 1949. The paper of Fermi \cite{fermi} was submitted in January 1949 but 
it contains no reference to the work of Hiltner and Hall. This indicates the Fermi was probably not aware 
of these optical measurements.}. 
\end{itemize}
According to the presented chain of events it is legitimate to conclude that
\begin{itemize}
\item{} the discoveries of Alfv\'en were essential in the Fermi proposal 
(who was already thinking of the origin of cosmic 
rays in 1938 before leaving Italy);
\item{} the idea that cosmic rays are in equilibrium with the galactic 
magnetic fields (and hence that the galaxy possess a magnetic field) 
was essential in the correct interpretation of the first, fragile, optical 
evidence of  galactic magnetization.
\end{itemize}
The origin of the galactic magnetization, according to \cite{fermi}, had to be somehow primordial.
This idea was further stressed in two subsequent investigations of Fermi and Chandrasekar \cite{fermi2,fermi3}
 who tried, rather ambitiously, to connect the magnetic field of the 
galaxy to its angular momentum.
 
In the fifties various observations on polarization
of Crab nebula suggested that the Milky Way is not the only magnetized 
structure in the sky. The 
effective new twist in the observations of large-scale magnetic fields 
was the development (through the fifties and sixties) of radio-astronomical 
techniques. From these measurements, the first unambiguous 
evidence of radio-polarization from the Milky Way (MW)
was obtained \cite{wiel}.  

It was also soon realized that the radio-Zeeman effect (counterpart of the optical 
Zeeman splitting employed to determine the magnetic field of the sun)
could offer accurate determination of (locally very strong) magnetic fields 
in the galaxy. The observation of Lyne and Smith \cite{lyne} 
that pulsars could be used to determine the column density 
of electrons along the line of sight opened 
the possibility of using not only synchrotron 
emission as a diagnostic of the presence of a large-scale magnetic field, but also 
Faraday rotation. In the seventies all the basic experimental tools 
for the analysis of galactic and extra-galactic magnetic fields 
were present. Around this epoch also 
extensive reviews on the experimental endeavors 
started appearing and a very nice account 
could be found, for instance, in the review of Heiles \cite{heiles}.

It became gradually clear in the early eighties, that measurements 
of large-scale magnetic fields in the MW and in the external galaxies 
are two complementary aspects of the same problem. While MW studies 
can provide valuable informations  concerning the {\em local} 
structure of the galactic magnetic field, 
the observation of external galaxies provides 
the only viable tool for the reconstruction
of the {\em global } features of the 
galactic magnetic fields. As it will be clarified 
in the following Sections, the complementary 
nature of global and local morphological 
features of large-scale magnetization may become, sometimes, 
a source of confusion. 

Since the early seventies, some relevant attention 
has been paid not only to the magnetic fields of the 
galaxies but also to the magnetic fields of the {\em clusters}.
A cluster is a gravitationally bound system of galaxies. 
The {\em local group} (i.e. {\em our} cluster containing the MW, Andromeda
together with other fifty galaxies) is an {\em irregular} cluster 
in the sense that it contains fewer galaxies than typical clusters 
in the Universe. Other clusters (like Coma, Virgo) are more typical 
and are then called {\em regular} or Abell clusters. As an order 
of magnitude estimate, Abell clusters can contain $10^{3}$ galaxies.

In the nineties magnetic fields have been measured in single Abell
clusters but around the turn of the century these estimates became 
more reliable thanks to improved experimental techniques.
In order to estimate magnetic fields in clusters, 
an independent knowledge of the electron density along 
the line of sight is needed (see Sec. 3). Recently Faraday rotation measurements
obtained by radio telescopes (like VLA \footnote{The Very Large Array Telescope, 
consists of 27 parabolic antennas spread over a surface of 20 ${\rm km}^2$ 
in Socorro (New Mexico)}) have been combined with independent measurements 
of the electron density in the intra-cluster medium. This was made possible
by the maps of the x-ray sky obtained with satellites measurements 
(in particular  ROSAT \footnote{The  ROegten SATellite (flying 
from June 1991 to  February 1999) provided maps of the x-ray sky in the 
range $0.1$--$2.5$ keV. A catalog of x-ray bright Abell  clusters was compiled.}).
This improvement in the experimental capabilities seems to have partially 
settled the issue confirming the measurements of the early nineties and implying 
that also clusters are endowed with a magnetic field of $\mu $G strength 
which is {\em not associated with individual galaxies}. 

The years to come are full of interesting experimental questions to be 
answered. These questions, as it will be discussed in the following, 
may have a rather important theoretical impact both on the 
theory of processes taking place in the local universe (like 
ultra-high energy cosmic ray propagation) and on the models 
 trying to explain the origin of large scale magnetic fields. 
Last but not least, these measurements may have an impact 
on the physics of CMB anisotropies and, in particular, on the 
CMB polarization. In fact, the same mechanism leading to the 
Faraday rotation in the radio leads to a Faraday  rotation 
of the CMB {\em provided} the CMB is linearly polarized.
One of the important questions to be answered is, for instance, 
the nature and strength of the  supercluster magnetic field and 
now more careful statistical studies are starting also 
along this important direction.
Superclusters are gravitationally bound systems of clusters. An 
example is the local supercluster formed by the local group 
and by the VIRGO cluster. Together ROSAT, various observations with the EINSTEIN, EXOSAT, 
and GINGA satellites showed
the presence of hot diffuse gas in the Local Supercluster.
 The estimated magnetic field 
in this system is, again, of the order of the $\mu$ G but 
the observational evidence is still not conclusive.  
Another puzzling evidence is the fact that Lyman-$\alpha$ 
systems (with red-shifts $z \sim 2.5$) 
are endowed with a magnetic field \cite{kro}.

From the historical development of the subject a series of 
questions arises naturally:
\begin{itemize}
\item{} what is the origin of large-scale magnetic 
fields? 
\item{} are magnetic fields primordial as
assumed by Fermi more than 50 years ago? 
\item{} even assuming that large-scale magnetic fields are primordial, is there a theory 
for their generation? 
\end{itemize}

\section{Large-scale magnetic fields observations}
\renewcommand{\theequation}{3.\arabic{equation}}
\setcounter{equation}{0}

Before describing the theory of something (not of everything) 
it is highly desirable, in natural sciences, to understand 
in some detail the empirical evidence of the subject under discussion.
While technical accounts of the various experimental techniques exist
already \cite{heiles,kro,vallee1,beck,battaner,han,zweibel}
it seems useful, in the present context, to give 
an account of the experimental situation by emphasizing 
the physical principles of the various measurements.
There are valuable books dealing directly with 
large-scale magnetism \cite{zeldovich,parker,ruzmaikin}. Furthermore 
excellent reviews of the morphological features 
of large scale magnetism can be found in \cite{vallee1,beck,battaner,han}.
Even if the present review will not be concerned with planetary magnetic fields, 
analogies with physical situations arising closer to the earth are often useful and, along this perspective, 
magnetic field observations in the solar system have been recently 
reviewed \cite{vallee2}.

Magnetic fields observed in the Universe 
have a homogeneous (or uniform) component and a non-uniform component. It is 
then useful, for the purposes of this Section, to write, in a schematic notation, that 
\begin{equation}
B_{\rm tot} = \overline{B} + \delta B,
\label{btot}
\end{equation}
where $\overline{B}$ is the uniform component and $\delta B$ the non-uniform
component. Both components are phenomenologically very relevant. Different 
experimental techniques can probe different components of large-scale magnetic fields 
(i.e. either the total magnetic field or its homogeneous component). Another 
important distinction is between $B_{\perp}$ and $B_{\parallel}$: there are measurements 
(like synchrotron emission) which are sensitive to $B_{\perp}$ i.e. the transverse 
component of the magnetic field; there are other measurements (like the Faraday Rotation 
measure or the Zeeman splitting of spectral lines) which are sensitive to $B_{\parallel}$ 
i.e. the magnetic field along the line of sight.

\subsection{Zeeman splitting}

In order to measure large scale magnetic fields, 
one of the first effects coming to mind,  is  the Zeeman effect.
The energy levels of  an hydrogen atom in the background of a magnetic field 
are not degenerate. The presence of a magnetic field
produces a well known splitting of the spectral lines:
\begin{equation}
\Delta \nu_{Z} = \frac{e \overline{B}_{\parallel}}{2 \pi m_{e}}.
\label{zee}
\end{equation}
where $\overline{B}_{\parallel}$ denotes the {\em  uniform component of the 
magnetic field along the line of sight}.
From the estimate of the 
splitting, the magnetic field intensity can be deduced. Indeed
this technique is the one commonly employed 
in order to measure the magnetic field of the sun \cite{vallee2}. 
The most common element in the interstellar medium 
is neutral hydrogen, emitting the celebrated $21$-cm 
line (corresponding to a frequency of $1420$ MHz). 
If a magnetic field of $\mu$ G strength is present 
in the interstellar medium, according to 
Eq. (\ref{zee}), an  induced splitting,
 $\Delta\nu_{Z} \sim 3 {\rm Hz}$, can be estimated. 
Zeeman splitting of 
the $21$-cm line generates two oppositely circular 
polarized spectral lines whose apparent splitting is 
however sub-leading if compared to the Doppler broadening.
In fact, the atoms and molecules in the interstellar medium 
are subjected to thermal motion and the 
amount of induced Doppler broadening is roughly given by 
\begin{equation}
\Delta \nu_{\rm Dop} \sim \biggl(\frac{v_{\rm th}}{c}\biggr) \nu,
\label{dop}
\end{equation}
where $ v_{\rm th}$ is the thermal velocity $\propto \sqrt{T/m}$ where 
$m$ is the mass of the atom or molecule. The amount of Doppler broadening 
is $\Delta \nu_{\rm Dop} \sim 30$ kHz which is much larger than the 
Zeeman splitting which should be eventually detected.
Zeeman splitting of molecules and recombination lines should however 
be detectable if the magnetic field strength gets larger with the density. 
Indeed in the interstellar medium there are molecules with an unpaired 
electron spin. From Eq. (\ref{zee}) it is clear that a detectable Zeeman splitting (i.e. comparable or possibly 
larger than the Doppler broadening) can be generically obtained for 
magnetic fields where $\overline{B}_{\parallel} \simeq 10^{-3}$ G, i.e.
magnetic fields of the order of the m G.
Molecules with an unpaired electron spin include OH, CN, CH and some other. In the past, for instance, 
magnetic fields have been estimated in OH clouds (see \cite{heiles} and references therein). 
Magnetic fields of the order of $10$ m G have been detected in interstellar ${\rm H}_2 {\rm O}$ maser 
clumps (with typical densities ${\cal O}(10^{10} {\rm cm}^{-3})$) \cite{fiebig}. More recently 
attempts of measuring magnetic fields in CN have been reported \cite{crutcher}. 
The possible 
caveat with this type of estimates is that the measurements can 
only be very {\em local}: the above mentioned molecules are much less common 
than neutral hydrogen and are localized in specific 
regions of the interstellar medium. In spite of this caveat, 
Zeeman splitting measurements  can provide reliable 
informations on the local direction of the magnetic field. This 
determination is important in order to understand the possible 
origin of the magnetic field. This aspect will be discussed, in more detail,
when describing the magnetic field of the Milky Way.

\subsection{Synchrotron emission}

The first experimental evidence of the existence of large-scale magnetic 
fields in external galaxies came, historically, from the synchrotron emission of Crab nebula \cite{shklovskij}.  
The emissivity formula (i.e. the energy emitted from a unit volume, per unit time, per unit frequency interval and 
per unit solid angle)
for the synchrotron is a function of  $B_{\perp}$ and and of the relativistic 
electron density, namely 
\begin{equation}
{\cal W}(B_{\perp}, \nu) = {\cal W}_{0}~ n_{0}~ B_{{\rm tot}, \perp}^{(1+\alpha)/2} \nu^{(1-\alpha)/2},
\end{equation}
where $\nu$ is the frequency, ${\cal W}_{0}$ is a proportionality constant depending 
only upon the spectral index $\alpha$ which also determines the (isotropic) relativistic electron number density distribution
for electrons with energies in the range $dE$:  
\begin{equation}
n_{\rm e} d E = n_{0} E^{-\alpha} d E.
\end{equation}
A useful estimate for the maximal emission from electrons of energy $E$ 
is given by \cite{ruzmaikin}
\begin{equation}
\biggl(\frac{\nu}{{\rm MHz}}\biggr) \simeq 15 \biggl( \frac{B_{{\rm tot},\perp}}{\mu {\rm G}} \biggr) \biggl( \frac{E}{{\rm GeV}}\biggr)^2.
\end{equation}
In generic terms the synchrotron emission is then sensitive to the 
{\em total} transverse magnetic field. 
In oversimplified terms the measurement proceeds in three steps:
\begin{itemize} 
\item{} estimate of the emission intensity;
\item{} from the estimate of the emission intensity, defining as $L$ the typical size of the source, the 
quantity $B_{{\rm tot},\perp}^{(\alpha + 1)/2} {\cal W}_0 L$ can be deduced;
\item{} the magnetic field can then be obtained by specifying the electron density 
distribution of the source.
\end{itemize}
While the second step is rather well defined ($\alpha$ can indeed be determined 
from the observed emission spectrum \footnote{Typical galactic values 
 of $\alpha$ are between $2.4$ and $2.8$. An important caveat to this 
statement, relevant for foreground extraction (for instance in studies 
of CMB anisotropies) is that the synchrotron spectrum may have a break.})
the third step has to be achieved in a model-dependent way.
The relativistic electron density is sometimes 
estimated using equipartition, i.e. the idea that 
magnetic and kinetic energy densities may be, after all, comparable. 
Equipartition is not always an empirical evidence, but can certainly
 be used as a working hypothesis which may or may not be 
realized in the system under consideration. For instance equipartition 
probably holds for the Milky Way but it does not seem to be valid in 
the Magellanic Clouds \cite{magcl}. The average equipartition 
field strengths in galaxies ranges from the $4 \mu$ G of M33 up to the 
$19 \mu$ G of NGC2276 \cite{eq,eq2}. 

In order to illustrate
the synchrotron emission take for instance the MW. Away 
from the region of the galactic plane synchrotron emission 
is the dominant signal of the galaxy for frequencies below $5$ GHz. Between 
$1.4$ and $5$ GHz the spectral dependence is ${\cal O}( \nu^{-0.9})$. It 
should be noticed that the spectral index of synchrotron emission may change 
in different frequency ranges and also in different regions of the sky 
due to the variation of the magnetic field \cite{lawson}. 
This aspect is crucial in the experimental 
analysis of CMB anisotropy experiments in the GHz region. In fact, in the 
context of CMB studies the synchrotron is a {\em foreground} 
which should be appropriately subtracted. The way this is done 
is by extrapolation of lower frequency measurements. The problem 
of proper subtraction of synchrotron foreground is particularly acute in the 
case of the PLANCK mission \footnote{The satellite mission PLANCK EXPLORER 
will provide, after 2008, full sky maps in nine frequency channels ranging from 
$30$ to $900$ GHz.}

The synchrotron has an intrinsic polarization 
which can give the orientation of the magnetic field, 
but not {\em the specific sign of 
the orientation vector}. The strength of synchrotron 
polarization is proportional to $|\overline{B}_{\perp}/B_{{\rm tot}, \perp}|^2$, i.e. 
the ratio between the magnetic energy densities of the uniform and total magnetic field. 
Thus, according to Eq. (\ref{btot}), the synchrotron polarization is sensitive 
to the {\em random component} of the large-scale magnetic field.

\subsection{Faraday Rotation Measurements}

To infer the magnitude of the 
magnetic field strength Faraday effect has been 
widely used. When a polarized radio wave passes 
through a region of space of size $\delta \ell$ containing a plasma with a magnetic field 
the polarization plane of the wave gets rotated by an amount
\begin{equation}
\delta \phi \propto \omega_{\rm B} \biggl(\frac{\omega_{\rm p}}{\omega}\biggr)^2 \delta \ell,
\end{equation} 
which is directly proportional to the square of the plasma 
frequency \footnote{See Eq. (\ref{pl}) in the following Section.}
 $\omega_{\rm p}$ (and hence to the electron density) and to the 
Larmor frequency $\omega_{\rm B}$ (and hence to the magnetic field intensity). 
A linear regression connecting the shift in the polarization
plane and the square of the wavelength $\lambda$, can be obtained:
\begin{equation}
\phi = \phi_{0} + {\rm RM} \lambda^2.
\label{regr}
\end{equation}
By measuring the relation expressed by Eq. (\ref{regr}) 
for two (or more) separate (but close) 
wavelengths, the angular coefficient of the regression can be 
obtained and it turns out to be
\begin{equation}
\frac{ \Delta \phi}{\Delta \lambda^2} = 811.9 \int \biggl(\frac{n_{e}}{
{\rm cm^{-3}}}\biggr) \biggl( \frac{ \overline{B}_{\parallel}}{\mu {\rm G}}\biggr) d 
\biggl(\frac{\ell}{ {\rm kpc}}\biggr), 
\label{FR}
\end{equation}
in units of ${\rm rad}/{\rm m}^2$ when all the quantities 
of the integrand are measured in the above units. The explicit 
dependence of the red-shift can be also easily included 
in Eq. (\ref{FR}). Notice, in general terms that the RM is an integral 
over distances. Thus the effect of large distances will reflect in high 
values of the RM. Furthermore, the Faraday effect occurs typically 
in the radio (i.e. ${\rm cm} <\lambda <{\rm m}$), however, some 
possible applications of Faraday effect in the microwave can be also expected 
(see Section 8 of the present 
review).
 
The shift in the polarization plane  should be 
determined with an accuracy greater than $\delta \phi \sim \pm \pi$. Otherwise
ambiguities may arise in the determination of the angular 
coefficient appearing in the linear regression of Eq. (\ref{regr}). 
\begin{figure}[tp]
\centerline{\epsfig{file=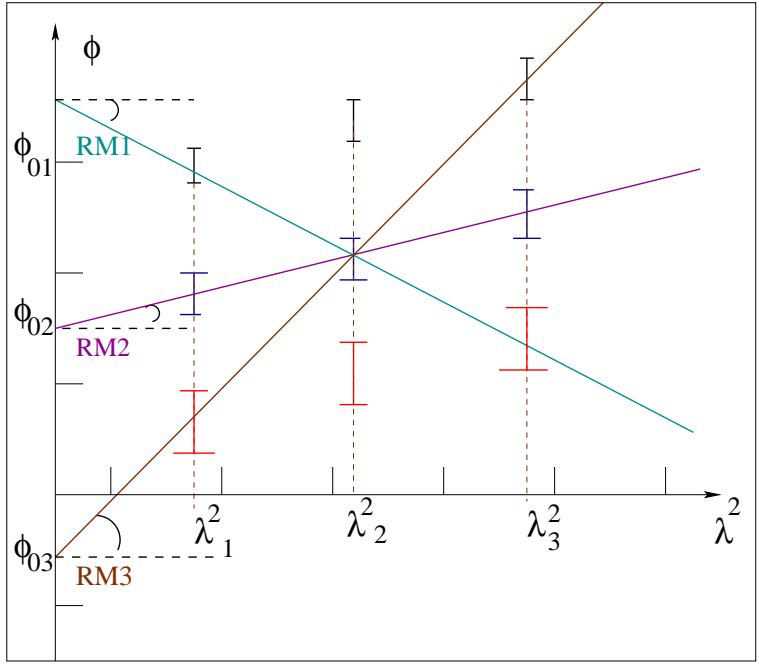, width=10cm}}
\vspace*{8pt}
\caption[a]{The possible ambiguities arising in practical determinations 
of the RM are illustrated. The RM is the angular coefficient of the linear 
regression expressed by Eq. (\ref{regr}). Clearly it is not necessary to know 
the initial polarization of the source to determine the slope of a straight line in the ($\phi,\lambda^2$) plane, 
but it is enough to measure 
$\phi$ at two separate wavelength. However, if the accuracy in the determination 
of $\phi$ is of the order of $\pi$ the inferred determination of the angular coefficient 
of the linear regression (\ref{regr}) is ambiguous.}
\label{FARFIG}
\end{figure}
This aspect is illustrated in Fig. \ref{FARFIG} 
which is rather standard but 
it is reproduced here 
in order to stress  the possible problems arising in the 
physical determination of the RM if the determination of the shift in the polarization
plane is not accurate. 

The RM defined in Eq. (\ref{FR}) 
not only the magnetic field (which should be 
observationally estimated), but also the column density 
of electrons. From the radio-astronomical 
observations, different techniques can be used in order 
to determine $n_{e}$. One 
possibility is to notice that in 
 the observed Universe there 
are pulsars. Pulsars are astrophysical objects emitting 
regular pulses of electromagnetic radiation with periods 
ranging from few milliseconds to few seconds. By comparing the arrival 
times of different radio pulses at different radio wavelengths, it is 
found that signals are slightly delayed as they pass through the 
interstellar medium exactly because electromagnetic waves travel 
faster in the vacuum than in an ionized medium. Hence, from pulsars 
the column density of electrons can be obtained in the 
form of the  dispersion measure, i.e. 
$DM \propto \int n_{e} d\ell$. Dividing the RM by DM,
an estimate of the magnetic field can be obtained. Due to their 
abundance, pulsars lead to the best 
determination of the magnetic field in the 
galactic disk \cite{han2}.
\begin{figure}[tp]
\centerline{\epsfig{file=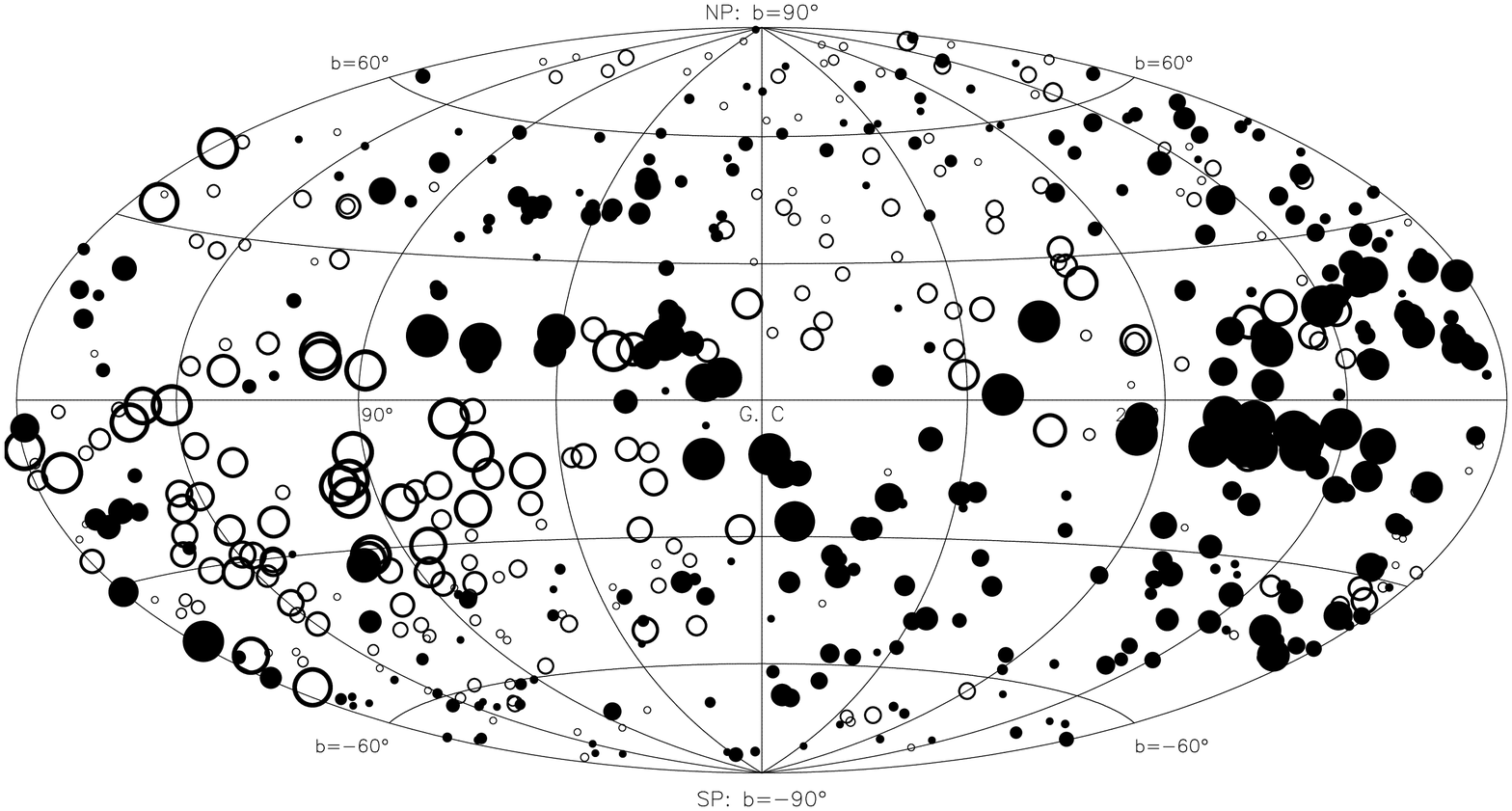, width=12cm}}
\vspace*{8pt}
\caption[a]{The filtered RM distribution of extragalactic radio sources. The antisymmetric 
distribution is clear especially from the inner galactic quadrant. This picture 
is adapted from \cite{han4}.}
\label{F1a}
\end{figure}
In Fig. \ref{F1a} (adapted from \cite{han4}) a map of 
the antisymmetric RM sky is reported. In the picture the open circles denote 
negative RM while filled circles denote positive RM. 
The size of the circle is proportional to the magnitude 
of the RM. The convention is, in fact, to attribute 
negative RM to a magnetic field directed away from the observer and positive 
RM if the magnetic field is directed toward the observer.

As in the case of synchrotron emission also Faraday rotation 
measurements can be used as a diagnostic for foreground 
contamination. The idea would be, in this context to 
look for cross-correlations in the Faraday rotation measure  
of extra-galactic sources and the measured microwave signal 
at the same angular position. A recent analysis has been 
recently reported \cite{coles}.

If magnetic field or the column density change considerably over 
the integration path of Eq. (\ref{FR}) one should probably 
define and use the two-point function of the RM, i.e. 
\begin{equation}
{\cal C}(\vec{r}) = \langle {\rm RM}(\vec{x} + \vec{r}) {\rm RM}(\vec{x})\rangle. 
\label{2ptRM}
\end{equation}
The suggestion  to study the mean-squared fluctuation of the RM was proposed
 \cite{goldshmidt,crusius}. More recently, using this statistical approach 
particularly appropriate in the case of magnetic fields in clusters (where both 
the magnetic field intensity and the electron density change 
over the integration path), Newmann, Newmann and Rephaeli \cite{rephaeli2} 
quantified the possible (statistical) uncertainty in the determination of cluster 
magnetic fields (this point will also be discussed later). 
The rather ambitious program of measuring the $RM$ power spectrum 
is also pursued \cite{vogt1,vogt2}. In \cite{kolatt} the analysis 
of correlations in the RM has been discussed.

\subsection{Magnetic fields in the Milky Way}

Since the early seventies \cite{heiles,mw1} the magnetic field 
of the Milky way (MW) was shown to be parallel to the galactic plane.
RM derived from pulsars allow  consistent determinations 
of the magnetic field direction and intensity \cite{mw2,mw3}. In the Milky Way,
the uniform component of the 
 magnetic field  is  in the plane of 
the galactic disk and it is thought to be directed approximately along the spiral 
arm. There is, though, a slight difference between 
the northern and southern hemisphere. While in the southern hemisphere 
the magnetic field is roughly $2 \mu$G, the  magnetic field 
in the northern hemisphere is three times smaller than the 
one of the southern hemisphere. 
The magnetic field of the MW is predominantly toroidal 
\footnote{Recall that 
the toroidal field is defined as the vector sum of the radial and 
azimuthal components, i.e. $\vec{B}_{\rm toroidal} 
= B_{\theta} \hat{e}_{\theta} + B_{r} \hat{e}_{r}$.}. 
However, looking at the center of the galaxy, 
a poloidal component (typically $1/3$ of the toroidal one) 
has been detected. In fact, in the central $100$ pc of the MW there are in fact pooloidal 
(dipole) 
fields whose origin is probably primordial 
\cite{morris}( see also \cite{yusef,yusef2}).
The reason for this statement is that through the usual plasma physics 
mechanisms (like the dynamo theory to be discussed later in Section 4) 
it is hard to account for a sizable poloidal component, even if 
localized in the central region of the MW. 

The common practice is to classify large-scale magnetic fields 
in spiral galaxies according to the parity of the toroidal field 
under reflections of the azimuthal angle \cite{ruzmaikin,vallee1}. According to some authors 
this distinction palys a crucial r\^ole in order to assess the 
primordial (or non-primordial ) nature of the observed field (see, for instance, \cite{beck,zweibel}). 
In short, the virtues and limitations of the previous distinction can be 
summarized as follows. Suppose to plot the toroidal component of the measured magnetic field 
(for instance of the MW) as a function of the azimuthal angle $\theta$. 
This is, in fact, equivalent, to plot the $RM(\theta)$ for instance near 
the equatorial plane of the galaxy. Then  two qualitatively different
situations can arise. In the first case $RM(\theta)$ upon rotation 
of $\pi$ around the galactic center is even and this corresponds 
to axisymmetric spiral galaxies (ASS). In the second case $ RM(\theta)$ upon rotation 
of $\pi$ around the galactic center is odd
and this corresponds to bisymmetric spiral galaxies (BSS). 
This sharp distinction is invalidated by two experimental facts. First of all 
there may be extra phases so that it is not clear, in practice, if $RM(\theta)$ 
is, predominantly, even or odd. In spite of possible ambiguities 
(which are, however, an experimental problem) the distinction may be very useful: 
 if the magnetic fields would originate through 
some dynamo amplification the preferred configuration will be of the ASS-type
and not of the BSS-type. There are strong indications that the MW is a BSS spiral
\cite{han2,han4}. A related issue is the {\em reversal } of the magnetic field 
from one spiral arm to the other. In \cite{han4} $63$ rotation 
measures from polarization observations of southern pulsars 
have been reported. For the galactic disk, convincing evidence of field 
reversal near the {\em Perseus} arm is presented. In the solar circle three reversals are 
observed: near the {\em Carina-Sagittarius} arm, near the {\em Crux-Scutum} arm, and possibly a 
third one in the {\em Norma} arm. These reversals are claimed to be consistent with BSS models.

\begin{figure}[th]
\centerline{\epsfig{file=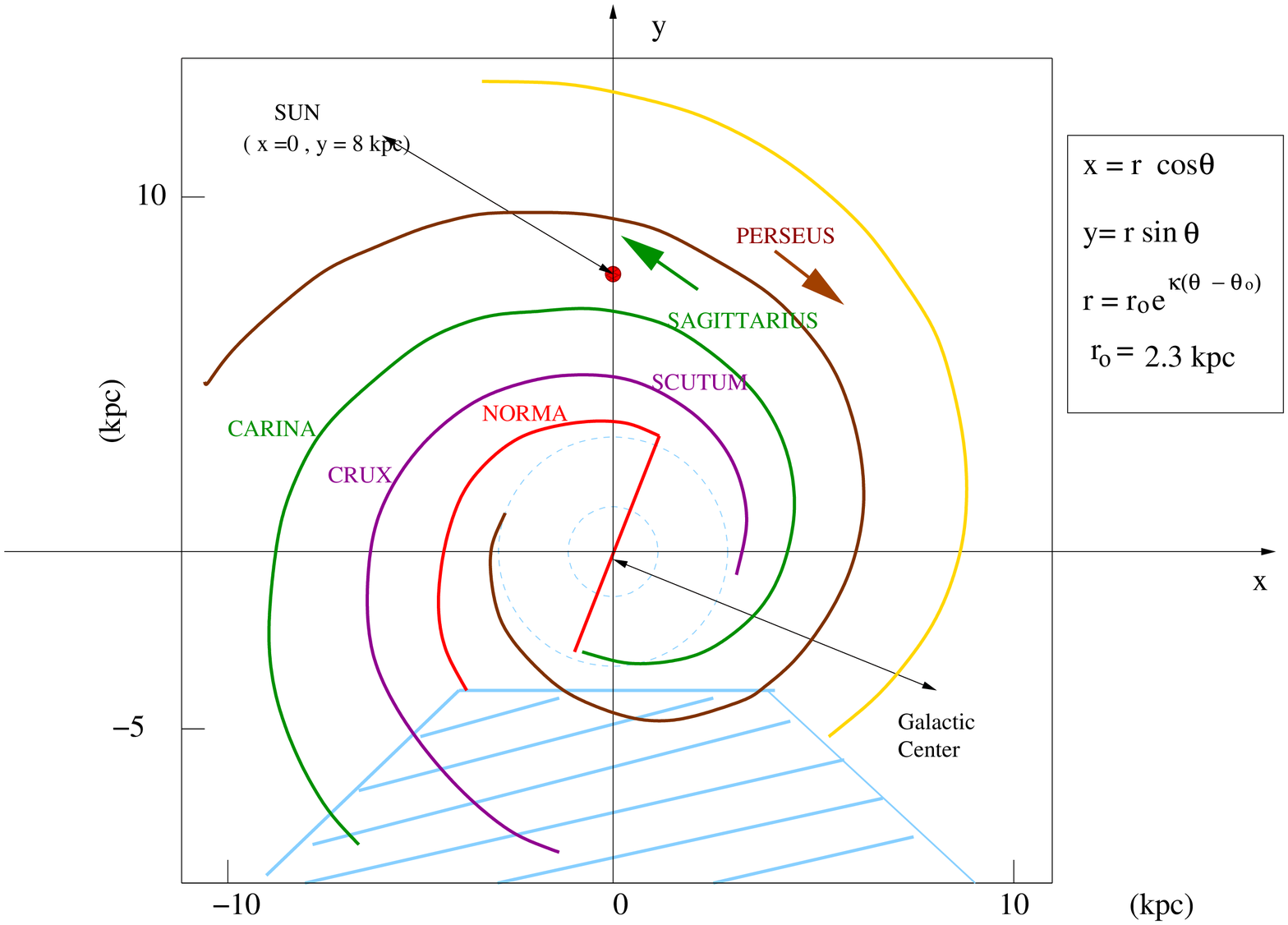,width=10cm}}
\vspace*{8pt}
\caption[a]{The map of the MW is illustrated. Following \cite{vallee3} the origin of the two-dimensional 
coordinate system are in the Galactic center. The two large arrows indicate 
one of the possible (3 or 5) field  reversals observed so far. The field reversal 
indicated in this figure is the less controversial one. }
\label{F2}
\end{figure}

Since this is a  relevant point, the possible 
controversies arising in the analysis of the magnetic field of the MW will now be swiftly 
mentioned. It is  appropriate, for this purpose, to give a geometrical characterization 
of the arm structure of the MW \cite{vallee3}.  The spiral structure 
of the galaxy can be described in terms of a two-dimensional 
coordinate system $(x,y)$ whose origin $(0,0)$ is at the galactic 
center. In this coordinate system the sun is located at $(0,8)$ where 
the coordinates are expressed in kpc. For each spiral 
arm the equation reduces to four curves, each rotated 
by $\pi/2$ of the form:
\begin{equation}
x = r \cos{\theta},~~~~~~y= r \sin{\theta},~~~~r=r_0 e^{\kappa(\theta - \theta_{0})}.
\label{spiralarms}
\end{equation}
In Eq. (\ref{spiralarms}) $r_0\sim 2.3 {\rm kpc}$, $\theta_0 \sim 0, -\pi/2, -\pi, -3/2\pi$, $\kappa \sim 0.21$.
In Fig. \ref{F2} the map of the spiral arms is illustrated according to the 
model of Eq. (\ref{spiralarms}). Fig. \ref{F2} is adapted from the recent paper of Vall\'ee \cite{vallee3}.
The region with dashed lines is poorly known from the observational point of view (``zona galactica incognita'' in the 
terminology of Vall\'ee \cite{vallee3}). In terms of 
Fig. \ref{F2} the basic question related to the magnetic field structure 
concerns the {\em relative} orientation of the 
magnetic field direction between one arm and the nearest one. As mentioned above\cite{han2,han4} 
some studies suggest that three reversals are present. Some other studies \cite{vallee3} based
on a statistical re-analysis of the most recent data from 1995 to 2001 suggest that 
{\em only one} reversal is observed. The two large arrows in Fig. \ref{F2} represent the field directions in the Carina-Sagittarius and in the Perseus arms. In \cite{vallee3} it has been argued 
that the magnetic field of the MW has the structure of axisymmetric spiral. However, the presence 
of one field reversal seems to be not fully consistent with this interpretation.  

Recently \cite{HMLQ}
Han, Manchester, Lyne and Qiao, using the Parkes multibeam pulsar survey, provided 
further observational support for the detection of a counterclockwise magnetic field in the 
Norma spiral arm. The morphological properties of the magnetic field in the Carina-Sagittarius and in the Crux-Scutum arms have been confirmed \cite{HMLQ}.
These results were obtained from the analysis of pulsar rotation measures.

If  the magnetic field of the MW flips 
its direction from one spiral arm to the other, then, as 
pointed out by Sofue and Fujimoto \cite{mw4} 
the galactic magnetic field should probably be associated with a BSS 
model. In the Sofue-Fujimoto model the radial and azimuthal components 
of the magnetic field in a bisymmetric logarithmic spiral configuration 
is given through
\begin{eqnarray}
&& B_{r} =  f(r) \cos{\biggl( \theta - \beta \ln{\frac{r}{r_{0}}} 
\biggr)} ~\sin{p},
\nonumber\\
&& B_{\theta} =  f(r) \cos{\biggl( \theta - \beta
\ln{\frac{r}{r_0}}\biggr)} \cos{p},
\label{BSS}
\end{eqnarray}
where $r_{0}\sim 10.5$ kpc is the galactocentric distance 
of the maximum of the field in our spiral arm, $\beta = 1/\tan{p}$  and 
$p$ is the pitch angle of the spiral. The smooth profile 
$f(r)$ can be chosen in different ways. A motivated choice is 
 \cite{mw5,mw6} (see also \cite{mw4})
\begin{equation}
f(r) = 3 \frac{r_1}{r} \tanh^3{\biggl(\frac{r}{r_2}\biggr)} \mu {\rm G},
\end{equation}
where $r_{1} = 8.5  $ kpc is the distance of the Sun to the 
galactic center and $r_2 = 2 $ kpc. The original 
model of SF does not have dependence  in the $z$ direction, however, 
the $z$ dependence can be included and also more complicated 
models can be built \cite{mw4}. Typically, along the $z$ axis, 
magnetic fields are exponentially suppressed as $\exp{[-z/z_0]}$ 
with $z_{0} \sim 4$ kpc.

In the BSS model there are then three issues to be determined in order 
to specify the parameters\cite{han5}:
\begin{itemize}
\item{} the number and location of field reversals;
\item{} the value of the pitch angle $p$ introduced in Eq. (\ref{BSS}); 
\item{} the field strength in the disk.
\end{itemize} 
 The answers to these three questions are, in short, the following: 
\begin{itemize}
\item{} 3 to 5 field reversals have been claimed; 
\item{} the pitch angle 
is determined to be $p = -8.2^{0} \pm 0.5^{0}$ \cite{han4,han5}; 
\item{}the strength  of the regular field is, as anticipated $1.8 \pm 0.3$ $\mu {\rm G}$.
\end{itemize}
It is interesting to notice that the {\em total} field strength, in the notation 
of Eq. (\ref{btot}) can reach even 6 $\mu$ G indicating, possibly, a strong stochastic 
component. Differently from 
other spirals, the Milky Way has also a large 
{\em radio halo}. The radio-halo indicates the large-scale height 
of the magnetic field and its origin is unclear 
As far as the stochastic component of the galactic magnetic field 
is concerned, the situation seems to be, according to the reported results,
still unclear \cite{fl1,fl2,fl3}. It is, at present, not fully understood if the stochastic 
component of the galactic magnetic field is much smaller than 
(or of the same order of) the related homogeneous part as implied by the estimates 
of the total field strength.

The structure of magnetic fields can be relevant when investigating the 
propagation of high-energy protons \cite{bier,far} as noticed already long ago \cite{wolf}. 
This aspect  leads 
to crucial (and structural) ambiguities in the analysis of the propagation of charged particles 
of ultra-high energy. 

The current observations of ultrahigh energy cosmic rays (UHECR) do not lead 
to a firm evidence of the existence of a cutoff of the cosmic ray spectrum between $10^{19}$ 
and $10^{20}$ eV. This is the celebrated Greisen, Zatsepin, Kuzmin (GZK) cutoff \cite{G,ZK}. 
If the source distribution of UHECR is isotropic and homogeneous because photoproduction interactions 
on the microwave background, then the GZK cutoff should be present. 
The isotropization of UHECR can be explained by scattering in large 
magnetic fields. This suggestion can be achieved in the presence of a sizable galactic 
halo. Magnetic fields may change not only the local intensity of UHECR but also 
their energy spectrum \cite{medina}. Various papers analyzed the acceleration and propagation of UHECR
in magnetized structures (see, for instance, \cite{bla,lem,stanev} and references therein).

\subsection{Magnetic fields in external galaxies}
An excellent presentation of the evidence of cosmic magnetism in nearby 
galaxies has been given by Beck et al.\cite{beck} and, in a more concise 
form, by Widrow \cite{widrow}.
The past few years have witnessed a long discussion on the primordial 
nature of magnetic fields in nearby galaxies whose features are partially 
reported in \cite{beck}. In analogy with the problem discussed in the case of the MW 
two different opinions have been confronted. Theorists believing the primordial 
nature of the galactic magnetic field supported the BSS model. On the contrary, dynamo 
theorists supported the conclusion that most of the nearby galaxies are ASS. As previously 
described, the variation of the $RM$ as a function of the azimuthal angles has been 
used in order to distinguish ASS from BSS \cite{krause1,krause2}.
The simplistic summary of the recent results is that 
there are galaxies where BSS structure dominates (like M81,M51) \cite{beck2}.
At the same time in some galaxies the ASS dominates \cite{krause3}. 

\subsection{Magnetic fields in Abell clusters}
Already the short summary of the main experimental techniques 
used for the detection of large scales magnetic fields shows that 
there may be problems in the determination of magnetic 
fields right outside galaxies. There magnetic fields 
are {\em assumed} to be often of ${\rm n G}$ strength. 
However, due to the lack of sources for the determination 
of the column density of electrons, it is hard 
to turn the assumption into an experimental evidence.
Magnetic fields in clusters have been recently reviewed by 
Giovannini \cite{chalonge} and, more extensively, by Carilli and Taylor \cite{carilli}.

Since various  theoretical speculations  
suggest that also  clusters are magnetized, it would be interesting 
to know if regular Abell clusters posses large scale magnetic fields. 
Different results in this direction have been reported 
\cite{cl1,cl2,cl4} (see also \cite{cl3}). Some studies during the past
decade \cite{cl1,cl2}
dealt mainly with the case of a single cluster (more specifically the 
Coma cluster). The idea was to target (with Faraday rotation 
measurements)  radio sources inside the
cluster. However, it was also soon realized that 
the study of many radio sources inside 
different clusters may lead to experimental 
problems due to the sensitivity limitations of 
radio-astronomical facilities. The strategy 
is currently to study a sample of clusters each with one or two 
bright radio-sources inside.

In the past it was shown that regular clusters have  cores with
a detectable component of RM \cite{cl3,cl4}. Recent results 
suggest that $\mu$ Gauss magnetic fields are indeed detected 
inside regular clusters \cite{cl5}. Inside the cluster means 
in the intra-cluster medium. Therefore, these magnetic fields
cannot be associated with individual galaxies.
 
Regular Abell clusters with strong x-ray emission were 
studied using a twofold technique \cite{cl5,cl6}. From the ROSAT 
full sky survey, the electron density has been determined \cite{roscat}.
Faraday RM (for the same set 
of 16 Abell clusters) has been estimated through observations at the VLA
radio-telescope.

The amusing result (confirming previous claims based only on one cluster 
\cite{cl1,cl2}) is that x-ray bright Abell clusters
 possess a magnetic field of $\mu$ Gauss 
strength. The clusters have been selected in order 
to show similar morphological features. All the 16 clusters 
monitored with this technique are at low red-shift ($z<0.1$) 
and at high galactic latitude ($|b|>20^{0}$).

These recent developments are rather promising 
and establish a clear connection between radio-astronomical 
techniques  and the improvements in the knowledge 
of  x-ray sky. There are 
various satellite missions mapping the
x-ray sky at low energies (ASCA, CHANDRA, NEWTON 
\footnote{ASCA is operating between $0.4$ AND $10$ keV and it is 
flying since February 1993. CHANDRA (NASA mission)  and NEWTON (ESA mission) 
have an energy range 
comparable with the one of ASCA and were launched, almost 
simultaneously, in 1999.}). There is 
the hope that a more precise knowledge of the surface brightness of regular
clusters will help in the experimental determination of large scale 
magnetic fields between galaxies. It should be however mentioned that evidence 
for the presence of relativistic electrons and magnetic fields in clusters 
was directly available even before,  from measurements of extended regions of radio synchrotron 
emission (for frequencies $ 10^{-2} < \nu < 1$ GHz) \cite{giofer}.

\begin{figure}[th]
\centerline{\psfig{file=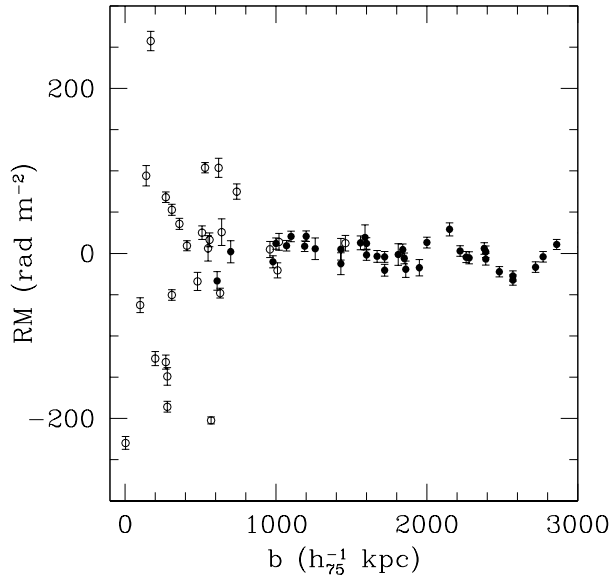,width=8cm}}
\vspace*{8pt}
\caption[a]{From Ref. \cite{cl5} the RM deduced from a 
sample of 16 X-ray bright Abell clusters is reported as a function 
of the source impact parameter.}
\label{F1}
\end{figure}

It is interesting to notice that intra-cluster magnetic fields 
of $\mu$ G strength can induce Faraday rotation on CMB polarization.
By combining informations from Sunyaev-Zeldovich effect and X-ray 
emission from the same clusters, it has been recently 
suggested that a richer information concerning electron column 
density can be obtained \cite{ohno}. 
In Fig. \ref{F1} the results reported in \cite{cl5} 
are summarized. In Fig. \ref{F1} the RM 
of the sample of x-ray bright Abell clusters is reported after 
the subtraction of the RM of the galaxy. At high 
galactic latitude (where all the observed clusters are) 
the galactic contribution is rather small and of the 
order of $9.5 {\rm rad}/{\rm m}^2$. 

The results reported in \cite{cl4} 
In Fig. \ref{F1} 
the open points represent sources viewed through 
the thermal cluster gas, whereas the full points 
represent control sources at impact parameters larger than the cluster gas.
The excess in RM attributed to clusters sources is clearly visible.

Using the described techniques large scale magnetic fields can be
observed and studied in external galaxies, in clusters and also in our own 
galaxy.  While the study of external galaxies 
and clusters may provide a global picture of magnetic fields, 
the  galactic observations performed within the Milky Way 
are more sensitive to the local spatial variations of the magnetic field.
For this reasons local and global observations are complementary. The flipped
side of the coin, as we will see in the second part of the present Section, 
is that the global structure of the magnetic field of our galaxy 
is not known directly and  to high precision but it is deduced from 
(or corroborated by)  the global knowledge of other spiral galaxies.

\subsection{Magnetic fields in superclusters}

Magnetic fields in the local supercluster are notoriously 
difficult to measure. In the absence of direct measurements 
it is the opinion of the author that these fields 
cannot be assumed to be small, i.e. $\ll \mu {\rm G}$. In
this situation it seems useful to consider the possibility 
that these fields are {\em large}. In \cite{far} this point of view 
has been taken in connection with the propagation of the most 
energetic cosmic rays. This approach is  desirable
since the (few) reliable claims of measurements of large-scale magnetic fields 
in the Local Supercluster seem to indicate that these fields are rather 
strong and more in the range of the $\mu$ G than in the range of the n G.

Recently \cite{vallee4} the possible existence, strength and structure 
of magnetic fields in the intergalactic plane, within the Local Supercluster, has been scrutinized. 
The local supercluster is centered approximately at the VIRGO cluster, about $18$ Mpc from the Local 
Group. A statistically significant Faraday screen acting on the radio-waves coming from the most distance sources
has been found. This analysis supports the existence of a regular magnetic field of $0.3 \mu$ G in the Local Supercluster.

In the past detection of radio emission in the Coma supercluster 
has been reported \cite{cl3}. The plane of the Coma supercluster is defined by  the Coma cluster and by the Abell cluster  1367.
The observed magnetic field has been extimated to be of the order of $0.5 \mu$ G.

In \cite{isola} using numerical simulations the spectrum of protons 
above $10^{19}$ eV has been determined under the assumption that 
the injection spectrum is determined by a discrete set of continuously emitting sources.
The sources follow the profile of the local Supercluster. It has been shown that 
if $|\vec{B}|\leq 0.05 \mu {\rm G}$  the source distribution assumed in the 
Supercluster is inconsistent with the observations. On the other hand, if 
$|\vec{B}|\leq 0.3 \mu {\rm G}$, 10 sources in the local Supercluster 
would lead to spectra consistent with the observations. This analysis 
seems to fit with \cite{vallee4} and with the general arguments 
put forward in \cite{far}. In a complementary perspective, the authors 
of \cite{igor} claim to be able to simulate the evolution of large-scale magnetic fields from reasonably high red-shifts.
The obtained results, according to \cite{igor}, correctly reproduce  the present large-scale 
structure of cluster and supercluster magnetic fields both in their observed
and unobserved features. 

\section{Globally Neutral Plasmas}
\renewcommand{\theequation}{4.\arabic{equation}}
\setcounter{equation}{0}
The simplified discussion of the plasma evolution 
is often related to magnetohydrodynamics (MHD) which 
is a one-fluid plasma description holding under very 
specific assumptions. The goal of the first part of the present 
Section (together with the related Appendix A) will be 
to present the reduction of the common two-fluids 
plasma dynamics to the simplified case of one-fluid 
description provided by MHD. 

\subsection{Qualitative aspects of plasma dynamics}
Consider, as a starting point, a globally neutral plasma of electrons and ions (in the simplest 
case protons) 
all at the same temperature $T_0$ and with mean particle density $n_{0}$.
Suppose, for simplicity, local thermodynamical equilibrium \footnote{This assumption is violated 
in the realistic situations.}.  
The charge densities will be given, respectively, by 
\begin{eqnarray}
n_{\rm e}(\phi) = n_0 e^{ e \phi/T_{0}} \simeq ( 1 + \frac{e \phi}{T_{0}}),
\nonumber\\
n_{\rm p}(\phi) =   n_0 e^{- e \phi/T_{0}} \simeq ( 1 - \frac{e \phi}{T_{0}}),
\label{weakly}
\end{eqnarray}
where $\phi$ is the electrostatic potential. In Eqs. (\ref{weakly}) it has been 
assumed that the plasma is weakly coupled, i.e. $|e\phi/T_0|\ll 1$.
A test charge $q_{\rm t}$ located in the origin will then experience  the electrostatic potential 
following from the  Poisson equation,
\begin{equation}
\nabla^2 \phi = 4\pi e [ n_{\rm e}(\phi) - n_{\rm p}(\phi)] - 4\pi q_{\rm t} \delta^{(3)}(\vec{x}),
\label{pois}
\end{equation}
or, using Eqs. (\ref{weakly}) into Eq. (\ref{pois})  
\begin{equation}
\nabla^2 \phi - \frac{1}{\lambda_{\rm D}^2} \phi = - 4 \pi\, q_{\rm t} \delta^{(3)}(\vec{x}),
\end{equation}
where 
\begin{equation}
\lambda_{\rm D} = \sqrt{\frac{T_0}{8 \pi n_0 e^2}},
\label{debye}
\end{equation}
 is the Debye length. 
For a test particle  the Coulomb potential 
will then have the usual Coulomb form but it will be suppressed, at large 
distances by a Yukawa term, i.e.  $e^{- r/\lambda_{\rm D}}$.

In the interstellar medium there are three kinds of regions which 
are conventionally defined, namely ${\rm H}_{2}$ regions (where the Hydrogen
is predominantly in molecular form), ${\rm H}^{0}$ regions (where Hydrogen is 
in atomic form) and ${\rm H}^{+}$ regions (where Hydrogen is ionized). 
Sometimes ${\rm H}^{+}$ regions are denoted with HI and ${\rm H}_{2}$ regions 
with HII. In the ${\rm H}^{+}$ regions the typical temperature $T_0$ is of the order 
of $10$--$20$ eV while for $n_{0}$ let us take, for instance, 
 $n_0 \sim 3 \times 10^{-2} {\rm cm}^{-3}$.
Then $\lambda_{\rm D} \sim  30 {\rm km}$.

For $r \gg 
\lambda_{\rm D}$ the Coulomb potential is screened by the global effect of 
the other particles in the plasma. Suppose now that particles 
exchange momentum through two-body interactions. Their cross 
section will be of the order of $\alpha_{\rm em}^2/T_0^2$ and the mean free 
path will be $\ell_{\rm m f p} \sim T_0^2/(\alpha_{\rm em}^2 n_0)$, i.e. recalling 
Eq. (\ref{debye}) 
$\lambda_{\rm D} \ll \ell_{\rm m f p}$. This means that the
plasma is a weakly collisional system which is, in general, {\em  not}
 in local thermodynamical 
equilibrium. This observation can be made more explicit by defining another 
important scale, namely the plasma frequency which, in the system 
under discussion, is given by 
\begin{equation}
\omega_{\rm p} = \sqrt{\frac{ 4 \pi n_0 e^2}{m_{\rm e}}},
\label{pl}
\end{equation}
where $m_{\rm e}$ is the electron mass. The plasma frequency is 
the oscillation frequency of the electrons when they are displaced 
from their equilibrium configuration in a background of approximately 
fixed ions. Recalling that $v_{\rm ther}\simeq \sqrt{T_0/m_{\rm e}}$ is the thermal 
velocity of the charge carriers, the collision frequency $\omega_{\rm c} \simeq v_{\rm ther}/\ell_{\rm mfp}$ 
is always much smaller than $\omega_{\rm p} \simeq v_{\rm ther}/\lambda_{\rm D}$.
Thus, in the idealized system described so far, the following 
hierarchy of scales holds:
\begin{equation}
\lambda_{\rm D} \ll \ell_{\rm mfp},~~~~~~\omega_{\rm c} \ll \omega_{\rm p}, 
\end{equation} 
which means that before doing one collision the system undergoes 
many oscillations, or, in other words, that the mean free path 
is not the shortest scale in the problem.
Usually one defines also the {\em plasma parameter} ${\cal N} = n_0^{-1} \lambda_{\rm D}^{-3}$, 
i.e. the number of particles in the Debye sphere. In the approximation of weakly coupled plasma  
${\cal N}\ll 1$ which also imply that the mean kinetic energy of the 
particles is larger than the mean inter-particle potential, i.e.  $|e \phi| \ll T_0$ in the 
language of Eq. (\ref{weakly}).

\subsection{Kinetic and fluid descriptions}

From the point of view of the evolution of large-scale magnetic fields, the galaxy 
is globally neutral system 
of charged particles with typical rotation period of $3\times 10^{8}$ yrs. 
Two complementary descriptions of the plasma can then be adopted. The first 
possibility is to study the full system of kinetic equations (the Vlasov-Landau description 
\cite{vla,lan,lif} for the one particle distribution function). 
In the kinetic description the ``observables'' of the plasma dynamics are 
related to the various moments of the distribution function. In a complementary 
perspective, it is also possible to study 
directly the evolution equations of the various moments of the distribution 
function (charge and matter densities, momentum transfer etc.). The resulting 
description is an effective one since not {\em all} the moments 
of the distribution function can be kept. The necessity 
of a consistent truncation in the hierarchy of the 
various moments is usually called {\em closure problem}. 
MHD is an effective description which holding for 
\begin{equation}
L \gg \lambda_{\rm D}, ~~~~~t \gg \omega_{\rm p}^{-1},
\end{equation} 
where $L$ and $t$ are the typical length and time scales of the problem.
The MHD approximation holds for typical length scales 
larger than the Debye scale and typical frequencies much smaller than the plasma 
frequency. In particular, the spectrum of plasma excitations obtained  
from the full kinetic description matches with the spectrum of MHD 
excitations but only at sufficiently small frequencies (typically 
of the order of the ionoacoustic waves) \cite{krall,chen}.

If the plasma is weakly collisional and out of 
thermodynamical equilibrium, an approximation where 
the evolution equations are still tractable is given in terms of the one-particle 
distribution function $f_{i}(\vec{x},\vec{v}, t)$  where the subscript denotes the given charge carrier (for instance 
electrons and ions). The one-particle distribution obeys the Vlasov-Landau equation for each 
particle species of charge $q_{\rm i}$ and of mass $m_{\rm i}$:
\begin{equation}
\frac{\partial f_{\rm i}}{\partial t} + \vec{v}\cdot \vec{\nabla}_{\vec{x}} 
f_{\rm i} + \frac{q_{i}}{m_{\rm i}} ( \vec{E} + \vec{v} \times \vec{B}) \cdot \vec{\nabla}_{\vec{v}} f_{\rm i} = \biggl(\frac{\partial f_{\rm i}}{\partial t}\biggr)_{\rm coll}.
\label{vlasov1}
\end{equation}
The term appearing at the right-hand side of Eq. (\ref{vlasov1}) is the binary collision term. 
In a weakly coupled plasma, the long-range force is given by the Coulomb interaction, 
while the short-range component arises thanks to binary collisions. In other words the electric and magnetic fields 
appearing in Eq. (\ref{vlasov1}) are {\em mean fields} obeying {\em mean} Maxwell's equations:
\begin{eqnarray}
&& \vec{\nabla}\cdot \vec{E} = 4 \pi \sum_{\rm i} \overline{n}_{\rm i} q_{\rm i} \int d^{3}v f_{\rm i}(\vec{x}, \vec{v}, t),
\label{gauss}\\
&& \vec{\nabla} \times \vec{E} + \frac{\partial \vec{B}}{\partial t} = 0,
\label{bianchi1}\\
&& \vec{\nabla} \times \vec{B} = \frac{\partial \vec{E}}{\partial t} + 4 \pi \sum_{\rm i} \overline{n}_{\rm i} q_{\rm i} \int d^{3} v
 \vec{v} f_{\rm i}( \vec{x}, \vec{v}, t), 
\label{mx}\\
&& \vec{\nabla}\cdot \vec{B} =0,
\label{transv}
\end{eqnarray}
where $\overline{n}_{\rm i}$ is the mean particle density. In the context of the kinetic approach 
the evolution equation should be solved self-consistently with the Maxwell's equations. Clearly, in a system 
where the density of particles is small the mean field given by the distant particles is more important 
than the the force produced by the closest particles. In practice the Vlasov-landau system 
can be linearized around some equilibrium value of the one-particle 
distribution function and specific examples will be discussed later on.

The one-particle distribution function is not directly observable. The directly observable 
quantities are the the various moments of the one-particle distribution, for instance, the particle
density, 
\begin{equation}
n_{\rm i}(\vec{x}, t) = \overline{n}_{\rm i} \int f_{\rm i}(\vec{x}, \vec{v}, t) d^{3} v,
\end{equation}
and the related matter and charge densities for each particle 
species (i.e.  $ \rho_{{\rm m},{\rm i}}(\vec{x}, t) = m_{\rm i} n_{\rm i}(\vec{x}, t)$ and 
$ \rho_{{\rm q},{\rm i}}(\vec{x}, t) = q_{\rm i} n_{\rm i}(\vec{x}, t)$) or summed over 
the different species (i.e. $\rho_{\rm m} = \sum_{\rm i} \rho_{{\rm m}, {\rm i}}$ and 
 $\rho_{\rm q} = \sum_{\rm i} \rho_{{\rm q}, {\rm i}}$). However,
the dynamical informations on the  plasma evolution are not only 
encoded in the moment of order zero but also in the higher moments (the first, the second 
and even higher).

For some class of problems the kinetic approach is  
to be preferred. For instance the kinetic approach is 
mandatory for all the high frequency phenomena 
(like, for instance, equilibration
of charge and current density fluctuations). As
far as the low-frequency phenomena are concerned one can imagine 
to obtain a reduced (fluid-like) description. 

\subsection{From one-fluid equations to MHD}

The evolution equations for the one-fluid variables are obtained from the two-species 
kinetic description by some algebra which is summarized in the first part of Appendix A. 
The bottom line of the derivation  
is that the moments of the one-particle distribution function obey a set of 
partial differential equations whose solutions can be studied in different approximations.
One of these approximations is the MHD description. 
 The set of one-fluid equations (derived in Appendix A) can be written as 
\begin{eqnarray}
&& \frac{\partial \rho_{\rm m}}{\partial t} + \vec{\nabla} \cdot( \rho_{\rm m} \vec{v}) =0,
\label{cm}\\
&& \frac{\partial  \rho_{\rm q}}{\partial t} + \vec{\nabla} \cdot \vec{j} =0.
\label{cq}\\
&&
\rho_{\rm m} \biggl[ \frac{\partial \vec{v}}{\partial t} + \vec{v}\cdot \vec{\nabla}\biggr] \vec{v} = 
\vec{J} \times \vec{B} - \vec{\nabla} P,
\label{mt}\\
&&
\vec{E} + \vec{v}\times\vec{B} = \frac{1}{\sigma} \vec{J} + \frac{1}{e n_{\rm q}} ( \vec{J} \times \vec{B} - \vec{\nabla} P_{e}).
\label{genohm1}
\end{eqnarray}
supplemented by the Maxwell's equations
\begin{eqnarray}
&& \vec{\nabla}\cdot \vec{E} = 4 \pi \rho_{\rm q}, ~~~~~~~~~~~\vec{\nabla}\cdot \vec{B}=0
\label{MX1}\\
&&  \vec{\nabla} \times \vec{E} + \frac{\partial \vec{B}}{\partial t} = 0,
\label{MX2}\\
&& \vec{\nabla} \times \vec{B} = \frac{\partial \vec{E}}{\partial t} + 4 \pi \vec{J}
\label{MX3}
\end{eqnarray}
Eqs. (\ref{cm}) and (\ref{cq}) are the mass and charge conservation, Eq. (\ref{mt}) is the equation for the momentum 
transfer and Eq. (\ref{genohm1}) is the generalized Ohm law. In Eqs. (\ref{cm})--(\ref{genohm1}) $\vec{v}$ is the 
one-fluid velocity field, $P$ is the total pressure and $P_{e}$ is the pressure of the electrons.

If the plasma is globally neutral the charge density can be  consistently neglected. 
However, global neutrality  is  plausible only for typical lengths much larger 
than $\lambda_{\rm D}$ where the collective properties of the plasma are the leading 
dynamical effect. Under this assumption, Eqs. (\ref{cq}) and (\ref{MX1}) imply
\begin{equation}
\vec{\nabla} \cdot \vec{J}=0,~~~~~~~~~~\vec{\nabla}\cdot \vec{E} =0,
\end{equation}
which means that  MHD currents are, in the first approximation, solenoidal. 

The second step to get to a generalized form of the MHD equations is 
to neglect the terms responsible for the high 
frequency plasma excitations, like the displacement current in Eqs. (\ref{MX3}) whose form becomes then 
\begin{equation}
\vec{\nabla} \times \vec{B} =  4 \pi \vec{J}.
\label{mhd1}
\end{equation}
Given the two sets of consistent 
approximations discussed so far, the form taken by the generalized Ohm law 
may not be unique and, therefore, different MHD descriptions 
arise depending upon which term at the right hand side of Eq. (\ref{genohm1}) 
is consistently neglected. The simplest (but often not realistic) approximation 
is to neglect all the terms at the right hand side of Eq. (\ref{genohm1}). This 
is sometimes called {\em ideal MHD approximation}. In this case 
the Ohmic electric field is given by $\vec{E} \simeq - \vec{v} \times \vec{B}$.
This ideal description is  also called sometimes {\em superconducting MHD approximation}
since, in this case, the resistive terms in Eq. (\ref{genohm1}), i.e. $\vec{J}/\sigma$ go to zero
(and the conductivity $\sigma \to \infty$). In the early Universe 
it is practical to adopt the superconducting approximation
 (owing to the large value of the conductivity) but it is also 
very dangerous since all the ${\cal O}(1/\sigma)$ effects are dropped. 

A more controllable approximation is the  {\em real (or resistive) MHD}.
The resistive approximation is more controllable since the small 
expansion parameter (i.e. $1/\sigma$) is not zero (like in the ideal case)
but it is small and finite. This allows to compute the various quantities 
to a a given order in $1/\sigma$ even if, for practical purposes, only the first 
correction is kept.
In this framework the Ohmic current is not neglected and the 
generalized Ohm law takes the form 
\begin{equation}
\vec{J} \simeq \sigma (\vec{E} + \vec{v}\times \vec{B}). 
\label{genohm2}
\end{equation}
Comparing Eq. (\ref{genohm2}) with Eq. (\ref{genohm1}), it is possible to argue   
that  the resistive MHD scheme is inadequate whenever the Hall and thermoelectric terms 
are cannot be neglected. Defining $\omega_{B} = e B/m$ as the Larmor 
frequency, the Hall term $\vec{J}\times \vec{B}$ can be neglected provided $L \omega_{\rm p}^2/\omega_{B} v_0 \gg 1$ where 
$L$ us the typical size of the system and $v_0$ the typical bulk velocity of the plasma. With analogous 
dimensional arguments it can be argued that the pressure gradient, i.e. the thermoelectric term,
 can be consistently 
neglected provided $L v_0 \omega_{B} m_{e}/T_e \gg 1$. Specific examples where 
the Hall term may become relevant will be discussed in the context of the Biermann
battery mechanism.

The set of MHD equations, as it has been presented so far, 
is not closed if a further relation among the different variables is not specified  \cite{bis}. Typically 
this is specified through an equation relating pressure and matter density (adiabatic or isothermal 
closures). Also the incompressible closure (i.e. $\vec{\nabla}\cdot\vec{v} =0$ ) is often justified 
in the context of the evolution of magnetic fields prior to recombination. It should be stressed 
that sometimes the adiabatic approximation may lead to paradoxes. It would correspond, in the 
context of ideal MHD, to infinite electrical conductivity and vanishingly small thermal conductivity (i.e. 
the electrons should be extremely fast not to feel the resistivity and, at the same time, extremely  slow
not to exchange heat among them).
 
\subsection{The Alfv\'en theorems}

The ideal and resistive MHD descriptions 
are rather useful  in order to illustrate some global properties 
of the evolution of the plasma which are relevant  both in the 
evolution of large-scale magnetic fields prior to recombination and 
in the discussion of the gravitational collapse in the presence 
of magnetic fields. These global properties of the 
plasma evolution go under the name of Alfv\'en theorems \cite{bis} (see also 
\cite{mgknot}).

In the ideal limit both the magnetic flux and the magnetic helicity 
are conserved. This means, 
\begin{eqnarray}
\frac{d}{d t} \int_{\Sigma} \vec{B} \cdot d\vec{\Sigma}=-
\frac{1}{4\pi \sigma} \int_{\Sigma} \vec{\nabla} \times\vec{\nabla}
\times\vec{B}\cdot d\vec{\Sigma},
\label{flux}
\end{eqnarray}
where $\Sigma$ is an arbitrary closed surface which moves with the
plasma \footnote{Notice that in Ref. \cite{mgknot}  
Eq. (\ref{flux}) has been derived without the $4\pi$ term at the right hand side because of different 
conventions.}.
If $\sigma \to \infty$  the expression appearing at the right hand side is
sub-leading
and the magnetic flux lines evolve glued to the plasma element.

The other quantity which is conserved in the superconducting limit
is the magnetic helicity
\begin{equation}
{\cal H}_{M} = \int_{V} d^3 x \vec{A}~\cdot \vec{B},
\label{h1}
\end{equation}
where $\vec{A}$ is the vector potential whose presence may lead to think that the 
whole expression (\ref{h1}) is not gauge-invariant.
In fact $\vec{A}\cdot\vec{B}$ is not gauge invariant but,
none the less, ${\cal H}_{M}$ is gauge-invariant since 
the integration volume is defined in such a way that the magnetic field
$\vec{B}$
is parallel to the surface which bounds $V$ and which we will call
$\partial V$. If $\vec{n}$ is the unit vector normal to $\partial V$
then $\vec{B}\cdot\vec{n}=0$ on $\partial V$ and the gauge dependent 
contribution to the integral appearing in Eq. (\ref{h1}) vanishes.
In physical terms  the integration may always be performed imagining 
that the  space is sliced in flux tubes of the magnetic field. This procedure 
 is correct provided the magnetic flux is, at least 
approximately, conserved as implied by Eq. (\ref{flux}).

The magnetic gyrotropy
\begin{equation}
\vec{B}\cdot\vec{\nabla} \times\vec{B}
\end{equation}
it is a gauge invariant measure of the diffusion rate of ${\cal H}_{M}$. In 
fact, in the resistive approximation  \cite{mgknot}
\begin{equation}
\frac{d}{d t} {\cal H}_{M} = - \frac{1}{4\pi \sigma} \int_{V} d^3 x
{}~\vec{B}\cdot\vec{\nabla} \times\vec{B}.
\label{h2}
\end{equation}
As in the case of Eq. (\ref{h1}), for $\sigma\to\infty$ the magnetic helicity 
is approximately constant.
The value of  the  magnetic gyrotropy allows  to distinguish 
different mechanisms for the magnetic field generation. Some 
mechanisms are able to produce magnetic fields whose 
flux lines have a non-trivial gyrotropy.
 The properties of turbulent magnetized plasmas may change 
depending upon the value of the gyrotropy and of the helicity
\cite{bis}.

The physical interpretation of the flux and magnetic 
helicity conservation is the following. In MHD the magnetic field 
has to be always solenoidal (i.e. $\vec{\nabla} \cdot \vec{B} =0$).
Thus, the magnetic flux conservation implies that, in the 
superconducting limit (i.e. $\sigma \to \infty$) the 
magnetic flux lines, closed because of the transverse nature of the field, evolve always 
glued together with the plasma element. In this 
approximation, as far as the magnetic field 
evolution is concerned, the plasma is a collection 
of (closed) flux tubes. The theorem of flux conservation 
states then  that the energetical properties of large-scale
magnetic fields are conserved throughout the plasma evolution. 

While 
the flux conservation concerns the 
energetical properties of the magnetic flux lines, the 
magnetic helicity concerns chiefly the 
{\em topological } properties of the 
magnetic flux lines. If the magnetic field is completely 
stochastic, the magnetic flux lines will be closed loops 
evolving independently in the plasma and the helicity 
will vanish. There could be, however, 
more complicated topological situations \cite{mgknot}
where a single magnetic loop is twisted (like some 
kind of M\"obius stripe) or the case where 
the magnetic loops are connected like the rings of a chain.
In both cases the magnetic helicity will not be zero 
since it measures, essentially, the number of links and twists 
in the magnetic flux lines. Furthermore, in the 
superconducting limit, the helicity will not change 
throughout the time evolution. 
The conservation of the magnetic flux and of the magnetic 
helicity is a consequence of the fact that, in ideal 
MHD, the Ohmic electric field is always orthogonal 
both to the bulk velocity field and to the magnetic 
field. In the resistive MHD approximation this is 
no longer true.

If the conductivity is very large (but finite), the 
resistive MHD approximation suggests, on one hand, that 
the magnetic flux is only approximately conserved. On the 
other hand the approximate conservation of the 
magnetic helicity implies that the closed magnetic loops 
may modify their topological structure. The  breaking of the flux lines, occurring at 
finite conductivity, is related to the possibility that 
bits of magnetic fluxes are  ejected from a galaxy 
into the inter-galactic medium. This 
phenomenon is called {\em magnetic reconnection} and it 
is the basic mechanism explaining, at least qualitatively, why,
during the solar flares, not only charged particles are emitted, 
but also magnetic fields. In the context of large 
scale magnetic fields the approximate (or exact) magnetic flux conservation
has relevant  consequences for the r\^ole of magnetic fields during 
gravitational collapse.

\subsection{Magnetic diffusivity equation}

From Eqs.  (\ref{mhd1}) and  (\ref{genohm2})
the Ohmic electric field can be expressed as 
\begin{equation}
\vec{E} = \frac{1}{4\pi \sigma} \vec{\nabla} \times \vec{B} - \vec{v} \times \vec{B},
\end{equation}
which inserted into Eq. (\ref{MX2}) leads to the magnetic diffusivity equation
\begin{equation}
\frac{\partial \vec{B}}{\partial t} = \vec{\nabla} \times (\vec{v} \times \vec{B}) + \frac{1}{4\pi \sigma} \nabla^2 \vec{B}.
\label{magndiff1}
\end{equation}
The first term of Eq. (\ref{magndiff1}) is the dynamo term. 
The second term of Eq. (\ref{magndiff1}) is the magnetic diffusivity term whose 
effect is to dissipate the magnetic field. By comparing the left and the right hand side of Eq. (\ref{magndiff1}), 
the typical time scale of resistive phenomena is
\begin{equation}
t_{\sigma} \simeq 4\pi \sigma L^2 ,
\label{difftime}
\end{equation}
where $L$ is the typical length scale of the magnetic field. In a non-relativistic plasma 
the conductivity $\sigma$ goes typically as $T^{3/2}$  \cite{krall}. In the case of planets, like 
the earth, one can wonder why a sizable magnetic field can still be present. One of 
the theories is that the dynamo term regenerates continuously 
the magnetic field which is dissipated by the diffusivity term \cite{parker}. 
In the case of the galactic disk the 
value of the conductivity \footnote{It is common use in the astrophysical applications 
to work directly with $\eta = (4\pi \sigma)^{-1}$. In the case of the galactic disks 
$\eta = 10^{26} {\rm cm}^{2}~{\rm Hz}$. The variable $\eta$ denotes, in the 
present review, the conformal time coordinate.}
is given by $\sigma \simeq 7\times 10^{-7} {\rm Hz}$. Thus, for  $L \simeq {\rm kpc}$ 
$t_{\sigma} \simeq 10^{9} (L/{\rm kpc})^2 {\rm sec}$. 

In Eq. (\ref{difftime}) the typical time of resistive phenomena has been 
introduced. Eq. (\ref{difftime}) can also give the typical 
resistive length scale once the time-scale 
of the system is specified. Suppose that the time-scale of the system is 
given by $t_{U} \sim H_{0}^{-1} \sim 10^{18} {\rm sec}$ where $H_0$ is the 
present value off the Hubble parameter. Then  
\begin{equation}
L_{\sigma} = \sqrt{\frac{t_{U}}{\sigma}},
\label{diffscale}
\end{equation} 
leading to $L_{\sigma } \sim {\rm AU}$. The scale (\ref{diffscale}) gives then the 
upper limit on the diffusion scale for a magnetic field whose lifetime is 
comparable with the age of the Universe at the present epoch. Magnetic fields with typical 
correlation scale larger than $L_{\sigma}$ are not affected by resistivity. On the other 
hand, magnetic fields with typical correlation scale $ L< L_{\sigma}$ are diffused. The
value $L_{\sigma} \sim {\rm AU}$ is consistent with the phenomenological 
evidence that there are no magnetic fields coherent over scales smaller than $10^{-5}$ pc.

The dynamo term may  be responsible for the origin of the magnetic field of the galaxy. 
The galaxy has a typical rotation period of 
$3 \times 10^{8}$ yrs and comparing this number with the typical age of the galaxy, ${\cal O}(10^{10} {\rm yrs})$, 
it can be appreciated that the galaxy performed about $30$ rotations since the time 
of the protogalactic collapse. 

From Eq.  (\ref{magndiff1}) the usual structure of the dynamo term may be derived
  by carefully averaging
over the velocity filed according to the procedure of \cite{vains,matt}.
By assuming that the motion of the  fluid is random and with zero mean
velocity the average is taken over the ensemble of the possible
velocity fields.
In more physical terms this averaging procedure of Eq. (\ref{magndiff1}) is
equivalent to average over scales and times exceeding the
characteristic correlation scale and time $\tau_{0}$ of the velocity
field. This procedure assumes that the correlation scale of the
magnetic field is much bigger than the correlation scale of the
velocity field which is required to be divergence-less
(${\vec{\nabla}}\cdot \vec{v}=0$).
In this approximation the magnetic diffusivity equation can be written
as:
\begin{equation}
\frac{\partial\vec{B}}{\partial t} =
\alpha(\vec{\nabla}\times\vec{B}) +
\frac{1}{4\pi\sigma}\nabla^2\vec{B} ,
\label{dynamored}
\end{equation}
where 
\begin{equation}
\alpha 
= -\frac{\tau_{0}}{3}\langle\vec{v}\cdot\vec{\nabla}
\times\vec{v}\rangle,
\label{alphafirst}
\end{equation}
 is the so-called dynamo term which vanishes
in the absence of vorticity. In Eqs. (\ref{dynamored})--(\ref{alphafirst}) $\vec{B}$ is
the magnetic field averaged 
over times longer that $\tau_{0}$ which is the typical correlation
time of the velocity field. 

It can be argued  that the essential
requirement for the consistence of the mentioned averaging procedure is that the
turbulent velocity field has to be ``globally'' non-mirror
symmetric \cite{zeldovich}. If the system would 
be, globally, invariant under parity transformations, then, the $\alpha$ term would 
simply vanish. This observation is related to the turbulent features 
of cosmic systems. In cosmic turbulence the systems are 
usually rotating and, moreover, they possess a gradient in the 
matter density (think, for instance, to the case of the galaxy). It is then 
plausible that parity is broken at the level of the galaxy since terms 
like $ \vec{\nabla} \rho_{\rm m} \cdot \vec{\nabla} \times \vec{v}$ 
are not vanishing \cite{zeldovich}.

The dynamo term, as it appears in Eq. (\ref{dynamored}),
 has a simple electrodynamical meaning,
namely, it can be interpreted as a mean ohmic current directed along
the magnetic field :
\begin{equation}
\vec{J} = - \alpha \vec{B}
\label{vort}
\end{equation}
This equation tells us that an ensemble of screw-like vortices with
zero mean helicity is able to generate loops in the magnetic flux
tubes in a plane orthogonal to the one of the original field.
Consider, as a simple application of Eq. (\ref{dynamored}), the case where the magnetic field 
profile is given by 
\begin{equation}
B_{x}(z,t) = f(t) \sin{k z}, ~~~~B_{y}=f(t) \cos{k z},~~~~ B_{z}(k,t) =0.  
\label{knot1}
\end{equation}
For this profile the magnetic gyrotropy is non-vanishing, i.e. 
$\vec{B}\cdot\vec{\nabla}\times\vec{B} = k f^2(t)$. From Eq. (\ref{dynamored}), using Eq. (\ref{knot1}) 
$f(t)$ obeys the following equation
\begin{equation}
\frac{d f}{dt} = \biggl( k \alpha - \frac{k^2}{4\pi\sigma}\biggr) f
\label{f1}
\end{equation}
admits exponentially growing solutions for sufficiently large scales, i.e. $k < 4\pi |\alpha| \sigma$.
Notice that in this naive example the $\alpha$ term is assumed to be constant. However, as the amplification proceeds, 
$\alpha$ may develop a dependence upon $|\vec{B}|^2$, i.e. $\alpha \to \alpha_0 ( 1 - \xi |\vec{B}|^2) \alpha_0 [ 1 - \xi f^2(t)]$. In the case 
of Eq. (\ref{f1}) this modification will introduce non-linear terms  whose effect will be to stop the growth of the magnetic field. 
This regime is often called saturation of the dynamo and the non-linear equations appearing in this context are sometimes 
called Landau equations \cite{zeldovich} in analogy with the Landau equations appearing in hydrodynamical 
turbulence.

In spite of the fact that in the previous example  the velocity field has been averaged, its evolution 
obeys  the Navier-Stokes equation 
\begin{equation}
\rho\biggl[ \frac{\partial \vec{v}}{\partial t} + (\vec{v}\cdot \vec{\nabla}) \vec{v} - \nu \nabla^2 \vec{v}\biggr] = - \vec{\nabla}p + 
\vec{J}\times \vec{B},
\label{NS1}
\end{equation}
where $\nu$ is the thermal viscosity coefficient. Since in MHD the matter current is solenoidal
(i.e. $\vec{\nabla}\cdot( \rho \vec{v}) =0$) the incompressible 
closure $\vec{\nabla} \rho =0$, corresponds 
to a solenoidal velocity field $ \vec{\nabla}\cdot\vec{v} =0$. 
Recalling Eq. (\ref{mhd1}), the Lorentz 
force term can be re-expressed through vector identities  and Eq. (\ref{NS1}) becomes 
\begin{equation}
\rho\biggl[ \frac{\partial \vec{v}}{\partial t} + (\vec{v}\cdot \vec{\nabla}) \vec{v} - \nu \nabla^2 \vec{v}\biggr] = - \vec{\nabla}\biggl[ p + \frac{|\vec{B}|^2}{8\pi} \biggr]
+ (\vec{B}\cdot\vec{\nabla}) \vec{B}.
\label{NS2}
\end{equation}
In typical applications to the evolution of magnetic fields prior to recombination the magnetic pressure term is always smaller 
than the fluid pressure \footnote{Recall that in fusion studies the quantity $\beta = 8\pi |\vec{B}|^2/p$ is usually defined. 
If the plasma is confined, then $\beta$ is of order 1. On the contrary, if $\beta \gg 1$, as implied 
by the critical density bound in the early Universe, then the plasma may be compressed at higher temperatures and densities.}
, i.e. $ p \gg |\vec{B}|^2$. Furthermore, there are cases where the Lorentz force term can be ignored. This 
is the so-called  force free approximation. Defining the kinetic helicity as $ \vec{\omega} = \vec{\nabla} \times \vec{v}$, 
the magnetic diffusivity and Navier-Stokes equations can be written in a rather simple and symmetric form 
\begin{eqnarray}
&& \frac{\partial \vec{B}}{\partial t} = \vec{\nabla} \times (\vec{v} \times \vec{B}) + \frac{1}{4\pi \sigma} \nabla^2 \vec{B},
\nonumber\\
&& \frac{\partial \vec{\omega}}{\partial t} = \vec{\nabla} \times (\vec{v} \times \vec{\omega}) + \nu \nabla^2 \vec{\omega}.
\label{symm}
\end{eqnarray}

In MHD various dimensionless ratios can be defined. The most frequently used are 
the magnetic Reynolds number, the kinetic Reynolds number and the Prandtl number:
\begin{eqnarray}
&& {\rm R}_{\rm m} = v L_{B} \sigma,
\label{magnrey}\\
&& {\rm R}= \frac{v L_{v}}{\nu},
\label{rey}\\
&& {\rm Pr} = \frac{ {\rm R}_{\rm m}}{{\rm R}} = \nu\sigma \biggl(\frac{L_{B}}{L_{v}}\biggr),
\label{Pr}
\end{eqnarray}
where $L_{B}$ and $L_{v}$ are the typical scales of variation of the magnetic and velocity fields. 
In the absence of pressure and density perturbations 
the combined system of Eqs. (\ref{mhd1}) and (\ref{NS2}) can be linearized easily. 
Using then the incompressible closure the propagating modes are the Alfv\'en 
waves whose typical dispersion relation is $\omega^2 = c_{\rm a}^2 k^2$ where 
$c_{\rm a} = |\vec{B}|/\sqrt{4\pi\rho}$. Often the Lundqvist number is called, in plasma 
literature \cite{krall,bis} magnetic Reynolds number. This confusion arises from the 
fact that the Lunqvist number, i.e. $c_{a} L \sigma$, is the magnetic Reynolds number 
when $v$ coincides with the Alfv\'en velocity. To have a very large Lundqvist number 
implies that the the conductivity is very large. In this sense the Lunqvist 
number characterizes, in fusion theory, the rate of growth of resistive instabilities
and it is not necessarily related to the possible occurrence of turbulent dynamics.
On the contrary, as large Reynolds numbers are related to the occurrence of 
hydrodynamical turbulence, large {\em magnetic} Reynolds numbers are related 
to the occurence of MHD turbulence \cite{bis}. 

\subsection{The dynamo mechanism}

According to the  naive description of the dynamo instability presented above 
the origin of large-scale  magnetic fields in spiral galaxies can be reduced to the 
following steps:
\begin{itemize}
\item{} during the $30$ rotations performed by the galaxy 
since the protogalactic collapse, the magnetic field should be amplified 
by about $30$ e-folds;
\item{}  if the large scale magnetic field of the 
galaxy is, today, ${\cal O}(\mu {\rm G})$ the magnetic field 
at the onset of galactic rotation might have been even $30$ e-folds smaller, i.e. 
${\cal O}(10^{-19} {\rm G})$;
\item{} assuming perfect flux freezing during the gravitational 
collapse of the protogalaxy (i.e. $\sigma \to \infty$) the 
magnetic field at the onset of gravitational collapse should 
be  ${\cal O}(10^{-23})$ G over a typical scale of 1 Mpc.
\end{itemize}
This picture is  oversimplified and each of the three steps mentioned above 
will be contrasted with the most recent findings.

The idea that a celestial body may acquire a magnetic field by differential 
rotation can be traced back to the paper of Larmor of 1919 \cite{larmor}.
One of the early ancestors of the dynamo mechanism, can be traced back to the model of
Fermi and Chandrasekar \cite{fermi2,fermi3}. In \cite{fermi2,fermi3} the attempt was to connect 
the existence off the galactic magnetic field with the existence of a galactic angular momentum.
Later on dynamo theory has been developed in greater detail (see  \cite{parker}) and 
its possible application to large-scale magnetic fields has been 
envisaged. 

The standard dynamo theory 
has been questioned in different ways.  Piddington \cite{pid1,pid2} 
pointed out that small-scale magnetic fields can grow large 
enough (until equipartition is reached) to swamp the dynamo action.
The quenching of the dynamo action has been numerically shown by
Kulsrud and Anderson \cite{ka}. More recently, it has been argued that 
if the large-scale magnetic field reaches the  critical value \footnote{  $v$ is the velocity
field at the outer scale of turbulence.}
$R_{\rm m}^{-1/2} v$  the dynamo action could also be quenched \cite{VC1,VC2}.
 
Eq. (\ref{magndiff1}) is exact, in the sense 
that both $\vec{v}$ and $\vec{B}$ contain long and short 
wavelength modes. The aim of the 
various attempts of the dynamo theory is to get 
an equation describing only the ``mean value'' of the magnetic field.
To this end the first step is to separate 
the exact magnetic and velocity fields as 
\begin{eqnarray}
&&\vec{B} = \langle \vec{B} \rangle + \vec{b},
\nonumber\\
&& \vec{v} = \langle \vec{V} \rangle + \vec{V},
\label{sep}
\end{eqnarray}
where $\langle \vec{B} \rangle$ and  $\langle \vec{V} \rangle$
are the  averages over an ensemble of many realizations 
of the velocity field $\vec{v}$. 
In order to derive the standard form of the dynamo equations
few important assumptions should be 
made. These assumptions can be summarized as follows:
\begin{itemize}
\item{} the scale of variation of the turbulent motion $\vec{V}$ 
should be smaller than the typical scale of variation of $\langle \vec{B}\rangle$.
In the galactic problem $\langle \vec{V}\rangle$ is the differential rotation of the 
galaxy, while $\vec{V}$ is the turbulent motion generated by 
stars and supernovae. Typically the scale of variation of $\vec{v}$ is 
less than $100$ pc while the interesting scales for $\langle \vec{B} \rangle$ 
are  larger than the kpc;

\item{} the field $\vec{b}$ is such that $|\vec{b}|\ll |\langle \vec{B} \rangle|$.

\item{} it should happen 
that $\langle \vec{V} \cdot \vec{\nabla} \times \vec{V} \rangle \neq 0$. 

\item{} magnetic flux is frozen into the plasma 
(i.e. magnetic flux is conserved).
\end{itemize}
From the magnetic diffusivity equation (\ref{magndiff1}), and using 
the listed assumptions, it is possible to
derive the typical structure of the dynamo term by carefully averaging
over the velocity field according to the procedure outlined in 
\cite{zeldovich,parker,kulsrud}. Inserting Eq. (\ref{sep}) into (\ref{magndiff1}) 
and breaking the equation into a mean part and a random part,
two separate induction equations can be obtained 
for the mean and random parts of the magnetic field
\begin{eqnarray}
&&\frac{\partial \langle \vec{B} \rangle}{\partial t} = 
\vec{\nabla} \times \biggl( \langle \vec{V}\rangle \times \langle 
\vec{B} \rangle\biggr) + 
\vec{\nabla} \times \langle 
\vec{V} \times \vec{b} \rangle,
\label{mean}\\
&& \frac{\partial \vec{b}}{\partial t} = \vec{\nabla} \times ( \vec{V} \times \langle \vec{B}\rangle) +
\vec{\nabla} \times( \langle \vec{V} \rangle \times \vec{b}) 
\nonumber\\
&&+ \vec{\nabla}\times( \vec{V} \times \vec{b} ) -
 \vec{\nabla} \times \langle 
\vec{V} \times \vec{b} \rangle,
\label{random} 
\end{eqnarray}
where the (magnetic) diffusivity terms have been neglected.
In Eq. (\ref{mean}), $ \langle \vec{V} \times \vec{b} \rangle$ is called
mean field (or turbulent) electromotive force and it is the average of the cross 
product of the small-scale velocity field $\vec{V}$ and of the small 
scale magnetic field $\vec{b}$ over a scale much smaller than the scale 
of $\langle \vec{B} \rangle$ but much larger than the scale of turbulence.
Sometimes, the calculation of the effect of $\langle \vec{V} \times \vec{b}\rangle$ 
is done in the case of incompressible and isotropic turbulence. In this case 
$\langle \vec{V} \times \vec{b}\rangle =0$. This estimate is, however, not realistic 
since $\langle \vec{B}\rangle$ is not isotropic. 
More correctly \cite{kulsrud}, $\langle \vec{V} \times \vec{b}\rangle$ should be 
evaluated by using Eq. (\ref{random}) which is usually written in a simplified form 
\begin{equation}
 \frac{\partial \vec{b}}{\partial t} =    \vec{\nabla} \times ( \vec{V} \times \langle \vec{B}\rangle),
\label{inran}
\end{equation}
where all but the first term of Eq. (\ref{random}) have been 
neglected. To neglect the term $ \vec{\nabla} \times (\langle\vec{V}\rangle \times \vec{b})$ 
does not pose any problem since it corresponds to choose a reference 
frame where $\langle \vec{V} \rangle$ is constant. 
However, the other terms, neglected in Eq. (\ref{inran}), 
are dropped because it is assumed that $|\vec{b}| \ll |\langle \vec{B}\rangle| $. 
This assumption may not be valid all the time and for all the scales. 
The validity of Eq. (\ref{inran}) seems to require that $\sigma$ is very large 
so that magnetic diffusivity can keep always $\vec{b}$ small \cite{KR}.
On the other hand \cite{kulsrud} one can argue that $\vec{b}$ is only present 
over very small scales (smaller than $100$ pc) and in this case 
the approximate form of eq. (\ref{inran}) seems to be more justified.

From Eqs. (\ref{mean})--(\ref{inran}) 
 it is possible to get to the final result 
for the evolution equation of $\langle \vec{B} \rangle$ \cite{kulsrud} as it
is usally quoted
\begin{equation}
\frac{\partial \langle \vec{B}\rangle}{\partial t} =
\vec{\nabla}\times (\alpha\langle \vec{B}\rangle) +
\beta\nabla^2\langle\vec{B}\rangle + \vec{\nabla} \times \biggl( \langle \vec{V} \rangle \times 
\langle \vec{B}\rangle \biggr), 
\label{fin}
\end{equation}
where 
\begin{eqnarray}
&&\alpha 
= -\frac{\tau_{0}}{3}\langle\vec{V}\cdot\vec{\nabla}
\times\vec{V}\rangle,
\label{at}\\
&& \beta = \frac{\tau_0}{3} \langle \vec{V}^2\rangle,
\label{bt}\end{eqnarray}
where $\alpha$ is the dynamo term, $\beta $ is the diffusion term 
and $\tau_0$ is the typical correlation time of the velocity field.
The term $\alpha$ is, in general, space-dependent. 
The standard lore is that the dynamo action stops when 
the value of the magnetic field reaches the equipartition value 
(i.e. when the magnetic and kinetic energy of the plasma are comparable).
At this point the dynamo ``saturates''. 
The mean velocity field can be expressed as  $\langle \vec{V} \rangle\simeq \vec{\Omega}\times \vec{r}$
where $|\vec{\Omega}(r)|$ is the angular velocity of differential rotation 
at the galactocentric radius $r$. In the case of flat rotation curve 
$|\vec{\Omega}(r)| = \Omega(r) \sim r^{-1}$ which also implies that $\partial |\Omega(r)|/\partial r <1$.

Eq. (\ref{fin}) can then be written 
in terms of the radial and azimuthal components of the mean 
magnetic field, neglecting, for simplicity, the diffusivity term:
\begin{eqnarray}
&& \frac{\partial  \langle B_{\phi}\rangle  }{\partial t} = - r \frac{\partial \Omega}{\partial r} \langle B_{r}\rangle,
\label{azim}\\
&& \frac{\partial  \langle B_{r}\rangle  }{\partial t} = \alpha \frac{\partial  \langle B_{\phi}\rangle }{\partial r} .
\end{eqnarray}

The second equation shows that the $\alpha$ effect amplifies the radial component 
of the large-scale field. Then, through Eq. (\ref{azim}) the amplification of the radial 
component is converted into the amplification of the azimuthal field, this is the $\Omega$ effect.

Usually the picture 
for the formation of galactic magnetic fields is related to the 
possibility of implementing the dynamo mechanism.  
By comparing 
the rotation period with the age of the galaxy (for a Universe with 
$\Omega_{\Lambda} \sim 
0.7$, $h \sim 0.65$ and $\Omega_{\rm m} \sim 0.3$) the number of rotations
performed by the galaxy since its origin is approximately  $30$. 
During these $30$ rotations the dynamo term of Eq. (\ref{fin}) 
 dominates against the magnetic diffusivity. As a 
consequence an instability develops. This instability can be used
in order to drive the magnetic field from some small initial condition
up to its observed value.
Eq. (\ref{fin}) is linear in the mean 
magnetic field. Hence, initial conditions for the mean magnetic field
 should be postulated at a given time and over a given scale. 
This initial mean field, postulated as initial 
condition of (\ref{fin}) is usually called seed. 

Most of the work in the context of the dynamo 
theory focuses on reproducing the correct features of the 
magnetic field of our galaxy.
The achievable amplification produced by the 
dynamo instability can be at most of $10^{13}$, i.e. $e^{30}$. Thus, if 
the present value of the galactic magnetic field is $10^{-6}$ Gauss, its value 
right after the gravitational collapse of the protogalaxy might have 
been as small as $10^{-19}$ Gauss over a typical scale of $30$--$100$ kpc.

There is a simple way to relate the value of the magnetic fields 
right after gravitational collapse to the value of the magnetic field 
right before gravitational collapse. Since the gravitational collapse 
occurs at high conductivity the magnetic flux and the magnetic helicity
are both conserved. Right before the formation of the galaxy a patch 
of matter of roughly $1$ Mpc collapses by gravitational 
instability. Right before the collapse the mean energy density  
of the patch, stored in matter, 
 is of the order of the critical density of the Universe. 
Right after collapse the mean matter density of the protogalaxy
is, approximately, six orders of magnitude larger than the critical density.

Since the physical size of the patch decreases from $1$ Mpc to 
$30$ kpc the magnetic field increases, because of flux conservation, 
of a factor $(\rho_{\rm a}/\rho_{\rm b})^{2/3} \sim 10^{4}$ 
where $\rho_{\rm a}$ and $\rho_{\rm b}$ are, respectively the energy densities 
right after and right before gravitational collapse. The 
correct initial condition in order to turn on the dynamo instability
would be $|\vec{B}| \sim 10^{-23}$ Gauss over a scale of $1$ Mpc, right before 
gravitational collapse. 

This last estimate  is
rather generous and has been presented just in order to 
make contact  with several papers (concerned with the origin 
of large scale magnetic fields) using such an estimate.
The estimates presented in the last paragraph are
 based on the (rather questionable) 
assumption that the amplification 
occurs over thirty e-folds while the magnetic flux is 
completely frozen in. In the real situation, the 
achievable amplification is much smaller. Typically a good 
seed would not be $10^{-19}$ G after collapse (as we assumed for 
the simplicity of the discussion) but rather \cite{kulsrud}
\begin{equation}
|\vec{B}| \geq  10^{-13} {\rm G}. 
\label{dynreq}
\end{equation}
The possible applications of dynamo mechanism to  clusters is still
under debate and it seems more problematic \cite{cl5,cl6}.  
The typical scale of the gravitational collapse of a cluster 
is larger (roughly by one order of magnitude) than the scale of gravitational
collapse of the protogalaxy. Furthermore, the mean mass density 
within the Abell radius ( $\simeq 1.5 h^{-1} $ Mpc) is roughly 
$10^{3}$ larger than the critical density. Consequently, clusters 
rotate much less than galaxies. Recall that clusters are 
formed from peaks in the density field. The present overdensity 
of clusters is of the order of $10^{3}$. Thus, in order to get 
the intra-cluster magnetic field, one could think that 
magnetic flux is exactly conserved and, then, from an intergalactic 
magnetic field $|\vec{B}| > 10^{-9}$ G  an intra cluster magnetic field
$|\vec{B}| > 10^{-7}$ G can be generated. This simple estimate 
shows why it is rather important to improve the accuracy of magnetic 
field measurements in the intra-cluster medium discussed in Section 3: the 
change of a single order of magnitude in the estimated magnetic field 
may imply rather different conclusions for its origin. Recent numerical
simulations seem to support the view that cluster magnetic fields 
are entirely primordial \cite{BJ}.
 
\section{Magnetic Fields in Curved Backgrounds}
\renewcommand{\theequation}{5.\arabic{equation}}
\setcounter{equation}{0}

As the temperature increases above $1$ MeV, a relativistic 
plasma becomes almost a perfect conductor.
There are two complementary effects associated with large-scale magnetic fields 
in curved backgrounds:
\begin{itemize}
\item{}  large-scale magnetic fields may evolve over a 
 rigid background space-time determined, for instance, by the dynamics of barotropic fluids and, in this 
case the energy density of the magnetic field must always be smaller than the energy 
density of the fluid sources;
\item{} magnetic fields may be so intense to modify the structure of the space-time and, in this case, their 
energy density is comparable with the energy density of the other sources of the geometry.
\end{itemize}
The first effect will be examined in the present Section, while the discussion
of the second effect, which is more speculative, will be confined to Section 9.

\subsection{The standard cosmological model}

The standard model of cosmological evolution rests on three important assumptions \cite{weinberg}. 
The first assumption 
 is  that over very large length scales (greater than $50$ Mpc) the Universe is described by 
a homogeneous and isotropic Friedmann-Robertson-Walker (FRW) line element:
\begin{equation}
ds^2 = G_{\mu\nu} d x^{\mu} d x^{\nu}= a^2(\eta)[d\eta^2 - d \vec{x}^2],
\label{metric}
\end{equation}
where $a(\eta)$ is the scale factor and $\eta$ the conformal-time coordinate (notice that
 Eq. (\ref{metric}) has been written, for simplicity, in the conformally flat case). The second hypothesis 
is that the sources of the evolution of the background geometry are perfect fluid sources. As 
a consequence the entropy of the sources is constant.
The third and final hypothesis is that the dynamics of the sources and of the geometry 
is dictated by the general relativistic FRW equations \footnote{Units $M_{\rm P} = (8\pi G)^{-1/2} = 1.72 \times 10^{18} {\rm GeV}$ 
will be used.}: 
\begin{eqnarray}
&& M_{\rm P}^2 {\cal H}^2 = \frac{a^2}{3} \rho,
\label{frw1}\\
&& M_{\rm P}^2 ({\cal H}^2 - {\cal H}') = \frac{a^2}{2}( \rho + p),
\label{frw2}\\
&& \rho' + 3 {\cal H} ( \rho + p) =0,
\label{frw3}
\end{eqnarray}
where ${\cal H} = (\ln{a})'$ and the prime \footnote{ The overdot will usually denote 
derivation with respect to the cosmic time coordinate $t$ related to conformal time as 
$dt = a(\eta) d\eta$.} denotes the derivation with 
respect to $\eta$. Recall also, for notational convenience, that $a H= {\cal H}$ where 
$H= \dot{a}/a$ is the conventional Hubble parameter.
The various tests of the standard cosmological model are well known \cite{peebles,peebles2}. Probably 
one of the most stringent one comes from the possible distortions, in the Rayleigh-Jeans region, 
of the CMB spectrum. The absence of these distortions clearly rules out steady-state 
cosmological models. 
In the standard cosmological model
one usually defines the proper distance of the event horizon at the time $t_1$ 
\begin{equation}
d_{\rm e} = a(t_1) \int_{t_1}^{t_{\rm max}} \frac{d t}{a(t)},
\label{evhor}
\end{equation}
this distance represents the maximal extension of the region over which causal connection is possible.
Furthermore, the proper distance of the particle horizon can also 
be defined:
\begin{equation}
d_{\rm p}(t_2) = a(t_2) \int_{t_{\rm min}}^{t_2}  \frac{d t}{a(t)}.
\label{parthor}
\end{equation}
If the scale factor is parametrized as $a(t) \sim t^{\alpha}$, for $0 <\alpha <1$ the Universe experiences 
a decelerated expansion, i.e. 
$\dot{a} >0$ and $ \ddot{a} <0$ while the curvature scale decreases, i.e. $\dot{H} <0$. This is the peculiar 
behaviour when the fluid sources are dominated either by dust ($p=0$) or by radiation ($p=\rho/3$).
In the case of the standard model, the particle horizon increases linearly in cosmic time (therefore 
faster than the scale factor). This implies that the CMB radiation, today observed with a temperature 
of $2.7 {\rm K}$ over the whole present horizon has been emitted from space-time 
regions which were not in causal contact. This problem is known as the horizon problem of the 
standard cosmological model. The other problem of the standard model 
is related to the fact that today the intrinsic (spatial) curvature $k/a^2$ is smaller than the extrinsic 
curvature, i.e. $H^2$. Recalling that $k/(a^2 H^2) = k/\dot{a}^2$ it is clear that if $\ddot{a} <0$, $1/\dot{a}^2$ 
increases so that the intrinsic curvature could be, today,  arbitrarily large. The third problem 
of the standard cosmological model is related to the generation of the large entropy of the 
present Universe. 
The solution of the kinematical problems of the standard model is usually discussed in the 
framework of a phase of accelerated expansion \cite{kolb}, i.e. $\ddot{a} >0$ and $\dot{a}>0$. In the case 
of inflationary dynamics the extension of the causally connected regions grows as the scale factor and hence 
faster than in the decelerated phase. This solves the horizon problem. Furthermore, during inflation 
the contribution of the spatial curvature becomes very small. The way inflation solves the 
curvature problem is by producing a very tiny spatial curvature at the onset of the radiation epoch taking 
place right after inflation.
The spatial curvature can well grow during the decelerated phase of expansion but is will be always subleading 
provided inflation lasted for sufficiently long time. In fact, the minimal requirement in order to solve these 
problems is that inflation lasts, at least, $60$-efolds. 
The final quantity which has to be introduced is the Hubble radius $H^{-1}(t)$. This quantity is local in time, however, 
a sloppy nomenclature often exchanges the Hubble radius with the horizon. Since this terminology 
is rather common, it will also be used here.
In the following applications it will be relevant to recall some of the useful thermodynamical relations.
In particular, in radiation dominated Universe, the relation between the Hubble 
parameter and the temperature is given by 
\begin{eqnarray}
&& H = 1.08 \times \biggl(\frac{g_{\ast}}{10.75}\biggr)^{1/2} \frac{T^2}{M_{\rm P}},
\nonumber\\
&& H^{-1} = 9 \times 10^{4}  \biggl(\frac{g_{\ast}}{10.75}\biggr)^{-1/2} \biggl(\frac{{\rm GeV}}{T}\biggr)^2 {\rm cm},
\label{therm1}
\end{eqnarray}
where $g_{\ast}$ is the effective number of relativistic degrees of freedom 
at the corresponding temperature. Eq. (\ref{frw1}), implying $H^2 M_{\rm P}^2 = \rho/3$, has been 
used in Eq. (\ref{therm1}) together with the known relations valid in a radiation 
dominated background
\begin{equation}
\rho(T) = \frac{\pi^2}{30} g_{\ast} T^4,~~~~~~n(T) = \frac{\zeta(3)}{\pi^2} g_{\ast} T^3
\label{therm2}
\end{equation}
where $\zeta(3) \simeq 1.2$.
In Eq. (\ref{therm2}) the thermodynamical relation for the 
number density $n(T)$ has been also introduced for future convenience.

\subsubsection{Inflationary dynamics and its extensions}

The inflationary dynamics can be realized in different ways. Conventional 
inflationary models are based either on one single inflaton field \cite{guth,albrecht,linde1,linde2} 
or on various fields \cite{linde3} (see \cite{kolb2} for a review). Furthermore, in the context 
of single-field inflationary models one oughts to distinguish between 
small-field models \cite{linde1} (like in the so-called new-inflationary models) and 
large-field models \cite{linde2} (like in the case of chaotic models).
 
In spite of their various quantitative differences, conventional  inflationary models 
are based on the idea that during the phase of accelerated expansion
the curvature scale is approximately constant (or slightly decreasing). After inflation, the radiation 
dominated phase starts. It is sometimes useful  for numerical estimates to assume that 
radiation suddenly dominates at the end of inflation. In this case   
the scale factor can be written as 
\begin{eqnarray}
&& a_{i}(\eta) = \biggl( - \frac{\eta}{\eta_1}\biggr)^{- \alpha}, 
~~~~\eta < - \eta_1,
\nonumber\\
&& a_{r}(\eta) = \frac{\alpha \eta + (\alpha + 1)\eta_1}{\eta_1},
~~~~\eta \geq - \eta_1, 
\label{a}
\end{eqnarray}
where $\alpha$ is some effective exponent parameterizing the  dynamics
of the primordial phase of the Universe.  Notice that if $\alpha = 1$
we have that the primordial  phase coincides with a de Sitter
inflationary epoch. The case $\alpha = 1$ is not completely realistic since 
it corresponds to the case where the energy-momentum tensor 
is simply given by a (constant) cosmological term. In this case 
the scalar fluctuations of the geometry are not amplified and the 
large-scale angular anisotropy in the CMB would not be reproduced. 
The idea is then to discuss more realistic energy-momentum tensors 
leading to a dynamical behaviour close to the one of pure de Sitter 
space, hence the name quasi-de Sitter space-times.  Quasi-de Sitter 
dynamics can be realized in different ways. One possibility is to 
demand that the inflaton slowly rolls down from its potential 
obeying the approximate equations 
\begin{eqnarray}
&& M_{\rm P}^2 H^2 \simeq \frac{V(\varphi)}{3},
\nonumber\\
&& 3 H \dot{\varphi} \simeq - \frac{\partial V}{\partial \varphi},
\label{sr}
\end{eqnarray}
valid provided $\epsilon_{1} = - \dot{H}/H^2 \ll 1$ and 
$\epsilon_{2} = \ddot{\varphi}/(H \dot{\varphi}) \ll 1$. There also exist 
exact inflationary solutions like the power-law solutions obtainable in the case 
of exponential potentials:
\begin{eqnarray}
&& V = V_0 e^{ - \sqrt{\frac{2}{p}} \frac{ \varphi}{M_{\rm P}}},~~~~~~~~ 
\dot{\varphi}= \frac{ \sqrt{2 p} M_{\rm P}}{t}, 
\nonumber\\
&&a(t) \simeq t^{p}.
\label{pot}
\end{eqnarray}
 Since, from the exact equations,
$ 2 M_{\rm P}^2 \dot{H} = - {\dot{\varphi}}^2$ the two slow-roll parameters can also be written as 
\begin{equation}
\epsilon_{1} = \frac{M_{\rm P}^2}{2} \biggl(\frac{\partial\ln{ V}}{\partial\varphi}\biggr)^2 , 
~~~~~~~~~~~~\epsilon_{2} =   -\frac{M_{\rm P}^2}{2} \biggl(\frac{\partial\ln{ V}}{\partial\varphi}\biggr)^2
+ \frac{M_{\rm P}^2 }{V} \frac{\partial^2 V}{\partial\varphi^2}.
\label{sr2}
\end{equation}
In the case of the exponential potential (\ref{pot}) 
the slow-roll parameters are all equal, $\epsilon_{1} = \epsilon_{2} = 1/p$. 
Typical potentials leading to the usual inflationary dynamics 
are power-law potentials of the type $V(\varphi) \simeq \varphi^{n}$, exponential 
potentials, trigonometric potentials and nearly any potential satisfying, in some 
region of the parameter space, the slow-roll conditions.

Inflation can also be realized in the case when the curvature scale is increasing, i.e. $\dot{H}>0$ and $\ddot{a}>0$.
This is the case of superinflationary dynamics. For instance the propagation of fundamental strings  in curved backgrounds 
\cite{dvs1,dvs2} may lead to superinflationary solutions \cite{gsv1,gsv2}. A particularly simple case 
of superinflationary solutions arises in the case when internal dimensions are present. 

Consider a homogeneous and anisotropic manifold 
whose line element can be written as 
\begin{eqnarray}
ds^2 = G_{\mu\nu} dx^{\mu} dx^{\nu} = 
a^2(\eta) [ d\eta^2 - \gamma_{i j} d x^{i} dx^{j}] - b^2(\eta) \gamma_{a b} 
dy^a d y^b,
\nonumber \\
\mu,\nu = 0,..., D-1=d+n , ~~~ i, j=1,..., d , ~~~
a,b = d+1,..., d+n.
\label{metric2}
\end{eqnarray}
[$\eta$ is the
conformal time coordinate related, as usual to the cosmic time $t=\int a(\eta)
d\eta$ ; $\gamma_{ij}(x)$, $\gamma_{ab}(y)$ are the metric
tensors of two maximally symmetric Euclidean 
manifolds parameterized,
respectively, by the ``internal" and the ``external" coordinates $\{x^i\}$ and
$\{y^a\}$]. 
The metric of Eq. (\ref{metric2})
 describes the situation in which the $d$ external dimensions 
(evolving
with scale factor $a(\eta)$) and  the $n$ internal ones (evolving with scale
factor $b(\eta)$) are dynamically decoupled from each other \cite{gio1}. 

A model of background evolution can be generically written as  
\begin{eqnarray}
&& a(\eta) = a_1 \biggl(-\frac{\eta}{\eta_1}\biggr)^{\sigma},\,\,\,\,
   b(\eta) = b_1 \biggl(-\frac{\eta}{\eta_1}\biggr)^{\lambda},\,\,\,\,\,\,
   \,\,\,\, \eta< - \eta_1,
\nonumber\\
&& a(\eta) = a_1\biggl(\frac{ \eta + 2 \eta_1}{\eta_1}\biggr),\,\,\,\,\, 
   b(\eta) = b_1,\,\,\,\,\,\,\,\,\,\,\,\,\,\,\,\,\,\,\,\,\,\,\,\,\,\,\,\,
-\eta_1 \leq \eta \leq \eta_2,
\nonumber\\
&& a(\eta) = a_1 \frac{ (\eta+ \eta_2 + 4 \eta_1)^2}{4 \eta_1 (\eta_2 +  
2 \eta_1)}, \,\,\,\,\, 
   b(\eta) = b_1,\,\,\,\,\,\,\,\,\,\,\,\,\,\,\,\,\,\,\,\,\,\,\,\,\,\,\,
\eta > \eta_2.
\label{back}
\end{eqnarray}
In the parameterization of Eq. (\ref{back})
 the internal dimensions grow (in conformal time) for $\lambda <0$ and they 
shrink for $\lambda > 0$ \footnote{To assume 
that the internal dimensions 
are constant during the radiation and matter dominated 
epoch is not strictly necessary. If 
the internal dimensions have a time variation during the radiation 
phase we must anyway impose the BBN bounds on their variation 
\cite{int1,int2,int3}. The tiny variation allowed by BBN implies that $b(\eta)$ must be
 effectively constant for practical purposes.}.

Superinflationary solutions are also common in the context 
of the low-energy string effective action \cite{lovelace,fradkin,callan}.
In critical superstring theory the
electromagnetic field $F_{\mu\nu}$ is coupled not only
to the metric ($g_{\mu\nu}$), but also to the dilaton background
($\varphi$).
In the low energy limit such interaction is represented by the
string effective action \cite{lovelace,fradkin,callan}, which reads, after
reduction from ten to four expanding dimensions,
\begin{eqnarray}
&&S=- \int d^4x\sqrt{-g}e^{-\varphi}\biggl[ R +
\partial_{\mu} \varphi \partial^{\mu} \varphi - \frac{1}{2} e^{2 \varphi}\partial_{\mu} \sigma \partial^{\mu} \sigma
\nonumber\\
&& + \frac{1}{4} F_{\mu\nu}F^{\mu\nu} -\frac{1}{8} e^{\varphi} \sigma F_{\mu\nu} \tilde{F}^{\mu\nu} \biggr]
\label{action4}
\end{eqnarray}
were  $\varphi = \phi - \ln{V_6} \equiv \ln (g^2)$ controls the
tree-level four-dimensional gauge coupling ($\phi$ being the
ten-dimensional dilaton field, and $V_6$ the volume of the
six-dimensional compact internal space). The field $\sigma$ is the Kalb-Ramond axion 
whose pseudoscalar coupling to the gauge fields may also be interesting.

In the inflationary models based on the above
effective action \cite{Veneziano1,Veneziano3}
the dilaton background is not at all constant, but
undergoes an accelerated evolution
from the string perturbative vacuum ($\varphi= -\infty$) towards the
strong
coupling regime, where it is expected to remain frozen at its
present value. The peculiar feature of this string cosmological scenario 
(sometimes called pre-big bang \cite{Veneziano3}) is that 
not only the curvature evolves but also the gauge coupling. Suppose, for 
the moment that the gauge fields are set to zero.
The phase
of growing  curvature and dilaton coupling
($\dot H>0$, $\dot\varphi>0$), driven by the kinetic energy of the
dilaton field,  is correctly described in terms of
 the lowest order string effective action only
up to the conformal time $\eta=\eta_{s}$  at which the curvature
reaches the string scale $H_{s}=\lambda_{s}^{-1}$ ($
\lambda_{s}\equiv
\sqrt{\alpha^{\prime}}$ is the fundamental length of string theory).
A first important parameter of this cosmological model is thus the
value $\varphi_s$ attained by the dilaton at $\eta=\eta_{s}$.
Provided such
 value is sufficiently negative (i.e. provided the coupling
$g=e^{\varphi/2}$ is sufficiently small to be still in the
perturbative region at $\eta=\eta_{s}$), it is also arbitrary, since
there
is no perturbative potential to break invariance under shifts of
$\varphi$.
 For $\eta >\eta_{s}$ high-derivatives terms (higher
orders in $\alpha^{\prime}$)
become important in the string effective action,
and the background enters a genuinely ``stringy" phase of
 unknown duration. An  assumption of string cosmology
is that the stringy phase eventually ends at some conformal time
$\eta_1$
in the strong coupling regime.  At this time the dilaton,
feeling a non-trivial potential, freezes to its present constant
value
$\varphi=\varphi_{1}$ and the standard radiation-dominated era starts.
The total duration $\eta_1/\eta_s$, or the total red-shift $z_s$
encountered during
 the stringy epoch (i.e. between $\eta_s$ and $\eta_1$),
will be the second crucial parameter besides $\varphi_{s}$
entering  our discussion. For the purpose of this paper, two
parameters are enough to specify completely our model of
background, if we accept that during the string phase the
curvature stays controlled by the string scale, that is $H\simeq g
M_{\rm P} \simeq \lambda_{s}^{-1}$ ($M_p$ is the Planck mass) for
$\eta_s<\eta <\eta_1$.

During the string era $\dot\varphi$ and $H$ are
 approximately constant, while, during the dilaton-driven epoch
\begin{equation}
a=(-t)^{\alpha},~~~~~ \alpha = - \frac {1}{\sqrt{3}}
\sqrt{1- \Sigma}, 
\end{equation}
Here
$\Sigma \equiv \sum_{i}{\beta_i^2}$ represents the
possible effect of internal dimensions whose radii $b_i$ shrink
like $(-t)^{\beta_i}$ for $t \rightarrow 0_-$ (for the sake of
definiteness we show in the figure the case $\Sigma=0$).
The shape of the coupling curve corresponds to the fact that the
dilaton
is constant during the radiation era, that $\dot\varphi$ is
 approximately constant during the string era, and that it evolves
like
\begin{equation}
g(\eta) = e^{\varphi/2}=a^{\lambda}, \,\,\,\,\,\,\,\,\,\,
\lambda =
\frac{1}{2}(3+\frac{\sqrt{3}}{\sqrt{1-\Sigma}}),
\end{equation}
during the dilaton-driven era.

An interesting possibility, in the pre-big bang context is that 
the exit to the phase of decelerated expansion and decreasing curvature 
takes place without any string tension correction. Recently a model of 
this kind has been 
proposed \cite{nonloc}. The idea consists in adding a non-local 
dilaton potential which is invariant under scale factor duality. The 
evolution equations of the metric and of the dilaton will 
then become, in 4 space-time dimensions,
\begin{eqnarray}
&&\dot{\overline{\varphi}}^2 - 3 H^2 -V=0,
\nonumber\\
&& \dot H -H\dot{\overline{\varphi}}=0,
\nonumber\\
&& 2 \ddot{\overline{\varphi}} - {\dot{\overline{\varphi}}}^2 -3 H^2 +V -{\partial V\over \partial \overline{\varphi}} = 0,
\label{nonlocdil}
\end{eqnarray}
where $\overline{\varphi} = \varphi - 3\log{a}$ is the shifted dilaton.
A particular solution to these equations will be given by 
\begin{eqnarray}
&& V (\overline{\varphi}) = - V_0 e^{ 4\overline{\varphi} },
\label{pot1}\\
&&a(t) = a_0 \biggl[ \tau + \sqrt{\tau^2 + 1}\biggr]^{1/\sqrt{3}},
\label{scalefactor1}\\
&& \overline{\varphi} = - \frac{1}{2} \log{ ( 1 + \tau^2)} + \varphi_{0},
\label{dilaton1}
\end{eqnarray}
where
\begin{equation}
\tau = \frac{t}{t_0}, ~~~~~~~~~~~~ t_0 = \frac{e ^{- 2 \varphi_0}}{ \sqrt{V_{0}}}.
\end{equation}
This solution interpolates smoothly between two self-dual solutions. For
$t \to -\infty$ the background superinflates while for $t\to +\infty$ the 
decelerated FRW limit is recovered.

\subsection{Gauge fields in FRW Universes}

In a FRW metric of the type (\ref{metric}) the evolution equations 
of the gauge fields are invariant under  Weyl  
rescaling of the four-dimensional metric. This property is often 
named  conformal invariance. 
Consider, for simplicity, the evolution equation of sourceless Maxwell fields:
\begin{equation}
\nabla_{\mu} F^{\mu\nu} \equiv \frac{1}{\sqrt{-G}} \partial_{\mu}\biggl[ \sqrt{-G} F^{\mu\nu}\biggr] =0.
\label{conformal} 
\end{equation} 
Since the metric (\ref{metric}) is conformally flat, the time dependence arising
in Eq. (\ref{conformal}) from $\sqrt{-G}$, is always compensated 
by the controvariant indices of the fields strength.
 Maxwell fields in conformally flat 
metrics obey the same equations obtained in the case of Minkowski space-time 
provided the conformal time coordinate is used and provided the 
fields are appropriately rescaled. This 
property holds not only for Maxwell 
fields but also for chiral fermions in FRW backgrounds. 

Using Weyl invariance, the generalization of MHD equations to the case of conformally flat FRW space-times can 
then be easily obtained:
\begin{eqnarray}
&& \frac{\partial \vcB}{\partial \eta} + \vec{\nabla} \times \vcE =0
\label{cmh1}\\
&& \vec{\nabla} \cdot \vcE =0, ~~~~~~~~~\vec{\nabla} \cdot \vcB =0,
\label{cmh2}\\
&& \vec{\nabla} \times \vcB = 4 \pi \vcJ + \frac{\partial \vcE}{\partial \eta},
\label{cmh3}\\
&& \vcJ = \sigma ( \vcE + \vec{v} \times \vcB),
\end{eqnarray}
where 
\begin{equation}
\vcB = a^2 \vec{B},~~~~\vcE = a^2 \vec{E},~~~~\vcJ = a^3 \vec{J}, ~~~~\sigma = \sigma_{c} a,
\label{curvflat}
\end{equation}
are the curved space fields expressed as a function of the corresponding 
flat space quantities.

In  the case 
of a radiation dominated background, the Navier-Stokes equation takes the form 
\begin{equation}
[(\overline{p}+ \overline{\rho}) \vec{v}]' + \vec{v}\cdot\vec{\nabla}[(\overline{p} +\overline{\rho})\vec{v}]  = - \vec{\nabla}\biggl[\overline{p} + 
\frac{ |\vcB|^2]}{8\pi} \biggr]
+ \frac{[\vcB\cdot\vec{\nabla}]\vcB}{4\pi} + (\overline{p} + \overline{\rho})\nu \nabla^2 \vec{v},
\label{cns}
\end{equation}
where the incompressible  closure, i.e. $\vec{\nabla}\cdot\vec{v}=0$, 
 has been already adopted and where the vector identity $\vec{\nabla}\times\vcB\times \vcB = 
- \frac{1}{2}\vec{\nabla} |\vcB|^2
 + [\vcB\cdot\vec{\nabla}] \vcB$ has been used. 
In Eq. (\ref{cns}) $\overline{p} = a^4 p$ and 
$\overline{\rho} = a^4 \rho$. In terms of the rescaled pressure and energy densities 
the continuity equation becomes 
\begin{equation}
\frac{\partial \overline{\rho}}{\partial\eta} + ( 3 \gamma -1) {\cal H} \overline{\rho} + (\gamma + 1) \vec{\nabla}\cdot[\overline{\rho} \vec{v}] =0,
\label{conteq}
\end{equation}
where a barotropic equation of state has been assumed for the background fluid, i.e. $\overline{p} = \gamma \overline{\rho} $. 
The energy density is homogeneous and hence the last term in Eq. (\ref{conteq}) can be consistently dropped within the incompressible 
closure. However, one may ought to linearize Eq. (\ref{conteq}) and this is why the last term may be kept for future 
considerations. If $\gamma = 1/3$, then Eq. (\ref{conteq}) implies  $\overline{\rho}'=0$. 
For generic barotropic index Eq. (\ref{cns}) becomes 
\begin{equation}
\frac{\partial \vec{v}}{\partial\eta} + ( 1 - 3 \gamma) {\cal H} \vec{v}  = \frac{[\vcB\cdot\vec{\nabla}]\vcB}{4\pi (\gamma + 1) \overline{\rho}} + \nu \nabla^2 \vec{v},
\label{cnsgen}
\end{equation}
where the magnetic pressure term has been dropped since it is negligible with respect to $\overline{p}$.

Recalling that for $L\gg L_{\sigma}$ the magnetic diffusivity equation can be written, using vector identities, as
\begin{equation}  
\frac{\partial \vcB}{\partial\eta}+ [\vec{v}\cdot\vec{\nabla}]
\vcB= 
[\vcB\cdot\vec{\nabla}]\vec{v} + \frac{1}{4\pi\sigma} \nabla^2 \vcB.
\label{dif}
\end{equation}

Eqs.  (\ref{cmh1})--(\ref{cns}) form a closed system sharing analogies  with 
their flat space counterpart. Thanks to conformal invariance, various solutions 
can be lifted from flat to curved space like, for instance \cite{bis}, the following 
 fully nonlinear solution of Eqs. (\ref{dif})
and (\ref{cns})
\begin{equation}
\vec{v} = \pm \frac{\vcB}{\sqrt{\overline{\rho} + \overline{p}}}\sim  
\pm \frac{\vcB}{\sqrt{\overline{\rho}}}.
\end{equation}
valid if thermal and magnetic diffusivities are 
neglected and if $ \overline{p} \gg |\vcB|^2$. 

Eqs. (\ref{cmh1})--(\ref{cns}) present also some differences with 
respect to the flat space case: since the Universe is expanding, the 
relative balance between diffusion scales may change as time goes by.
All the propagation of MHD excitations can be then  generalized to the curved-space case.
In various papers \cite{vlcur1,vlcur2,vlcur3,vlcur3} this program 
has been achieved (see also \cite{holcomb,tajima} for earlier attempts). 
In \cite{mgknot} the evolution of magnetic fields  with non-vanishing magnetic helicity 
has been studied using the resistive MHD approximation 
in curved space. In \cite{subra} the evolution of non-linear 
Alfv\'en waves with the purpose of deriving possible observable 
effects \cite{subra2}. 
 
In order to analyze the high-frequency spectrum of plasma excitations, the 
Vlasov-Landau approach can be generalized to curved spaces \cite{ber}.
There are two options in order to achieve this goal. The first option 
is to assume that the magnetic field breaks the isotropy 
of space-time. This approach leads to the study of plasma effects 
in a magnetic field in full analogy with what is done 
in terrestrial tokamaks. The second choice is to discuss 
magnetic fields which are stochastic and which do not break the 
isotropy. Later in the present review the effects of magnetic fields breaking 
the isotropy of space-time will be considered but, for the moment, the attention will 
be concentrated on the second approach.

A typical problem which should be discussed within the kinetic approach is the
relaxation of charge and current density 
fluctuations in a relativistic plasma \cite{mginf1}. The problem 
is then to compute what magnetic field is induced from charge and 
current density fluctuations.

Consider an equilibrium homogeneous and
isotropic conducting plasma, characterized by a distribution function
\begin{equation}
f_{0}(p) = \frac{n_{q}}{8 \pi T^3} e^{-\frac{p}{T}},
\label{dist0}
\end{equation}
common for both positively and negatively charged
ultrarelativistic particles (for example, electrons and positrons).
The normalization is chosen in such a way 
that $\int d^3 p f_0(p)= n_{q}$. 
Within the kinetic approach,  the initial charge and current density 
fluctuations are related to the  perturbations of the 
initial value of the distribution function around the equilibrium 
distribution (\ref{dist0}), i.e 
\begin{eqnarray}
&& f_{+}(\vec{x}, \vec{p}, \eta) = f_0(p) + 
\delta f_{+}(\vec{x},\vec{p},\eta),
\nonumber\\
&& f_{-}(\vec{x},\vec{p}, \eta) =  
f_0(p) + \delta f_{-}(\vec{x},\vec{p},\eta),
\end{eqnarray}
where $+$ refers to positrons and $-$ to electrons, and $\vec{p}$ is
the conformal momentum.  The Vlasov equation   defining the
curved-space evolution of the perturbed distributions can  be written
as  
\begin{eqnarray} 
&&\frac{\partial f_{+}}{\partial \eta} + 
\vec{v} \cdot \frac{\partial f_{+}}{\partial\vec{x}} + e ( \vec{\cal E} + 
\vec{v}\times \vec{\cal B}) \cdot\frac{\partial f_{+}}{\partial \vec{p}} = 
\biggl( 
\frac{\partial f_{+}}{\partial \eta}\biggr)_{\rm coll}
\label{Vl+},\\
&&\frac{\partial f_{-}}{\partial \eta} + 
\vec{v} \cdot \frac{\partial f_{-}}{\partial\vec{x}} - e ( \vec{\cal E} + 
\vec{v}\times \vec{\cal B})\cdot \frac{\partial f_{-}}{\partial \vec{p}} = 
\biggl(\frac{\partial f_{-}}{\partial\eta}\biggr)_{\rm coll}
\label{Vl-},
\end{eqnarray}
where the two terms appearing at the right hand side of each 
equation are the collision terms. This system of equation represents 
the curved space extension of the Vlasov-Landau approach  to plasma
fluctuations \cite{vla,lan}. All particle number densities here are
related to the comoving volume.

Notice that, in general  $\vec{v} = \vec{p}/\sqrt{\vec{p}^2 + m_{e}^2
a^2}$. In the  ultra-relativistic limit $\vec{v}= \vec{p}/|\vec{p}|$
and the  Vlasov equations are conformally invariant.  This  implies
that, provided we use conformal  coordinates and rescaled gauge
fields, the system of equations which we would have in flat space
\cite{lif} looks exactly the same  as the one we are discussing in a
curved FRW (spatially flat) background \cite{ber}.

The evolution equations of the gauge fields coupled to the two
Vlasov  equations can be written as 
\begin{eqnarray}
&& \vec{\nabla} \cdot \vec{\cal E} = 4 \pi e \int d^3 p [f_{+}(\vec{x}, \vec{p},\eta) - 
f_{-}(\vec{x}, \vec{p},\eta)],
\nonumber\\
&& \vec{\nabla}\times \vec{\cal E} + \vec{B}' =0,
\nonumber\\
&& \vec{\nabla}\cdot \vec{\cal B} =0,
\nonumber\\
&& \vec{\nabla}\times \vec{\cal B} -\vec{\cal E}'= 4\pi e \int d^3 p \vec{v} 
[f_{+}(\vec{x}, \vec{p},\eta) - f_{-}(\vec{x}, \vec{p},\eta)].
\label{max}
\end{eqnarray}
In order to illustrate the physical relevance of this description, Eqs. 
(\ref{max}) will be used later in this Section to study
the fate of charge and current density fluctuations. 

\subsection{Physical scales in the evolution of magnetic fields}

When MHD is discussed in curved spce-times, the evolution of 
the geometry reflects in the time evolution of the physical scales 
of the problem.

\subsubsection{Before matter-radiation equality}

From Eq. (\ref{dif}), the comoving momentum corresponding to the magnetic 
diffusivity scale is given by the relation
\begin{equation}
\frac{k^2}{4 \pi \sigma} \sim \frac{1}{\eta}\sim {\cal H} = a H.
\label{ksigma}
\end{equation}
In the  case of a relativistic plasma at typical temperatures  $T \gg 1~{\rm MeV}$, the flat-space conductivity
(see Eq. (\ref{curvflat})) is given by  
\begin{equation}
\sigma_{c} \simeq \frac{\alpha_{\rm em}}{ T \overline{\sigma}},
\end{equation}
where $\overline{\sigma} \sim \alpha_{\rm em}^2/T^2$ 
is the cross section \footnote{The cross section is denoted with $\overline{\sigma}$ while the conductivity is $\sigma$.}. 
Then, it can be shown \cite{enqvist1} that 
\begin{equation}
\sigma_{c} = \sigma_{0} \frac{T}{\alpha_{\rm em}}
\end{equation}
where $\sigma_0$ is a slowly increasing function of the temperature.
In \cite{enqvist1} it has been convincingly argued that $\sigma_0 \sim 0.06$ 
at $T\sim 100 $ MeV while it becomes $0.6$ at $T\sim 100~ {\rm GeV}$ (for a recent analysis 
of conductivity in high temperature QED see \cite{bd1}).
Recalling Eq. (\ref{curvflat}) the physical diffusivity scale $\omega_{\sigma}= k_{\sigma}/a(\eta)$ 
will be, in units of the Hubble parameter, 
\begin{equation}
\frac{\omega_{\sigma}}{H} \simeq \sqrt{\frac{4 \pi \sigma_0}{\alpha_{\rm em}}}\biggl(\frac{T}{H}\biggr)^{1/2} = 72.2 \times 
\sigma_0^{1/2} g_{\ast}^{-1/4} \biggl(\frac{T}{M_{\rm P}}\biggr)^{-1/2}.
\label{omsig}
\end{equation}
Discounting for the effect due to the possible variations of the 
effective relativistic degrees of freedom, the (physical) magnetic diffusivity length scale 
is much smaller than the Hubble radius at the corresponding epoch. For instance, if $T\sim 10~{\rm MeV}$, then
 $L_{\sigma} \sim \omega_{\sigma}^{-1} \sim 10^{-10} H^{-1}$. This  means that there are ten decades 
length-scales where the magnetic field spectrum will not experience diffusion due to the finite 
value of the conductivity. 

The  estimate of the magnetic diffusivity scale will now be repeated in  a different range 
of temperatures, namely for  $T< 0.2 m_{\rm e}$ when weak interactions have fallen out of thermal 
equilibrium. In this case the relevant degrees of freedom 
are the three neutrino species and the photon, i.e. $g_{\ast} = 3.36$. In this case the conductivity will be 
given by 
\begin{equation}
\sigma \simeq \frac{\alpha_{\rm em}}{m_{\rm e} v_{\rm th} \overline{\sigma}},
\end{equation}
which leads to 
\begin{equation}
\sigma_{\rm c} = \frac{1}{\alpha_{\rm em}} \biggl(\frac{T}{m_{\rm e}}\biggr)^{1/2} T,
\label{NR}
\end{equation}
if $v_{\rm th} \sim (T/m_{\rm e})^{1/2}$ and $\overline{\sigma} \sim \alpha_{\rm em}^2/T^2$.
Eq. (\ref{NR}) leads to the known $T^{3/2}$ dependence of the conductivity 
from the temperature, as expected from a non-relativistic plasma.
The physical magnetic diffusivity momentum is, in this case
\begin{equation}
\frac{\omega_{\sigma}}{H} \simeq \sqrt{\frac{4 \pi}{\alpha}} \biggl(\frac{T}{m_{\rm e}}\biggr)^{1/4} \sqrt{\frac{T}{H}} \sim  
\frac{11}{\alpha_{\rm em}} g_{\ast}^{-1/4} \biggl(\frac{T}{m_{\rm e}}\biggr)^{1/4} 
\biggl(\frac{M_{{\rm P}}}{T} \biggr)^{1/2}.
\label{NRdiffscale}
\end{equation}
 Eqs. (\ref{omsig}) and (\ref{NRdiffscale}) refer to two different ranges 
of temperatures. The magnetic diffusivity scale in units of the Hubble parameter obtained  in Eq. (\ref{NRdiffscale}) is 
smaller than the one given by in Eq. (\ref{omsig}). Hence, by lowering the temperature across the weak equilibration 
temperature makes the conductivity length scale larger in units of the Hubble radius :
 {\em inertial range} \footnote{In more general terms the inertial range is the interval of scales (either in momentum 
or in real space) where the dynamics is independent on the scales of dissipation so that, in this range, the difffusivities 
can be taken to zero.} of length scales (i.e. the region where conductivity effects are negligible) is  larger 
above the weak equilibration temperature.
Summarizing
\begin{eqnarray}
&& L_{\sigma}(T) = \frac{\sigma_0^{-1/2} g_{\ast}^{1/4}}{72.2} \biggl( \frac{T}{M_{\rm P}}\biggr)^{1/2} ~~L_{H}(T),~~~~~~~~~~~~~T> {\rm MeV}
\nonumber\\
&& L_{\sigma}(T) = \frac{\sigma_0^{-1/2} g_{\ast}^{1/4}}{72.2} \biggl( \frac{T}{M_{\rm P}}\biggr)^{1/2} \biggl(\frac{T}{m_{\rm e}}\biggr)^{-1/4}~~
L_{H}(T),~~~~~T< {\rm MeV},
\label{difflength}
\end{eqnarray}
where $L_{H}(T) = H^{-1}(T)$.

In the evolution of the plasma it is sometimes relevant to estimate the  thermal diffusivity scale setting 
 a bound on the coherence 
of the velocity field and not of the magnetic field. There are however phenomena, like 
the propagation of MHD waves, where the  excitations of a background magnetic field are coupled 
to the excitations of the velocity field.  Generally 
speaking the typical correlation scale of the velocity field is always 
smaller, in various  physical situations, than the typical correlation 
scale of the magnetic fields. 
The range of physical momenta affected by thermal diffusivity is bounded from above 
by the thermal diffusivity scale which can be obtained from the following relations
\begin{equation}
k^2 \nu \simeq \frac{k^2}{5} \ell_{\gamma}(\eta) \simeq \frac{1}{\eta} = {\cal H} = a H,
\label{thdiffmom}
\end{equation}
were $\ell_{\gamma}$ is the photon mean free path, i.e.
\begin{equation}
\ell_{\gamma} \simeq \frac{1}{n_{\rm e} \overline{\sigma}}.
\label{MFP}
\end{equation}
The mean free path measures the efficiency of a given process to transfer 
momentum. From Eqs. (\ref{cns}) and (\ref{thdiffmom}) the typical thermal diffusivity momentum 
is:
\begin{equation}
\frac{\omega_{\rm diff}}{H} \simeq \sqrt{\frac{5}{H(T) \ell_{\gamma}(T)}}.
\label{thdif2}
\end{equation}
where $\omega_{\rm diff} = k_{\rm diff}/a$ is the physical momentum as opposed to 
the comoving momentum $k_{\rm diff}$.  The scale defined in Eq. (\ref{thdif2}) is 
sometimes called Silk scale. Notice that  the mean free path given in Eq. (\ref{MFP}) 
increases as $a^3(\eta)$. Since the physical scale of an inhomogeneous region 
increases as $a(\eta)$, the photon mean free path can become larger than the physical 
size of an inhomogeneous region. This is the so-called free streaming regime.
 
If  $T < 0.2 m_{\rm e}$, 
the photon mean free path is given by Eq. (\ref{MFP}) 
where $\overline{\sigma}_{\rm T}= 1.7 \times 10^{3} {\rm GeV}^{-2} \simeq \alpha_{\rm em} m_{\rm e}^{-2}$ is the Thompson 
cross section and 
\begin{equation}
n_{\rm e} \simeq 6.6 \times 10^{-9}\, (\Omega_{\rm b}\, h_0^2)\,\, x_{\rm e}\, T^3
\label{ne1}
\end{equation}
is the electron number density depending both on the critical fractions of baryons, $\Omega_{\rm b}$ and on 
the ionization fraction $x_{\rm e}$. The ionization fraction is the ratio between the number density 
of protons and the total number density of baryons. The value of $x_{\rm e}$ drops from $1$ to $10^{-5}$ in a short 
time around recombination. Eq. (\ref{ne1}) is simply derived 
by noticing that, after electron positron annihilation, 
$n_{\rm e} \simeq \overline{\eta} n_{\gamma}$ where 
$\overline{\eta}= 2.74 \times 10^{-8} \Omega_{\rm b} h_0^2$  
is the baryon-to-photon ratio, the free parameter 
of big-bang nucleosynthesis 
calculations \cite{BBNobs1,BBNobs2,BBNobs3,hannu}. Thus 
\begin{equation}
\ell_{\gamma}(T) \simeq 8.8 \times 10^{12} \biggl(\frac{x_{\rm e}}{0.5}\biggr)^{-1} \biggl(\frac{\Omega_{\rm b} h_0^2}{0.02}\biggr)^{-1} 
\biggl(\frac{ {\rm MeV}}{T}\biggr)^{2} T^{-1}.
\label{mfp1}
\end{equation}

For temperatures larger than the MeV the photon mean free path  can be estimated 
by recalling that, above the MeV, the Thompson cross section is replaced 
by the Klein-Nishina cross section (i.e. $\overline{\sigma} \sim T^{-1} {\rm GeV}^{-1}$). 
Thus, from Eq. (\ref{MFP}),  the photon mean free path is 
\begin{equation}
\ell_{\gamma}(T) \simeq 0.1 ~\biggl(\frac{{\rm GeV}}{T}\biggr)^2 {\rm GeV}^{-1}.
\label{mfp2}
\end{equation}
As before, it is useful to refer $\omega_{\rm diff}$ to the 
Hubble rate and to compute the typical diffusion distances for $T > 1~{\rm MeV}$
\begin{equation}
 \frac{L_{\rm diff}^{(\gamma)}(T)}{L_{H}(T)} \simeq 1.1 \times 10^{-10} \biggl(\frac{g_{\ast}}{10.75}\biggr)^{1/4} 
\label{difM}
\end{equation}
and for  $T < 1~{\rm MeV}$
\begin{equation}
\frac{L_{\rm diff}^{(\gamma)}(T)}{L_{H}(T)}\simeq 1.03 \times 10^{-5} 
\biggl(\frac{\Omega_{\rm b} h_0^2}{0.02}\biggr)^{-1/2} \biggl(\frac{x_{\rm e}}{0.5}\biggr)^{-1/2} \biggl(\frac{g_{\ast}}{10.75}\biggr)^{1/4}
\sqrt{\frac{{\rm MeV}}{T}}.
\label{difm}
\end{equation}
At temperatures ${\cal O}({\rm MeV})$,  corresponding
roughly to $e^{+}$ $e^{-}$ annihilation, Eqs. (\ref{difM}) and (\ref{difm}) 
differ by a factor $10^{5}$. This drop of five orders of magnitude of the thermal diffusivity scale
reflects the drop (by ten orders of magnitude) of the photon mean free path as it can be appreciated by comparing 
Eqs. (\ref{mfp1}) and (\ref{mfp2}) at $T \simeq {\cal O}({\rm MeV})$.

For temperatures 
smaller than the electroweak temperature (i.e. $T\sim 100$ GeV) down to the 
temperature of neutrino decoupling, neutrinos are the species with 
the largest mean free path and, therefore, the most 
efficient momentum transporters. Neglecting possible variations of the number of relativistic
species the neutrino mean free path will be simply given by $\ell_{\nu} = {\rm G}_{\rm F}^{-2} T^{-2} (n_{\rm lep} + n_{\rm qua})^{-1}$
where $n_{\rm lep}$ and $n_{\rm qua}$ are the leptons and quark number densities. Using this estimate the diffusion scale 
will be
\begin{equation}
L_{\rm diff}^{(\nu)} \simeq 10^{-5} \biggl(\frac{{\rm MeV}}{T}\biggr)^{3/2} L_{H}
\end{equation}
which should be compared with Eq. (\ref{difM}). Clearly $L_{\rm diff}^{(\nu)}>L_{\rm diff}^{(\gamma)}$.

It seems then  difficult to get 
observable effects from inhomogeneities induced by MHD 
excitations. In \cite{olinto1} the damping rates of  Alfv\'en and 
magnetosonic waves have been computed and it has been shown that 
MHD modes suffer significant damping (see, however, \cite{subra,subra2}). 
To avoid confusion, the results 
reported in Ref. \cite{olinto1} apply to the damping of MHD waves and not 
to the damping of magnetic fields themselves as also correctly pointed out in \cite{son}.

\subsubsection{After matter-radiation equality}

In standard big bang cosmology there are three rather close (but rather 
different) epochs,
\begin{itemize}
\item{} the matter-radiation equality taking place
at $T_{\rm eq}\simeq 5.5~(\Omega_0~ h_0^2)$ eV, 
i.e. $t_{\rm eq} \simeq 4 \times 10^{10}~ (\Omega_0~ h_0^2)^{-2}$ sec;
\item{} the  decoupling of radiation from matter occurring at $T_{\rm dec} \simeq 0.26~ {\rm eV} $ i.e. 
$t_{\rm dec} \simeq 5.6 \times 10^{12}~ (\Omega_{0}~ h_0^2)^{-1/2}~ {\rm sec}$
\item{} the  recombination occurring, for a typical value of the critical baryon fraction, 
 at $T_{\rm rec} \simeq 0.3~ {\rm eV}$, i.e. $t_{\rm rec} \simeq 4.3~ \times 10^{12} 
(\Omega_0~ h_0^2)^{-1/2}~ {\rm sec}$.
\end{itemize}
 After plasma recombines, the baryons are 
no longer prevented from moving by viscosity and radiation pressure. Thus, the only term distinguishing 
the evolution of the baryons from the evolution of a cold dark matter fluid is the presence 
of the Lorentz force term in the Navier-Stokes equation. In this situation, it is interesting
to investigate the interplay between the forced MHD regime and the generation 
of gravitational inhomogeneities. It was actually argued long ago by Wasserman \cite{wasserman}
and recently revisited by Coles \cite{coles2} that large-scale magnetic fields 
may have an important impact on structure formation.
 Assuming that, for illustrative purposes, 
the matter fluid consists only of baryons and cold dark matter, for 
$t > t_{\rm rec}$ the evolution of the system is described by the following set 
of non relativistic equations
\begin{eqnarray}
&& \frac{\partial \vec{v}_{\rm b}}{\partial \eta} + {\cal H} \vec{v}_{\rm b} = - \vec{\nabla} \Phi + \frac{\vec{\nabla} \times \vcB \times \vcB}{4 \pi  a^4(t) \rho_{\rm b}},
\label{vbaryon}\\
&& \frac{\partial \vec{v}_{\rm c}}{\partial \eta} + {\cal H} \vec{v}_{\rm c} = - \vec{\nabla}\Phi,
\label{vcold}\\
&& \nabla^2 \Phi = \frac{a^2}{2 M_{\rm P}^2} \biggl[ \rho_{\rm b} \delta_{\rm b} + \rho_{\rm c} \delta_{\rm c} \biggr],
\label{phi}
\end{eqnarray}
where $\Phi$ is the Newtonian potential. Eqs. (\ref{vbaryon})  come from Eq. (\ref{cnsgen}) for $\gamma=0$ in the case of homogeneous 
pressure density. Eq. (\ref{vcold}) can be obtained, in the same limit,  from Eq.  Eq. (\ref{cnsgen}) by setting 
to zero the Lorentz force.

From Eq. (\ref{conteq}), by linearizing the continuity equations for the two species, the relation between the divergence of the velocity field
and the density contrast can be obtained
\begin{equation}
\delta_{\rm b}' = - \vec{\nabla}\cdot\vec{v}_{\rm b},~~~~~~~~\delta_{\rm c}' = - \vec{\nabla}\cdot\vec{v}_{\rm c},
\label{deltav}
\end{equation}
where $\delta_{\rm b} = \delta\rho_{\rm b}/\rho_{\rm b}$ and  $\delta_{\rm c} = \delta\rho_{\rm c}/\rho_{\rm c}$
Thus, taking the divergence of Eqs. (\ref{vbaryon}) and (\ref{vcold}) and using, in the obtained equations, Eqs. (\ref{phi}) and 
(\ref{deltav}) the following system can be obtained
\begin{eqnarray}
&& \delta_{\rm b}'' + {\cal H}\delta_{\rm b} - \frac{a^2}{2 M_{\rm P}^2} \rho_{{\rm m}} \delta = 
\frac{\vec{\nabla} \cdot[ (\vec{\nabla}\times\vcB )\times \vcB]}{4\pi \rho_{\rm b} a^4(t)},
\label{deltab}\\
&&  \delta_{\rm c}'' + {\cal H}\delta_{\rm c} - \frac{a^2}{2 M_{\rm P}^2} \rho_{{\rm m}} \delta = 0,
\label{deltac}
\end{eqnarray}
where $\delta= \Omega_{\rm b} \delta_{\rm b} + \Omega_{\rm c} \delta_{\rm c}$ and $\rho_{\rm m}= \rho_{\rm b} + \rho_{\rm c}$ is the total matter density. 
Multiplying Eq. (\ref{deltab}) by $\Omega_{\rm b}$ and Eq. (\ref{deltac}) by $\Omega_{\rm c}$, the sum of the two resulting equations can be written as 
\begin{equation}
\delta'' + {\cal H} \delta - \frac{a^2}{2 M_{\rm P}^2} \rho_{{\rm m}} \delta = 
\Omega_{\rm b} \frac{\vec{\nabla} \cdot[(\vec{\nabla}\times \vcB )\times \vcB]}{4\pi \rho_{\rm b} a^4(t)},
\label{deltadec}
\end{equation}
where it has been used that the background geometry is spatially flat. Going to cosmic time, Eq. (\ref{deltadec}) reads 
\begin{equation}
\ddot{\delta} + \frac{2}{3 t} \dot{\delta} - \frac{2}{3 t^2} \delta = \frac{\vec{\nabla} \cdot[ (\vec{\nabla}\times\vcB) \times \vcB]}{4\pi \rho_{\rm b} a^6(t)}.
\label{finalcontrast}
\end{equation}
The homogeneous term of Eq. (\ref{finalcontrast})  can be easily solved. The source term strongly depends upon the 
magnetic field configuration \cite{mgknot}. However, as far as time evolution 
is concerned, $\vcB$ is constant in the inertial range of scales, i.e. $L > L_{\sigma}$. Hence, $\vcB = \vcB_0$ and $\vec{B}(t) = \vcB_0/a^2(t)$.
Assuming a stochastic magnetic field 
the solution of equation (\ref{finalcontrast}) can be obtained integrating from $t_{\rm rec}$ and recalling that $\rho_{\rm b}(t) \sim \rho_{\rm b}(t_{\rm rec}) a^{-3}$.
The result is 
\begin{equation}
\delta(t) = \Omega_{\rm b}\frac{t_{\rm rec}^2 |\vec{B}_{\rm rec}|^2}{4\pi \rho_{\rm b}(t_{\rm rec}) a^2(t_{\rm rec})
\lambda_{\rm B}^2}\biggl[ \frac{9}{10} 
\biggl(\frac{t}{t_{\rm rec}}\biggr)^{2/3} + \frac{3}{5} \biggl(\frac{t_{\rm rec}}{t}\biggr) - \frac{3}{2}\biggr].
\end{equation}
where $\lambda_{\rm B}$ is the typical length scale of variation of the magnetic field.
The total density contrast grows as $t^{2/3}$ with an initial condition set by the Lorentz force, i. e.
\begin{equation}
\delta(t_{\rm rec}) \simeq \Omega_{\rm b} \frac{|\vec{B}_{\rm rec}|^2 t_{\rm rec}^2}{4 \pi  \rho_{\rm b}(t_{\rm rec}) a^2(t_{\rm rec}) \lambda_{\rm B}^2}.
\end{equation}
To seed galaxy directly the effective $\delta(t_{\rm rec})$ should be of the order off $10^{-3}$ for a flat Universe, implying 
that 
\begin{equation}
|\vec{B}_{\rm rec}| \simeq 10^{-3} \biggl( \frac{\lambda_{\rm B}}{\rm Mpc}\biggr) (\Omega_{\rm b} h^2)^{1/2} ~~{\rm G}.
\label{sform}
\end{equation}
Also the converse is true. Namely one can read Eq. (\ref{sform}) from left to right. In this case Eq. (\ref{sform}) will define the 
typical scale affected by the presence of the magnetic field. This scale is sometimes called magnetic Jeans scale \cite{peebles}: 
it can be constructed from the Jeans length 
\begin{equation}
\lambda_{\rm J} = c_{\rm s} \biggl(\frac{\pi}{G\rho_{\rm b}}\biggr)^{1/2},
\end{equation}
where $c_{\rm s}$ is replaced by $c_{\rm a}$, i.e. the Alfv\'en velocity.

\subsection{Charge and current density fluctuations}

As a simple application of kinetic techniques, it is 
useful to understand which magnetic field can be obtained from 
an initial inhomogeneous distribution of charge and current 
density fluctuations.

Consider then the  system of Eqs. (\ref{max}) under the assumption that, 
initially, electric and magnetic fields 
vanish:
\begin{equation}
\delta f_{\pm}(\vec{x},\vec{p},\eta_0)\neq 0, ~~~~~~~\vec{\cal E}(\vec{x},\eta_0)=\vcB(\vec{x},\eta_0)=0.
\end{equation}
Using  Eqs. (\ref{Vl+})--(\ref{max}) the induced magnetic fields can be obtained..
This problem is discussed, in greater detail, in  \cite{mginf1}.

Subtracting  Eqs. (\ref{Vl+}) and (\ref{Vl-}),  and defining 
\begin{equation}
f(\vec{x},\vec{p},\eta)=  \delta f_{+}(\vec{x},\vec{p},\eta) - \delta f_{-}(\vec{x},\vec{p},\eta)
\end{equation}
the Vlasov-Landau system can be written as 
\begin{eqnarray}
&& - g_{\vec{k}}(\vec{p}) + i (\vec{k}\cdot \vec{v} - \omega) 
f_{\vec{k}\omega}(\vec{p}) + 2 e \vec{\cal E}_{\vec{k} \omega} \cdot 
\frac{\partial f_0}{\partial\vec{p}} = -\tilde{\nu} f_{\vec{k},\omega},
\label{vl1}\\
&& i \vec{k} \cdot \vec{\cal E}_{\vec{k} \omega} = 4\pi e 
\int f_{\vec{k} \omega}(\vec{p}) d^3 p,
\label{vl2}\\
&& i \vec{k} \cdot \vec{\cal B}_{\vec{k} \omega} =0,
\label{vl3}\\
&& \vec{\cal B}_{\vec{k} \omega} = \frac{1}{\omega} \vec{k}\times 
\vec{E}_{\vec{k} \omega},
\label{vl4}\\
&&i \omega \bigl( 1 - \frac{k^2}{\omega^2}\bigr) \vec{\cal E}_{\vec{k}
\omega}   + \frac{i}{\omega} \vec{k} (\vec{k}\cdot \vec{\cal E}_{\vec{k}
\omega})  = 4\pi e \int d^3 p ~\vec{v} ~f_{\vec{k} \omega}(\vec{p}),
\label{vl5}
\end{eqnarray}
where $g_{\vec{k}}(\vec{p})$ is the initial profile of the distribution
function and where $\tilde{\nu}$ is the typical collision frequency. 
The subscripts $\vec{k}\omega$ denote the Fourier transform with respect to space 
and the Laplace transform with respect to time of the corresponding 
quantity.

After enforcing the Gauss constraint at the initial 
time, the  electric field can be separated in its polarizations parallel
and transverse to the  direction of propagation of the fluctuation. 
The transverse current provides a source for the evolution 
of transverse electric field fluctuations 
\begin{equation}
i \omega \bigl( 1 - \frac{k^2}{\omega^2} \bigr)  \vec{{\cal E}}_{\vec{k}
\omega}^{\perp} = 4 \pi e \int d^3 p f_{\vec{k} \omega}(\vec{p})
\vec{v}_{\perp},
\label{perp}
\end{equation}
whereas the charge fluctuations provide a source for 
 the evolution of longitudinal  electric field fluctuations
\begin{equation}
i \vec{k} \cdot \vec{\cal E}_{\vec{k} \omega}^{\parallel} =4\pi e \int d^3 p 
f_{\vec{k} \omega} (\vec{p}). 
\label{parallel}
\end{equation}
In Eqs. (\ref{perp}) and (\ref{parallel})  the 
longitudinal and transverse part of the electric field fluctuations 
have been defined:
\begin{equation}
\vec{\cal E}_{\vec{k} \omega}^{\perp} = \vec{\cal E}_{\vec{k} \omega}^{\perp}
-  \frac{\vec{k}}{|\vec{k}|^2}  (\vec{k} \cdot \vec{{\cal E}}_{\vec{k}
\omega}),\,\,\,\,\,\, \vec{\cal E}_{\vec{k} \omega}^{\parallel} =
\frac{\vec{k}}{|\vec{k}|^2}  (\vec{\cal E}_{\vec{k} \omega} \cdot
\vec{k}).
\end{equation}
The solution of  Eq. (\ref{vl1}) is given by  
\begin{equation}
f_{\vec{k} \omega}(\vec{p}) =  \frac{1}{i( \vec{k} \cdot \vec{v} -
\omega -i\tilde{\nu})} \bigl[ g_{\vec{k}}(\vec{p}) - 2 e  \vec{v}\cdot
\vec{{\cal E}}_{\vec{k} \omega} \frac{\partial f_0}{\partial p} \bigr],
\label{f}
\end{equation}
where we used that $\partial f_0/\partial \vec{p} \equiv 
\vec{v} \partial f_0/\partial p$. 
The longitudinal  and transverse components of the electric
fluctuations can be obtained  by inserting Eq. (\ref{f}) into Eqs.
(\ref{perp})-(\ref{parallel})
\begin{eqnarray}
&& |\vec{\cal E}_{\vec{k} \omega}^{\parallel}| = \frac{e}{ 
i k \,\,\, \epsilon_{\parallel} } \int d^3 p 
\frac{g_{\vec{k}}(\vec{p})}{i (\vec{k}\cdot \vec{v} - \omega -i\tilde{\nu})},
\label{par2}\\
&& \vec{\cal E}_{\vec{k} \omega}^{\perp} = 
\frac{e \,\,\,\omega}{\omega^2 \epsilon_{\perp} - k^2} 
\int d^3 p \vec{v}^{\perp} 
 \frac{ g_{\vec{k}}(\vec{p})}{(\vec{k}\cdot \vec{v} - \omega-i\tilde{\nu})}, 
\label{perp2}
\end{eqnarray}
where $\epsilon_{\parallel}$ and $\epsilon_{\perp}$ are,
respectively,  the longitudinal and transverse part of the
polarization tensor
\begin{eqnarray}
&&\epsilon_{\parallel}(k, \omega) = 1 - \frac{ 2 e^2 }{k^2} \int d^3 p 
\frac{\vec{k}\cdot \vec{v}}{ (\vec{\vec{k}} \cdot \vec{v} - \omega - i \tilde{\nu})}
 \frac{\partial f_0}{\partial p},
\label{polpar}\\
&& \epsilon_{\perp} (k, \omega) = 1 - \frac{e^2}{ \omega} \int d^3 p 
\frac{ \vec{v}_{\perp}^2 }{(\vec{k}\cdot \vec{v} - \omega - i \tilde{\nu})} 
\frac{\partial 
f_0}{\partial p}.
\label{polperp}
\end{eqnarray}
Now, the general expression for the generated magnetic field is 
\begin{equation}
\vec{\cal B}_{\vec{k} \omega} = \frac{ e }{ \omega^2 
\epsilon_{\perp}(k,\omega) - k^2} 
\int d^3 p [\vec{v} \times \vec{k}] \frac{g_{\vec{k}}(\vec{p}) }{(\vec{k} 
\cdot \vec{v} - \omega -i\tilde{\nu})}.
\label{Bg}
\end{equation}

The space-time evolution of the magnetic fluctuations can be
determined by performing the inverse Laplace and  Fourier transforms:
\begin{equation}
\vec{{\cal B}}(\vec{x},\eta) = 
\int e^{- i \omega\eta}\frac{e~d\omega }{\omega^2 
\epsilon_{\perp}(k,\omega) - k^2 } \int d^3 k 
e^{ i \vec{k}\cdot \vec{x} } [\vec{v}\times \vec{k}]\int d^3 p 
\frac{ g_{\vec{k}}(\vec{p})}{ (\vec{v}\cdot \vec{k} - \omega -i\tilde{\nu})}.
\label{bf}
\end{equation}

In order to perform this integral, the explicit relations for the
polarization tensors should be given. They depend on the equilibrium
distribution function $f_0(p)$, 
which we take to be the expression given in Eq. (\ref{dist0})\footnote{ 
Notice that most of  our considerations  can be easily
extended to the case of a Bose-Einstein or Fermi-Dirac  distribution.
What is important, in our context, is the analytical  structure of
the polarization tensors and this is the same  for different
distributions \cite{vlcur1}.} 

Then we have for transverse polarization
\begin{equation}
\epsilon_{\perp}(k, \omega) = 1 +   \frac{e^2 \,\,
n_q}{2 \omega k T} \biggl\{  \bigl[ 1 - \frac{(\omega+i\tilde{\nu})^2}{k^2} \bigr]
\ln{\frac{k - \omega-i\tilde{\nu}}{k + \omega + i\tilde{\nu}}} - 2 \frac{\omega+i\tilde{\nu}}{k}\biggr\},
\label{perpapp}
\end{equation}
and for the longitudinal polarization:
\begin{equation}
\epsilon_{\parallel}(k, \omega) = 1 +  
\frac{e^2 \,\, n_q}{ k^2 T} \biggl\{ 
 2 +\frac{\omega+i\tilde{\nu}}{k} \ln{\frac{k - \omega-i\tilde{\nu}}{k + \omega +i\tilde{\nu}}}
\biggr\}.
\label{parallapp}
\end{equation}

Consider now the case of very small momenta $k\ll \omega$ and $\omega
\ll \tilde{\nu}$, relevant for long-ranged magnetic fields. Then the
computation of the integral (\ref{bf}) in the large time limit and with
the use of explicit form of the transverse polarization tensor in
(\ref{perpapp}) gives \footnote{For small $k$,  the equation  $\omega^2
\epsilon_{\perp}(k,\omega) - k^2=0$ defining the poles of the 
inverse Laplace transform implies $\omega \sim i k^2/\sigma$.}: 
\begin{equation}
\vcB(\vec{x},\eta) \simeq \frac{T}{4\pi\alpha_{\rm em} n_q} \exp{(-k^2 \eta/\sigma)}
\vec{\nabla} \times \vec{{\cal J}},
\label{BF}
\end{equation}
where $\sigma$ is the plasma conductivity in the relaxation time
approximation,
\begin{equation}
\sigma = \frac{4\pi e^2 n_q}{\tilde{\nu} T} 
\end{equation}
and initial electric current is given by
\begin{equation}
\vec{{\cal J}}(\vec{x}) = \int d^3 p \,\vec{v}\, g_{\vec{k}}(\vec{p}) ~.
\end{equation}
 
 The obtained  results
assumed that the linearization of the Vlasov equation is consistent 
with the physical assumptions of our problem. This is indeed the
case.  In order to safely linearize the Vlasov equation we have to 
make sure that the perturbed distribution function of the charge 
fluctuations is always smaller than the first order of the 
perturbative expansion (given by the distribution of Eq.
\ref{dist0}).  In other words we have to make sure that
\begin{equation}
|\delta f_{+}(\vec{x},\vec{p},\eta)| \ll f_{0}(\vec{p}),\,\,\,
|\delta f_{-}(\vec{x},\vec{p},\eta)| \ll f_{0}(\vec{p}).
\end{equation}
These conditions imply that
\begin{equation}
\frac{e \vcE_{\vec{k} \omega}}{|\vec{k}\cdot \vec{v} - \omega-i\tilde{\nu}|}\cdot
\frac{\partial f_0(\vec{p})}{\partial \vec{p}} \ll f_0(\vec{p}).
\label{cond}
\end{equation}
If  the relativistic plasma frequency is defined,
\begin{equation}
\omega^2_{p} = \frac{4\pi \,e^2\, n_q}{3 \,T}, 
\end{equation}
the condition expressed by Eq. (\ref{cond})  can
be restated, for modes $ k \leq \omega_{p}$, as  $|
\vec{{\cal E}}_{\vec{k},\omega}|^2 < n_q T$ (where we essentially took the
square modulus of Eq. (\ref{cond})). This last inequality expresses
the fact that the energy  density associated with the gauge field
fluctuations  should always be smaller than the critical energy
density  stored in radiation. The linear treatment of the Vlasov
equation is certainly  accurate provided the typical modes of the the
field are  smaller than the plasma frequency and provided  the energy
density in electric and magnetic fields is smaller  than $T^4$, i.e.
the energy density stored in the radiation  background.

\subsection{Inverse cascades}

A relevant topic in the evolution of large scale 
magnetic fields concerns the concepts of {\em direct}
and {\em inverse} cascades. A direct cascade is a process, occurring in a plasma, 
where energy is transferred from large to small length scales. An inverse cascade is a process where 
the energy transfer goes from small scales to large length scales. One can also generalize the 
the concept of energy cascade to the cascade of any conserved quantity in the plasma, like, for instance, 
the helicity. Thus, in general terms, the transfer process of a conserved 
quantity is a cascade.

The concept of cascade (either direct or inverse) is related with the concept 
of turbulence, i.e. the class of phenomena taking place in fluids and plasmas
at high Reynolds numbers, as anticipated in Section 4.
Non-magnetized fluids become turbulent at high {\em kinetic} 
Reynolds numbers. Similarly, magnetized fluids
 should become turbulent for sufficiently large {\em magnetic }
Reynolds numbers. The experimental evidence of this occurrence is still 
poor. One of the reasons is
that it is very difficult to reach, with terrestrial plasmas, the physical situation
where the magnetic  and the kinetic Reynolds numbers are both large 
but, in such a way that their ratio is also large i.e. 
\begin{equation}
{\rm R}_{\rm m} \gg 1, ~~~~~{\rm R}\gg1, ~~~~~~ {\rm Pr} = \frac{{\rm R}_{\rm m}}{{\rm R}} \gg 1.
\label{ratio}
\end{equation}
The physical regime expressed through Eqs. (\ref{ratio}) rather common 
in the early Universe. Thus, MHD turbulence is probably one of the key aspects 
of magnetized plasma dynamics at very high temperatures and densities.
Consider, for instance, the plasma at the electroweak 
epoch when the temperature was of the order of $100$ GeV. One can compute the Reynolds 
numbers and the Prandtl number from their definitions given in Eqs. (\ref{magnrey})--(\ref{Pr}).
In particular, 
\begin{equation}
{\rm R}_{\rm m} \sim 10^{17}, ~~~~~~~~~~{\rm  R} = 10^{11}, ~~~~~~{\rm  Pr} \simeq 10^{6},
\end{equation}
which can be obtained from  Eqs. (\ref{magnrey})--(\ref{Pr}) using as fiducial 
parameters $ v \simeq 0.1$, $\sigma T/\alpha$, $\nu\simeq ( \alpha T)^{-1}$ and $L \simeq 0.01 ~H_{\rm ew}^{-1} 
\simeq 0.03~{\rm cm}$ for $T \simeq 100 ~{\rm GeV}$.

At high Reynolds and Prandtl numbers the evolution of the large-scale magnetic fields 
is not  obvious since the inertial regime is  large. There 
are then two scales separated by a huge gap: the scale of the horizon ($3$ cm at the electroweak epoch) and 
the diffusivity scale (about $10^{-8}$ cm). One would like to know, given an 
initial spectrum of the magnetic fields, what is the evolution of the spectrum, i.e. in what 
modes the energy has been transferred. The problem is that it  is difficult to simulate systems with 
huge hierarchies of scales.
 
Suppose, for instance, that large scale 
magnetic fields are generated at some time $t_0$ and with a given spectral dependence. 
The energy spectrum of magnetic fields 
may receive, for instance, the dominant contribution for large comoving momentum $k$ (close 
to the cut-off provided by the magnetic diffusivity scale). The ``initial'' 
spectrum at the time $t_0$ is sometimes called injection spectrum. The problem 
will then be to know the how the spectrum is modified at a later time $t$.
If inverse cascade occurs, the magnetic energy density present in the 
ultra-violet modes may be transferred to the infra-red modes. In spite 
of the fact that inverse cascades are a powerful dynamical principle, it is 
debatable under which conditions they may arise. 

The possible occurrence  of  inverse cascades in a magnetized Universe 
has been put forward, in a series of papers,
by Brandenburg, Enqvist and Olesen \cite{enqvist2,enqvist3,olesen1,olesen2} (see also \cite{enqvist4} for 
an excellent review covering also in a concise way different mechanisms for the generation of large-scale
magnetic fields).

MHD simulations in $2+1$ dimensions  \cite{enqvist2,enqvist3} seem to 
support  the idea of an inverse 
cascade in a radiation dominated Universe.  
Two-dimensional MHD simulations are very interesting but also 
peculiar. In $2+1$ dimensions, the magnetic helicity does not
 exist while the magnetic flux is well defined 
and conserved in the ideal MHD limit. It is therefore 
impossible to implement $2+1$-dimensional MHD simulations
where the magnetic flux and the magnetic helicity are constrained to be 
constant in the ideal limit. This technical problem is related to a deeper 
physical difference between magnetized plasmas with and without helicity.
Consider then, for the moment, the situation where the helicity is strictly 
zero. It is indeed possible to  argue that, in this case, $2+1$-dimensional MHD 
simulations can capture some of the relevant features of $3+1$-dimensional MHD 
evolution.

In \cite{olesen1,olesen2} the scaling properties 
of MHD equations have been investigated in the force-free regime (i.e. in the approximation of negligible 
Lorentz force).
In the absence of helical components at the onset of the MHD evolution a rather 
simple  argument has been proposed by Olesen \cite{olesen1} in order to decide
under which conditions inverse cascade may arise. Consider the magnetic diffusivity and 
Navier-Stokes equations in the limit where $\vec{\cal J}\times \vcB  \simeq 0$. Then, the evolution 
equations are invariant under the following similarity transformations:
\begin{eqnarray}
&&\vec{x} \to L \vec{x},~~~ \eta \to L^{1 - \epsilon} \eta,~~~\vec{v}\to L^{\epsilon} \vec{v},~~~~
\nonumber\\
&& \vcB \to L^{\epsilon} \vcB,~~~~
\nu \to L^{1 +\epsilon} \nu,~~~\sigma\to  L^{-1-\epsilon} \sigma.
\label{ol0}
\end{eqnarray}
For simplicity it is also useful to concentrate the attention 
on the inertial range of momenta  where the dynamics 
is independent on the scale of dissipation.

In order to understand correctly the argument under discussion the 
volume average of the magnetic energy density has to be 
introduced (see \cite{everett} for a lucid discussion on the 
various possible averages of large-scale magnetic fields and see \cite{dim1} for a complementary 
point of view).
A generic rotationally and parity invariant two-point 
function for the magnetic inhomogeneities can be written as 
\begin{equation}
{\cal G}(\vec{r},\eta) =  \langle \vcB(\vec{x}, \eta) \vcB(\vec{y},\eta) \rangle, ~~~~~~~\vec{r} = \vec{x} - \vec{y}.
\label{G11}
\end{equation}
Defining now
\begin{equation}
{\cal G}(\vec{r},\eta) = \int d^{3} k {\cal G}(k,\eta) e^{-i \vec{k} \cdot \vec{r}}.
\label{G22}
\end{equation}
From Eqs. (\ref{G11})--(\ref{G22}) the following relation can be obtained
\begin{equation}
V\int d k {\cal Q}_{k}( \eta) = \int d^{3} x d^{3} y \langle \vcB(\vec{x}, \eta) \vcB(\vec{y},\eta) \rangle,
\label{defQ}
\end{equation}
where ${\cal Q}_{k}(\eta)$ is related to ${\cal G}(k,\eta)$ and $V$ is the averaging volume.

The idea is now  to study the scaling properties of ${\cal Q}_{k}(\eta)$ under the 
similarity transformations given in Eq. (\ref{ol0}).
Following \cite{olesen1}, it is possible to show that 
\begin{eqnarray}
a^4(\eta) {\cal Q}_{k}(\eta) &=& k^{\alpha} F(x)
\nonumber\\
                           x &=& k^{(3 + \alpha)/2} \eta.
\label{ol1}
\end{eqnarray}
with $\alpha = - 1 - 2 \epsilon$.
In Eq. (\ref{ol1})  ${\cal Q}_{k}(\eta)$ is related to the magnetic energy density and 
$F(x)$ is a function of the {\em single argument} $x =(k^{(3 + \alpha)/2} \eta)$. If, initially, the spectrum 
of the magnetic energy density is $k^{\alpha}$, then at later time the evolution will be dictated by $F(x)$.
Since $a^4(\eta) {\cal Q}_{k}(\eta)$ is approximately constant, because of flux conservation, then the 
comoving momentum will evolve, approximately, as 
\begin{equation}
k \simeq \eta^{-2/( 3 + \alpha)} \simeq t^{-1/( 3 + \alpha)},
\label{ol2}
\end{equation}
while the physical momentum will scale like 
\begin{equation}
\omega \simeq \eta^{-(5 +\alpha)/(\alpha + 3)} \sim t^{-(5 +\alpha)/(2\alpha + 6)}.
\label{ol3}
\end{equation} 
The conclusion of this argument is the following:
\begin{itemize}
\item{} if $\alpha < -3 $ the cascade will be direct (or forward);

\item{}  if $\alpha > -3 $ the comoving momentum moves toward the infra-red and the cascade
will be inverse;

\item{} if $\alpha =-3$, then, 
 $\epsilon = 1$ and, from Eqs. (\ref{ol0})-(\ref{ol1}), the momentum 
and time dependence will be uncorrelated.
\end{itemize}

The  Olesen scaling argument has interesting implications:
\begin{itemize}
\item{} if the initial spectrum 
of the magnetic energy density is concentrated at high momenta then, 
an inverse 
cascade is likely;
\item{}
from Eqs. (\ref{ol2}) and (\ref{ol3}) 
the typical correlation scale of the magnetic field 
may, in principle, evolve faster than in the case where 
the inverse cascade is not present. 
\end{itemize}

In a FRW Universe the correlation scale evolves as $\ell(\eta) = \ell(\eta_0) a(\eta)$ 
from a given initial time $\eta_0$ where magnetic energy density is injected with 
spectral dependence $k^{\alpha}$. If $\alpha >- 3$, Eqs. (\ref{ol2})--(\ref{ol3}) 
imply that 
\begin{equation}
\ell(\eta)  = \ell(\eta_0) a(\eta) \biggl(\frac{\eta}{\eta_0}\biggr)^{\frac{2}{\alpha + 3}} \simeq 
\ell(\eta_0)\biggl(\frac{\eta}{\eta_0}\biggr)^{\frac{\alpha + 5}{\alpha + 3}}
\label{corr1}
\end{equation}
where the second equality follows in a radiation dominated phase of expansion.
In flat space, the analog of 
 Eqs. (\ref{ol2})--(\ref{ol3}) would imply that $\ell(t) \sim \ell(t_0) (t/t_0)^{2/(\alpha + 3)}$.
Finally, in the parametrization of Olesen \cite{olesen1} the case of a Gaussian 
random field corresponds to $\alpha =2$, i.e. $\epsilon= -3/2$. 
In this case the correlation scale will evolve as 
\begin{equation}
\ell(\eta) = \ell(\eta_0) a(\eta) \biggl(\frac{\eta}{\eta_0}\biggr)^{2/5}.
\label{corr2}
\end{equation}
In the context of hydrodynamics the scaling law expressed through
Eq.(\ref{corr2}) is also well known \cite{saffman}. 
It is appropriate, at this point, to recall that 
Hogan \cite{hogan} was the first one to discuss the 
evolution of the correlation scale produced by a Gaussian random
injection spectrum (i.e. $\alpha=2$). Hogan was also the first one to 
think of the first order phase transitions as a possible source of large scale 
magnetic fields \cite{hogan}.

An argument similar to the one discussed in the case 
of MHD can be discussed in the case of ordinary hydrodynamics.
In this case, the analog of ${\cal Q}_{k}(\eta)$ will be 
related to the kinetic energy density whose spectrum, in the case 
of fully developed turbulence goes as $k^{-5/3}$, i.e. the usual 
Kolmogorov spectrum.

The  evolution of the correlation scale derived in Eq. (\ref{corr2}) 
can be obtained through another class of arguments 
usually employed in order to discuss inverse cascades in MHD: the mechanism 
of  selective decay \cite{bis} which can be triggered in the context 
of magnetic and hydrodynamical turbulence. The basic idea is that modes with 
large wave-numbers decay faster than modes whose wavenumber is 
comparatively small. Son \cite{son} applied these considerations 
to the evolution of large-scale magnetic fields at high temperatures 
and his results (when the magnetic field configuration has, initially, vanishing 
helicity) are in agreement with the ones obtained in \cite{olesen1,olesen2}.
In \cite{shiro} the topics discussed in \cite{olesen1,son} have been revisited and 
in \cite{berera} the possible occurrence of inverse cascades 
has been criticized using renormalization group techniques.

The main assumption of the arguments presented so far 
is that the evolution of the plasma occurs in an inertial regime.
As discussed above in this Section, at high temperatures the inertial
regime is rather wide both from the magnetic end thermal points of view.
However, in the case where the magnetic and thermal diffusivity scales
are {\em finite} the arguments relying on  the inertial regime 
cannot be, strictly speaking, applied. In this situation 
numerical simulations should be used in order to understand if 
the diffusive regime either precedes or follows the inverse cascade.

\begin{figure}[th]
\centerline{\epsfig{file=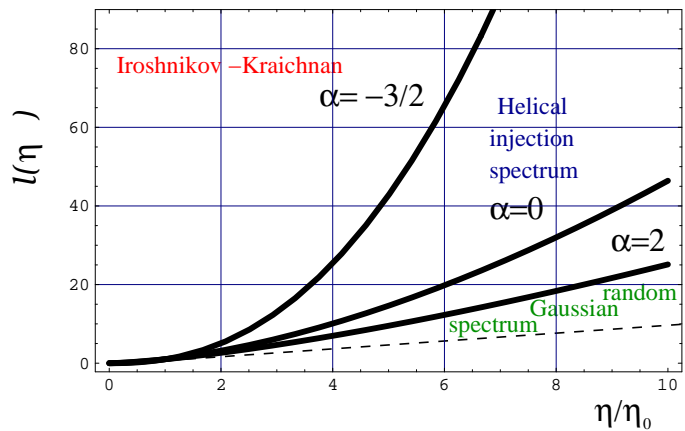,width=9cm}}
\vspace*{8pt}
\caption[a]{The time evolution of the correlation scale is illustrated 
in the case of a FRW Universe dominated by radiation and in units $\ell_0 =1$. The case $\alpha =2$ corresponds 
to the behaviour obtained in (\ref{corr2}). The case $\alpha = 2$, corresponding 
to Eq. (\ref{corr3}), refers to the situation of inverse cascade with an initial helical spectrum. For comparison 
the case $\alpha =-3/2$ is also reported. This last evolution takes place if 
the injection spectrum is of Iroshnikov-Kraichnan type.}
\label{F3}
\end{figure}
 
If the magnetic and thermal diffusivity scales are 
finite, two conceptually  different possibilities can be envisaged. One possibility is that 
the inverse 
cascade terminates with a diffusive regime. The other possibility  is that the onset of diffusion 
prevents the occurrence of the  inverse cascade. This dilemma
 can be solved by  numerical simulations 
along the lines of the ones already mentioned \cite{enqvist2,enqvist3}.
The numerical simulations presented in \cite{enqvist2,enqvist3} 
are in $(2+1)$ dimensions. The results support the occurrence of inverse 
cascade in the presence of large kinetic viscosity. The moment at which the 
cascade stops depends on the specific value of the viscosity coefficient.
 In \cite{hindmarsch1,hindmarsch2} 
$(3+1)$-dimensional MHD simulations have shown  evidence of inverse cascade.
Another possible limitation is that in the early Universe the magnetic and kinetic 
Reynolds numbers are both large and also their ratio, the Prandtl number, is very large.
This feature seems to be difficult to include in simulations.

The considerations presented so far do not take into account the possibility
that the initial magnetic field configuration has non vanishing helicity.
If magnetic helicity is present the typical features of MHD turbulence 
can qualitatively change \cite{bis} and it was argued that the occurrence 
of inverse cascades may be even more likely \cite{pouquet1,pouquet2} (see also \cite{cornwall}).
The conservation of the magnetic flux is now supplemented, in the magnetic 
sector of the evolution, by the conservation of the helicity obtained in Eqs. 
(\ref{h1}) and (\ref{h2}). The conservation of the helicity is extremely important 
in order to derive the MHD analog of Kolmogorov spectrum. In fact, the heuristic derivation 
of the Kolmogorov spectrum is based on the conservation of the kinetic energy density.
If the conservation of the magnetic helicity is postulated, the same 
qualitative argument leading to the Kolmogorov spectrum will lead in the case of 
fully developed MHD turbulence to the so-called Iroshnikov-Kraichnan spectrum 
\cite{bis,iro1,iro2} whose specific form, is, in terms of the quantity introduced in Eq. 
(\ref{ol1}) ${\cal Q}_{k}(\eta_0)\sim k^{-3/2}$.

Suppose, for simplicity, that the injection spectrum has a non-vanishing helicity and suppose 
that the helicity has the same sign everywhere. In this case $\vec{\cal A}\cdot\vcB \sim \ell |\vcB|^2$.
Since the helicity is conserved the magnetic field will scale according to Eqs. (\ref{ol0}) but 
with $\epsilon = -1/2$ corresponding  to $\alpha =0$. Hence, the 
correlation scale will evolve, in this case, as
\begin{equation}
\ell(\eta)  = \ell(\eta_0) a(\eta) \biggl(\frac{\eta}{\eta_0}\biggr)^{\frac{2}{3}} \simeq 
\ell(\eta_0)\biggl(\frac{\eta}{\eta_0}\biggr)^{\frac{5}{3}},
\label{corr3}
\end{equation}
i.e. faster than in the case of Eq. (\ref{corr2}).
It is difficult to draw general conclusions without specifying 
the nature and the features of the mechanism producing the injection
configuration of the magnetic field. The results 
critically summarized in this Section will be important 
when discussing specific ideas on the origin of large-scale magnetic fields.
In Fig. \ref{F3} the cases of different injection spectra are compared.
It is clear that an injection spectrum 
given by a Gaussian random field leads to a rather inefficient inverse cascade.
Since the magnetic helicity vanishes at the onset of the evolution, the rate
of energy transfer from small to large scales is rather slow. On the contrary 
if the injection spectrum is helical, the rate increases significantly.
For comparison the situation where 
the injection spectrum corresponds to the Iroshnikov-Kraichnan case, i.e. the case 
of fully developed MHD turbulence, is also reported. The dashed line of Fig. \ref{F3} 
illustrates the evolution of the correlation scale dictated by the the Universe expansion.

\subsection{Hypermagnetic fields}
At small temperatures and small densities of the  different 
fermionic charges the $SU_{L}(2) \otimes U_{Y}(1)$  is 
broken down to the $U_{\rm em}(1)$ and the long range fields which can 
survive in the plasma are the ordinary magnetic fields. 
For sufficiently high temperatures 
(and for sufficiently high values of the various fermionic charges) the 
$SU_{L}(2) \otimes U_{Y}(1)$ is restored and
non-screened vector modes correspond to hypermagnetic fields. 
In fact, Abelian electric fields decay within a typical time  scale 
$1/\sigma$ where $\sigma$ is the conductivity.
The long-ranged non-Abelian magnetic fields (corresponding, for instance,
to the color $SU(3)$  or to the weak $SU(2)$) cannot exist 
because at high temperatures the non-Abelian interactions 
induce a ``magnetic'' mass gap $\sim g^2 T$ where $g$ is the gauge coupling 
constant.    
Also the non-Abelian electric fields decay because of the finite value
of the conductivity as it occurs for Abelian electric fields.
Therefore, the only long scale field that  can
exist in the plasma for enough time must be associated with some
Abelian U(1) group. This statement, valid to all orders in
perturbation theory, has been confirmed non-perturbatively for the
electroweak theory by recent lattice studies in \cite{ms1}.
Under normal conditions (i.e.
small temperatures and small densities of the different fermionic
charges) the SU(2)$\times$U(1)$_Y$ symmetry is ``broken" down to
U(1)$_{EM}$, the massless field corresponding to U(1)$_{EM}$ is the
ordinary photon and the only long-lived field  in the
plasma is the ordinary magnetic one. At sufficiently high
temperatures, $T > T_c$, the SU(2)$\times$U(1)$_Y$ symmetry is
``restored", and non-screened vector modes $Y_\mu$ correspond to the
U(1)$_Y$ hypercharge group. Hence, if primordial fields existed at $T >
T_c$, they did correspond to hypercharge rather than to U(1)$_{EM}$. 

At the electroweak epoch the typical size of the Hubble radius is of the 
order of $3$ cm . The typical diffusion scale is of the order of $10^{-9}$ cm.
Therefore, over roughly eight orders of magnitude hypermagnetic fields can 
be present in the plasma without being dissipated \cite{mg3}. 

\subsubsection{Anomalous MHD equations}

The evolution of hypermagnetic fields can be obtained from the  anomalous 
magnetohydrodynamical (AMHD) 
equations. The AMHD equations generalize the treatment 
of plasma effects involving hypermagnetic fields to the case 
of finite fermionic density\cite{mg4}. 

There are
essential differences between the interactions of magnetic fields and
the ones of hypermagnetic fields with matter. 
The ordinary electromagnetic field
has a vector-like coupling to the fermions, while the coupling of the
hypercharge fields is chiral. Thus, if hyperelectric ($\vec{
E}_{Y}$) and hypermagnetic ($\vec{H}_{Y}$) fields are present
simultaneously, they cause a variation of the fermionic number
according to the anomaly equation, $\partial_\mu j_\mu \sim
\frac{g'^2}{4\pi^2} \vec{H}_{Y}\cdot \vec{ E}_{Y}$ (here
$g'$ the hypercharge gauge coupling constant). Now, the presence of
{\em non-homogeneous} hypermagnetic fields in the EW plasma with 
(hyper)conductivity $\sigma_c$ always implies the existence of a related
electric field, $\vec{E}_{Y}\sim \frac{1}{\sigma_c} \vec{\nabla}
\times \vec{ H}_{Y}$. Since for a general stochastic magnetic
background $\langle(\vec{H}_{Y}\cdot \vec{\nabla}\times
\vec{H}_{Y})^2\rangle \neq 0$, the non-uniform hypermagnetic
field may absorb or release fermions and produce, ultimately, 
baryon and lepton density perturbations because of the anomaly equation. 

The behaviour of {\em cold} fermionic matter with non-zero anomalous Abelian
charges was considered in \cite{rub1} where it  was pointed out  that the
anomalous fermionic matter is unstable against the creation of 
Abelian gauge field
with non-zero Chern-Simons number, which eats up fermions because of
the anomaly. The right electron number
density may be converted to the hypercharge field because of a
similar effect. Also the opposite effect is possible: hypercharge
fields may be converted into fermions in a {\em hot} environment.

The electroweak plasma in {\em complete} thermal equilibrium at a temperature
$T$ can be characterized by $n_f$ chemical potentials
$\mu_i,~i=1,..., n_f$ corresponding to the exactly conserved global charges  
\begin{equation}
N_{i} = L_{i} - \frac{B}{n_{f}}
\label{2.11}
\end{equation}
($L_{i}$ is the lepton number of the $i$-th generation, $B$  is the baryon
number, and $n_{f}$ is the number of fermionic generations). One should
also introduce a chemical potential $\mu_Y$ corresponding to weak
hypercharge, but its value is fixed from the requirement of the
hypercharge neutrality of the plasma, $\langle Y \rangle =0$.

It is interesting to study this plasma slightly out of thermal equilibrium, 
for instance in the situation  where a  non-uniform distribution of the
hypermagnetic field is present. Because of the  anomaly,
 this field is coupled to the fermionic number densities. In
principle, different chemical potentials can be assigned to all the 
fermionic degrees of freedom of the electroweak theory ($45$ if
$n_{f}=3$) and the coupled system of Boltzmann-type 
equations for these chemical potentials and the hypercharge fields may be
written. Since we are interested in the slow processes in the
plasma, this is not necessary. If the coupling, corresponding to some
slow process, is switched off, then the  electroweak theory acquires an
extra conserved charge and  a further  chemical potential should be added
to the system  given in Eq.  (\ref{2.11}). 

An interesting observation (which turns out to be quite important in our
 context) has been made in
\cite{campbell,ibanez,rummukainen}, where it was noticed that perturbative reactions with
right-handed electron chirality flip are out of thermal equilibrium at
temperatures higher than some temperature $T_R$.\footnote{ This
temperature depends on the particle physics model, see also the 
 discussion reported in Section 5. In the MSM $T_R \simeq 80$ TeV 
\cite{campbell,ibanez,rummukainen}.} 
Thus, the number of right electrons  is
perturbatively conserved at temperatures $T>T_R$ 
and the  chemical potential $\mu_R$ can be introduced for
it. On the other hand, this charge  is not conserved 
because of the  Abelian anomaly,
\begin{equation}
\partial_\mu j^\mu_R = -\frac{g'^2 y_R^2}{64 \pi^2} {\cal Y}_{\mu\nu}
\tilde{{\cal Y}}^{\mu \nu},
\end{equation}
and it is therefore coupled  to the hypermagnetic field.
Here  ${\cal Y}$ and $\tilde{{\cal Y}}$ are, respectively, the $U_Y(1)$ 
hypercharge field strengths and their duals, $g'$ is the associated gauge
coupling and $y_R=-2$ is the hypercharge of the right electron. 

Now we are ready to derive the anomalous MHD equations in flat space \cite{mg3,mg4}.
The effective Lagrangian density describing
the dynamics of the gauge fields at finite fermionic density is \cite{redlich}:
\begin{equation}
{\cal L}_{Y, e_{R}} = -\frac{1}{4}\sqrt{-g}
Y_{\alpha\beta}Y^{\alpha\beta} - \sqrt{-g} J_{\alpha}Y^{\alpha} 
+ \mu \epsilon_{ijk} Y^{ij} Y^{k},~~~\mu = \frac{g'^2}{4\pi^2}\mu_{R}
\label{2.15}
\end{equation}
($g$ is the determinant of the metric defined in (\ref{metric});
$Y_{\alpha\beta} = \nabla_{[\alpha}Y_{\beta]}$; 
$\nabla_{\alpha}$ is the covariant
derivative with respect to the metric (\ref{metric})[notice that in the
metric (\ref{metric}) 
$\nabla_{[\alpha}Y_{\beta]} = \partial_{[\alpha}Y_{\beta]}$]; $g'$ is the
Abelian coupling constant).
The first and the last terms in Eq. (\ref{2.15}) are nothing but the
curved space generalization of the flat-space effective Lagrangian for
the hypercharge fields at finite fermion density \cite{mg3,mg4},
$J_{\alpha}$ is the ohmic current.
The equations of motion for the
hyperelectric and hypermagnetic fields are then
\begin{eqnarray}
\frac{\partial\vcH_{Y}}{\partial\eta} = -\vec{\nabla}
\times {\vcE}_{Y}
,~~~~~
\frac{\partial\vcE_{Y}}{\partial\eta}+ {\vcJ}_{Y} =
\frac{g'^2}{\pi^2} \mu_{R} a {\vcH}_{Y}+
{\vec{\nabla}}\times \vcH_{Y}, 
\nonumber\\
{\vec{\nabla}}\cdot\vcH_{Y}=0,~~~~~~~~ ~~~~~~~~~~~~
{\vec{\nabla}}\cdot \vcE_{Y}=0,
\nonumber\\
{\vec{\nabla}}\cdot \vcJ_{Y}=0,~~~ ~~~~
\vcJ_{Y}=\sigma (\vcE_{Y} +
\vec{v}\times\vcH_{Y}),~~\sigma=\sigma_{c} a(\eta),
\label{2.16}
\end{eqnarray}
with the same notations introduced  in the case of the conventional MHD equations.

To Eqs. (\ref{2.16}), the evolution equation of the right electron chemical
potential, accounting for the anomalous and perturbative 
non-conservation of the right electron number density ($n_R$), must be added:
\begin{equation}
 \frac{\partial n_{R}}{\partial t} = - \frac{g'^2
}{4\pi^2} 
{\vec{{\cal E}}}_{Y}\cdot{\vec{{\cal H}}}_{Y} - \Gamma (n_{R}-n_R^{eq}),
\label{nr}
\end{equation}
where $\Gamma$ is the perturbative chirality-changing rate, 
$\Gamma = T\frac{T_R}{M_0}$, $n_R^{eq}$ is the equilibrium value of
the right electron number density,
and the term proportional to ${\vec{E}}_{Y}\cdot{\vec{H}}_{Y}$ is 
the  right electron anomaly contribution. 

Finally, the relationship between the right electron number density
and the chemical potential must  be specified. This relation depends
upon  the particle content of the theory. In the case of the Minimal 
Standard Models (MSM) the evolution equation of the chemical potential becomes 
\cite{mg4}
 \begin{equation}
\frac{1}{a} \frac{\partial (\mu_{R} a)}{\partial \eta} = - \frac{g'^2
}{4\pi^2} \frac{783 }{88} \frac{ 1}{a^3 T^3}
{\vcE}_{Y}\cdot{\vcH}_{Y} - \Gamma (\mu_{R} a).
\label{mur}
\end{equation}

At finite hyperconductivity (in what
we would call, in a MHD context, resistive approximation)
we have that from Eq. (\ref{2.16}) the induced
hyperelectric field is not exactly orthogonal to the hypermagnetic one and,
moreover, an extra ``fermionic'' current comes in the game thanks to
the fact that we are working at finite chemical potential. 
Therefore in our context the resistive Ohm law can be written as
\begin{eqnarray}
&& {\vcE}_{Y} = \frac{{\vec{{\cal J}}}_{Y}}{\sigma}
-{\vec{v}}\times \vcH_{Y}
 \simeq 
\nonumber\\
&& \frac{1}{\sigma}\left( 
\frac{4 \alpha'}{\pi} \mu_{R} a\vcH_{Y}+
{\vec{\nabla}}\times \vcH_{Y}\right)
-{\vec{v}}\times \vcH_{Y}, ~~~\alpha' =\frac{g'^2}{4\pi}~.
\label{ohm}
\end{eqnarray}

In the bracket appearing in 
 Eq. (\ref{ohm}) we can identify two different contributions. One is
associated with the curl of the magnetic field. We will call this the MHD
contribution, since it appears in the same way in ordinary plasmas.
The other contribution contains the chemical potential and it is
directly proportional to the magnetic field and to the chemical
potential. This is a typical finite density effect. In fact the extra
Ohmic current simply describes the possibility that the energy sitting
in real fermionic degrees of freedom  can be transferred to the 
 hypermagnetic field. 
It may be of some interest to
notice the analogy between the first term of Eq. (\ref{ohm}) and the
typical form of the ohmic current discussed in Eq. (\ref{genohm1})
appearing in the context of the dynamo mechanism. In the latter case the
presence of a current (proportional to the vorticity through the
$\alpha$ dynamo term)  was indicating that large length scales 
magnetic fields could grow by eating up fluid vortices. 
By inserting ${\vec{E}}_{Y}$ obtained from the generalized Ohm law
(\ref{ohm}) in the evolution equations (\ref{2.16}) of the
hypercharge fields, we obtain the generalized form of the magnetic 
diffusivity equation:
\begin{equation}
\frac{\partial{\vcH_{Y}}}{\partial \eta} =- \frac{4 a
\alpha'}{\pi\sigma}  
\vec{\nabla}\times\left({\mu_{R} \vcH}_{Y}\right) 
+{\vec{\nabla}}\times(\vec{v}\times{\vcH})
+ \frac{1}{\sigma}
 \nabla^2 {\vcH}_{Y}.
\label{hyperdiffusivity}
\end{equation}
In order to be consistent with our resistive approach
 we neglected terms
containing time derivatives of the electric field, which are 
sub-leading provided the conductivity is finite. In our considerations
 we will also make a further simplification, namely we will assume
 that the EW plasma is (globally) parity-invariant and that,
 therefore, no global vorticity is present. Therefore, since the
 length scale of variation of the bulk velocity field is much shorter
 than the correlation distance of the hypermagnetic field, the
 infrared modes of the hypercharge will be practically unaffected by
 the velocity of the plasma, which will be neglected when the large-scale part
 of the hypercharge is concerned.  This corresponds to the usual MHD
 treatment of a mirror symmetric plasma (see, e.g. Eq. 
(\ref{dynamored})--(\ref{vort}), when $\alpha =0$).

Eqs. (\ref{hyperdiffusivity}) and (\ref{mur}) form a set of
AMHD equations for the hypercharge magnetic field and right electron
chemical potential in the expanding Universe.

\subsubsection{Hypermagnetic knots}

The Abelian nature of the hypercharge field does not forbid  that 
the hypermagnetic flux lines should have a trivial topological 
structure. This situation is similar to what 
already encountered in the case of conventional 
electromagnetic fields with the important 
difference that the evolution equations of hypermagnetic fields 
are different from the ones of ordinary magnetic fields.
After a swift summary of the properties of
hypermagnetic knots (based on \cite{hypmg1,hypmg2}),
some interesting applications of these hypercharge 
profiles will be reviewed.

In
 the gauge $Y_0=0$, $\vec{\nabla}\cdot\vec{Y}=0$, 
an example of 
 topologically non-trivial configuration of the hypercharge field is 
the Chern-Simons wave \cite{cswave,hyp1,hyp2}
\begin{eqnarray}
Y_{x} (z, t) &=& Y(t) \sin{k_0 z},
\nonumber\\
Y_{y} (z, t) &=& Y(t) \cos{k_0 z}, 
\nonumber\\
Y_{z} (z,t) &=& 0.
\label{conf1}
\end{eqnarray}
This particular configuration is not homogeneous but it
 describes a hypermagnetic knot with {\em homogeneous} helicity and 
 Chern-Simons number  density
\begin{eqnarray}
&&\vec{ H}_{Y} \cdot \vec{\nabla} \times\vec{H}_{Y} = 
k_0 H^2(t),
\nonumber\\
&&n_{CS} = - \frac{g'^2}{32 \pi^2} \vec{ H}_{Y}\cdot \vec{Y} = 
\frac{ g'^2 }{32 \pi^2 k_0} H^2(t),
\end{eqnarray}
where $\vec{H}_{Y} 
=\vec{\nabla} \times \vec{Y}$, $
H(t) = k_0 Y(t)$; $g'$ is the $U(1)_{Y}$ coupling. 

It is  possible to construct hypermagnetic knot configurations 
with finite energy and helicity which are localized in space and within 
typical distance scale  $L_{s}$. 
Let us consider in fact the following configuration
in spherical coordinates \cite{hypmg2}
\begin{eqnarray}
Y_{r}({\cal R},\theta) &=& - \frac{2 B_0}{ \pi L_{s}}
\frac{\cos{\theta} }{\bigl[{\cal R}^2 +1\bigr]^2},
\nonumber\\
Y_{\theta}({\cal R},\theta) &=& \frac{2 B_0}
{ \pi L_{s}} \frac{ \sin{\theta}}{\bigl[ {\cal
R}^2 + 1\bigr]^2},
\nonumber\\
Y_{\phi}({\cal R},\theta) &=& - \frac{ 2 B_0}{ \pi L_{s}} \frac{ n
{\cal
R}\sin{\theta}}{\bigl[{\cal R}^2 + 1\bigr]^2},
\label{conf2}
\end{eqnarray}
where ${\cal R}= r/L_{s}$ is the rescaled radius and $B_{0}$ is some 
dimensionless amplitude and $n$ is just an integer number 
whose physical interpretation will become clear in a moment. 
The hypermagnetic field can be easily computed 
from the previous expression and it is 
\begin{eqnarray}
&&H_{r}({\cal R},\theta) = - \frac{4 B_{0}}{\pi~ L_{s}^2}\frac{n
\cos{\theta}}{\bigl[  {\cal
R}^2 + 1\bigr]^2},
\nonumber\\
&&H_{\theta}({\cal R}, \theta) = - \frac{4 B_{0}}{\pi~
L_{s}^2}\frac{{\cal R}^2 -1}{\bigl[
{\cal R}^2 + 1\bigr]^3}n \sin{\theta},
\nonumber\\
&&H_{\phi}({\cal R}, \theta) = 
- \frac{8 B_0}{ \pi~ L_{s}^2}\frac{ {\cal R}
\sin{\theta}}{\bigl[
{\cal R}^2 + 1\bigr]^3}.
\label{knot}
\end{eqnarray}
The poloidal and toroidal components of $\vec{{\cal H}}$ can be usefully 
expressed as $\vec{H}_{p} = 
H_{r} \vec{e}_{r} + H_{\theta} \vec{e}_{\theta} $ 
and $\vec{\cal H}_{t}= {\cal H}_{\phi} \vec{e}_{\phi}$.
The Chern-Simons number is finite and it is given by 
\begin{equation}
N_{CS} =\frac{g'^2}{32\pi^2}
\int_{V} \vec{{\cal Y}} \cdot \vec{H}_{Y} d^3 x=
\frac{g'^2}{32\pi^2} \int_{0}^{\infty}
\frac{ 8 n
B^2_0}{\pi^2} \frac{ {\cal R}^2 d {\cal R}}{\bigr[ {\cal R}^2 +
1\bigl]^4} = \frac{g'^2 n B^2_0}{32 \pi^2}.
\label{CS}
\end{equation}
The total helicity of the configuration can also be computed 
\begin{equation}
\int_{V} \vec{{\cal H}}_{Y}
 \cdot \vec{\nabla} \times \vec{{\cal H}}_{Y} d^3 x=
\frac{256~B^2_0~n}{\pi L^2 } \int_{0}^{\infty} \frac{ {\cal R}^2 d
{\cal R}}{(1 + {\cal R}^2)^5} = \frac{5 B^2_0 n}{L_s^2}.
\label{helic}
\end{equation}
We can compute also the total energy of the field
\begin{equation}
E = \frac{1}{2}\int_{V} d^3 x |\vec{{\cal H}}_{Y}|^2 = \frac{B^2_0}{2~L_{s}}
(n^2 + 1).
\end{equation}
and we discover that it is proportional to $n^2$.
 This means that one way of increasing the total energy of
 the field is to increase the number of knots and twists in the flux lines.

This type of configurations can be also obtained by projecting a 
non-Abelian SU(2) (vacuum) gauge field on a fixed electromagnetic 
direction \cite{JPI} \footnote{ In order to interpret these solutions it is 
very interesting to make use of the Clebsh decomposition. The 
implications of this  decomposition (beyond the hydrodynamical context, where 
it was originally discovered) have been recently discussed (see \cite{JPI2} 
and references therein).}. The resulting profile of the knot 
depends upon an arbitrary function of the radial distance.

These configurations have been also 
studied in \cite{ad1,ad2}.  In particular, in \cite{ad2}, the relaxation 
of HK has been investigated with a technique different from the one employed
 in \cite{hypmg1,hypmg2} but with similar results. The problem of scattering of 
fermions in the background of hypermagnetic fields has been also studied in 
\cite{ay1,ay2}.

Hypermagnetic knots with large correlation scale 
can be also generated dynamically provided an unknotted 
hypermagnetic background is 
already present.

Let us assume that  dynamical pseudoscalar
particles are  evolving in the background geometry given by Eq. (\ref{metric}). 
The pseudoscalars are {\em not} a source of the background (i.e. they do
not affect the time evolution of the scale factor) but, nonetheless,
they evolve according to their specific dynamics and can excite 
other degrees of freedom.

The  action describing the interaction of a dynamical 
pseudoscalar with hypercharge fields can be written as
\begin{equation}
S= \int d^{4} x \sqrt{-g} \biggl[ \frac{1}{2}g^{\alpha\beta}
\partial_{\alpha}\psi \partial_{\beta}\psi  - V(\psi) -
\frac{1}{4}Y_{\alpha\beta}Y^{\alpha\beta} + 
c\frac{\psi}{4 M}
Y_{\alpha\beta}\tilde{Y}^{\alpha\beta}\biggr].
\label{action}
\end{equation}
This action is quite generic. In the case $V(\psi) = (m^2/2) \psi^2$
Eq. (\ref{action}) is  nothing but the curved 
space generalization of the model usually employed in direct searches
of axionic particles \cite{sikivie,maiani,gasperini}. The constant in front of the anomaly 
is a model-dependent factor. For example, in the case of axionic particles
, for large temperatures $T \geq m_{W}$,
 the Abelian gauge fields present in Eq. (\ref{action}) will be 
hypercharge fields and $c = c_{\psi Y} \alpha'/(2 \pi)$ where $\alpha'=
g'^2/4\pi$ and $c_{\psi Y}$ is a numerical factor of order $1$ 
which can be computed (in a specific axion scenario) by knowing the
 Peccei-Quinn charges of all the fermions present in the model \cite{kim,cheng}. 
For small temperatures 
 $T\leq m_{W}$ we have that the Abelian fields present in the 
action (\ref{action}) will coincide with ordinary  electromagnetic 
fields and
$c = c_{\psi\gamma} \alpha_{{\rm em}}/2\pi$ where $\alpha_{{\rm em}}$ 
is the fine structure constant and $c_{\psi\gamma}$ 
is again a numerical factor.

The coupled system of equations describing the evolution of the 
pseudoscalars 
and of the Abelian gauge fields can be easily derived by varying the action 
with respect to $\psi$ and $Y_{\mu}$,
\begin{eqnarray}
&& \frac{1}{\sqrt{- g}} \partial_{\mu}\biggl[ \sqrt{-g} g^{\mu\nu} \partial_{\nu} \psi \biggr] 
+ \frac{\partial V}{\partial\psi}= \frac{c}{4 M} Y_{\alpha\beta} \tilde{Y}^{\alpha\beta},
\nonumber\\
&&\nabla_{\mu}  Y^{\mu\nu} = 
\frac{c}{M}\nabla_{\mu}\psi\tilde{ Y}^{\mu\nu},~~~
\nabla_{\mu} \tilde{ Y}^{\mu\nu}=0,
\label{eqmotion}
\end{eqnarray}
where,
\begin{equation}
\nabla_{\mu} Y^{\mu\nu} = \frac{1}{\sqrt{-g}}
 \partial_{\mu}\biggl[ \sqrt{-g} Y^{\mu\nu}\biggr],~~~\nabla_{\mu}
 \tilde{Y}^{\mu\nu} = \frac{1}{\sqrt{-g}}\partial_{\mu} \biggr[
 \sqrt{-g} \tilde{Y}^{\mu\nu}\biggr],
\end{equation}
are the usual covariant derivatives defined from the background FRW metric
Eqs. (\ref{eqmotion}) can be written in terms of the physical gauge  fields
\begin{eqnarray}
&&\psi'' + 2 {\cal H} \psi' - \nabla^2\psi 
+ a^2 \frac{\partial V}{\partial \psi} 
= - \frac{1}{a^2}\frac{c}{M} \vcE_{Y}\cdot \vcH_{Y},
\nonumber\\
&&\vec{\nabla}\cdot\vcH_{Y} =0,
~~~\vec{\nabla}\times\vcE_{Y} +\vcH_{Y}' =0,~~~
\vec{\nabla}\cdot\vcE_{Y} = \frac{c}{M}\vec{\nabla}\psi \cdot\vcH_{Y},
\nonumber\\
&&\vec{\nabla}\times\vcH_{Y} = \vcE_{Y}' 
- \frac{c}{M} \biggl[ \psi' \vcH_{Y} 
+ \vec{\nabla}\psi\times \vcE_{Y}\biggr].
\label{system}
\end{eqnarray}

We want now to study the amplification of gauge field 
fluctuations induced by 
the time evolution of $\psi$. Then, the evolution equation for the 
hypermagnetic fluctuations $\vcH_{Y}$ can be obtained by
 linearizing  Eqs. (\ref{system}). We will assume that any background 
gauge field is absent.
 In the linearisation procedure we will also assume that the pseudoscalar 
field can be treated as completely homogeneous 
(i.e.$ |\vec{\nabla}\psi|\ll \psi'$). This seems to be natural if, 
prior to the radiation dominated epoch, an inflationary phase diluted 
the gradients of the pseudoscalar.

In this approximation, the result of the linearization can be simply written
in terms of the vector potentials in the gauge  
$Y^0 =0$ and $\vec{\nabla}\cdot\vec{Y} =0$:
\begin{eqnarray}
&&\vcY'' - \nabla^2  \vcY +\frac{c}{M} \psi' \vec{\nabla} \times 
\vcY =0,
\label{vectorp}\\
&& \ddot{\psi} + 3 H\dot{\psi} + \frac{\partial V}{\partial\psi} =0,
\label{psi}    
\end{eqnarray}
By combining the evolution equations for the gauge fields we can  find a
decoupled evolution equation for $\vcH_{Y}$,
\begin{equation}
\vcH_{Y}'' - \nabla^2\vcH_{Y} + \frac{c}{M} \psi' \vec{\nabla}
 \times\vcH_{Y} =0.
\end{equation}

From this equation is already apparent that the pseudo-scalar vertex 
induces an interaction in the two physical 
polarizations of the hypermagnetic field. Giving initial 
conditions which are such that   $\vcH_{Y} \neq 0$ with 
$\vcH_{Y}\cdot \vec{\nabla}\times \vcH_{Y} =0$ a profile 
with $\vcH_{Y}\cdot \vec{\nabla}\times \vcH_{Y} \neq 0$ can be 
generated provided $\psi' \neq 0$.

\section{Two approaches to the Origin}
\renewcommand{\theequation}{6.\arabic{equation}}
\setcounter{equation}{0}
Since the flux is conserved the ratio between the magnetic energy 
density, $\rho_{\rm B}(L,\eta)$ 
 and the energy density sitting in radiation, $\rho_{\gamma}(\eta)$
is almost constant and therefore, in terms of this quantity (which is only scale 
dependent but not time dependent), the most simplistic
 dynamo requirement can be rephrased (see Section 3) as
\begin{equation}
r_{\rm B}(L) = \frac{\rho_{\rm B}(L,\eta)}{\rho_{\gamma}(\eta)} 
\geq 10^{-34},\,\,\,\, L\sim 1\,{\rm Mpc},
\label{dyn}
\end{equation}
to be compared with the value $r_{\rm B} \sim 10^{-8}$ which would lead to
the galactic magnetic field only thanks to the collapse 
and without the need of dynamo action (this would be  
the case when the magnetic field is fully primordial). Notice that 
Eq. (\ref{dyn}) assumes that the magnetic flux is exactly frozen into the plasma 
element.

As previously pointed out during the discussion of the 
dynamo mechanism, the requirement expressed 
by Eq. (\ref{dyn}) is unrealistic: it would 
correspond to thirty e-folds of amplification 
during galactic rotation and perfect flux freezing 
during the collapse of the protogalaxy. In a more realistic situation 
situation, taking into account the effectively achievable 
amplification of the dynamo action the requirement would 
rather be, prior to collapse, 
\begin{equation}
r_{\rm B}(L) = \frac{\rho_{\rm B}(L,\eta)}{\rho_{\gamma}(\eta)} 
\geq 10^{-23},\,\,\,\, L\sim 1\,{\rm Mpc}.
\label{dyn2}
\end{equation}
In spite of the richness of the theoretical 
models, the mechanisms for magnetic field generation can be divided,
broadly speaking, into two categories: astrophysical  and 
cosmological. The cosmological mechanisms can be divided, in turn,
into {\em causal} mechanisms (where the magnetic seeds are produced at a given time inside the 
horizon) and {\em inflationary} mechanisms where correlations in the magnetic field 
are produced outside the horizon.  

\subsection{Inhomogeneous MHD equations}

The first attempts in this direction have been made by Biermann \cite{biermann} 
and Harrison \cite{harrison1,harrison2,harrison3}.
The Biermann battery mechanism is easy to understand from the 
form of the generalized Ohm law discussed in Eq. (\ref{genohm1}).
MHD equations are linear and homogeneous in the magnetic field 
intensities. The idea 
is then to look for  a natural source term  which seemed 
to be provided by the thermoelectric current already introduced 
in the context of the generalized Ohm law . Consider, indeed, Eq.
 (\ref{genohm1}) in the approximation where the Hall term
is neglected (the argument can be also generalized to the case of non 
vanishing Hall term). Then, the Ohmic electric field will not be the one 
simply obtained in resistive MHD but it will get a contribution 
from the thermoelectric term:
\begin{equation}
\vec{E} = - \vec{v}\times \vec{B} + \frac{\vec{\nabla}\times \vec{B}}{4\pi \sigma} - \frac{1}{e n_{\rm e}} \vec{\nabla}P_{\rm e}.
\label{bier1}
\end{equation}
Using the other MHD equations in the incompressible approximation the new form of 
the magnetic diffusivity equation can be derived:
\begin{equation}
\frac{\partial \vec{B}}{\partial t} - \vec{\nabla}\times( \vec{v} \times \vec{B}) - \frac{1}{4\pi \sigma} \nabla^2 \vec{B} = - \frac{1}{e n_{\rm e}^2}\vec{\nabla} 
n_{\rm e} \times \vec{\nabla} P_{\rm e}. 
\label{bier2}
\end{equation}
The magnetic diffusivity equation has now a source terms which does not vanish provided 
the gradients of the charge density of the electrons is not parallel to the pressure 
gradients. As a consequence of the presence of the source term, the magnetic field will 
grow linearly in time until the thermoelectric source term is comparable 
with the dynamo term. The estimate of the magnetic field intensity prior to the 
onset of galactic rotation will then be 
\begin{equation}
|\vec{B}| \simeq \frac{t}{e n_{\rm e}^2}| \vec{\nabla} 
n_{\rm e} \times \vec{\nabla} P_{\rm e}|,
\end{equation}
leading to magnetic fields ${\cal O}(10^{-20})$ G over typical scales of 
the order of $10$ kpc (see also \cite{LPA} for a recent analysis in a similar framework).

In a complementary perspective, Harrison\cite{harrison1,harrison2,harrison3}  
discussed the possibility that the pressure gradients are strictly vanishing,
i.e. $\vec{\nabla}p_{e} =0$. This is a reasonable assumption, for instance, 
in a radiation-dominated stage of expansion where pressure gradients are 
expected to be negligible by the global homogeneity 
of the background. In this situation, an evolution equation 
for the vorticity can be derived
\begin{equation}
\frac{\partial}{\partial \eta} ( a^2 \vec{\omega} + \frac{e}{m_{\rm p}} \vcB) = \frac{e}{4\pi \sigma m_{\rm p}} \nabla^2\vcB,
\label{harrison2}
\end{equation}
where $\omega = \vec{\nabla} \times \vec{v}$ and $m_{\rm p}$ is the ion mass. 
If we now postulate that some vorticity was present prior to decoupling, then Eq. (\ref{harrison2}) 
can be solved  and the magnetic field can be related to the initial vorticity as 
\begin{equation}
\vec{B} \sim - \frac{m_{\rm p}}{e} \vec{\omega}_{\rm i} \biggl(\frac{a_i}{a}\biggr)^2.
\label{harrison3}
\end{equation}  

It is clear that if the estimate of the vorticity is made prior to equality (as originally 
suggested by Harrison\cite{harrison1}) of after decoupling as also suggested, a bit later\cite{mishustin}
can change even by two orders of magnitude. Prior to equality $|\vec{\omega}(t) \simeq 0.1/t$ and, therefore,
$|\vec{B}_{\rm eq}| \sim 10^{-21}$G. If a similar estimate is made after decoupling the typical 
value of the generated magnetic field is of the order of $10^{-18}$ G. So, in this context, the 
problem of the origin of magnetic fields is circumvented by postulating an appropriate 
form of vorticity whose origin must be explained.

The idea was employed, later on, in the context of the physics of topological defects.
Vachaspati and Vilenkin \cite{vilenkin1} have suggested that cosmic strings with small scale 
structure may be a source of the wanted vorticity. The argument is, in short, that since 
matter flow in baryonic wakes is turbulent, velocity gradients will be induced in the flow by the small-scale
wiggles which produced the required vorticity. Furthermore, dynamical friction between cosmic strings and matter
may provide a further source of vorticity \cite{shellard1}. There have been also studies 
trying to generate large-scale magnetic fields in the context of superconducting cosmic strings (see, for instance,
\cite{dm2} and references therein).

Recently the possible generation of large-scale magnetic fields prior to hydrogen 
recombination has been discussed in \cite{dolr1,dolr2}. The vorticity required in order 
to produce the magnetic fields is generated, according to \cite{dolr1}, by the photon 
diffusion at second order in the temperature fluctuations. In a similar 
perspective Hogan \cite{hoganr} got much less optimistic estimates which, according to
\cite{dolr1,dolr2}, should be attributed to different approximation schemes employed 
in the analysis.

\subsection{Inflationary mechanisms}

Different astrophysical objects, of different physical sizes, have 
comparable magnetic fields. This coincidence is hard to explain in 
the context of causal mechanisms 
of generation. The  analogy with structure formation, already presented in the Introduction,
 is here useful. In the 
late seventies, prior to the formulation of the inflationary paradigm, 
the initial conditions for the density contrast were taken as 
primordial input. Later on, in the context of inflationary models, 
the primordial spectrum of curvature and density fluctuations could be calculated.

During inflation, in fact, fields of various spin are present and they 
can be excited by the dynamical evolution of the geometry. 
In the context of inflationary models of particular relevance are the 
scalar and tensor fluctuations of the geometry, corresponding 
respectively, to fluctuations of the scalar curvature and to 
gravitational waves. Gravitational waves and curvature perturbations 
obey evolution equations which are not invariant under Weyl rescaling of the 
four-dimensional metric. Then the quantum mechanical 
fluctuations in the corresponding fields will simply be amplified 
by the evolution of the background geometry.

It is interesting to speculate that large scale magnetic fields could be 
produced thanks to a similar mechanism. The major obstruction 
to this type of models is that gauge fields are not amplified
thanks to the evolution of the background geometry the reason being that 
the evolution equations of gauge fields are invariant under Weyl 
rescaling of the metric. 

When conformal invariance is broken, by some means, one is often led to estimate 
the amplitude of gauge field fluctuations arising as a result of the breaking.
Clearly the detailed amount of amplified gauge fields will be specific of the 
particular model. In the following various ways of 
breaking conformal invariance will be listed without getting through 
the details of the calculations. 
Before doing so it is anyway instructive to introduce 
some general considerations stressing the analogy between the production 
of magnetic inhomogeneities and the production of gravitational inhomogeneities 
whose late time evolution leads the anisotropies in the CMB.

The tensor modes of the geometry are described by a rank-two (transverse and traceless) 
tensor in three spatial dimensions \footnote{The theory of cosmological perturbations is assumed in the following considerations. 
Comprehensive accounts of the relevant topics can be found in \cite{gi1,gi2} (see  also \cite{gi3}).}, i.e.
\begin{equation}
ds^2 = a^2(\eta)[ d\eta^2 - (\delta_{i j} + h_{i j}) dx^{i} dx^{j}],~~~~~h_{i}^{i} = \partial_{i} h_{j}^{i} =0,
\label{perttens}
\end{equation}
 obeying the equation, in Fourier space,
\begin{equation}
\mu_{k}'' + \biggl[ k^2 - \frac{a''}{a} \biggr] \mu_{k} =0,
\label{tens}
\end{equation}
where $\mu_{k} = a h_{k}$ and $h_{k}$ is the Fourier mode 
of each polarization. In this equation the ``pump field'' is simply given by the 
scale factor. When $k^2 \gg |a''/a|$ the mode is said to be adiabatically 
damped: in fact, in this regime, $ |h_{k}| \simeq a^{-1}$, i.e. decreasing 
in an expanding Universe. In the opposite regime, i.e. $k^2 \ll |a''/a|$ 
the mode is super-adiabatically amplified. In fact, during a de Sitter or quasi-de Sitter 
$ \mu_{k} \sim a(\eta)$ and $ h_{k}$ is constant. Hence, for the whole time 
the given mode spends under the ``potential barrier'' of Eq. (\ref{tens}), 
it is amplified. 

A similar equation  holds for the canonical 
normal mode for the scalar fluctuations of the geometry, i.e. 
\begin{equation}
v_{k}'' + \biggl[ k^2 - \frac{z''}{z} \biggr] v_{k} =0.
\label{vvar}
\end{equation}
The variable $v_{k}$ is defined as 
\begin{equation}
v_{k} = a \delta\varphi_{k} + z \psi_{k}, ~~~~~~~~~~~~~{\cal R}_{k} = - \frac{v_{k}}{z},
\label{defv}
\end{equation}
where $\delta\varphi_{k}$  is the fluctuation 
of the inflaton, ${\cal R}_{k}$ is the curvature perturbation and $\psi$ is the scalar fluctuation 
of the geometry in the conformally Newtonian gauge \cite{gi1,gi2}. If the inflaton has an exponential potential
$z(\eta) \simeq a(\eta)$.

In the case of gauge fields, each of the two polarization of the appropriately rescaled vector potential obeys, the following equation 
\begin{equation}
{\cal A}_{k}'' + \biggl[k^2 - g(g^{-1})''\biggr] {\cal A}_{k} =0.
\label{scheq}
\end{equation}
In specific models, 
$g(\eta) $ may be associated with the gauge coupling. However $g^{-1}(\eta)$ can also be 
 viewed as a generic pump field arising as a result of the breaking 
of conformal invariance. The time dependence of the potential term is 
rather common to different models: it goes to zero for $|\eta| \to \infty$ 
as $(\nu^2 - 1/4)\eta^{-2}$. The numerical coefficient appearing in the potential 
determines the strength and spectrum of the amplified gauge fields whose subsequent 
evolution has to be however computed at finite conductivity.
\begin{figure}[th]
\centerline{\epsfig{file=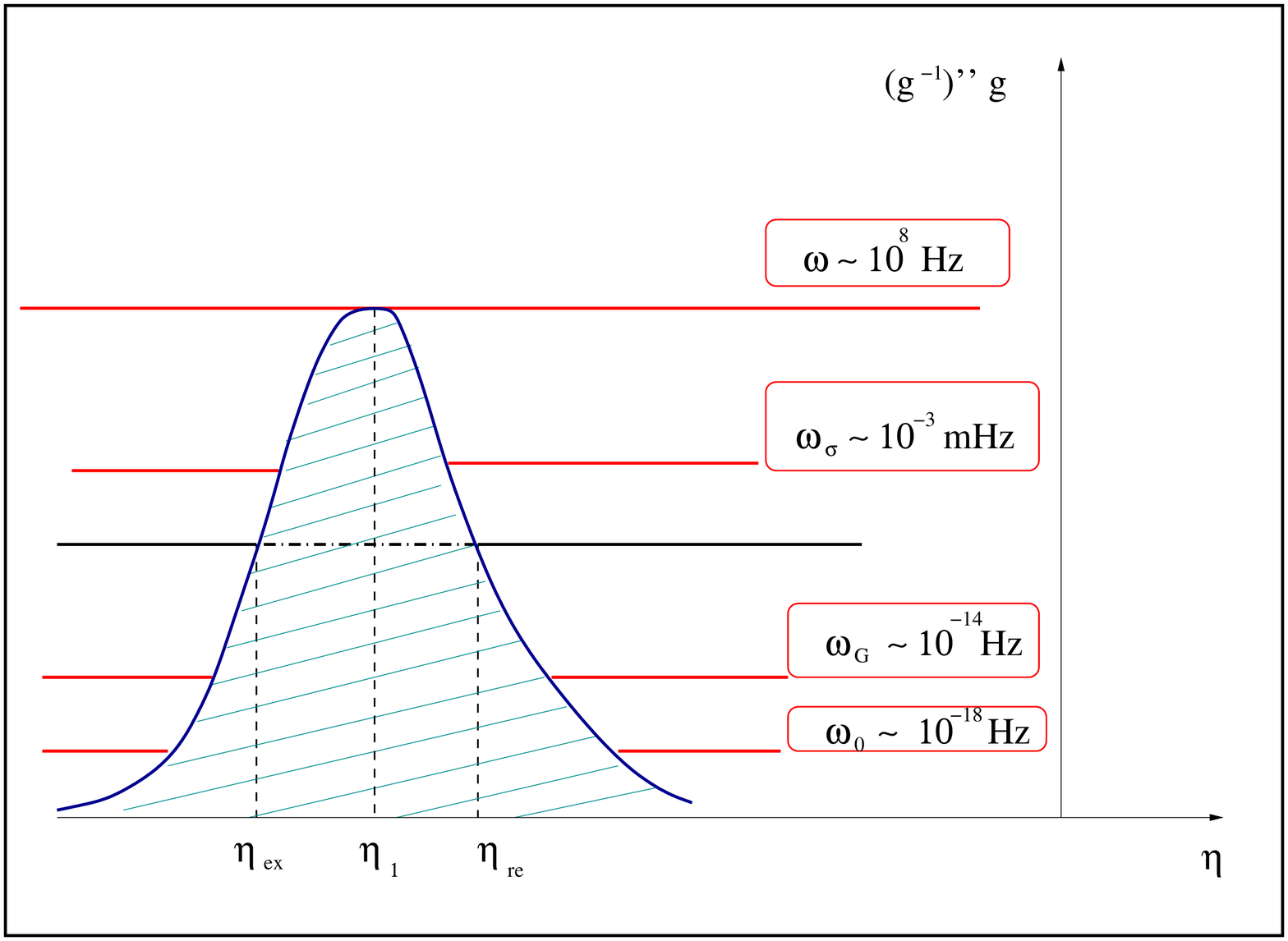,width=9cm}}
\vspace*{8pt}
\caption[a]{The effective potential appearing in Eq. (\ref{scheq}) is illustrated in general terms. In this 
example the pump field leading to amplification of electromagnetic 
quantum  fluctuations is assumed to be constant prior to the onset of BBN. On the vertical axis 
few relevant physical frequencies (to be compared with the height of the potential) have been 
reported.}
\label{F4}
\end{figure}
In Fig. \ref{F4} the typical form of the potential term appearing in Eq.  (\ref{scheq})
is illustrated. Different comoving frequencies go under the barrier at different times 
and the amount of amplification is roughly proportional to the time spent under 
the barrier. Clearly, given the generic form of the barrier, smaller frequencies are 
more amplified than the frequencies comparable with the height of the barrier at $\eta_{1}$.
In Fig. \ref{F4} the explicit numerical value of the height of the barrier corresponds 
to a (present) frequency of $10^{8}$ Hz which can be realized if the pump field
goes to a constant right after the end of a conventional inflationary phase 
followed by a radiation-dominated stage of expansion.

\begin{figure}[th]
\centerline{\epsfig{file=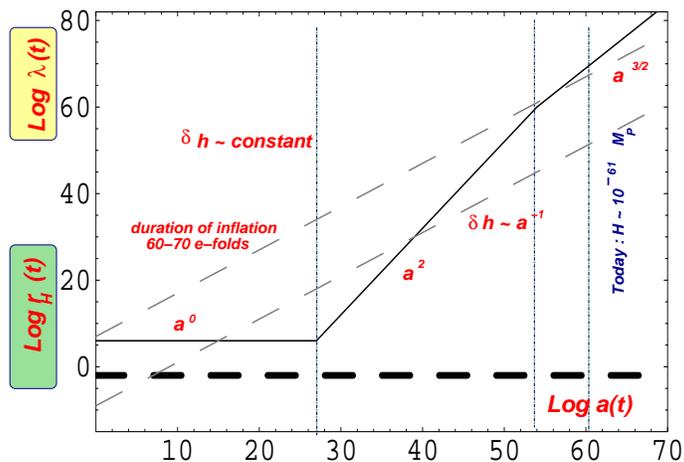,width=9cm}}
\vspace*{8pt}
\caption[a]{The evolution of a given physical wavelength is 
illustrated in the case of an inflationary model with minimal duration. The region 
above the full curve (denoting the Hubble radius  $r_{H})= H^{-1}$) corresponds, for the tensor and scalar modes of the geometry,
to the region where a given comoving wavenumber $k$
is below the ``potential barrier'' appearing in Eqs. (\ref{tens}) and (\ref{vvar}).}
\label{HR}
\end{figure}

In the case of gravitational waves, and, with some caveats, in the case of scalar 
metric fluctuations, the ``potential barriers'' appearing, respectively, in Eqs. (\ref{tens}) and (\ref{vvar}) 
may be  related with the inverse Hubble radius. Hence, in the case of metric fluctuations, 
a mode which is under the barrier is also, with a 
swift terminology, outside the horizon (see Fig. \ref{HR}). This is the reason why, following 
the conventional nomenclature, in Fig. \ref{F4} the moment 
when a given scale gets under the barrier has been denoted by $\eta_{\rm ex}$ (i.e. 
horizon exit). According to the same convention, 
 the moment when a given scale gets out the potential  barrier is 
labeled by $\eta_{\rm re}$ (i.e. horizon re-entry).
In Fig. \ref{F4} few relevant frequencies have been compared with the height 
of the barrier. Consider, for instance, $\omega_{\rm G}$ i.e. the typical scale 
of gravitational collapse which is of the order of $1$ Mpc, i.e. $10^{-14} $ Hz. 
In Fig. \ref{F4} the physical frequency has been directly reported. 
Another interesting frequency is $\omega_{\sigma} \sim {\rm m Hz}$ corresponding 
to the present value of the magnetic diffusivity momentum. 
The amplification caused by the parametric amplification of 
the vacuum fluctuations can be computed by solving Eq. (\ref{scheq}) in the different regimes.
\begin{eqnarray}
{\cal  A}_{k} &=&{e^{-ik\eta}\over \sqrt{k}}\;\;~~~,~~~~~
\,\,\,\,\,\,\,\,\,\,\,\,\,\,\,\,\,\,\,\,\,\,\,\,\,\,\,\,\,\,\,\,\,\,\,\,\,\,\,
\eta < \eta_{ex} ,
 \nonumber \\
{\cal A}_{k} &=& g^{-1}(\eta)
 [C_k + D_k \int^{\eta} d\eta'~~ g^2(\eta^{\prime})] \; \;
{}~~~~,~~~~~\eta_{ex} < \eta < \eta_{re} \; \nonumber \\
{\cal A}_{k} &=& {1\over \sqrt{k}}[ c_+(k) e^{-ik\eta} +  c_-(k)
e^{ik\eta}]
\;\;~~~~~,~~~~~~~~ \eta > \eta_{re},
\label{solution}
\end{eqnarray}
where  ( $C_{k}$, $D_{k}$, and 
$c_{\pm}(k)$ are integrations constants).

The mixing coefficients $c_{\pm}(k)$, determining the
parametric amplification of a mode $k^2<|V(\eta_1)|$,  computed 
 by matching these various branches of the solution reported in Eq. (\ref{solution}). One finds:
\begin{eqnarray}
&&{2ik}e^{- ik(\eta_{ex} \mp \eta_{re})} c_\pm(k) =
\mp 
\frac{g_{ex}}{g_{re}}(-\frac{{g_{re}}^{\prime}}{g_{re}} \mp ik) \pm
\frac{g_{re}}{g_{ex}}(-\frac{{g_{ex}}^ {\prime}}{g_{ex}} +ik) 
 \nonumber \\
&\pm &\frac{1}{g_{ex} g_{re}} (-\frac{{g_{ex}}^{\prime}}{g_{ex}}
+ik)  (-\frac{{g_{re}}^{\prime}}{g_{re}} \mp ik)
\int_{\eta_{ex}}^{\eta_{re}} g^2 d\eta.
\label{Bog}
\end{eqnarray}
Similar calculations can be performed in order to obtain the 
spectrum of scalar and tensor fluctuations of the geometry \cite{gr1,gr2,gr3}.

Suppose now to make a simple estimate. Assume, for instance,
that $g$ is evolving prior to $\eta_{1}$ according to the 
dynamics dictated by a given model. Suppose also that after $\eta_{1}$ the Universe 
is suddenly dominated by radiation,  and $g'\sim 0$ for $\eta > \eta_1$.
In this situation all the modes reenter during radiation and the 
amplification will be roughly given, to leading order by
\begin{equation}
|c_{-}|\simeq\frac{g_{re}}{g_{ex}}.
\label{amplification}
\end{equation}
If the function $g(\eta)$ is identified with the evolving gauge coupling this result 
suggest that in order to have large amplification, $g(\eta)$ has to grow from smaller 
to larger values. This is what happens, for instance, in the case of pre-big bang models where 
$g\sim e^{\varphi/2}$ and $\varphi$ is the for-dimensional dilaton field \cite{Veneziano3}.

Since we ought to estimate the amplification of an initial quantum mechanical fluctuations, a fully 
quantum mechanical treatment is certainly appropriate also in view of the  discussion of  the correlation 
properties of the obtained fluctuations. This analysis has been performed in \cite{sqmg} where the 
squeezing properties of the obtained photons have also been discussed. 

The perturbed effective Lagrangian density 
\begin{equation}
{\cal L}(\vec{x}, \eta) = 
\frac{1}{2}\sum_{\alpha} \biggl[{{\cal A}'}_{\alpha}^2 +
 2 \frac{g'}{g} {\cal A}'_{\alpha} {\cal A}_{\alpha} + \biggl(\frac{g'}{g}\biggr)^2 {\cal A}_{\alpha}^2 - 
(\partial_i {\cal A}_{\alpha})^2 \biggr], ~~L(\eta) 
= \int d^3 x {\cal L}(\vec{x},\eta),
\label{action2}
\end{equation}
describes the evolution of the two ( $\alpha= \otimes,~\oplus$) 
transverse degrees of 
freedom defined by  the Coulomb gauge condition $A_0=0$ and 
$\vec{\nabla}\cdot\vec{A} =0$ (the prime denotes 
differentiation with respect to conformal time). The fields
$A_{\alpha}= g {\cal A}_{\alpha}$  have kinetic terms 
with canonical normalization and the time evolution given 
in Eq. (\ref{scheq}) stems from the Euler-Lagrange equations derived from 
Eq. (\ref{action2}).
By functionally deriving the 
 the action  the canonically conjugated momenta can be obtained 
leading 
to the Hamiltonian density and to the associated Hamiltonian
\begin{eqnarray}
&&{\cal H}(\vec{x},\eta) = \frac{1}{2}\sum_{\alpha} 
\biggl[ \pi_{\alpha}^2 + (\partial_{i} {\cal A}_{\alpha} )^2 - 
2 \frac{g'}{g} {\cal A}_{\alpha} \pi_{\alpha}\biggr],
\nonumber\\
&& H(\eta) =\int d^3 x {\cal H}(\vec{x}).
\end{eqnarray}
The operators corresponding to the classical polarizations
appearing in the Hamiltonian density 
\begin{eqnarray}
&&\hat{\cal A}_{\alpha} (\vec{x}, \eta) = \int{ d^3 k}\frac{1}{(2 \pi)^{3/2}}  
\hat{\cal A}_{\alpha}(k,\eta) e^{i \vec{k}\cdot{\vec{x}}},
\nonumber\\
&&\hat{\cal A}_{\alpha}(k,\eta) = \frac{1}{ \sqrt{ 2 k}}
\bigl[ \hat{a}_{k,\alpha}(\eta)+ \hat{a}^{\dagger}_{-k,\alpha}(\eta)\bigr],
\nonumber\\
&&\hat{\pi}_{\alpha} (\vec{x}, \eta) = \int{ d^3 k}
\frac{1}{(2 \pi)^{3/2}}  \hat{\pi}_{\alpha}(k,\eta) 
e^{i \vec{k}\cdot{\vec{x}}},
\nonumber\\
&&\hat{\cal \pi}_{\alpha}(k,\eta) 
= i \sqrt{\frac{k}{2}}\bigl[
\hat{a}_{k,\alpha}(\eta)- \hat{a}^{\dagger}_{-k,\alpha}(\eta)\bigr],
\end{eqnarray}
obey canonical commutation relations and the associated 
creation and annihilation operators satisfy
$[\hat{a}_{k,\alpha}, \hat{a}^{\dagger}_{p,\beta}] = 
\delta_{\alpha\beta}\delta^{(3)} (\vec{k} - \vec{p})$.

The (two-modes) Hamiltonian 
contains  a free part and the effect of the variation 
of the coupling constant is  encoded in the 
(Hermitian) interaction term which is quadratic in the creation 
and annihilation operators whose evolution equations, read,
in the Heisenberg picture
\begin{eqnarray}
&&\frac{ d \hat{a}_{k,\alpha}}{d \eta} = - i k 
\hat{a}_{k,\alpha} - \frac{ g'}{g} 
\hat{a}^{\dagger}_{-k,\alpha},
\nonumber\\
&&
\frac{ d \hat{a}^{\dagger}_{k,\alpha}}{d \eta} = 
i k \hat{a}^{\dagger}_{k,\alpha} - \frac{ g'}{g} 
\hat{a}_{-k,\alpha}.
\label{heiseq}
\end{eqnarray}
The general solution of the previous system of equations can be written 
in terms of  a Bogoliubov-Valatin transformation, 
\begin{eqnarray}
&& \hat{a}_{k,\alpha}(\eta) = \mu_{k,\alpha}(\eta) \hat{b}_{k,\alpha} + 
\nu_{k,\alpha}(\eta)\hat{b}^{\dagger}_{-k,\alpha}
\nonumber\\
&& \hat{a}^{\dagger}_{k,\alpha}(\eta) 
= \mu^{\ast}_{k,\alpha}(\eta) \hat{b}^{\dagger}_{k,\alpha} + 
\nu_{k,\alpha}^{\ast}(\eta)\hat{b}_{-k,\alpha}
\label{heis}
\end{eqnarray}
where $\hat{a}_{k,\alpha}(0) = 
\hat{b}_{k,\alpha}$ and $\hat{a}_{-k,\alpha}(0) = \hat{b}_{-k,\alpha}$. 
Notice that the Bogoliubov coefficients are the quantum analog of the 
mixing coefficients discussed in the semiclassical approach to the problem.

Unitarity requires that  
 the two complex functions $\mu_{k}(\eta)$ and $\nu_{k}(\eta)$ 
are subjected to the condition $|\mu_{k}(\eta)|^2 - |\nu_{k}(\eta)|^2 =1$ 
which also implies that
$\mu_{k}(\eta)$ and $\nu_{k}(\eta)$ can be parameterized in terms of 
one real amplitude and two real phases
\begin{equation}
\mu_k = e^{i \theta_{k}} \cosh{r_{k}},
~~~~\nu_k = e^{ i(2\phi_{k} -  \theta_{k})} 
\sinh{r_{k}},
\label{sq1}
\end{equation}
($r$ is sometimes called squeezing parameter and $\phi_{k}$
is the squeezing phase; from now on we will drop the subscript 
labeling each polarization if not strictly necessary).
The total number of produced photons  
\begin{equation}
\langle 0_{-k} 0_{k}| 
\hat{a}^{\dagger}_{k}(\eta) \hat{a}_{k}(\eta) + 
\hat{a}_{-k}^{\dagger} \hat{a}_{-k} |0_{k} 0_{-k}\rangle= 
2 ~\overline{n}_k.
\label{num}
\end{equation}
is expressed in terms of
 $\overline{n}_{k} =\sinh^2{r_{k}}$, i.e. the  mean number 
of produced photon pairs in the mode $k$.
Inserting  Eqs. (\ref{heis})--(\ref{num}) 
into Eqs. (\ref{heiseq})  
we can derive a closed system involving only the 
$\overline{n}_k$  and the related phases:
\begin{eqnarray}
&&\frac{d  \overline{n}_{k}}{d \eta} = 
-2 f(\overline{n}_{k}) \frac{g'}{g} 
\cos{2 \phi_{k}},
\label{I}\\
&& \frac{d \theta_{k}}{d\eta} = - k + \frac{g'}{g} 
\frac{\overline{n}_{k}}{f(\overline{n}_{k})}
\sin{2 \phi_{k}} ,
\label{II}\\
&& \frac{ d \phi_{k}}{ d \eta} = - k + \frac{g'}{ g} 
\frac{d f(\overline{n}_{k})}{d \overline{n}_{k}}
\sin{ 2 \phi_{k}},
\label{III}
\end{eqnarray}
where $f(\overline{n}_{k}) = \sqrt{ \overline{n}_{k}(\overline{n}_{k} + 1)}$.

In quantum optics \cite{lou,mandel} the coherence properties of light fields have 
been  a subject of intensive investigations for nearly half a century. 
In the present context the multiparticle states described so fare are nothing 
but squeezed states of the electromagnetic field \cite{lou,mandel}.
In fact, up to now the Heisenberg description has been adopted. In the Schr\"odinger picture 
the quantum mechanical states obtained as a result of the time evolution are 
exactly squeezed vacuum states \cite{lou,mandel}.
 
Magnetic fields over galactic scales have typical frequency 
of the order $10^{-14}$--$10^{-15}$ Hz which clearly fall well outside the 
optical range. Thus, the analogy with quantum optics is only technical.
The same quantum optical analogy has been successfully exploited 
in particle \cite{car} and heavy-ions physics \cite{baym2}  
of pion correlations (in order to measure the size of the 
strongly interacting region) and in the phenomenological 
analysis of hadronic multiplicity distributions. 

The interference between the amplitudes of the magnetic fields (Young 
interferometry \cite{lou2}, in a quantum optical language) estimates 
 the first order coherence of the magnetic background at different spatial 
locations making use of the two-point correlation function  
whose trace  over the physical polarizations and for coincidental 
spatial locations is  related to the magnetic energy density.
Eqs. (\ref{I})--(\ref{III}) can be solved once the pump field
is specified but general expressions can be also obtained \cite{sqmg}.

\subsubsection{Conventional inflationary models}

In conventional inflationary models it is very difficult to produce 
large scale magnetic fields with phenomenologically 
relevant strength. This potential difficulty has been 
scrutinized in various 
investigations \cite{mgst1,turner,ratra,dolgov}.

Turner and Widrow \cite{turner} listed a series of field theory models 
in de Sitter space with the purpose of finding a natural way of breaking 
conformal invariance. The first suggestion was that 
conformal invariance may be broken, at an effective level, 
through the coupling of photons to the geometry \cite{drummond}. Typically, the 
breaking of conformal invariance occurs through  products 
of gauge-field strengths and curvature tensors, i.e.
\begin{equation} 
\frac{1}{m^2}F_{\mu\nu}F_{\alpha\beta} R^{\mu\nu\alpha\beta},~~~~~~~\frac{1}{m^2} R_{\mu\nu} F^{\mu\beta} F^{\nu\alpha} g_{\alpha\beta},~~~~~~
\frac{1}{m^2} F_{\alpha\beta}F^{\alpha\beta} R
\label{curv}
\end{equation}
where $m$ is the appropriate mass scale; $R_{\mu\nu\alpha\beta}$ and  $R_{\mu\nu}$ are the Riemann and Ricci tensors 
and $R$ is the Ricci scalar. If the evolution of gauge fields is studied during a phase of de Sitter (or quasi-de Sittter) 
expansion, then the amplification of the vacuum fluctuations induced by the 
couplings listed in Eq. (\ref{curv}) is minute. The price in order to get large amplification 
should be, according to \cite{turner}, an explicit breaking of gauge-invariance 
by direct coupling of the vector potential to the Ricci tensor or to the Ricci scalar, i.e. 
\begin{equation}
R A_{\mu} A^{\mu},~~~~~~~~~R_{\mu\nu} A^{\mu} A^{\nu}.
\label{vector}
\end{equation}
In \cite{turner} two other different models were proposed (but not scrutinized in detail) 
namely  scalar electrodynamics and the axionic coupling to the Abelian field strength.

Dolgov \cite{dolgov} considered the possible breaking of conformal invariance due 
to the trace anomaly. The idea is that the conformal invariance of 
gauge fields is broken by the triangle diagram where two photons in the external 
lines couple to the graviton through a loop of fermions. The local 
contribution to the effective action 
leads to the vertex $(\sqrt{-g})^{1+ \epsilon} F_{\alpha\beta}F^{\alpha\beta}$ where 
$\epsilon$ is a numerical coefficient depending upon the number of scalars and fermions present 
in the theory. The evolution 
equation for the gauge fields, can be written, in Fourier space, as 
\begin{equation}
{\cal A}_{k}'' + \frac{\epsilon}{8} {\cal H} {\cal A}_{k}' + k^2 {\cal A}_{k} =0,
\end{equation}
and it can be shown that only if $\epsilon >0$ the gauge fields are amplified.
Furthermore, only 
is $\epsilon \sim 8$ substantial amplification of gauge fields is possible.

In a series of papers \cite{carroll1,carroll2,carroll3} the possible 
effect of the axionic coupling to the amplification of gauge fields has been investigated.
The idea is here that conformal invariance is broken through 
the explicit coupling of a pseudo-scalar field to the gauge field (see Section 5), i.e. 
\begin{equation}
\sqrt{-g} c_{\psi\gamma}\alpha_{\rm em} \frac{\psi}{8\pi M} F_{\alpha\beta}\tilde{F}^{\alpha\beta},
\label{pseudo}
\end{equation}
where $\tilde{F}^{\alpha\beta}$ is the dual field strength and where 
$c_{\psi\gamma}$ is a numerical factor of order one. Consider now the case of  a standard 
pseudoscalar potential, for instance $m^2 \psi^2$, evolving in a de Sitter (or quasi-de Sitter space-time).
It can be shown, rather generically, that the vertex given in Eq. (\ref{pseudo}) leads to negligible 
amplification at large length-scales. The coupled system of evolution equations 
to be solved in order to get the amplified field is similar to Eqs. (\ref{system}) 
already introduced in the duscussion of hypermagnetic fields
\begin{eqnarray}
&& \vcB'' - \nabla^2 \vcB - \frac{\alpha_{\rm em}}{2 \pi M} \psi' \vec{\nabla} \times \vcB =0,
\label{bpsi}\\
&& \psi'' + 2 {\cal H} \psi' + m^2 a^2 \psi =0.
\label{psib}
\end{eqnarray}
From Eq. (\ref{bpsi}),  there is a maximally amplified physical frequency 
\begin{equation}
\omega_{\rm max} \simeq \frac{\alpha_{{\rm em}}}{2 \pi M} \dot{\psi}_{\rm max} \simeq  \frac{\alpha_{{\rm em}}}{2 \pi } m
\end{equation}
where the second equality follows from $\psi \sim a^{-3/2} M \cos{m t}$ (i.e. $\dot{\psi}_{\rm max} \sim m M$).
The amplification for $\omega \sim \omega_{\rm max}$ is of the order of $\exp{[m \alpha_{\rm em}/(2 \pi H)]}$ 
where $H$ is the Hubble parameter during the de Sitter phase of expansion. From the above expressions one can 
argue that the modes which are substantially amplifed are the ones for which $\omega_{\rm max} \gg H$. The modes 
interesting for the large-scale magnetic fields are the ones which are in the opposite range, i.e.  $\omega_{\rm max} \ll H$.
Clearly, by lowering the curvature scale of the problem the produced seeds may be larger and the 
conclusions much less pessimistic \cite{carroll3}.

Another interesting idea pointed out by Ratra \cite{ratra} is that the electromagnetic field may be 
directly coupled to the inflaton field. In this case the coupling is specified 
through a parameter $\alpha$, i.e. $e^{\alpha \varphi} F_{\alpha\beta}F^{\alpha\beta}$ where $\varphi$ is the 
inflaton field in Planck units. In order to get 
sizable large-scale magnetic fields the effective gauge coupling must be larger than one during 
inflation (recall that $\varphi$ is large, in Planck units, at the onset of inflation).

In \cite{variation} it has been suggested that the evolution of the Abelian gauge 
coupling during inflation induce the growth of the two-point function of magnetic inhomogeneities.
This model is different from the one previously discussed \cite{ratra}. Here 
the dynamics of the gauge coupling is not related to the dynamics of the 
inflaton which is not coupled to the Abelian field strength. In particular, 
$r_{B}({\rm Mpc}) $ can be as large as $10^{-12}$. In \cite{variation}
the MHD equations have been generalized to the case of evolving gauge coupling.
Recently a scenario similar to \cite{variation} has been discussed in \cite{bamba}.

In the perspective of generating large scale magnetic fields Gasperini \cite{gravphot} 
 suggested to consider the possible mixing between the photon and the 
graviphoton field appearing in supergravity theories (see also, in a related context \cite{okun}). 
The graviphoton is 
the massive vector component of the gravitational supermultiplet and its 
interaction with the photon is specified by an interaction term of the 
type $ \lambda F_{\mu\nu} G^{\mu\nu}$ where $G_{\mu\nu}$ is 
the filed strength of the massive vector. Large-scale magnetic fields 
with $r_{B}({\rm Mpc})\geq 10^{-34}$ can be obtained 
if $\lambda \sim {\cal O}(1)$ and for a mass of the vector $m \sim 10^{2} {\rm TeV}$.

Bertolami and Mota \cite{bertolami} argue that if Lorentz invariance 
is spontaneously broken, then  photons acquire naturally a coupling 
to the geometry which is not gauge-invariant and which is similar to the 
coupling considered in \cite{turner}.

Finally Davis and Dimopoulos \cite{acdavis} considered the possibility 
of phase transitions taking place during inflation. They found that 
sizable large-scale magnetic fields can be generated provided the phase 
transition occurs in the last 5 e-foldings of the inflationary stage 
of expansion. 

\subsubsection{Abelian Higgs model}

 While the coupling
of electromagnetic field to the metric and to the charged fields is
conformally invariant, the coupling of the charged 
scalar field to
gravity is not. Thus, vacuum fluctuations of the charged scalar field
can be amplified during inflation at super-horizon scales, leading
potentially to non-trivial correlations of the electric currents and
charges over cosmological distances. The  fluctuations of
electric currents, in turn, may induce magnetic fields through
Maxwell equations at the corresponding scales. The role of the
charged scalar field may be played by the Higgs boson which couples to
the hypercharge field above the electroweak phase transition; the
generated hypercharged field is converted into ordinary magnetic
field at the temperatures of the order of electroweak scale. 

No detailed computations were carried out in \cite{turner} in order to
substantiate this idea. The suggestion of \cite{turner} was further
developed quite recently in \cite{cal} for the standard electroweak
theory with an optimistic conclusion that large scale  magnetic
fields can be indeed generated. In \cite{cal1} a supersymmetric model
was studied. In \cite{mginf1}, previous treatments have been further 
scrutinized by computing, with higher accuracy, the amplification 
of the charged scalar field and the damping induced by the conductivity. 
It turns out that the resulting magnetic fields are insufficient 
in order to provide reasonable seeds for the dynamo amplification.

Introducing appropriately rescaled fields the action of the Abelian-Higgs 
model in a conformally flat FRW space-time can be written as 
\begin{equation}
S = \int d^3 x d\eta\bigl[ 
\eta^{\mu\nu}D_{\mu} \tilde{\phi}^{\ast} D_{\nu} \tilde{\phi} + 
\bigl( \frac{a''}{a}- 
m^2 a^2\bigr)\tilde{\phi}^{\ast} \tilde{\phi} - \frac{1}{4} 
F_{\alpha\beta}F^{\alpha\beta}].
\label{actionab}
\end{equation}
Now, since the evolution equation of the charged scalar is not conformally invariant, 
current density and charge density fluctuations will be induced. Then, by 
employing a Vlasov-Landau description similar to the one introduced 
in Section 5, the resulting magnetic field will be of the order of 
$B_{\rm dec} \sim 10^{-40} T_{\rm dec}^2$ which is, by far, too small. Later 
it has been proposed that much larger magnetic fields may be obtained in the 
context of the Abelian-Higgs model \cite{davdim} (see however \cite{mgs2} for a detailed 
criticism of this proposal).

\subsubsection{Internal dimensions}

If internal dimensions are dynamical, then Weyl invariance 
may be naturally broken \cite{mgint}.
Consider  a pure electromagnetic fluctuation decoupled 
from the sources, representing an electromagnetic wave propagating 
in the $d$-dimensional external space such that $A_{\mu} \equiv A_{\mu}(\vec{x}, 
\eta)$, 
$A_{a} =0$. In the metric given in Eq. (\ref{metric}) the evolution 
equation of the gauge field fluctuations can be written as 
\begin{equation}
\frac{1}{\sqrt{-G}} \partial_{\mu}\biggl( \sqrt{-G} G^{\alpha\mu} 
G^{\beta\nu} F_{\alpha\beta} \biggr) =0,
\end{equation}
where $F_{\alpha\beta} = \nabla_{[\alpha}A_{\beta]}$ 
is the gauge field strength and 
$G$ is the determinant of the $D$ dimensional metric. Notice that 
if $n=0$ the space-time is isotropic and, therefore, the Maxwell's 
equations can be reduced (by trivial rescaling) to the flat space equations. 
If $n \neq 0$ we have that 
the evolution equation of the electromagnetic fluctuations propagating in the 
external $d$-dimensional manifold will receive a contribution from the internal
 dimensions which cannot be rescaled away \cite{berg}
\footnote{Notice that the electromagnetic field couples only to 
the internal dimensions through the determinant of the $D$-dimensional 
metric. In string theories, quite generically, the one-form fields are also 
coupled to the dilaton field. This case has been already analyzed in the 
context of string inspired cosmological scenarios and will be discussed later. }.
In the radiation gauge ($A_0 =0$ and $\nabla_{i} A^{i} =0$) \footnote{For a 
discussion of gauges in curved spaces see \cite{ford}.}
the evolution  the vector potentials can be written as 
\begin{equation}
A_{i}'' + n {\cal F} A_{i}' - \vec{\nabla}^2 A_{i} =0, \,\,\,\,\,
 {\cal F} = \frac{b'}{b}.
\label{vec1}
\end{equation}
The vector potentials $A_{i}$ are already rescaled with respect 
to the (conformally flat) $d+1$ dimensional metric.  
In terms of the canonical normal modes of oscillations ${\cal A}_{i} = b^{n/2}
 A_{i}$ 
the previous equation can be written in a simpler form, namely 
\begin{equation}
{\cal A}_{i}''  - V(\eta) {\cal A}_{i} -\vec{\nabla}^2 {\cal A}_{i}  =0,\,\,\,
\,\, 
V(\eta) 
= \frac{n^2}{4} {\cal F}^2 + \frac{n}{2}{\cal F}'.
\label{eq1}
\end{equation}
In order to estimate the amplification of the gauge fields induced by the 
evolution 
of the internal geometry we shall consider the background metric of Eq. 
(\ref{metric2})
in the case of maximally symmetric subspaces $ \gamma_{ij} = \delta_{ij}$, 
$\gamma_{a b} = \delta_{a b}$. 

Suppose now that the background geometry evolves along three different epochs. 
During the first phase (taking place for $] -\infty, -\eta_1]$) 
the evolution is truly  multidimensional. At $\eta = -\eta_1$ the 
multidimensional dynamics is continuously matched 
to  a radiation dominated phase turning, after decoupling, into 
a matter dominated regime of expansion. 
During the radiation and matter dominated stages 
the internal dimensions are fixed to their (present) constant size 
in order not 
to conflict with possible bounds arising both at the BBN time and 
during the matter-dominated epoch. 
The evolution of the external dimensions does not affect the amplification
of the gauge fields as it can be 
argued from Eq. (\ref{eq1}) :in the limit 
$n \rightarrow 0$ (i.e. conformally invariant background) Eq. (\ref{eq1}) 
reduces to the flat space equation. A background with the features described 
above has been introduced in Eq. (\ref{back}).
 
Defining, respectively, $b_{BBN}$ 
and $b_0$ as the size of  the internal dimensions at the BBN 
time and at the present epoch, the maximal variation 
allowed to the internal scale factor from the BBN time 
can be expressed as $b_{BBN}/b_0 \sim 1 
+ \epsilon$ where $ |\epsilon | < 10^{-2} $ \cite{int1,int2,int3}. 
The bounds on the  variation 
of the internal dimensions during the matter dominated epoch
are even stronger. Denoting with an over-dot the derivation with 
respect to the cosmic time coordinate, we have that 
$ |\dot{b}/b| < 10^{-9} H_0$ where 
$H_0$ is the present value of the Hubble parameter \cite{int1}.
The fact that the time evolution of internal dimensions
is so tightly constrained for temperatures lower of $1$ MeV
does not forbid that they could have been dynamical 
prior to that epoch \cite{star}.

In the parameterization of Eq. (\ref{back})
 the internal dimensions grow (in conformal time) for $\lambda <0$ and they 
shrink \footnote{To assume 
that the internal dimensions 
are constant during the radiation and matter dominated 
epoch is not strictly necessary. If 
the internal dimensions have a time variation during the radiation 
phase we must anyway impose the BBN bounds on their variation 
\cite{int1,int2,int3} . 
The tiny variation allowed by BBN implies that $b(\eta)$ must be
 effectively constant for practical purposes. } for $\lambda > 0$.

By inserting this background into Eq. (\ref{eq1}) we obtain that 
for $\eta < - \eta_1$
\begin{equation}
V(\eta) = \frac{ n \lambda}{4 \eta^2}( n\lambda - 2),
\label{V}
\end{equation}
whereas $ V(\eta) \rightarrow 0$ for $ \eta > - \eta_1$.
Since $V(\eta)$ goes to zero for $\eta\rightarrow \pm \infty$ 
 we can define, in both limits,  
a Fourier expansion of ${\cal A}_{i}$ in terms of two 
distinct orthonormal sets of modes. The amplification of the quantum mechanical 
fluctuations of the gauge fields can then be computed using the standard techniques (see, for instance, 
\cite{bir,gar,gio2}).
In \cite{mgint} it has been shown that in simple models of dimensional decoupling 
magnetic fields can be generated with strength compatible 
with the bound of Eq. (\ref{dyn2}). In particular there is the 
interesting possibility that large-scale magnetic fields are produced 
in the case when internal dimensions also expand while the external ones also expand 
but at a different rate.

It should be mentioned that the interplay between gauge fields and large (or infinite) 
extra-dimensions is still under investigation. There is, at the moment, no definite 
model leading to the generation of large-scale magnetic field 
in a framework of infinite internal dimensions.  The main reason 
is that it is difficult to build reasonable cosmological models 
with large extra-dimensions. More specifically, one would like to have a model 
where gauge fields arise naturally in the bulk. In six- dimensions 
interesting models can be constructed \cite{infintdim1} where localized gauge zero modes 
may be naturally present (in analogy with what happens in the case of  six-dimensional 
warped models  \cite{infintdim2,infintdim3}).

\subsubsection{String cosmological models}

Large scale magnetic fields may also be produced in the context of string cosmological models \cite{mgst1,mgst2} (see also 
\cite{mgst0} for the analysis of the amplification of magnetic fields during a phase of coherent dilaton oscillations\footnote{The
possible production of (short scale) magnetic fields by parametric resonance explored in \cite{mgst0} , has been 
subsequently analyzed in the standard inflationary contex \cite{FG}}).
In many respects the case of string cosmological models is related to the one where 
the gauge coupling is dynamical. However, in the string cosmological context, the 
dynamics of the gauge coupling and of the geometry are connected 
by the solutions of the low-energy $\beta$-functions. The basic evolution of the background and of the gauge 
coupling has been discussed in Section 5.  In order to achieve a large 
amplification of the quantum fluctuations of the gauge fields, the gauge coupling should be 
sufficiently small  when the typical scale 
of the gravitational collapse hits the ``potential barrier''. In particular, 
$g_{\rm ex}(\omega_{\rm G}) < 10^{-33}$ is the dynamo requirement has to be satisfied. 
It is difficult to produce large scale magnetic fields with reasonable amplitudes if 
the pre-big bang phase matches immediately to the 
post-big bang evolution \cite{lemoine}. However, phenomenologically 
consistent pre-big bang models lead to a sufficiently long stringy 
phase when the dilaton and the curvature scale are 
roughly constant. If this phase is included the conditions expressed by Eqs. 
(\ref{dyn}) and (\ref{dyn2}) are easily satisfied. Furthermore, there are regions in 
the parameter space of the model where $r_{\rm B}(\omega_{\rm G}) \sim 10^{-8}$ 
which implies that the galactic magnetic field may be fully primordial \cite{mgst1,mgst2},
i.e. no dynamo action will be required.

\subsection{Inside the Hubble radius}

The experimental evidence concerning large-scale magnetic fields 
suggests that magnetic fields should have similar strength 
over different length-scales. In this 
sense inflationary mechanisms seem to provide 
a rather natural explanation for the largeness of the correlation scale.
At the same time, the typical amplitude of the obtained seeds is, in 
various models, rather minute.

Primordial magnetic fields can also 
be generated through physical mechanisms operating 
inside the Hubble radius at a given physical time.
Particularly interesting moments in the life of the 
Universe are the epoch of the electroweak phase transition 
(EWPT) or the QCD phase transition  where 
magnetic fields may be generated according to different 
physical ideas. In the following 
the different proposals emerged so far will be reviewed.

\subsubsection{Phase transitions}

At the time of the EWPT the typical size of the Hubble 
radius is of the order of $3$ cm and the 
temperature is roughly $100$ GeV. Before 
getting into the details of the possible electroweak 
origin of large-scale magnetic fields 
it is useful to present a
kinematical argument based on the evolution of the correlation scale 
of the magnetic fields \cite{son}.

Suppose that, thanks to some mechanism, sufficiently 
large magnetic fields compatible with the critical density 
bound are generated inside the Hubble radius at the electroweak epoch. 
Assuming that the typical coherence length of the 
generated magnetic fields is maximal,
the present correlation scale will certainly be much larger but, unfortunately, it does 
not seem to be as large as the Mpc scale at the epoch of the 
gravitational collapse. As already mentioned, the 
growth of the correlation scale 
may be enhanced, by various processes such as 
inverse cascade and helical inverse  cascade.
For instance, if the injection spectrum 
generated at the electroweak epoch is 
Gaussian and random a simple estimate shows 
that the present correlation scale is of the order 
of $100$ AU which is already larger than what 
one would get only from the trivial expansion
of length-scales (i.e. $1$ AU) \cite{son}. 
If, in a complementary 
perspective, the injection spectrum is strongly 
helical, then the typical correlation scale 
can even be of the order of $100$ pc but still 
too small than the typical scale of the gravitational 
collapse. 

Large-scale magnetic fields can be generated 
at the electroweak epoch in various ways. Consider 
first the case when the phase transition 
is strongly first order.

Hogan \cite{hogan} originally suggested the idea that 
magnetic fields can be generated during first-order phase 
transitions. Since during the phase transition there are gradients in the 
radiation temperature, similar thermoelectric 
source terms of MHD equations (which were discussed in the context 
of the Biermann mechanism) may arise. The magnetic fields, initially 
concentrated on the surface of the bubbles, are expelled 
when bubbles collide thanks to the finite 
value of the conductivity.

The idea that charge separation can be generated during first-order 
phase transitions has been exploited in \cite{baym3}. The suggestion 
is again that there are baryon number gradients at the phase boundaries 
leading to thermoelectric terms. In the process of bubble 
nucleation and collisions turbulence is then produced. In spite 
of the fact that the produced fields are sizable, the correlation scale, as 
previously pointed out, is constrained to be smaller than $100$ pc.

In a first-order phase transition the phases of the complex order parameter of the 
nucleated bubbles are not correlated. When the bubbles undergo 
collisions a phase gradient arises leading to a source terms for the 
evolution equation of the gauge fields. Kibble and Vilenkin \cite{kibblevil}
proposed a gauge-invariant difference between the phases of the 
Higgs field in the two bubbles. This idea has been investigated in the context of the Abelian-Higgs 
model \cite{copeland1,copeland2,ahonen2}. The collision of two spherical bubbles 
in the Abelian-Higgs model leads to a magnetic field which is localized in the 
region at the intersection of the two bubbles. The estimate of the strength of the field 
depends crucially upon the velocity of the bubble wall. 
The extension of  this idea to the case of the standard model $SU(2)\times U(1)$ 
has been discussed in \cite{torn2}. A relevant aspect to be mentioned is 
that the photon field in the broken phase of the electroweak 
theory should be properly defined. In \cite{kocian} 
it has been shown that the definition employed in \cite{torn2} 
is equivalent to the one previously discussed in \cite{vlasov2}.

It has been argued by Vachaspati \cite{vachaspati1} that magnetic fields can be 
generated at the electroweak time even if the phase transition 
is of second order. The observation is that, provided the Higgs field 
fluctuates, electromagnetic fields may be produced since the gradients of
the Higgs field appear in the definition of the photon field 
in terms of the hypercharge and $SU(2)$ fields.
Two of the arguments proposed in \cite{vachaspati1}  have been scrutinized in subsequent 
discussions. 
The first argument is related to the averaging which should 
be performed in order to get to the magnetic field relevant 
for the MHD seeds. Enqvist and Olesen \cite{enqvistolesen}  
noticed that if line averaging is relevant the obtained magnetic field 
is rather strong. However \cite{enqvistolesen} (see also \cite{everett}) 
volume averaging is the one relevant for MHD seeds. 

The second point is related to the fact that the discussion of \cite{vachaspati1}
was performed in terms of gauge-dependent quantities. The problem is then to give a 
gauge-invariant definition of the photon field in terms of the standard model fields. 
As already mentioned this problem has been addressed in \cite{torn2} and the proposed  
gauge-invariant is equivalent \cite{kocian} to the one proposed in \cite{vlasov2}.

In \cite{shap5} a mechanism for the generation of magnetic fields at the 
electroweak epoch has been proposed in connection with 
the AMHD equations. The idea is to study the conversion of the 
right-electron chemical potential into hypercharge fields. In this 
context the baryon asymmetry is produced at some epoch prior to 
the electroweak phase. The obtained magnetic fields are 
rather strong (i.e. $|\vec{B}| \sim 10^{22} ~{\rm G}$ at the EW epoch) but over a small scale , i.e. 
$10^{-6} H_{\rm ew}^{-1}$ dangerously close to the diffusivity scale.

The final point to be mentioned is that, probably, the electroweak 
phase transition is neither first order nor second order but it is 
of even higher order at least in the context of the minimal standard 
model \cite{PT1,PT2}. This conclusion has been reached 
using non-perturbative techniques and the relevant point, in the present context, 
is that for Higgs boson masses larger than $m_{W}$ the phase transition seems 
to disappear and it is possible to pass from the symmetric to the broken phase without 
hitting any first or second order phase transition.

There have been also ideas concerniing the a possible generation 
of magnetic fields at the time of the QCD phase transition occurring 
roughly at $T\sim 140 {\rm MeV}$, i.e. at the moment when free 
quarks combine to form colorless hadrons. The mechanism here 
is always related to the idea of Biermann with thermoelectric 
currents developed at the QCD time.Since the strange quark is heavier than the
up and down quarks there may be the possibility that the quarks develop a 
net positive charge which is compensated by the electric charge in the leptonic sector.
Again, invoking the dynamics of a first-order phase transition, it is argued that 
the shocks affect leptons and quarks in a different way so that electric currents 
are developed as the bubble wall moves in the quark-gluon plasma.  
In \cite{spergel} the magnetic field has been estimated to be $|\vec{B}|\sim {\rm G}$ 
at the time of the QCD phase transition and with 
typical scale of the order of the meter at the same epoch.

In \cite{cheng1,sigl2} it has been pointed out that, probably, 
the magnetic fields generated at the time of QCD phase transition may be 
much stronger than the ones estimated in \cite{spergel}.
The authors of \cite{cheng1,sigl2} argue that strong magnetic fields 
may be generated when the broken and symmetric phase of the theory 
coexist. The magnetic fields generated at the boundaries between quark and hadron phases 
can be, according to the authors, as large as $10^{6}$ G over scales of the 
order of the meter at the time of the QCD phase transition.

Recently, in a series of papers, Boyanovsky, de Vega and Simionato \cite{bd2,bd3,bd4} 
studied the generation of large scale magnetic fields during a phase transition
taking place in the radiation dominated epoch. The setting is a theory 
of $N$ charged scalar fields coupled to an Abelian gauge field that 
undergoes a phase transition at a critical temperature much larger than the 
electroweak scale. Using non-equilibrium field theory techniques the authors argue that 
during the scaling regime (when the back-reaction effects are dominant) 
large scale magnetogenesis is possible. The claim is that the minimal dynamo requirement of Eq. (\ref{dyn}) 
is achievable at the electroweak scale. Furthermore, much larger magnetic fields 
can be expected if the scaling regime can be extended below the QCD scale. 
\subsection{Mixed mechanisms}

From the above discussion it is apparent that the mechanisms 
producing magnetic fields at some moment in the life of the Universe 
(and inside the Hubble radius) have to justify 
the large correlation scale of the magnetic seeds required by subsequent MHD 
considerations. The possibility of turbulent behaviour in the early 
Universe, even though physically justified, seems to be 
difficult to quantify even if rather reasonable quantitative 
estimates appeared so far. 
On the other hand, inflationary scenarios can 
lead to magnetic seeds with large correlation scale.

A useful point to be stressed is that there may exist 
``mixed'' mechanisms. Consider the following example
which has been proposed in \cite{hypmg1} and \cite{hypmg2}.
Suppose that during the inflationary phase a 
stochastic background of gauge fields is produced thanks 
to the breaking of conformal invariance. 
The  spectral amplitude of the generated field will be 
 constrained by the 
critical density bound and by all the other dynamical
considerations related to the finite values of the 
diffusion scales.

The generated magnetic field spectrum will be 
distributed over a large interval of frequencies. In particular, 
the energy spectrum of the magnetic field will be 
non negligible at some intermediate scale, like, for instance, the
electroweak scale, i.e. the physical frequency
$\omega_{\rm ew} \sim H_{\rm ew}$. In other words, if magnetic fields 
are generated thanks to some inflationary mechanism, the amplified magnetic 
inhomogeneities will be reentering at different epochs during the radiation 
dominated phase and the scales which left the horizon later during inflation will reenter earlier.

Given the background of abelian gauge fields, the dynamics of the 
electroweak epoch may add interesting 
physical features. For instance the magnetic field of inflationary origin
will be, as discussed, topologicaly trivial: its magnetic gyrotropy will vanish.
However, during the EW epoch, thanks to a pseudoscalar vertex due either 
to a pseudo goldstone boson or to the chemical pootential, the topologically trivial 
hypermagnetic background may be turned into a topologically non-trivial
background and large magnetic helicity may be generated.

\section{Effects of primordial magnetic fields}
\renewcommand{\theequation}{7.\arabic{equation}}
\setcounter{equation}{0}
In the context of various mechanisms, the magnetic fields generated at large
at the scale of the protogalactic 
collapse may be rather small. In this Section 
the possible effects of magnetic present over much smaller 
length-scales will be analyzed. Two important 
scales are the Hubble radius at the electroweak epoch and at the 
big-bang nucleosynthesis epoch.

\subsection{Electroweak epoch}
Hypermagnetic fields present for temperatures $ T \geq 100 {\rm GeV}$ 
have a twofold effect:
\begin{itemize}
\item{} they may affect the phase diagram of the electroweak theory;
\item{} they may offer a mechanism to seed the baryon asymmetry 
of the Universe (BAU).
\end{itemize}

\subsubsection{EW phase diagram}
The physical
picture  of the possible effects of magnetic fields 
on the electroweak phase diagram is exactly the same as 
the macroscopic description of  superconductors in the
presence of an external magnetic field. The normal-superconducting phase
transition, being of second order in the absence of magnetic
fields, becomes  of first order if a  magnetic field is present. The reason
for this is the Meissner effect, i.e.  the fact that the magnetic field cannot
propagate inside a superconducting cavity, and, therefore, creates
 an extra pressure acting on the normal-superconducting boundary. 
Our consideration below explores this simple picture.

Consider the plain domain wall that separates broken and symmetric
phase at some temperature $T$, in the presence of a uniform 
hypercharge magnetic field $Y_{j}$. Far from the domain wall, in
the symmetric phase, the non-Abelian SU(2)  field strength 
($W_{j}^{3}$) is equal to
zero, because of a non-perturbative mass gap generation. Inside the
broken phase, the massive $Z_{j}$ combination of $Y_{j}$  and $
W^{3}_{j}$,
\begin{equation}
Z_{j} = \cos{\theta_{W}} W_{j}^{3} - \sin{\theta_{W}}Y_{j}
\end{equation} 
must be equal to zero, while the massless combination, corresponding
to photon $A_{j}^{em}$, survives. The matching of the fields on the
boundary gives $A^{em}_{j} = {\cal Y}_{j} \cos{\theta_{W}}$. Thus, an
extra pressure $\frac{1}{2}|\vec{{\cal H}}_{Y}|^2\sin^2\theta_{W}$
acts on the domain wall from the symmetric side. At the critical
temperature it must be compensated by the vacuum pressure of the
scalar field. If we neglect loop corrections associated with the
presence of magnetic fields, then the condition that determines the
critical temperature is:
\begin{equation}
\frac{1}{2}|\vec{ H}_{Y}|^2\sin^2\theta_{W} = 
V(0,T_c) - V(\varphi_{\rm min},T_c)~~,
\label{pressure}
\end{equation}
where $ V(\varphi,T)$ is the effective potential in the absence of
magnetic field, $\varphi_{\rm min}$ is the location of the minimum of the
potential at temperature $T$.

The above consideration was dealing with the uniform magnetic
fields. Clearly, it remains valid when the typical distance scale of
inhomogeneities of  the magnetic field are larger than the typical bubble
size. This is the case for bubbles smaller than the magnetic
diffusion scale, and, in particular, at the onset of the bubble
nucleation. Thus, the estimate of the critical temperature coming from
(\ref{pressure}) is applicable. For bubbles larger than the
diffusivity scale, the presence of a stochastic magnetic field will
considerably modify their evolution. In particular, the spherical form
of the bubbles is very likely to be spoiled. 

These considerations were presented in \cite{mg3}. Later \cite{kainul} 
perturbative estimates were performed in order to corroborate the proposed 
picture. In \cite{magnpt} a full non-perturbative analysis 
of the phhase diagram of the electroweak theory in the presence of an hyermagnetic
background has been performed. As previously discussed \cite{PT1,PT2} 
for values of the Higggs boson mass larger than the W boson mass 
the electroweak phase diagram seems to exhibit a cross-over regime. 
The inclusion of a constant hypermagnetic background with typical 
strength $|\vec{ H}_{Y}|/T^2 \leq 0.3$ does modify the electroweak 
phase diagram but does not seem to make the phase transition 
strongly first order for $m_{H} \geq m_{W}$ as expected from perturbative considerations. 
Furthermore for  $|\vec{H}_{Y}|/T^2 > 0.3$  (but still compatible with the critical density bound) 
a new (inhomogeneous) phase has been observed. 
The analysis performed in \cite{magnpt} included a net hypermagnetic flux. It would be 
interesting to repeat the same calculation in the presence of a non-vanishing hypermagnetic helicity (or gyrotropy).

\subsubsection{Baryon asymmetry}
Depending upon the topology 
of the flux lines, hypermagnetic fields can have two distinct effects:
\begin{itemize}
\item{} topologically non-trivial 
configurations of the hypermagnetic flux lines lead to the formation 
of hypermagnetic knots  whose decay 
might seed the Baryon Asymmetry of the Universe (BAU);
\item{} even if the electrowak plasma has  vanishing hypermagnetic 
gyrotropy, i.e.  $\langle\vec{H}_{Y} \cdot\vec{\nabla} 
\times\vec{H}_{Y}\rangle =0$, matter--antimatter fluctuations may be generated.
\end{itemize}
Consider first the situation where the electroweak plasma contains, for $T > T_{c}$ 
a network of hypermagnetic knots of the type of the ones described in 
Eqs. (\ref{conf1})--(\ref{knot}). In Section 5 these configurations 
have been named hypermagnetic knots (HK) and they are 
Chern-Simons condensates carrying a non-vanishing (averaged) 
hypermagnetic helicity. As discussed, HK can be dynamically generated 
from a background of hypermagnetic fields with trivial topology provided 
a (time-dependent) pseudo-scalar is present in the plasma (see, for instance,
Eq. (\ref{action})).

In order to seed the BAU a network of HK should be present at high
temperatures. In fact
for temperatures larger than $T_{c}$
 the fermionic number is stored both in HK 
and in real fermions.  For $T<T_{c}$, 
the HK should release real fermions 
since the ordinary magnetic fields (present {\em after} EW 
symmetry breaking) do not carry fermionic number.
If the EWPT is strongly first order the decay of the HK 
can offer some seeds for the BAU generation.
This last condition can be met in the 
minimal supersymmetric standard model (MSSM).

The integration of the $U(1)_{Y}$ 
anomaly equation 
gives the CS number density carried by the HK
which is in turn related to the density of baryonic number $n_{B}$
for the case of $n_{f}$ fermionic generations. In fact, using 
Eqs. (\ref{nr})--(\ref{ohm}) and after some algebra \cite{mg3,hypmg2}
it can be shown that 
\begin{equation}
\frac{n_{B}}{s}(t_{c})=
\frac{\alpha'}{2\pi\sigma_c}\frac{n_f}{s}
\frac{\langle{\vec{H}}_{Y}\cdot \vec{\nabla}\times
\vec{H}_{Y}\rangle}{\Gamma + \Gamma_{H}}
\frac{M_{0}\Gamma}{T^2_c},~~\alpha' = \frac{g'^2}{4\pi}
\label{BAU}
\end{equation}
($g'$ is the $U(1)_{Y}$ coupling and $s = (2/45) \pi^2 g_{\ast}T^3$ 
is the entropy density; $g_{\ast}$, at $T_{c}$,
 is $106.75$ in the MSM;
$M_{0}= M_{P}/1.66 \sqrt{g_{\ast}}$).
In Eq. (\ref{BAU}) $\Gamma$ is the perturbative rate of the 
right electron chirality 
flip processes  (i.e. 
scattering of right electrons with the Higgs and gauge bosons and with 
the top quarks because of their large Yukawa coupling) which 
are the slowest reactions in the plasma and 
\begin{equation}
\Gamma_{H} = \frac{783}{22} \frac{\alpha'^2}{\sigma_{c} \pi^2} 
\frac{|\vec{H}_{Y}|^2}{T_{c}^2}
\end{equation}
is the rate of right electron dilution induced by the presence of a
 hypermagnetic field. In the MSM we have that 
$\Gamma < \Gamma_{H}$  
whereas in the MSSM $\Gamma$ can naturally 
be larger than $\Gamma_{H}$. 
Unfortunately, in the MSM 
a hypermagnetic field can modify the phase diagram of the phase transition 
but cannot make the phase transition strongly first order for large masses of
the Higgs boson.
Therefore, we will concentrate on the case $\Gamma > \Gamma_{H}$ and we
 will show that in the opposite limit the BAU will be anyway small 
even if some (presently unknown) mechanism would make the EWPT strongly 
first order in the MSM.

It is interesting to notice that, in this scenario, the value 
of the BAU is determined by various particle physics parameters but also 
by the ratio of the hypermagnetic energy density over the 
energy density sitting in radiation during 
the electroweak epoch.

Consider now the complementary situation where the electroweak plasma, for $T> T_{c}$, 
is filled with topologically trivial hypermagnetic fields.  In this 
case, fluctuations in the baryon to entropy ratio
will be induced since, in any case 
$\langle(\vec{H}_{Y} \cdot \vec{\nabla} 
\times \vec{H}_{Y})^2 \rangle \neq 0$.
These fluctuations are of {\em isocurvature}
type and can be  related to the spectrum 
of hypermagnetic fields at the EWPT. 
Defining as 
\begin{equation}
\Delta(r,t_{c}) = \sqrt{\langle\delta
\left(\frac{n_{B}}{s}\right)(\vec{x}, t_{c})
\delta\left(\frac{n_{B}}{s}\right)(\vec{x}+\vec{r},
t_{c})\rangle}, 
\label{13}
\end{equation}
the fluctuations in the ratio of baryon number density $n_{B}$  to the entropy density $s$ 
at $t={t_c}$ \cite{mg4}, the value of $\Delta(r,t_{c})$
can be related to the hypermagnetic spectrum
which is determined in terms of its amplitude 
$\xi$ and its slope $\epsilon$.
A physically realistic situation corresponds to the case in
which the Green's functions of the magnetic hypercharge fields decay at
large distance (i. e. $\epsilon> 0$) and this would
imply either ``blue''( $\epsilon \geq 0$ ) or ``violet''
($\epsilon \gg 1$) energy spectra. 
The fluctuations of the baryon to entropy ratio generated 
at the electroweak epoch may survive until the BBN epoch \cite{mg4}. 
The possibility of survival of these fluctuations is related 
to their typical scale which must exceed the neutron 
diffusion scale appropriately blue-shifted at the 
electroweak epoch. The implications of these fluctuations will 
be discussed in a moment.

\subsection{Big-bang nucleosynthesis epoch}

Large scale magnetic fields possibly present at the BBN epoch 
can have an impact on the light nuclei formation. By reversing 
the argument, the success of BBN can be used in order 
to bound the magnetic energy density possibly present at 
the time of formation of light nuclei.
Magnetic fields, at the BBN epoch may have four distinct  effects:
\begin{itemize}
\item{} they can enhance the rate of the Universe expansion 
at the corresponding epoch in a way proportional to $\rho_{\rm B}$;

\item{} they can affect the reaction rate in a way proportional
to $\alpha_{\rm em} \rho_{\rm B}$;

\item{} they can affect the phase space of the electrons;

\item{} they can induce, prior to the formation of the light 
nuclei, matter--antimatter fluctuations.
\end{itemize}
While the first three effects are direct, the fourth effect 
is mainly caused by hypermagnetic fields 
present even before BBN.

The idea that magnetic fields may increase the Universe 
expansion and, consequently, affect directly the abundance 
of $^{4} {\rm He}$ (which is 
directly sensitive to the expansion rate) has been 
pointed out long ago by Greenstein \cite{bbn1} and 
by Matese and O'Connel \cite{bbn2}. From a qualitative 
point of view the effects over the expansion 
should be leading in comparison with the 
effects over the rate of interactions (which are suppressed by $\alpha_{\rm em}$).

Thanks to detailed numerical simulations performed 
indipendently by different groups (i.e. Cheng et al. \cite{bbn3,bbn3a}; Kernan, Starkman and 
Vachaspati \cite{bbn4,bbn4a}; Grasso and Rubinstein \cite{bbn5,bbn5a}), the common opinion is that 
the Universe expansion is the leading effect even if this conclusion has been reached 
only recently and not without some disagreements on the details of the calculations (see, for instance,
\cite{bbn4a}).
 
In order to prevent the Universe 
from expanding too fast at the BBN epoch 
$\rho_{B} < 0.27 \rho_{\nu}$  where  $\rho_{\nu}$ is the 
energy density contributed by the standard three light neutrinos with masses much smaller 
than the MeV. The BBN bound  is physically relevant in many situations since it is 
a bit more constraining than the critical density bound.  

The BBN bounds discussed so far are derived assuming a stochastic magnetic field at the BBN 
epoch. This means that the isotropy of the geometry is not affected. However, 
it is not unreasonable to think of the possibility 
that a magnetic field with a preferred direction may break the isotropy 
of the background. Thus bounds on the isotropy of the expansion at the BBN epoch 
may be turned into bounds on the shear induced by the presence of a magnetic field (see, for instance, 
\cite{iso1,iso2} and \cite{iso3,iso4}). 

\begin{figure}[htb]
    \centering
    \includegraphics[height=2.5in]{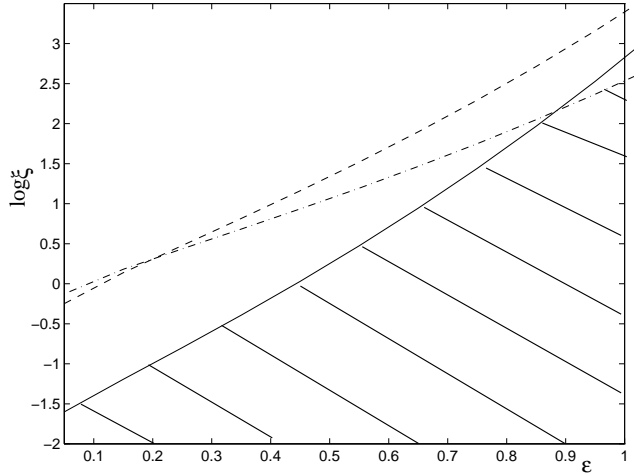}
    \caption{The parameter space of the hypermagnetic 
background in the case 
$\Delta(r, t_c) < n_{B}/s$ for $r> r_{\rm n}$ (full line).}
    \label{F5}
\end{figure}

Let us now come to the last point. The matter--antimatter fluctuations 
created at the electroweak epoch thanks to the presence of hypermagnetic fields 
may survive down to the epoch of BBN.
Since the fluctuations in the baryon to entropy
ratio are, in general, not positive definite, they will induce 
fluctuations in the baryon to photon ratio,
 at the BBN epoch. The possible effect of 
matter--antimatter fluctuations on BBN 
depends on the typical scale of of the 
baryon to entropy ratio at the electroweak epoch.
Recalling that for $T\sim T_{c}\sim 100 $ GeV 
the size of the electrowek horizon is 
approximately $3 $ cm, fluctuations whose 
scale is well inside the EW horizon at 
$T_{c}$ have dissipated by the BBN 
time through (anti)neutron diffusion.
The neutron diffusion scale at $T_{c}$ is 
\begin{equation}
 r_{\rm n}(T_{c})  =0.3 ~~{\rm cm}.
\end{equation}
 The neutron diffusion scale
at $T=1 ~{\rm keV}$ is  $10^{5}$ m, while, today 
it is $10^{-5}$ pc, i.e. of the order of the astronomical unit.
Matter--antimatter fluctuations smaller than $10^{5}$ m annihilate 
before neutrino decoupling and have no effect on BBN.
\begin{figure}
\begin{center}
\begin{tabular}{|c|c|}
      \hline
      \hbox{\epsfxsize = 6 cm  \epsffile{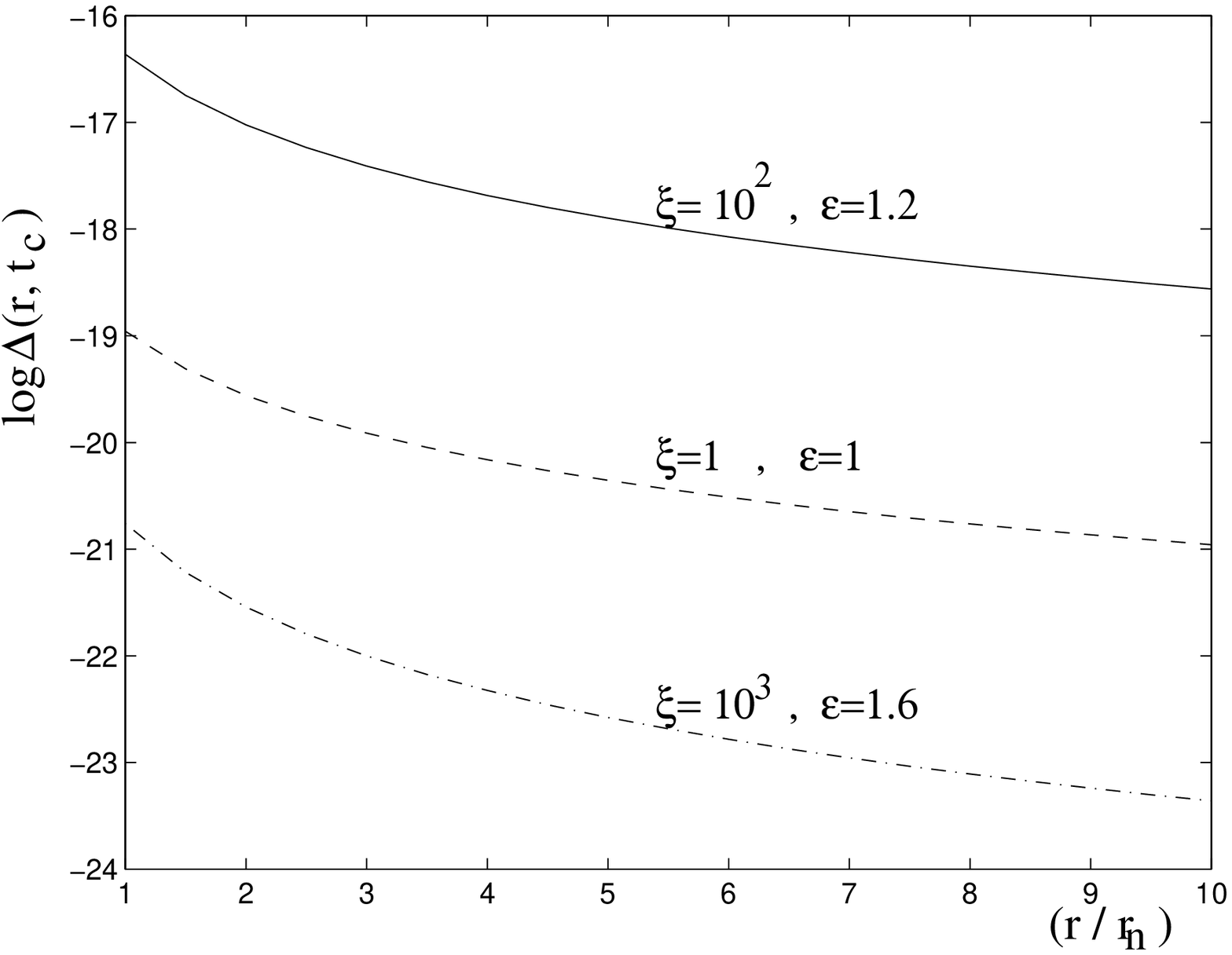}} &
      \hbox{\epsfxsize = 6 cm  \epsffile{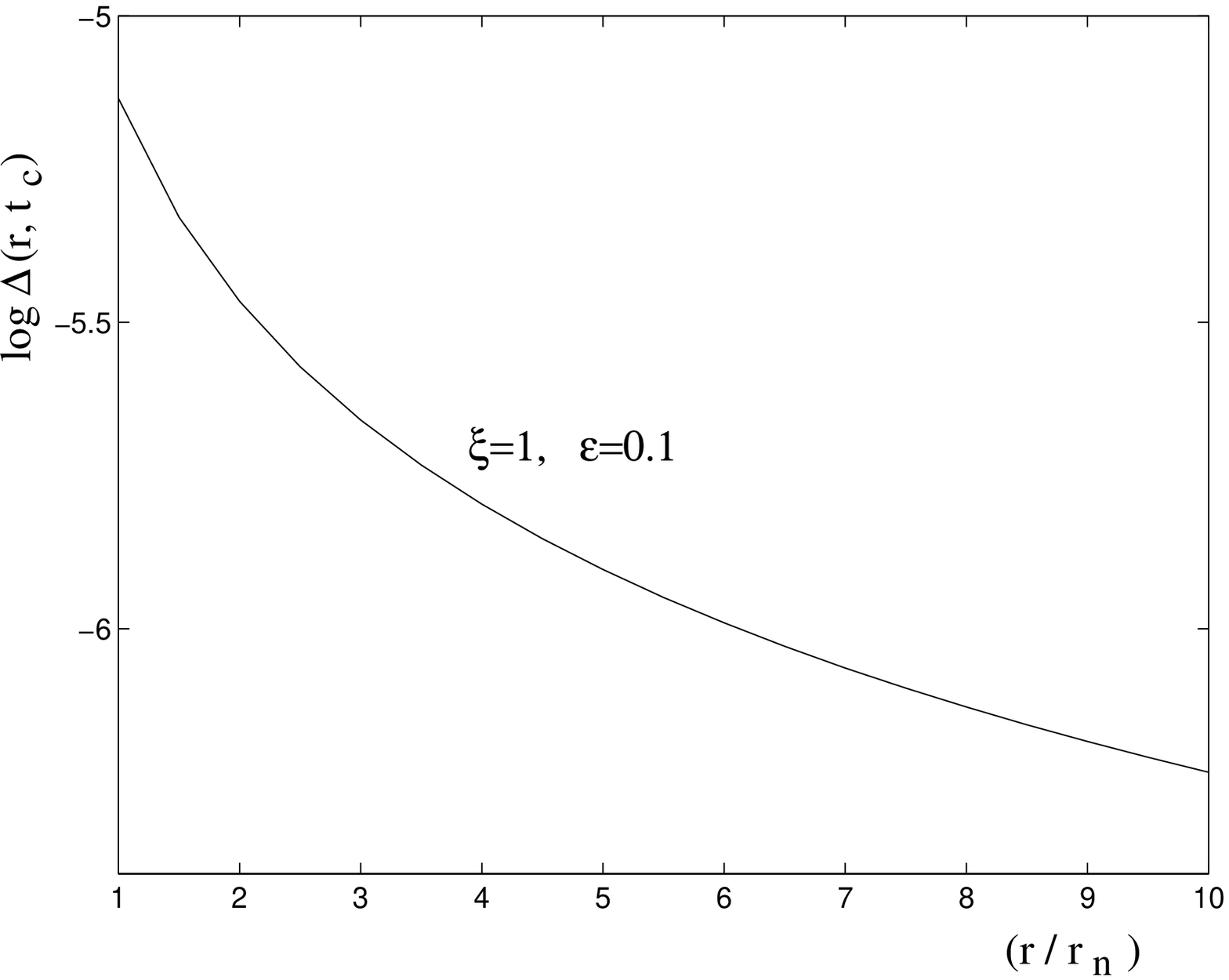}}\\
      \hline
\end{tabular}
\end{center}
\caption{The value of the baryon number 
fluctuations for different parameters of the hypermagnetic
background $\xi$ and $\epsilon$ is reported. }
\label{ad}
\end{figure}
Two possibilities can then be envisaged. 
We could require that the matter--antimatter 
fluctuations (for scales $r\geq r_{\rm n}$) 
are small. This will then imply a 
bound, in the ($\xi$,$\epsilon$) plane on the strength 
of the hypermagnetic background. 
In Fig. \ref{ad} some typical baryon number fluctuations 
have been reported for different choices of the parameters $\xi$ and $\epsilon$.
In Fig. \ref{F5} 
such an exclusion plot is reported with the full
line. With the dashed line the bound implied by 
the increase in the expansion rate 
(i.e. $\rho_{H} < 0.2 \rho_{\nu}$) is reported. 
Finally with the dot-dashed line the 
critical density bound is illustrated 
for the same hypermagnetic background.
The second possibility is to study the  effects 
of large matter--matter domains. 
These studies led to a slightly different 
scenario of BBN \cite{mam1}, namely BBN 
with matter--antimatter regions.
This analysis has been 
performed in a series of papers by Rehm and Jedamzik \cite{mam2,mam5} and by 
Kurki-Suonio and Sihvola \cite{mam3,mam4} (see also \cite{mam4a,mam4b}). 
The idea is to discuss BBN in the presence 
of spherically symmetric regions 
of anti--matter characterized by their 
radius $r_{\rm A}$ and by the parameter $R$, i.e. 
the matter/antimatter ratio.
Furthermore, in this scenario the net baryon-to-photon
ratio,  $\overline{\eta}$, is positive definite and non zero. 
Antimatter domains larger than $10^{5}$ m at $1$ keV may survive 
until BBN and their dissipation has been 
analyzed in detail \cite{mam4}.
Antimatter domains  in the range 
\begin{equation}
10^{5} ~~{\rm m}~ \leq 
r_{\rm A} ~\leq~ 10^{7} ~~{\rm m}
\label{range}
\end{equation}
 at 1 keV annihilate 
before BBN for temperatures between 70 keV and $1$ MeV.
Since the antineutrons annihilate on neutrons, the neutron 
to proton ratio gets smaller. As a consequence, the 
$^{4}{\rm He}$ abundance gets reduced if compared to the 
standard BBN scenario. The maximal scale of matter--antimatter
fluctuations is determined by the constraints following 
from possible distortions of the CMB spectrum.
The largest scale is of the order of $100$ pc (today), corresponding 
to $10^{12}$ m at $1$ keV.  
\begin{figure}[htb]
    \centering
    \includegraphics[height=2.5in]{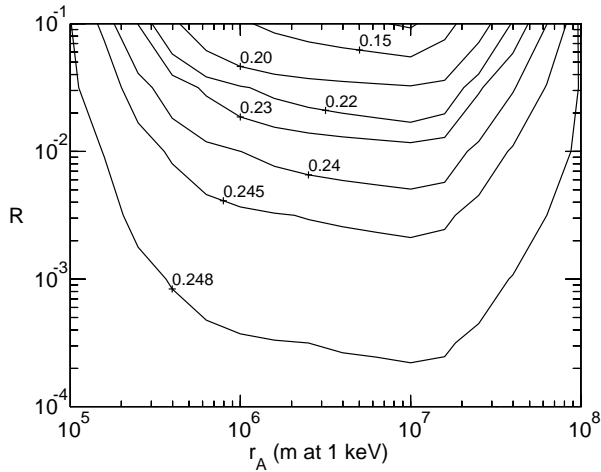}
    \caption{From \cite{mam6} the $^{4}{\rm He}$ yield is illustrated 
in the $(R,r_{\rm A})$ plane for $\overline{\eta} = 6\times 10^{-10}$. As the 
matter/antimatter ratio decreases, we recover the 
standard $^{4}{\rm He}$ yield.}
    \label{F6}
\end{figure}
Suppose that matter--antimatter regions are present
in the range of Eq. (\ref{range}). Then the abundance 
of $^{4}{\rm He}$ get reduced. The yield of $^{4} {\rm He}$ are 
reported as a function of $R$, the matter--antimatter ratio
and $r_{\rm A}$. Now, we do know that by adding 
extra-relativistic species the $^{4} {\rm He}$ can be 
increased since the Universe expansion gets larger. 
Then the conclusion is that BBN with 
matter--antimatter domains allows for a larger number 
of extra-relativistic species if compared to the 
standard BBN scenario. This observation 
may have implications for the upper bounds on the stochastic 
GW backgrounds of cosmological origin \cite{mam6} since 
the extra-relativistic species present at the BBN
epoch can indeed be interpreted as relic gravitons.

\section{CMB physics an magnetic fields}
\renewcommand{\theequation}{8.\arabic{equation}}
\setcounter{equation}{0}
The vacuum fluctuations of the gauge fields present during 
the inflationary stage of expansion may be 
amplified if conformal invariance is broken. It is 
then plausible that magnetic inhomogeneities are amplified not 
only at the scale of protogalactic collapse but also at larger length-scales.
agnetic fields can be  generated over all physical scales compatible 
Large-scale magnetic fields may then have various interesting 
implications on the physics of the CMB and of its anisotropies. 
The following possible effects have been 
discussed through the years:
\begin{itemize}
\item{} distortion of the Planckian spectrum;
\item{} shift of the polarization plane (provided the CMB is linearly polarized);
\item{} shift in the position of the first Doppler peak;
\item{} generic increase of the amount of the (primary)  anisotropy.
\end{itemize}
On top of these effects, magnetic fields 
can also modify the evolution of the tensor fluctuations of the geometry 
for typical length scales much smaller than the ones 
probed by CMB anisotropy experiments. 

The possible distortions of the Planckian spectrum of CMB are 
usually discussed in terms of a chemical potential 
which is bounded, by experimental data, to me $|\mu| < 9\times 10^{-5}$. 
Magnetic field dissipation 
at high red-shift implies the presence of a chemical potential.
Hence bounds on the distorsion of the Planckian spectrum of the CMB can be turned into bounds
on the magnetic field strength at various scales. In particular \cite{jed} the 
obtained bound are such that $B < 3\times 10^{-8}$ G for comoving 
coherence lengths between $0.4$ kpc and $500$ kpc.

Large scale magnetic fields can also afffect the poosition of the Doppler peak.
In \cite{rg} this analysis has been performed in a non-relativistic approximation
where the scalar perturbations of the geometry obey linearized Newtonian 
equations of motion. It has been found that, in this approximation, the 
effect of the presence of the magnetic fields is an effective 
renormalization of the speed of sound of the baryons.

\subsection{Large-scale magnetic fields and CMB anisotropies}

Large scale magnetic fields present prior to 
the recombination epoch may act as a source term in the evolution 
equation of the cosmological perturbations. If 
magnetic fields are created over large length scales 
(possibly even larger than the Mpc) it is rather plausible that they 
may be present after matter-radiation equality  (but before decoupling)
when  primordial fluctuations are imprinted on the CMB 
anisotropies. In fact, the relevant modes determining 
the large scale temperature anisotropies are the ones that are outside the horizon 
for $\eta_{\rm eq} < \eta < \eta_{\rm dec}$ where $\eta_{\rm eq}$ and $\eta_{\rm dec}$ are, respectively, the equality 
and the decoupling times. After equality there are two complementary 
approaches which can be used in order to address phenomenological implications
of large scale magnetic fields:
\begin{itemize}
\item{} large scale magnetic fields are completely {\em homogeneous} outside 
the horizon after equality;
\item{} large scale magnetic fields are {\em inhomogeneous} after equality.
\end{itemize}
The idea that {\em homogeneous} magnetic fields 
may affect the CMB anisotropies was originally pointed out by Zeldovich \cite{zelm1,zelm2} and 
further scrutinized by Grishchuk, Doroshkevich and Novikov \cite{GDN}.
Since this idea implies that large-scale magnetic fields are 
weakly gravitating the details of the discussion (and 
of the possible generalization of this idea) will be postponed to the following Section.

The suggestion that {\em inhomogeneous} magnetic fields  created during inflation may 
affect the CMB anisotropies was originally proposed in  \cite{mgst2}.
In \cite{mgst2} it was noticed that large scale 
magnetic fields produced during inflation may also lead to 
large fluctuations for modes which are 
outside the horizon after equality. In \cite{mgst2} it was noticed that 
either the produced magnetic fields may be used to seed directly the CMB 
anisotropies, or, in a complementary persepective, the CMB data can be used 
to put constraints on the production mechanism. 

If large-scale magnetic fields are inhomogeneous, their energy momentum tensor 
will have, in general terms, scalar, vector and tensor modes
\begin{equation}
\delta {\cal T}_{\mu\nu} = \delta {\cal T}_{\mu\nu}^{(S)}+ \delta {\cal T}_{\mu\nu}^{(V)} + \delta {\cal T}_{\mu\nu}^{(T)},
\label{decT}
\end{equation}
which are decoupled
and which act as source terms for the scalar, vector and tensor modes of the geometry
\begin{equation}
\delta g_{\mu\nu} = \delta g_{\mu\nu}^{(S)} + \delta g_{\mu\nu}^{(V)} + \delta g_{\mu\nu}^{(T)},
\end{equation}
whose gauge-invariant description is well explored \cite{gi1,gi2}(see also \cite{gi3}).
The effect of the scalar component of large scale magnetic fields should be 
responsible, according to the suggestion of \cite{mgst2}, of the 
large scale temperature anisotropies. The direct 
seeding of large-scale temperature anisotropies 
seems  unlikely. In fact, the 
spectrum of large scale gauge fields leading to 
plausible values of the magnetic energy density over the scale of protogalactic 
collapse is typically smaller than the value required by experimental data.
However, even if the primordial spectrum of magnetic fields would be 
tailored in an appropriate way, the initial conditions 
for the plasma evolution compatible with the presence of a sizable (but undercritical) 
magnetic field are of {\em isocurvature} type \cite{durgas}. 

In order to illustrate this point consider
the evolution equations for the gauge-invariant system of scalar 
perturbations of the geometry  expressed in terms 
of the two Bardeen potential $\Phi$ and $\Psi$ corresponding, in the Newtonian 
gauge, to the fluctuations of the temporal and spatial component 
of the metric \cite{gi2,gi3}. The linearization of the $(00)$, $(0i)$ and $(i,j)$ components 
of the Einstein equations leads, respectively, to:
\begin{eqnarray}
&& \nabla^2 \Psi - 3 {\cal H} ( {\cal H} \Phi + \Psi') = \frac{a^2}{2 M_{\rm P}^2} \rho \delta + 
\frac{a^2}{2 M_{\rm P}^2} \delta {\cal T}_{0}^{0},
\label{00}\\
&& \partial_{i}[ {\cal H} \Phi + \Psi'] =  \frac{a^2}{2 M_{\rm P}^2} (1 + w) \rho \partial_{i} u+
 \frac{a^2}{2 M_{\rm P}^2} \delta {\cal T}_{i}^{0},
\label{0i}\\
&& \biggl\{ \Psi'' + {\cal H}( \Phi' + 2 \Psi') + ( {\cal H}^2 + 2 {\cal H}') \Phi + 
\frac{1}{2} \nabla^2(\Phi - \Psi) \biggr\}\delta_{i}^{j} 
\nonumber\\
&& - 
\frac{1}{2} \partial_{i}\partial^{j} (\Phi - \Psi) = -\frac{a^2}{2 M_{\rm P}^2} \delta {\cal T}_{i}^{j} 
+ w \frac{a^2}{2 M_{\rm P}}^2 \rho \delta,
\label{ij}
\end{eqnarray}
where $\delta = \delta\rho/\rho$ and $w=p/\rho$ is the usual barotropic index, and 
\begin{eqnarray}
&& \delta {\cal T}_{0}^{0} = \frac{1}{8 \pi a^4} |\vcB(\vec{x})|^2,
\nonumber\\
&& \delta {\cal T}_{i}^{j} = \frac{1}{4\pi a^4} \biggl[ - {\cal B}_{i} {\cal B}^{j} + \frac{1}{2} |\vcB(\vec{x})|^2 \delta_{i}^{j} \biggr],
\nonumber\\
&&  \delta {\cal T}_{i}^{0} = \frac{\vcB \times \vec{\nabla} \times \vcB}{16 \pi^2 a^4\sigma}.
\label{pertB}
\end{eqnarray}
are the fluctuations of the energy-momentum tensor of the magnetic field.
Eqs. (\ref{00an})--(\ref{ij}) have to be supplemented by the 
perturbation of the covariant conservation equations for the 
fluid sources:
\begin{eqnarray}
&& \delta' - ( 1 + w) \nabla^2 u - 3 \Psi' ( 1 + w) =0,
\label{delta}\\
&& u' + {\cal H} ( 1 - 3 w) u - \frac{w}{w + 1} \delta - \Phi=0.
\label{u}
\end{eqnarray}
If the magnetic field is force-free, i.e. $\vcB \times \vec{\nabla} \times \vcB$ the above system simplifies. Notice that 
Eq. (\ref{u}) has been already written in the case where the Lorentz force is absent. However, the forcing 
term appears, in the resistive MHD approximation, in Eq. (\ref{pertB}). The term  $\vcB \times \vec{\nabla} \times \vcB$
is suppressed by the conductivity, however, it can be also set exactly to zero. In this case, vector 
identities imply
\begin{equation}
(\vcB \times \vec{\nabla} \times \vcB)_{i} = {\cal B}_{m} \partial_{i} {\cal B}^{m} - {\cal B}_{m}\partial^{m} {\cal B}_{i}=0,
\end{equation}
i.e. $  {\cal B}_{m} \partial_{i} {\cal B}^{m} ={\cal B}_{m}\partial^{m} {\cal B}_{i}$. 
With this identity in mind the system of Eqs. (\ref{00})--(\ref{ij}) and (\ref{delta})--(\ref{u}) 
can be written, in Fourier space, as 
\begin{eqnarray}
&& k^2 \Psi_{k} + 3 {\cal H} ( {\cal H} \Phi_{k} + \Psi_{k}') = - \frac{a^2}{2 M_{\rm P}^2} - \frac{ {\cal B}_{0}^2(k)  }{ 8 \pi a^2 M_{\rm P}^2 },
\label{00k}\\
&& \Psi_{k}'' + {\cal H} ( \Phi_{k}' + 2 \Psi_{k}') + ( 2 {\cal H}' + {\cal H}^2) \Psi 
\nonumber\\
&& - \frac{k^2}{3} ( \Phi_{k} - \Psi_{k}) = 
 \frac{a^2}{2 M_{\rm P}^2} w \rho \delta_{k} - \frac{{\cal B}_{0}^2(k)}{48 \pi a^2 M_{\rm P}^2},
\label{ijk}\\
&& k^2 ( \Phi_{k} - \Psi_{k}) = \frac{{\cal B}_{0}^2(k)}{16 \pi a^2 M_{\rm P}^2} {\cal B}_{0}^2(k),
\label{shk}\\
&& \delta_{k}' + k^2 (1 +w) u_{k} - 3 \Psi_{k}' (w +1) =0,
\label{deltak}\\
&& u_{k}' + {\cal H} ( 1 - 3 w) u_{k} - \frac{w}{w+1} \delta_{k} - \Phi_{k}= 0,
\label{uk}
\end{eqnarray}
where 
\begin{equation}
 {\cal B}_{0}^2(k) = \frac{1}{(2 \pi)^3 }\int d^{3}p \vcB(\vec{p}) \cdot \vcB( \vec{k} - \vec{p}) .
\end{equation}

Combining Eqs. (\ref{00k})--(\ref{ijk}) with Eq. (\ref{shk}) the following decoupled equation 
can be obtained
\begin{equation}
\Psi_{k}'' + 3 {\cal H}(1 + w) \Psi_{k}' + k^2 w \Psi_{k} = \frac{{\cal B}_{0}^2(k) }{8\pi a^2 M_{\rm P}^2} \biggl[ \frac{ {\cal H}^2 ( 1 - 3 w) - 2 {\cal H}' }{2 k^2} - w
\biggr].
\label{genpsi}
\end{equation}

Consider, for simplicity, the dark-matter radiation fluid after equality and in the presence of a 
force-free magnetic field. In this case Eq. (\ref{genpsi}) can be immediately integrated:
\begin{eqnarray}
\Psi_{k}(\eta) &=& \Psi_{0}(k) - \frac{\Psi_{1}(k)}{5} \biggl(\frac{\eta_{\rm eq}}{\eta}\biggr)^{5} - \frac{{\cal B}_{0}^2(k)}{8\pi k^2 M_{\rm P}^2} 
\biggl(\frac{\eta_{\rm eq}}{\eta}\biggr)^{4},
\nonumber\\
\Phi_{k}(\eta) &=& \Psi_{0}(k) - \frac{\Psi_{1}(k)}{5} \biggl(\frac{\eta_{\rm eq}}{\eta}\biggr)^{5} - \frac{{\cal B}_{0}^2(k)}{16\pi k^2 M_{\rm P}^2} 
\biggl(\frac{\eta_{\rm eq}}{\eta}\biggr)^{4}.
\label{solpsieq}
\end{eqnarray}
The solutions given in Eq. (\ref{solpsieq}) determine the source of the density and velocity 
fluctuations in the plasma, i.e. 
\begin{eqnarray}
&& \delta_{\rm r}' - \frac{4}{3} \nabla^2 u_{\rm r} - 4 \Psi' =0,
\label{dr1}\\
&& u_{\rm r}' - \frac{1}{4} \delta_{\rm r} - \Phi = 0,
\label{ur1}\\
&& \delta_{\rm m}' - \nabla^2 u_{\rm m} - 3 \Psi' =0,
\label{dm1}\\
&& u_{\rm m}' + {\cal H} u_{\rm m} - \Phi =0,
\label{udm}
\end{eqnarray}
The solution for the  velocity fields and density 
contrasts can be easily obtained by integrating Eqs. (\ref{dr1})--(\ref{udm}) with the source terms 
determined by Eq. (\ref{solpsieq}). The purpose is not to integrate here this
system (see for instance \cite{ggvcmb} for the standard case without magnetic fields). 
Rather it is important to notice that there 
are two qualitatively different situation. The first situation is the one where, 
\begin{equation}
|\Psi_{0}(k) | \ll  \frac{{\cal B}_{0}^2(k)}{8\pi k^2 M_{\rm P}^2} 
\biggl(\frac{\eta_{\rm eq}}{\eta}\biggr)^{4},
\end{equation}
namely the case where the constant mode of the Bardeen potential is negligible if compared to 
the contribution of the magnetic field. The derivative of the Bardeen potential will be, however, non vanishing.
In this case the CMB anisotropies are said to be seeded by isocurvature initial conditions for 
the fluid of radiation and dark-matter present after equality.  
In the  opposite case, nemely 
\begin{equation}
|\Psi_{0}(k) | \gg  \frac{{\cal B}_{0}^2(k)}{8\pi k^2 M_{\rm P}^2} 
\biggl(\frac{\eta_{\rm eq}}{\eta}\biggr)^{4},
\label{secondq}
\end{equation}
the constant mode of the Bardeen potential is provided, for instance, by inflation and 
the magnetic field represent a further parameter to be taken into account in the  analysis 
of CMB anisotropies. Assuming, roughly, that $|\Psi_{0}(k) | \sim 10^{-6}$, Eq. (\ref{secondq}) implies that for the typical 
scale of the horizon at decoupling the critical fraction of magnetic energy density should be smaller than $10^{-3}$.
In the case of Eq. (\ref{secondq}) the leading order relations among the different hydrodynamical 
fluctuations are well known and can be obtained from Eqs.(\ref{dr1})--(\ref{udm})  if the magnetic field 
is approximately negligible, i.e. 
\begin{equation}
k u_{\rm r} \simeq  k u_{\rm m},~~~~ \delta_{\rm r} \simeq (4/3) \delta_{\rm m}.
\label{dmrad}
\end{equation}
This result has a simple physical interpretation, and 
implies the adiabaticity of the fluid perturbations. 
The entropy per matter particle is indeed proportional to
 $S=T^3/n_{\rm m}$, where $n_{\rm m}$ is the number density of 
 matter particles and $T$ is the radiation temperature.
The associated entropy fluctuation, $\delta S$, satisfies  
\begin{equation} 
\frac{\delta S}{S} = \frac{3}{4} \delta_{\rm r} - \delta_{\rm m},
\end{equation}
where we used the fact that $\rho_{\rm r} \sim T^4 $ and that 
$\rho_{\rm m} = m n_{\rm m}$, where $m$ is the typical mass 
of the particles in the matter fluid. Equation (\ref{dmrad}) thus implies  $\delta S/S
=0$, in agreement with the adiabaticity  of the fluctuations.

After having computed the corrections induced by the magnetic field on the (leading) 
adiabatic initial condition the temperature anisotropies can be computed using 
the Sachs-Wolfe effect \cite{sw}.
In terms of the
gauge-invariant variables introduced in the present analysis, the
various contributions  to the Sachs--Wolfe effect, along the $\vec{n}$
direction, can be written as   
\begin{equation}
\frac{\delta T}{T}(\vec{n},\eta_0,x_0) = \biggl[ \frac{\delta_{\rm r}}{4} 
+ \vec{n} \cdot \vec{\nabla} v_{\rm b} 
+ \Phi\biggr](\eta_{\rm dec}, \vec{x}(\eta_{\rm dec})) 
- \int_{\eta_0}^{\eta_{\rm dec}} (\Phi' + \Psi')(\eta , \vec{x}(\eta))
d\eta, 
\label{SW}
\end{equation}
where $\eta_0$ is the present time, and $\vec{x}(\eta)=\vec{x}_0-
\vec{n}(\eta-\eta_0)$ is the 
 unperturbed photon position 
at the time $\eta$ for an observer 
 in $\vec{x}_0$. The term  $\vec{v}_{\rm b}$ is the peculiar velocity 
of the baryonic matter component.
\begin{figure}[ht]
\centerline{\epsfxsize = 8 cm  \epsffile{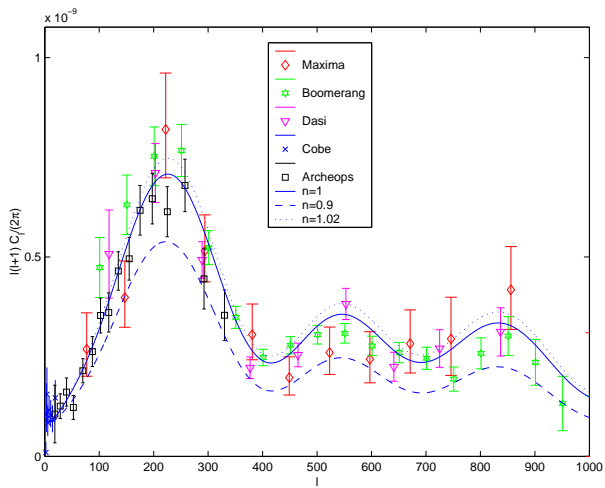}}
\vskip 3mm
\caption[a]{The  spectrum of $C_{\ell}$ is illustrated 
for a fiducial set of parameters ($h_0 =0.65$, $\Omega_{\rm b}=
0.04733$,  $\Omega_{\Lambda} = 0.7$, $\Omega_{\rm m} = 0.25267$) 
and for  flat (full line, $n =1$), slightly red (dashed line, $n =0.9$)
 and slightly blue (dotted line, $n =1.02$) spectral indices for the constant (adiabatic) mode.}
\label{FCL1}
\end{figure}
In Fig. \ref{FCL1} the  $C_{\ell}$ are plotted in the case of models with adiabatic 
initial conditions. 
\begin{figure}
\centerline{\epsfxsize = 8 cm  \epsffile{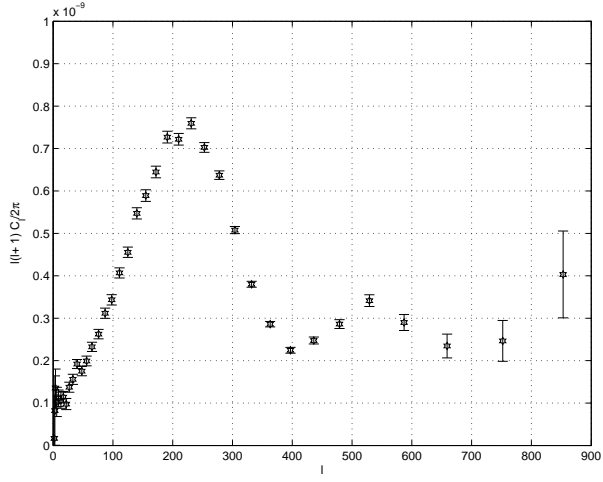}}
\vskip 3mm
\caption[a]{The WMAP data are reported.}
\label{FCL2}
\end{figure}
The data points reported in Figs. \ref{FCL1} and \ref{FCL2} 
are those from COBE \cite{cobe,cob2}, BOOMERANG \cite{boom},
DASI \cite{dasi}, MAXIMA \cite{maxima} and ARCHEOPS \cite{archeops}.
Notice that the data reported in \cite{archeops} fill 
the ``gap'' between the last COBE points and the points of 
\cite{boom,dasi,maxima}. In  Fig. \ref{FCL2} the recent WMAP data are reported \cite{WMAP1,WMAP2,WMAP3}.

Recall,  that the temperature fluctuations are customarily 
discussed in terms of their Legendre transform 
\begin{equation} 
\frac{\delta T}{T}(\vec{x}_0,\vec{n},\eta_0)
=\sum_{\ell,m} a_{\ell m}(\vec{x}_0) Y_{\ell m}(\vec{n}), 
\end{equation}
where the coefficients $a_{\ell m}$ define the angular power spectrum 
$C_\ell$ by 
\begin{equation}
\biggl\langle a_{\ell m}\cdot a_{\ell'm'}^*\biggr\rangle 
= \delta_{\ell\ell'}\delta_{mm'}C_\ell,
\label{anps}
\end{equation}
and determine the 
two-point (scalar)  correlation function of the temperature fluctuations,
namely
\begin{eqnarray}
\left\langle{\delta T\over T}(\vec{n}){\delta T\over T}(\vec{n}') \right\rangle_{{~}_{\!\!(\vec{n}\cdot
\vec{n}'=\cos\vartheta)}}&=&
\sum_{\ell \ell' m m'}\biggl\langle a_{\ell m}
 a_{\ell'm'}^*\biggr\rangle  Y_{\ell m}(\vec{n})  Y_{\ell'  m'}^*(\vec{n'})
\nonumber \\
& =& {1\over 4\pi}\sum_\ell(2\ell+1)C_\ell P_\ell(\cos\vartheta).
\end{eqnarray}
In Fig. \ref{FCL1} and \ref{FCL2} the experimental points are presented in terms of the 
 $C_{\ell}$ spectra.

In Fig. \ref{FCL1} and \ref{FCL2} the curves fitting the data are obtained by imposing 
adiabatic initial conditions.
If magnetic fields would seed directly the CMB anisotropies 
a characteristic ``hump'' would appear for $\ell <100$ which is 
not compatible with the experimental data. If, on the contrary, the
condition (\ref{secondq}) holds, then the modifications due to a magnetic 
field can be numerically computed for a given magnetic spectral index and for a given 
amplitude. This analysis has been performed by Koh and Lee \cite{koh}. The claim is that
by modifying the amplitude of the magnetic field in a way compatible with the cosmological 
constraint the effect on the scalar $C_{\ell}$ can be as large as the $10\%$ for a given magnetic spectral 
index. 

The main problem is order to detect large-scale magnetic fields from the spectrum of temperature 
fluctuations is the parameter space. On top of the usual parameters common in the CMB analysis 
at least two new parameters should be introduced, namely the magnetic spectral index and the 
amplitude of the magnetic field. The proof of this statement is that there is no available analysis of the WMAP
data including also the presence of large scale magnetic fields in the initial conditions 
according to the lines presented in this Section.

To set bounds on primordial magnetic fields from 
CMB anisotropies it is often assumed that the magnetic field is fully 
homogeneous. In this case the magnetic field amplitude 
has been bounded to be $ B \leq 10^{-3}$ G at the decoupling time \cite{ferreirabarr} (see also \cite{kan1} 
for a complementary analysis always in the case of a homogeneous magnetic field).

There is the hope that since magnetic fields contribute not only to the scalar
fluctuations, but also to the vector and tensor modes of the geometry useful informations 
cann be also obtained from the analysis of vector and tensor power spectra. In \cite{mack}
temperature and polarization power spectra induced by vector and tensor perturbations 
have been computed by assuning a power-law spectrum of magnetic inhomogeneities.
In \cite{kan0,kan2,caprdur} the tensor modes of the geometry have 
been specifically investigated.
In \cite{subrapol,subrabarrow2,subrabarrow1} it is argued that magnetic fields can induce 
anisotropies in the polarization for rather small length scales, i.e. $\ell >1000$. 

As already pointed out in the present Section, magnetic fields may be 
assumed to be force free in various cases. This is an approximation 
which may also be relaxed. However, even if magnetic fields are assumed to be 
force-free there is no reason to assume that their associated 
magnetic helicity (or gyrotropy) vanishes. In \cite{pogosian} the posssible 
effects of helical magnetic fields on CMB physics 
have been investigated. If helical flows are present at recombination, they would 
produce parity violating temperature-polarization correlations. The magnitude 
of helical flows induced by helical magnetic fields 
turns out to be unobservably small, but better prospects of constraining 
helical magnetic fields come from maps of Faraday rotation measurements
of the CMB \cite{pogosian}.

\subsection{Faraday rotation of CMB}

Large scale magnetic fields present at the decoupling epoch 
can also depolarize CMB \cite{FR}.
The polarization of the CMB
represents a very interesting observable which has been extensively
investigated in the past both from the theoretical \cite{1a} and
experimental points of view \cite{2a,2ab}. Forthcoming satellite missions
like PLANCK \cite{3a} seem to be able to achieve a level
of sensitivity which will enrich decisively our experimental knowledge
of the CMB polarization with new direct measurements. 

If the background geometry of the universe is homogeneous but not
isotropic the CMB is naturally polarized \cite{1a}. 
This phenomenon occurs, for example, in  Bianchi-type I models \cite{4a}.
On the other hand if the background geometry is homogeneous and
isotropic (like in the Friedmann-Robertson-Walker  case) it seems very
reasonable that the CMB acquires a small degree of linear
polarization provided the radiation field has a non-vanishing
quadrupole component at the moment of last scattering \cite{5a,5ab}.

Before decoupling photons, baryons and electrons form a unique fluid
which possesses only monopole and dipole moments, but not
quadrupole. Needless to say, in a homogeneous and isotropic model of
FRW type a possible source of linear polarization for the CMB becomes
efficient only at the decoupling and therefore a small degree of linear
polarization seems a firmly established theoretical option  which will
be (hopefully) subjected to direct tests in the near future.
The linear polarization of the CMB is a very promising
laboratory in order to directly probe the speculated existence of a
large scale magnetic field (coherent over the horizon size at the
decoupling) which might actually rotate (through the Faraday
effect) the polarization plane of the CMB. 

Consider, for
instance, a linearly polarized electromagnetic wave of physical 
frequency $\omega$
travelling along the $\hat{x}$ direction in a  plasma of ions and
electrons together with  a magnetic field ($ \vec{B}$)
 oriented along an arbitrary direction ( which might coincide with
$\hat{x}$ in the simplest case). 
If we let the polarization vector at the origin ($x=y=z=0$, $t=0$) 
be directed along the $\hat{y}$ axis, after the
wave has traveled a length $\Delta x$, the corresponding angular shift
($\Delta\alpha$) in the polarization plane will be :
\begin{equation}
\Delta\alpha= f_{e} \frac{e}{2m}
\left(\frac{\omega_{\rm p}}{\omega}\right)^2 (\vec{B}\cdot\hat{x})~ \Delta x.
\label{Faraday1}
\end{equation}
From Eq. (\ref{Faraday1})
by stochastically averaging over all the possible orientations 
of $\vec{B}$  and
by assuming that the last scattering surface is infinitely thin
(i.e. that  $\Delta x f_{e} n_{e} \simeq \sigma_{T}^{-1}$ where
$\sigma_{T}$ is the Thompson cross section) we
get an expression connecting the RMS of the rotation angle to the
magnitude of $\overline{B}$ at $t\simeq t_{dec}$
\begin{eqnarray}
&&\langle(\Delta\alpha)^2 \rangle^{1/2} \simeq 1.6^{0} 
\left(\frac{B(t_{dec})}{B_{c}} \right)
\left(\frac{\omega_{M}}{\omega}\right)^2,
\nonumber\\
&&B_{c} =
10^{-3}~{\rm G},~~~\omega_{M} \simeq  3\times10^{10}~Hz
\label{Faraday2}
\end{eqnarray}
(in the previous equation we implicitly assumed that the frequency of
the incident electro-magnetic radiation is centered around the maximum
of the CMB).
We can easily argue from Eq. (\ref{Faraday2}) that if $B(t_{dec}) \geq
B_c$ the expected rotation in the polarization plane of the CMB is
non negligible.
Even if we are not interested, at this level, in a precise estimate of
$\Delta\alpha$, we point out that more refined determinations of the
expected Faraday rotation signal (for an incident frequency
$\omega_{M}\sim 30~{\rm GHz}$) were recently carried out \cite{6b,6b2}
leading to a result fairly consistent with (\ref{Faraday1}).

Then, the statement is the following. {\em If} the CMB is linearly 
polarized and {\em if} a large scale magnetic field is 
present at the decoupling epoch, {\em then} the polarization plane of the 
CMB can be rotated \cite{FR}. The predictions of different 
models  can then be confronted 
with the requirements coming from a possible detection of depolarization 
of the CMB \cite{FR}.

\subsection{Relic gravitational waves}

Magnetic fields can source the evolution equations of the 
fluctuations of the geometry also over length-scales much smaller than the ones 
where CMB anisotropy experiments are conduced. This suggests 
that magnetic inhomogeneities may leave an imprint 
on the relc background of gravitational waves.

If a hypermagnetic background is present for $T> T_c$, then, as discussed
in Section 5 and 6,  the energy momentum tensor 
will acquire a small anisotropic component which will source the evolution 
equation of the tensor fluctuations of the metric. 
 Suppose now,  that 
$|\vec{{\cal H}}_{Y}|$ has constant amplitude and that it is also 
homogeneous. Then 
as argued in  \cite{griru} we can easily deduce 
the critical fraction of energy density  present today in relic gravitons 
of EW origin 
\begin{equation}
\Omega_{\rm gw}(t_0) = \frac{\rho_{\rm gw}}{\rho_c} 
\simeq z^{-1}_{{\rm eq}}
r^2,~~\rho_{c}(T_{c})\simeq N_{\rm eff} T^4_{c }
\end{equation}
($z_{\rm eq}$ is the redshift from the time of matter-radiation,
 equality). Because of the structure of the AMHD equations, stable 
hypermagnetic fields will be present not only for 
$\omega_{\rm ew}\sim k_{\rm ew}/a$ but 
for all the range $\omega_{{\rm ew}} <\omega< \omega_{\sigma}$ where 
$\omega_{\sigma}$ is the diffusivity frequency. Let us assume, 
 for instance, that $T_{c} \sim 100 $ GeV and $g_{\ast} = 106.75$. 
Then, the (present) values of 
$\omega_{\rm ew}$ is 
\begin{equation}
\omega_{\rm ew } (t_0) \simeq 2.01 \times 10^{-7} \biggl( \frac{T_{c}}{1 {\rm GeV}} \biggr) 
\biggl(\frac{ g_{\ast}}{100} \biggr)^{1/6} {\rm Hz} .
\end{equation}
Thus, $\omega_{\sigma}(t_0) \sim 10^{8} \omega_{\rm ew} $. Suppose now that 
$T_{c} \sim 100$ GeV; than we will have that $\omega_{\rm ew}(t_0) \sim 10^{-5}$ Hz. 
Suppose now,  that 
\begin{equation}
|\vec{{\cal H}}_{Y}|/T_{c}^2 \geq 0.3,
\end{equation} 
as, for instance, implied by the analysis of the electroweak 
phase diagram in the presence of a magnetized background.
This requirement imposes $ r \simeq 0.1$--$0.001$ and, consequently, 
\begin{equation}
h_0^2 \Omega_{\rm GW} \simeq 10^{-7} - 10^{-8}.
\end{equation}
Notice that this signal would occurr in a (present) frequency 
range between $10^{-5}$ and $10^{3}$ Hz. This signal 
satisfies the presently available phenomenological 
bounds on the graviton backgrounds of primordial origin.
The pulsar timing bound ( which applies for present 
frequencies $\omega_{P} \sim 10^{-8}$ Hz and implies 
$h_0^2 \Omega_{\rm GW} \leq 10^{-8}$) is automatically satisfied
since our hypermagnetic background is defined for $10^{-5} {\rm Hz} 
\leq \omega \leq 10^{3} {\rm Hz}$. The large scale bounds would imply 
$h_0^2 \Omega_{\rm GW} < 7 \times 10^{-11}$ but a at much lower frequency 
(i.e. $10^{-18 }$ Hz). The signal discussed here is completely 
absent for frequencies $\omega < \omega_{\rm ew}$. Notice that 
this signal is clearly distinguishable from other stochastic 
backgrounds occurring at much higher frequencies (GHz region) 
like the ones predicted by quintessential inflation \cite{gw10,gw1,gw1a}.
It is equally distinguishable from signals due to 
pre-big-bang cosmology (mainly in the window of
ground based interferometers \cite{gw2}).
The frequency of operation of the interferometric devices 
(VIRGO/LIGO) is located between few Hz and 10 kHz \cite{gw2}.
 The frequency of operation 
of LISA is well below the Hz (i.e. $10^{-3} $Hz, approximately). 
 In this model the signal 
can be located both in the LISA window and in the VIRGO/LIGO window
due to the hierarchy between the hypermagnetic diffusivity scale and the 
horizon scale at the phase transition.
\begin{figure}[htb]
    \centering
    \includegraphics[height=2.5in]{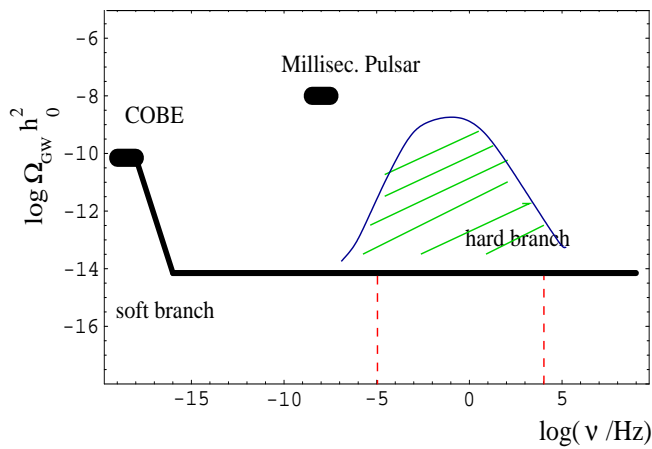}
    \caption{The stochastic background of GW produced by 
inflationary models with flat logarithmic energy spectrum, 
illustrated together with the GW background of hypermagnetic origin. The 
frequencies marked with dashed lines correspond to the electroweak frequency and to the 
hypermagnetic diffusivity frequency.}
    \label{F7}
\end{figure}
In Fig. \ref{F7} the full thick line illustrates the spectrum of relic gravitational waves 
produced in a conventional model for the evolution of the universe. The flat 
plateau corresponds to modes which left the horizon during the inflationary 
stage of expansion and re-entered duriing the radiation dominated phase. The 
decreasing slope between $10^{-16}$ and $10^{-18}$ Hz is 
due to modes leaving the horizon during inflation and re-entering 
during the maatter dominated stage of expansion. Clearly, the signal provided 
by a background of hypermagnetic fields can be even 7 order of magnitude larger than the 
inflationary prediction.
The interplay between gravitational waves and large-scale 
magnetic fields has been also the subject of recent interesting 
investigations \cite{tsagas1,tsagas2,papadop}.

\section{Gravitating magnetic fields}
\renewcommand{\theequation}{9.\arabic{equation}}
\setcounter{equation}{0}
Up to now large scale magnetic fields have been treated in different frameworks
but always within a perturbative approach. However, magnetic fields can also 
break explicitely the isotropy (but not the homogeneity) of the background geometry

Suppose that at some time $t_1$ the Universe becomes transparent to radiation 
and suppose that, at the same time, the four-dimensional background geometry 
(which we assume, for simplicity, spatially flat) has a very tiny 
amount of anisotropy in the scale factors associated with different spatial
 directions, namely
\begin{equation}
ds^2 = dt^2 - a^2(t)dx^2 - b^2(t) [ dy^2 + dz^2],
\label{1}
\end{equation}
where $b(t)$, as it will be clear in a moment, has to be only 
slightly different from $a(t)$. The electromagnetic radiation 
propagating along the $x$ and $y$ axes will have a  different
 temperature, namely   
\begin{equation}
T_{x}(t) = T_1 \frac{a_1}{a} = T_1 e^{- \int H(t) dt},~~~
T_{y}(t) = T_1 \frac{b_1}{b} = T_1 e^{- \int F(t) dt},
\label{2}
\end{equation}
where $H(t)$ and $F(t)$ are the two Hubble factors 
associated, respectively with $a(t)$ and $b(t)$. 
The temperature anisotropy associated with the electromagnetic radiation 
propagating along two directions with different expansion rates can be 
roughly estimated, in the limit $H(t)-F(t)\ll 1$, as 
\begin{equation}
\frac{\Delta T}{T} \sim \int [H(t) - F(t)] dt= \frac{1}{2} 
\int \varepsilon(t) d\log{t}
\end{equation}
where, in the second equality, we assumed that the deviations from the 
radiation dominated expansion can  be written as 
$F(t) \sim ( 1 - \varepsilon(t))/2t$ with $|\varepsilon(t)|\ll 1$. 
The function $\varepsilon$ can be connected with the shear 
parameter, i.e. $[H(t) - F(t)]/[H(t) + 2 F(t)]$ which 
measures the anisotropy in the expansion. In the standard context \cite{zelm1,zelm2}
Einstein equations are solved in the metric (\ref{1}) with fluid sources and in the 
presence of a magnetic field. The shear parameter is then connected with 
the magnetic field intensity.

The dynamical origin of the primordial
anisotropy in the expansion 
 can be connected with the existence of a primordial
magnetic field (not dynamically generated but postulated 
as an initial condition) or with some other sources 
of anisotropic pressure 
 and, therefore, the possible bounds on the temperature
 anisotropy can be translated into bounds on the time evolution of the 
anisotropic scale factors \cite{ferreirabarr}.

Today the amount of anisotropy in the expansion must be very small because 
of the effect we just described. In the present Section some highly 
speculative considerations on the possible r\^ole of magnetic fields in the
early Universe will be intrtoduced. In spite of the fact that 
magnetic fields must be sufficiently small today, in the past history of the Universe they might have been very large
even modifying the dynamics of the geometry. During the magnetic phase the anisotropy 
in the expansion will be constant (or even grow). However, after the magnetic phase 
a sufficienly long radiation dominated phase may isotropize the background leading 
to a tiny amount of anisotropy in the expansion. This is the basic scenario 
invoked long ago by Zeldovich \cite{zelm1,zelm2}. 

In modern approaches to cosmology, the dynamics of the Universe 
at very high densities is often discussed in terms of the low-energy string 
effective action. It is then useful to analyze the Zeldovich proposal 
in terms of the modern perspective. 

In the low energy limit, the dilaton field is directly coupled to the kinetic 
term of the Maxwell field \cite{anis1}
\begin{equation}
S= - \frac{1}{2 \lambda_{s}^2} \int d^4 x \sqrt{- g} e^{-\varphi} \biggl[ R + 
g^{\alpha\beta} \partial_{\alpha}\varphi \partial_{\beta} \varphi - 
\frac{1}{12} H_{\mu\nu\alpha} H^{\mu\nu\alpha} 
+ \frac{1}{4} F_{\alpha\beta}F^{\alpha\beta} +...\biggl]
\label{actionan}
\end{equation}
where $F_{\alpha\beta}= \nabla_{[\alpha} A_{\beta]}$ is the Maxwell 
field strength and $\nabla_{\mu}$ is the covariant derivative with respect 
to the String frame  metric $g_{\mu\nu}$. Notice that $H_{\mu\nu\alpha}$ is the antisymmetric field strength.
In Eq. (\ref{actionan}) 
the ellipses stand for a possible (non-perturbative) dilaton potential 
and for the string tension corrections parameterized by 
$\alpha' = \lambda_{s}^2$. 
In Eq. (\ref{actionan}) $F_{\mu\nu}$ can be thought as the Maxwell field 
associated with a $U(1)$ subgroup of $E_{8}\times E_{8}$.

Consider a spatially flat background configuration with vanishing 
antysimmetric field strangth 
($H_{\mu\nu\alpha} =0$) and vanishing dilaton potential. 
The dilaton depends only on time and the metric will be taken fully 
anisotropic since we want to study possible solutions with a homogeneous 
magnetic field which is expected to break the isotropy of the background:
\begin{equation} 
ds^2 = g_{\mu\nu} dx^{\mu} dx^{\nu} = dt^2 - a^2(t) dx^2 - b^2(t) dy^2 
- c^2(t) dz^2 .
\label{metrican}
\end{equation}

By varying the effective action with respect to $\varphi$, $g_{\mu\nu}$ and the 
vector potential $A_{\mu}$  we get, respectively 
\begin{eqnarray}
&&R - g^{\alpha\beta} \partial_{\alpha}\varphi \partial_{\beta}\varphi + 
2 g^{\alpha\beta} \nabla_{\alpha} \nabla_{\beta} \varphi = 
- \frac{ 1}{4} F_{\alpha\beta} F^{\alpha\beta}, 
\label{phian}\\
&& R_{\mu}^{\nu} - \frac{1}{2} \delta_{\mu}^{\nu} R = \frac{1}{2} 
\biggl[ \frac{1}{4} \delta_{\mu}^{\nu} F_{\alpha\beta} F^{\alpha\beta} 
- F_{\mu\beta} F^{\nu\beta}  \biggr]
\nonumber\\
&&- \frac{1}{2} \delta_{\mu}^{\nu}  g^{\alpha\beta}\partial_{\alpha}\varphi 
\partial_{\beta} \varphi - \nabla_{\mu} \nabla^{\nu} \varphi + \delta_{\mu}^{\nu} 
\frac{1}{\sqrt{-g}}\partial_{\alpha}[ \sqrt{-g} g^{\alpha\beta}\partial_{\beta} \varphi],
\label{g}\\
&& \nabla_{\mu}\biggl[ e^{-\varphi} F^{\mu\nu}\biggr] =0.
\label{b}
\end{eqnarray}
where $\nabla_{\mu}$ denotes covariant differentiation with respect 
to the metric 
of Eq. (\ref{metrican}). Inserting Eq. (\ref{phi}) into Eq. (\ref{g}) we 
get that Eq. (\ref{g}) 
can  be expressed as 
\begin{equation}
R_{\mu}^{\nu} + \nabla_{\mu}\nabla^{\nu} \varphi 
+ \frac{1}{2} F_{\mu\alpha}F^{\nu\alpha} =0.
\label{g2}
\end{equation}
Consider now a homogeneous magnetic field directed along the $x$ axis. 
The generalized Maxwell equations (\ref{b}) 
and the associated Bianchi identities 
can then be trivially solved by  the field strength $F_{yz}= -F_{zy}$.
The resulting system of equations (\ref{phi}), (\ref{g2}) can than be 
written, in the metric of Eq. (\ref{metrican}),  as
\begin{eqnarray}
&&\ddot{\varphi} = H^2 + F^2 + G^2 + \dot{H} + \dot{F} + \dot{G},
\label{00an}\\
&&H\dot{\varphi} = H F + H G +  H^2 + \dot{H},
\label{xx}\\
&&F\dot{\varphi} = H F + F G + F^2 + \dot{F} - \frac{B^2}{2 b^2c^2},
\label{yy}\\
&&G \dot{\varphi} = G H + F G + G^2 +\dot{G} - \frac{B^2}{2 b^2 c^2}, 
\label{zz}
\end{eqnarray}
where  Eqs. 
(\ref{00an})--(\ref{zz}) correspond, respectively to the 
$(00)$ and $(ii)$  components of Eq. (\ref{g2}).

The solution of Eqs. (\ref{00an})--(\ref{zz})
\begin{eqnarray}
&&a(\eta)= a_0 e^{{\cal H}_0 \eta},~~~b(\eta) = b_0 e^{\varphi_0} 
e^{({\cal F}_0 + {\cal H}_0 + \Delta_0)\eta}\biggl[e^{-2 \Delta_0 \eta}
 + e^{\varphi_1} \biggr],~~~
\nonumber\\
&&c(\eta) = c_0 e^{\phi_0} 
e^{({\cal G}_0 + {\cal H}_0 + \Delta_0)\eta}\biggl[e^{-2 \Delta_0 \eta}
 + e^{\varphi_1} \biggr],
\\
&& \varphi(\eta) = \varphi_0 + ({\cal H}_0 + \Delta_0)\eta + 
\log{\biggl[ e^{\varphi_1} + e^ {- 2 \Delta_0 \eta}\biggr]},
\label{s1}
\end{eqnarray}
where 
\begin{equation}
\Delta_{0} \equiv\sqrt{{\cal H}_0^2 + \Lambda_0}= \frac{a_0}{2 \sqrt{2}}
e^{- (\varphi_0 + \varphi_1)} B.
\label{c1}
\end{equation}
The solution given in Eq.(\ref{c1}) s given in terms of a new variable (a generalized 
conformal time) defined by the following differential relation
\begin{equation}
e^{-\overline{\varphi}} d \eta = dt,~~~~\overline{\varphi} = \varphi - \log{a b c }.
\label{time}
\end{equation}

These solutions represent the generalization of pre-big bang solutions \cite{Veneziano3} to the case
where a constant magnetic field is included.
Generalizations of the solution (\ref{c1}) 
can be obtained in various models anisotropic (but homogeneous) models of the Bianchi class \cite{anis2} like 
Bianchi I, II, III, VI$_{-1}$ and VII$_{0}$ according to the usual classification which can be 
found, for instance, in \cite{ryan}.
Further generalizations of the solution (\ref{c1}) are possible in the Kantowski-Sachs \label{ks} metric  with 
a magnetic field aligned alond the radial coordinate \cite{bark1,bark2}.

Magnetized cosmological solutions can be used in order to describe 
some early phase in the life of the universe. This highly speculative 
idea may be implemented in semi-realistic models provided the large 
anisotropy in the expansion decays at later epochs.
In the standard context \cite{zelm1,zelm2} the only source of isotropization
is the evolution in a radiation dominated background since the shear 
parameter decays if the background is dominated by radiation. If quadratic corrections are 
present, an initially anisotropic solution  becomes isotropic as a consequence of the dynamical 
properties of the evolution equations as discussed in detail in \cite{anis1}. 

Gauge fields have also been studied as a source of inflationary evolution by 
Ford \cite{fordv} and later by Lidsey \cite{lidseyv1,lidseyv2}. The gauge 
kinetic term is supplemented by a potential which is a function of $A_{\mu}A^{\mu}$.
Both exponential inflation \cite{fordv} and power-law inflation \cite{lidseyv1} can be realized 
in these models. 

\section{Concluding Remarks}

The topics reported in the present review suggest various open problems
on the nature of large-scale magnetic fields in the present and in the 
early Universe:
\begin{itemize}
\item{} in principle the morphology of magnetic fields 
in spiral galaxies could be used in order to understand the origin
of galactic magnetism but in practice the observations offer 
answers which are often contradictory;
\item{}  recent observations of magnetic fields 
inside Abell clusters confirm the presence 
of strong magnetic fields in the $\mu$ G range;
\item{} there is evidence that also superclusters 
are magnetized;
\item{} in spite of the fact that magnetic fields may modify the patterns of the CMB anisotropies 
and, eventually, induce anisotropies in the polarization, the theoretical analysis 
has been rarely corroborated by the comparison with the available 
experimental data;
\item{} Faraday rotation of the CMB polarization would 
be a powerful tool to study the possible exsitence of large scale magnetic fields 
prior to recombination;
\item{} the imprints of magnetic fields on the relic background of gravitational waves 
(in the window of wide band interferometers)
allows to test the existence of a background of Abelian gauge fields at the 
electroweak epoch.
\end{itemize} 
It would be important, for the theorist, a reasonable accuracy (within  one order of magnitude) in the experimental
determination of large-scale magnetic fields. For instance
even one order of magnitude in the intensity  of the intra-cluster magnetic field  (i.e. 
$10^{-6}$ G rather than $10^{-7} G$) 
makes a  difference as far as the problem of the origin is concerned. 
In the same perspective, more accurate determinations of the typical correlation scales of the intra-cluster 
fields would be desirable. 

There is, at the moment, no compelling reason why large-scale magnetic fields 
should not be primordial, at least to some degree:
\begin{itemize}
\item{} in the early Universe 
magnetic fields are easily produced during phase transitions but their typical correlation scales are small;
\item{} inflation greatly helps in producing correlations over very large scales 
and the (early) variation of gauge couplings during the inflationary phase 
allows the generation of intense large-scale fields;
\item{} primordial magnetic fields may have an impact  on the thermodynamical 
history of the Universe, but, in practice, the 
obtained constraints improve only marginally, in various interesting cases, the 
critical density bound imposed on the magnetic energy density;
\item{} if magnetic fields are generated during phase transitions, a growth in their correlation scale 
cannot be excluded because of turbulent phenomena whose existence 
can be reasonably expected (but not firmly predicted)  on the basis 
of our terrestrial knowledge of magnetized plasmas.
\end{itemize}

In the next decade a progress of the empirical 
knowledge is expected in apparently unrelated areas like $x$ and $\gamma$-ray astronomy, 
 radio-astronomy, CMB physics, detection of relic gravitational waves. All these 
areas are connected with the existence and with the properties of large-scale magnetic fields 
in the early and in the present Universe. This connection is sometimes rather fragile  
since it is  mediated by various theoretical assumptions. However, 
there is the hope that, in a not too distant future, some of the puzzles 
related to the origin and existence of large-scale magnetic fields may be 
resolved and some of the current theoretical guessworks firmly ruled out.

\section*{Acknowledgements}
Some of the ideas  presented in this review have been 
elaborated and assembled on the occasion of different sets of lectures delivered through the years.
The author wishes to express his gratitude to 
D. Babusci, H. de Vega,  M. Gasperini,
 H. Kurki-Suonio, E. Keih\"anen, N. Sanchez,  M. Shaposhnikov, G. Veneziano, A. Vilenkin.
A special thank is due to G. Cocconi for important remarks which improved the first draft.

\newpage

\appendix

\renewcommand{\theequation}{A.\arabic{equation}}
\setcounter{equation}{0}
\section{Complements on MHD description}

\subsection*{Two fluids and one-fluid MHD equations}
In a  two-fluid plasma description  
the charge carriers are the ions (for simplicity we can think of them as protons)
 and the electrons. The 
two fluid equations treat the ions and the electrons as two 
conducting fluids which are coupled as in Eqs. (\ref{vlasov1})--(\ref{transv}).
Given the two- fluid description, one-fluid variables can be defined directly in terms of the 
two-fluid variables
\begin{eqnarray}
&& \rho_{\rm m}(\vec{x}, t) = m_{e} n_{e}(\vec{x}, t) + m_{p} n_{p}(\vec{x}, t),
\nonumber\\
&& \rho_{\rm q}(\vec{x}, t) = e[ n_{p}(\vec{x}, t) - n_{e}(\vec{x}, t)],
\nonumber\\
&& \vec{J} = e[ n_{p} \vec{v}_{p} - n_{e} \vec{v}_{e}].
\end{eqnarray}
In the case of a globally neutral plasma $n_{e} \sim n_{p} = n_{q}$ and $\rho_{q} =0$. The ion and electron
equations then become 
\begin{eqnarray}
&& m_{p} n_{q}\biggl[ \frac{\partial}{\partial t} + \vec{v}_{\rm p}\cdot \vec{\nabla} \biggr]\vec{v}_{\rm p} 
= e n_{q} ( \vec{E} + \vec{v}_{p} \times \vec{B}) - \vec{\nabla} p_{p} + {\cal C}_{p e},
\label{prot}\\
&& m_{e} n_{q}\biggl[ \frac{\partial}{\partial t} + \vec{v}_{\rm e}\cdot \vec{\nabla} \biggr]\vec{v}_{\rm e}  
= -e n_{q}( \vec{E} + \vec{v}_{e} \times \vec{B}) - \vec{\nabla} p_{e} + {\cal C}_{e p},
\label{elec}
\end{eqnarray}
where   ${\cal C}_{p e}$ and $ {\cal C}_{p e}$ are the collision terms.
In the globally neutral case the  center of mass velocity 
becomes 
\begin{equation}
\vec{v} = \frac{m_{p} \vec{v}_{p} + m_{e} \vec{v}_{e}}{m_{p} + m_{e}},
\end{equation}
and the one-fluid mass and charge density conservations
become
\begin{eqnarray}
&& \frac{\partial \rho_{\rm m}}{\partial t} + \vec{\nabla} \cdot( \rho_{\rm m} \vec{v}) =0,
\label{consm}\\
&& \frac{\partial  \rho_{q}}{\partial t} + \vec{\nabla} \cdot \vec{J} =0.
\label{consq}
\end{eqnarray}
Summing up Eqs. (\ref{prot}) and (\ref{elec}) leads, with some algebra involving the continuity equation, to the 
momentum transport equation in the one-fluid theory:
\begin{equation}
\rho_{\rm m} \biggl[ \frac{\partial \vec{v}}{\partial t} + \vec{v}\cdot \vec{\nabla}\biggr] \vec{v} =  \vec{J} \times \vec{B} - \vec{\nabla} P,
\label{momt}
\end{equation}
where $P= P_{e} + P_{i}$.
In Eq. (\ref{momt}) the collision term vanishes if there are no neutral particles, i.e. if the plasma is fully ionized. 
The final equation of the one-fluid description is obtained by taking the difference of Eqs. (\ref{prot}) and (\ref{elec})
after having multiplied Eq. (\ref{prot}) by $m_{e}$ and Eq. (\ref{elec}) by $m_{p}$. This procedure is more tricky and it is 
discussed in standard textbooks of plasma physics\cite{krall,chen}. The key points in the derivation are that 
the limit for $m_{e}/m_{p} \to 0$ must be taken. The problem with this procedure is that the subtraction of the two 
mentioned equations does not guarantee that viscous and collisional effects are negligible. 
The result of this procedure is the so-called one-fluid generalized Ohm law:
\begin{equation}
\vec{E} + \vec{v}\times\vec{B} = \frac{1}{\sigma} \vec{J} + \frac{1}{e n_{q}} ( \vec{J} \times \vec{B} - \vec{\nabla} P_{e}).
\label{genohm}
\end{equation}
The term $\vec{J}\times B$ is nothing but the {\em Hall current} and $\vec{\nabla} P_{e}$ is often called
thermoelectric term. Finally the term $\vec{J}/\sigma$ is the resistivity term and $\sigma$ is 
the conductivity of the one-fluid description. In Eq. (\ref{genohm})  the pressure has been taken to be isotropic.
This is, however, not a direct consequence of the calculation presented in this Appendix but it is an assumption 
which may (and should) be relaxed in some cases. In the plasma physics literature \cite{krall,chen} 
the anisotropic pressure contribution is neglected for the simple reason that 
experiments terrestrial plasmas show that this terms is often negligible.

\subsection*{Conservation laws in resistive MHD}

Consider and arbitrary closed surface $\Sigma$ which moves with the
plasma. Then, by definition of the bulk velocity of the plasma
($\vec{v}$ we can also write  $d\vec{\Sigma} = \vec{v} \times
d\vec{l}~d\eta$. The (total) time derivative of the flux can
therefore
be expressed as
\begin{equation}
\frac{d}{d t} \int_{\Sigma} \vec{B}\cdot d\vec{\Sigma}=
\int_{\Sigma} \frac{\partial \vec{B}}{\partial t} \cdot
d\vec{\Sigma} + \int_{\partial\Sigma} \vec{B}\times\vec{v} \cdot
d\vec{l},
\label{first}
\end{equation}
where $\partial \Sigma$ is the boundary of $\Sigma$. Using now the
properties of the vector products
(i.e. $\vec{B}\times\vec{v}\cdot d\vec{l}= - \vec{v}\times\vec{B} \cdot
d\vec{l} $) we can express $\vec{v} \times \vec{B}$ though the Ohm
law
given in Eq. (\ref{genohm2}) and we obtain that
\begin{equation}
\vec{v} \times \vec{B} \simeq - \vec{E} + \frac{1}{4\pi \sigma} \vec{\nabla}
\times \vec{B}
\end{equation}
Using now Eq. (\ref{first}) together with the Stokes theorem, the following 
expression can be obtained
\begin{equation}
\frac{d}{d t} \int_{\Sigma} \vec{B}\cdot d\vec{\Sigma}=\int_{\Sigma} \biggl[
\frac{\partial \vec{B}}{\partial t} + \vec{\nabla} \times
\vec{E}\biggr]\cdot d\vec{\Sigma} - \frac{1}{4\pi \sigma}
\int_{\Sigma} \vec{\nabla} \times\vec{\nabla}
\times\vec{B}\cdot d\vec{\Sigma}.
\label{second}
\end{equation}
From the Maxwell's  equations the first part at the right
hand side of Eq. (\ref{second}) is zero and Eq. (\ref{flux}),
expressing the
Alfv\'en theorem, is recovered.

With similar algebraic manipulations (involving the use of various vector identities),
 the conservation of the
magnetic helicity can be displayed. Consider a closed volume in the plasma,
then we can write that $dV = d^3 x= \vec{v}_{\perp}\cdot
d\vec{\Sigma}~
d\eta \equiv \vec{n} \cdot \vec{v}_{\perp}~ d\Sigma~ d\eta$ where
$\vec{n}$ is the unit vector normal to $\Sigma$ (the boundary of V, i.e.
$\Sigma = \partial V$) and $\vec{v}_{\perp}$ is the component of the
bulk velocity orthogonal to $\partial V$. The (total) time derivative
of the magnetic helicity can now be written as
\begin{equation}
\frac{d}{d\eta} {\cal H}_{M} = \int_{V} d^3 x
\frac{\partial}{\partial\eta} \bigl[\vec{A} \cdot\vec{B}\bigr] +
\int_{\partial V=\Sigma}\vec{A}\cdot\vec{B}
\vec{v}_{\perp}\cdot\vec{n}
d\Sigma.
\label{ddthel}
\end{equation}
The partial derivative at the right
hand side of Eq. (\ref{ddthel})  can be made explicit. Then  the 
one-fluid  MHD equations should be used recalling that the relation between the 
electromagnetic fields and the vector potential, for instance in the Coulomb gauge. 
Finally using again the Ohm
law and transforming the obtained surface integrals into volume
integrals (through the divergence theorem)  Eq. (\ref{h2}), expressing the conservation of the 
helicity, can be obtained.

In spite of the fact that the conservation of the magnetic helicity can be 
derived in a specific gauge, the magnetic
helicity is indeed a gauge invariant quantity. Consider a gauge transformation
\begin{equation}
\vec{A} \rightarrow \vec{A}  + \vec{\nabla} \xi,
\end{equation}
then the magnetic helicity changes as
\begin{equation}
\int_{V} d^3x \vec{A} \cdot \vec{B} \rightarrow
\int_{V} d^3x \vec{A} \cdot \vec{B} + \int_{V} d^3 x
\vec{\nabla}\cdot\bigl[
\xi \vec{B}\bigr]
\end{equation}
(in the second term at the right hand side we used the fact that the
magnetic field is divergence free). By now using the divergence
theorem we can express the volume integral as
\begin{equation}
 \int_{V} d^3 x  \vec{\nabla}\cdot\bigl[ \xi \vec{B}\bigr] =
 \int_{\partial V=\Sigma } \xi\vec{B} \cdot\vec{n} d\Sigma.
\end{equation}
Now if, as we required, $\vec{B}\cdot\vec{n}=0$ in $\partial V$, the
integral is exactly zero and ${\cal H}_{M} $ is gauge invariant. The 
condition  $\vec{B}\cdot\vec{n}=0$ is not specific of a particular 
profile of the magnetic field. It can be always 
achieved by slicing the volume of integration in 
small flux tubes where, by definition, the magnetic field 
is orthogonal to the walls of the flux tube.


\newpage

\begin{thebibliography}{0}

\bibitem{alv1} H. Alfv\'en, {\it Arkiv. Mat. F. Astr., o. Fys.} {\bf 29 B}, 2 (1943).

\bibitem{fermi} E. Fermi, {\it Phys. Rev.}{\bf 75}, 1169 (1949).

\bibitem{alv2}  H. Alfv\'en, {\it Phys. Rev.}{\bf 75}, 1732 (1949).

\bibitem{hiltner} W. A. Hiltner, {\it Science}{\bf 109}, 165 (1949).

\bibitem{hall} J. S. Hall, {\it Science}{\bf 109}, 166 (1949).

\bibitem{davis} L. J. Davis and J. L. Greenstein, {\it Astrophys. J.}{\bf 114}, 206 (1951).

\bibitem{fermi2} E. Fermi and S. Chandrasekar {\it Astrophys. J.}{\bf 118}, 113 (1953).

\bibitem{fermi3}E. Fermi and S. Chandrasekar {\it Astrophys. J.}{\bf 118}, 116 (1953).

\bibitem{alv3}  R. D. Richtmyer and E. Teller  {\it Phys. Rev.} {\bf 75}, 1729 (1949).

\bibitem{wiel} R. Wielebinski and J. Shakeshaft, {\it Nature}{\bf 195}, 982 (1962).

\bibitem{lyne} A. Lyne and F. Smith, {\it Nature}{\bf 218}, 124 (1968).

\bibitem{heiles} C. Heiles, {\it Annu. Rev. Astron. Astrophys.}{\bf 14}, 1 (1976).

\bibitem{kro} P. P. Kronberg, {\it Rep. Prog. Phys.}{\bf 57}, 325 (1994).

\bibitem{vallee1} J. Vall\'ee, {\it Fundam. Cosmic Phys.}{\bf 19}, 1 (1997).

\bibitem{beck} R. Beck, A. Brandenburg, D. Moss, A. Skhurov, and D. Sokoloff
{\it Annu. Rev. Astron. Astrophys.}{\bf 34}, 155 (1996).

\bibitem{battaner} E. Battaner and E. Florido, {\it Fund. of Cosm. Phys.}{\bf 21}, 1 (2000).

\bibitem{han} J.-L. Han and R. Wieblinski, {\it Chin. J. Astron. Astrophys.}
{\bf 2} 293 (2002).

\bibitem{zweibel} E. Zweibel and C. Heiles, {\it Nature}{\bf 385}, 131 (1997).

\bibitem{zeldovich} Ya. B. Zeldovich, A. A. Ruzmaikin, and D.D. Sokoloff, 
{\it Magnetic Fields in Astrophysics} (Gordon and Breach Science, New York, 
1983).

\bibitem{parker} E. N. Parker, {\it Cosmical Magnetic Fields} (Clarendon Press, 
Oxford, 1979).

\bibitem{ruzmaikin} A. A. Ruzmaikin, A. M. Shukurov, and D. D. Sokoloff {\it Magnetic 
Fields of Galaxies}, (Kluwer Academic Publisher, Dordrecht, 1988).

\bibitem{vallee2} J. Vall\'ee, {\it Fundam. Cosmic Phys.}{\bf 19}, 319 (1998).

\bibitem{fiebig} D. Fiebig and R. Guensten, {\it Astron. and Astrophys.} {\bf 214}, 333 (1989).

\bibitem{crutcher} R. Crutcher et al. {\it Astrophys. J.}{\bf 514}, L121 (1999).

\bibitem{shklovskij} I. Shklovskij, {\em Dokl. Akad. Nauk. USSR}{\bf 90}, 983 (1953).

\bibitem{magcl} X. Chi and A. W. Wolfendale, {\it Nature}{\bf 362}, 610 (1993).

\bibitem{eq} V. R. Buczilowski and R. Beck, {\it Astron. Astrophys.}{\bf 241}, 46 (1991)

\bibitem{eq2}  E. Hummel and R. Beck, {\it Astron. Astrophys.}{\bf 303}, 691 (1995).

\bibitem{lawson} K. D. Lawson et al., {\it MNRAS}{\bf 225}, 307 (1987).

\bibitem{han2} J.-L. Han and G. J. Qiao, {\it Astron. Asrtophys.}{\bf 288} 759 (1994).

\bibitem{han4} J.-L. Han, R. Manchester, and G. Qiao {\it Mon. Not. R. Astron. Soc.}{\bf 306}, 371 (1999).

\bibitem{coles} P. Dineen and P. Coles, e-print Archive [astro-ph/0306529] (to appear in Mon. Not. R. Astron. Soc.).

\bibitem{goldshmidt}  O. Goldshmidt and Y. Rephaeli, {\it Astrophys. J.}{\bf 411}, 518 (1993).

\bibitem{crusius} A. Crusius-W\"atzel, P. Biermann, R. Schlickeiser, and I. Lerche, {\it Astrophys. J.}{\bf 360}, 417 (1990).

\bibitem{rephaeli2} W. Newmann, A. Newmann, and Y. Rephaeli, {\it Astrophys. J.}{\bf 575}, 755 (2002).

\bibitem{vogt1} C. Vogt and T. Ensslin,  e-print Archive [astro-ph/0302426] (to appear in Astronomy and Astrophysics).

\bibitem{vogt2} C. Vogt and T. Ensslin,  e-print Archive [astro-ph/0309441] (to appear in Astronomy and Astrophysics).

\bibitem{kolatt} T. Kolatt, {\it Astrophys. J}{\bf 495}, 564 (1998).

\bibitem{mw1} D. Mathewson and V. Ford, {\it Mem. R. Astron. Soc.}{\bf 74}, 139 (1970).

\bibitem{mw2} R. N. Manchester, {\it Astrophys. J.}{\bf 188}, 637 (1974).

\bibitem{mw3} R. Rand and A. Lyne, {\it Mon. Not. R. Astron. Soc.}{\bf 268}, 497 (1994).

\bibitem{morris} F. Yusef-Zadeh and M. Morris, {\it Astrophys. J.}{\bf 320}, 545 (1987).

\bibitem{yusef}  F. Yusef-Zadeh, D. Roberts, and M. Wardle, {\it Astrophys. J. }{\bf 490}, L83 (1997).

\bibitem{yusef2}  F. Yusef-Zadeh, M. Wardle, and P. Parastaran, {\it Astrophys. J.}{\bf 475}, L119 (1997).

\bibitem{vallee3} J. Vall\'ee, {\it Astrophys. J.}{\bf 566}, 261 (2002).

\bibitem{HMLQ} J. L. Han, R. N. Manchester, A. G. Lyne, and G. J. Qiao, {\it Astrophys. J. }{\bf 570}, L17 (2002).

\bibitem{mw4} Y. Sofue and M. Fujimoto, {\it Astrophys. J.}{\bf 265}, 722 (1983).

\bibitem{mw5} D. Harari, S. Mollerach and E. Roulet, {\it JHEP}{\bf 08} 022 (1999).

\bibitem{mw6} T. Stanev, {\it Astrophys. J.}{\bf 479}, 290 (1997).

\bibitem{han5} J. Han, {\it Astrophys. and Space Sci.} {\bf 278}, 181 (2001).

\bibitem{fl1} R. Rand and S. Kulkami, {\it Astrophys. J.} {\bf 343}, 760 (1989)

\bibitem{fl2} H. Ohno and S. Shibata, {\it Mon. Not. R. Astron. Soc.} {\bf 262}, 953 (1993).

\bibitem{fl3} T. Jones, D. Klebe, and J. Dickey, {\it Astrophys. J.} {\bf 458}, 194 (1992).

\bibitem{bier} P. Biermann, {\it J. Phys. G} {\bf 23}, 1 (1997).

\bibitem{far}  G. Farrar and T. Piran, {\it Phys. Rev. Lett.} {\bf 84}, 3527 (2000). 

\bibitem{wolf} M. Giler, J. Wdowczyk and A. W. Wolfendale, {\it J. Phys. G} {\bf 6}, 1561 (1980).

\bibitem{G} K. Greisen, {\it Phys. Rev. Lett.} {\bf 16}, 748 (1966).

\bibitem{ZK} G. Zatsepin and V. Kuzmin, {\it JETP Lett.}{\bf 4}, 78 (1966).

\bibitem{medina} G. Medina Tanco, {\it Astrophys. J. Lett.}{\bf 505}, L79 (1998).

\bibitem{bla} P. Blasi and A. Olinto, {\it Phys. Rev. D}{\bf 59}, 023001 (1999).

\bibitem{lem} G. Sigl, M. Lemoine, and P. Biermann, {\it Astropart. Phys.}{\bf 10}, 141 (1999).

\bibitem{stanev} T. Stanev, D. Seckel, and R. Engel {\it Phys. Rev. D}{\bf 68}, 103004 1.

\bibitem{widrow} L. Widrow, {\em Rev. Mod. Phys.} {\bf 74}, 775 (2002).

\bibitem{krause1} M. Krause, R. Beck and E. Hummel {\it Astron. and Astrophys.} {\bf 217}, 4 (1989).

\bibitem{krause2} M. Krause, R. Beck and E. Hummel {\it Astron. and Astrophys.} {\bf 217}, 17 (1989).

\bibitem{beck2} R. Beck, {\it Phil. Trans. R. Soc. Lond. A} {\bf 358}, 777 (2000).

\bibitem{krause3} F. Krause and R. Beck, {\it Astron. and Astrophys.} {\bf 335}, 789 (1998).

\bibitem{chalonge} M. Giovannini,  
{\it Proc. of 7th Paris Cosmology Colloquium on High Energy Astrophysics for and from Space} 
(eds. N. Sanchez and H. de Vega) e-Print Archive [hep-ph/0208152]. 

\bibitem{carilli} C. Carilli and G. Taylor  {\it Ann.Rev.Astron.Astrophys.} {\bf 40}, 319 (2002). 

\bibitem{cl1} K.-T. Kim, P.P. Kronberg, P. D. Dewdney and T. L. Landecker, {\it Astrophys. J.} {\bf 355}, 29 (1990).

\bibitem{cl2} L. Ferretti, D. Dallacasa, G. Giovannini, and A. Tagliani, {\it Astronom. and Astrophys.} {\bf 302}, 680 (1995).

\bibitem{cl4} O. Goldshmidt and Y. Raphaeli, {\it Astrophys. J.} {\bf 411}, 518 (1993).

\bibitem{cl5} T.E. Clarke, P.P. Kronberg and H. B\"ohringer, 
{\it Astrophys. J.} {\bf 547}, L111 (2001).

\bibitem{cl6}  H. B\"ohringer, Rev. Mod. Astron. {\bf 8}, 295 (1995).

\bibitem{roscat} H. Hebeling, W. Voges, H. B\"ohringer, 
A. C. Edge, J. P. Huchra, and U. G. Briel {\it Mon. Not. Astron. Soc.} 
{\bf 281}, 799 (1996).

\bibitem{giofer} G. Giovannini and L. Ferretti, {\it New Astronomy}{\bf 5}, 335 (2000).

\bibitem{ohno} H. Ohno et al., {\it Astrophys.J.} {\bf 584}, 599 (2003). 

\bibitem{vallee4} J. Vall\'ee, {\it Astron. J.} {\bf 124}, 1322 (2001).

\bibitem{cl3} K.-T. Kim, P. C. Tribble, and P.P. Kronberg, 
{\it Astrophys. J. } {\bf 379}, 80 (1991).

\bibitem{isola} C. Isola and G. Sigl, {\it Phys. Rev. D}{\bf 66}, 083002 (2002).

\bibitem{igor} K. Dolag, D. Grasso, V. Springel, and Igor Tkachev, e-Print Archive [astro-ph/0310912]. 

\bibitem{vla} A. Vlasov, Zh. \'Eksp. Teor. Fiz. {\bf 8}, 291 (1938);
J. Phys. {\bf 9}, 25 (1945).

\bibitem{lan} L. D. Landau, J. Phys. U.S.S.R. {\bf 10}, 25 (1945).

\bibitem{lif} E. M. Lifshitz and L. P. Pitaevskii, {\em Physical Kinetics},
(Pergamon Press, Oxford, England, 1980).

\bibitem{krall} N. A. Krall and A. W. Trivelpiece, {\it Principles of
Plasma Physics}, (San Francisco Press, San Francisco 1986).

\bibitem{chen} F. Chen, {\it Introduction to Plasma Physics}, (Plenum 
Press, New York 1974).

\bibitem{bis}  D. Biskamp, {\it Non-linear Magnetohydrodynamics}
(Cambridge University Press, Cambridge, 1994).

\bibitem{mgknot} M. Giovannini, {\it Phys. Rev. D}{\bf 58}, 124027 (1998). 

\bibitem{vains} S. I. Vainshtein and Ya. B. Zeldovich, 
{\it Usp. Fiz. Nauk.} {\bf 106}, 431 (1972).

\bibitem{matt} W. H. Matthaeus, M. L. Goldstein, and S. R. Lantz, {\it Phys. Fluids }
{\bf 29}, 1504 (1986).

\bibitem{larmor} J. Larmor, {\it Rep. 87th Meeting Brit. Assoc. Adv. Sci.} (John Murray, London 1919).

\bibitem{pid1} J. H. Piddington, {\it Aust. J. Phys.} {\bf 23}, 731 (1970).

\bibitem{pid2} J. H. Piddington, {\it Ap. and Sp. Sci.} {\bf 35}, 269 (1975).

\bibitem{ka} R. Kulsrud and S. Anderson, {\it Astrophys. J.} {\bf 396}, 606
(1992).

\bibitem{VC1} S. Vainshtein, {\it Sov. Phys. JETP} {\bf 34}, 327 (1972).

\bibitem{VC2} F. Cattaneo and S. Vainshtein, {\it Astrophys. J.} {\bf 376}, 
L21 (1991).

\bibitem{kulsrud} R. M. Kulsrud, {\it Annu. Rev. Astron. Astrophys.} {\bf 37}, 37 (1999).

\bibitem{KR} F. Krause and K. H. R\"adler, {\it Mean field Magnetohydrodynamics and Dynamo Theory}, (Pergamon Press, Oxford, England 1980).

\bibitem{BJ}  R. Banerjee and K. Jedamzik, {\it Phys. Rev. Lett}{\bf 91} , 251301 (2003)

\bibitem{weinberg} S. Weinberg, {\em Gravitation and cosmology} (Wiley, New York, 1971). 

\bibitem{peebles} P. Peebles, {\it The Large Scale Structure of the Universe}, (Princeton University Press, Princeton, New Jersey 1980).

\bibitem{peebles2} P. Peebles, {\it Principles of Physical Csomology}, (Princeton University Press, Princeton, New Jersey 1993).

\bibitem{kolb} E. W.  Kolb and M. S. Turner, {\em The Early Universe}  
(Addison-Wesley, Redwood City, CA, 1990).

\bibitem{guth} A. Guth, {\it Phys.Rev.D}{\bf 23}, 347 (1981).

\bibitem{albrecht} A. Albrecht and P. Steinhardt,  {\it Phys.Rev.Lett.}{\bf 48}, 1220 (1982).

\bibitem{linde1} A. Linde, {\it Phys.Lett.B} {\bf 108}, 389 (1982).

\bibitem{linde2} A. Linde, {\it Phys.Lett.B}{\bf 129}, 177 (1983).

\bibitem{linde3} A. Linde, {\it  Phys.Rev.D} {\bf 49}, 748 (1994).

\bibitem{kolb2} J. Lidsey, {\it et al.}, {\it Rev.Mod.Phys.}{\bf 69}, 373 (1997).

\bibitem{dvs1} H. de Vega  and N. Sanchez, {\it Phys. Lett. B} {\bf 197}, 320 (1987).

\bibitem{dvs2} H. de Vega and N. Sanchez, {\it Phys. Rev. D} {\bf 50}, 7202 (1994).

\bibitem{sv} N. Sanchez and G. Veneziano, {\it Nucl. Phys. B} {\bf 333}, 253 (1990).

\bibitem{gsv1} M. Gasperini, N. Sanchez and G. Veneziano, {\it Nucl. Phys. B} {\bf 364}, 365 (1991).

\bibitem{gsv2}  M. Gasperini, N. Sanchez and G. Veneziano, {\it Int. J. Mod. Phys. A}  {\bf 6}, 3853 (1991).

\bibitem{gio1} M. Giovannini, {\it Phys. Rev. D} {\bf 55}, 595 (1997).

\bibitem{int1} J. D. Barrow, {\it Phys. Rev. D} {\bf 35}, 1805 (1987).

\bibitem{int2} E. W. Kolb, M. J. Perry, and T.P. Walker, {\it Phys. Rev. D} {\bf 33}, 869 (1986).

\bibitem{int3} F. S. Accetta, L. M. Krauss, and P. Romanelli, {\it Phys. Lett. B} {\bf 248}, 146 (1990).

\bibitem{lovelace} C. Lovelace, {\it Phys. Lett. B} {\bf 135}, 75 (1984).

\bibitem{fradkin} E. Fradkin and A. Tseytlin, {\it Nucl. Phys. B} {\bf 261}, 1 (1985).

\bibitem{callan} C. Callan et al., {\it Nucl. Phys. B} {\bf 262},  593 (1985).

\bibitem{Veneziano1} G. Veneziano, {\it Phys. Lett. B} {\bf 265}, 287 (1991).

\bibitem{Veneziano3} M. Gasperini and G. Veneziano, {\it Phys.Rept.} {\bf 373}, 1 (2003).

\bibitem{nonloc} M. Giovannini, M. Gasperini and G. Veneziano, {\it Phys.Lett.B} {\bf 569}, 113 (2003).

\bibitem{vlcur1} C. P. Dettmann, N. E. Frankel, and V. Kowalenko, 
{\i Phys. Rev. D} {\bf 48}, 5655 (1993).

\bibitem{vlcur2} R. M. Gailis, C. P. Dettmann, N. E. Frankel, and V. 
Kowalenko, {\it Phys. Rev. D} {\bf 50}, 3847 (1994).

\bibitem{vlcur3} R. M. Gailis, N. E. Frankel, and C. P. Dettmann, {\it Phys. Rev. D} {\bf 52}, 6901 (1995). 

\bibitem{vlcur4} R. M. Gailis and N. E. Frankel {\it Phys. 
Rev. D} {\bf 56}, 6901 (1995).

\bibitem{holcomb} K. Holcomb and T.Tajima, {\it Phys. Rev. D} {\bf 40}, 3809 (1989).

\bibitem{tajima} T. Tajima and Taniuti, {\it Phys. Rev. A} {\bf 42}, 3587 (1990).

\bibitem{subra} K. Subramanian and J. Barrow, {\it Phys. Rev. D} {\bf 58}, 083502 (1998).

\bibitem{subra2} K. Subramanian and J. Barrow, {\it Phys.Rev.Lett.} {\bf 81}, 3575 (1998).

\bibitem{ber} J. Bernstein, {\em Kinetic Theory in the Expanding 
Universe}, (Cambridge University Press, Cambridge, England, 1988).

\bibitem{mginf1} M. Giovannini and M. Shaposhnikov, {\it Phys. Rev. D} {\bf 62}, 103512 (2000).

\bibitem{enqvist1} J. Ahonen and  K. Enqvist, {\it Phys. Lett. B} {\bf 382}, 40 (1996).

\bibitem{bd1} D. Boyanovsky, H. de Vega, and  S. Wang {\it Phys.Rev.D}{\bf 67}, 065022 (2003)

\bibitem{olinto1} K. Jedamzik, V. Katalinic, and A. Olinto, {\it Phys. Rev. D} {\bf 57}, 3264 (1998).

\bibitem{BBNobs1} D.N. Schramm and M.S. Turner, {\it Rev. Mod. Phys. } {\bf 70}, 303  (1998).

\bibitem{BBNobs2} D. Tytler, J.M. O'Meara, N. Suzuki, and D. Lubin,
{\it Physica Scripta}  {\bf T85}, 12 (2000).

\bibitem{BBNobs3}  D. B. Fields and S. Sarkar, Particle Data Book {\it Phys. Rev. D}  {\bf 66} , 162 (2002).

\bibitem{hannu} H. Kurki-Suonio, e-Print Archive [astro-ph/0112182]. 

\bibitem{son} D.T. Son, {\it Phys.Rev.D} {\bf 59} 063008 (1999).

\bibitem{wasserman} I. Wasserman, {\it Astrophys. J.} {\bf 224}, 337 (1978)

\bibitem{coles2}  P. Coles, {\it Comments Astrophys.} {\bf 16}, 45 (1992).

\bibitem{enqvist2} A. Brandenburg, K. Enqvist and P. Olesen  {\it Phys.Rev.D} {\bf 54}, 1291  (1996). 

\bibitem{enqvist3} A. Brandenburg, K. Enqvist and P. Olesen  {\it Phys.Lett.B} {\bf 392} , 395 (1997).

\bibitem{olesen1}  P. Olesen,  {\it Phys.Lett.B} {\bf 398}, 321 (1997). 

\bibitem{olesen2} P. Olesen, {\it  NATO ASI Series B} {\bf  366}, 159 (1998). 

\bibitem{enqvist4} K. Enqvist, {\it Int. J. Mod. Phys. D} {\bf 7}, 331 (1998).

\bibitem{everett} M. Hindmarsh and A. Everett, {\it Phys. Rev. D}, {\bf 58} 103505 (1998).

\bibitem{dim1} K. Dimopoulos and A. Davis, {\it Phys. Lett. B}, {\bf 390}, 87 (1996).

\bibitem{hogan} C. Hogan, {\it Phys. Rev. Lett.}, {\bf 51} 1488 (1983).

\bibitem{saffman} P. Saffman, Phys. Fluids {\bf 10}, 1349 (1967).

\bibitem{shiro} T. Shiromizu,  {\it Phys.Lett.B} {\bf 443}, 127 (1998).

\bibitem{berera}  A. Berera and David Hochberg, e-Print Archive [cond-mat/0103447] 

\bibitem{hindmarsch1} M. Christensson, M. Hindmarsh, and  A. Brandenburg, {\it Phys. Rev. E} {\bf 64} 056405 (2001).

\bibitem{hindmarsch2} M. Christensson, M. Hindmarsh, and  A. Brandenburg, e-Print Archive [astro-ph/0209119].

\bibitem{pouquet1} A. Pouquet, U. Frisch and J. L\'eorat, {\it J. Fluid Mech.} {\bf 77}, 321 (1976).

\bibitem{pouquet2} J. L\'eorat, A. Pouquet, and U. Frisch, {\it J. Fluid Mech.} {\bf 104}, 419 (1981).

\bibitem{cornwall} J. Cornwall, {\it Phys. Rev. D}{\bf 56}, 6146 (1997).

\bibitem{iro1} P. Iroshnikov, {\it Sov. Astron. } {\bf 7}, 566 (1964).

\bibitem{iro2} R. Kraichnan, {\it Phys. Fluids} {\bf 8}, 1385 (1965).

\bibitem{mg3} M. Giovannini and M. Shaposhnikov, {\it Phys. Rev. D} {\bf 57}, 2186 (1998).

\bibitem{mg4} M. Giovannini and M. Shaposhnikov, {\it Phys. Rev. Lett.} {\bf 80}, 22 (1998).

\bibitem{ms1} K. Kajantie et al., {\it Nucl. Phys. B} {\bf 493}, 413 (1997).

\bibitem{rub1} V. Rubakov and A. Tavkhelidze, Phys. Lett. {\bf B 165}, 109 (1985).

\bibitem{campbell}  B. Campbell, S. Davidson, J. Ellis and K. Olive, {\it Phys. Lett. B} {\bf 297}, 118 (1992).

\bibitem{ibanez} L.~E. Ibanez and F.~Quevedo, {\it  Phys. Lett.B}  {\bf 283}, 261 (1992).

\bibitem{rummukainen} J. M. Cline, K. Kainulainen and K. A. Olive, {\it Phys. Rev. Lett.} {\bf 71}, 2372 (1993).

\bibitem{redlich} A. N. Redlich and L. C. R. Wijewardhana, {\it Phys. Rev. Lett.} {\bf 54}, 970 (1984).

\bibitem{hypmg1} M. Giovannini, {\it Phys.Rev.D} {\bf 61}, 063004 (2000).

\bibitem{hypmg2} M. Giovannini, {\it Phys.Rev.D} {\bf 61}, 063502 (2000). 

\bibitem{cswave} S. Deser, R. Jackiw, and S. Templeton, {\it Ann. Phys.} {\bf 140}, 372 (1982).

\bibitem{hyp1} M. Shaposhnikov, {\it Nucl. Phys. B} {\bf 287}, 757 (1987).

\bibitem{hyp2}  M. Shaposhnikov, {\it Nucl. Phys. B} {\bf 288}, 757 (1988).

\bibitem{JPI} R. Jackiw and  S.-Y. Pi,  {\it Phys.Rev. D} {\bf 61}, 105015 (2000).

\bibitem{JPI2} R. Jackiw, V.P. Nair, S-Y. Pi, {\it Phys.Rev.D}{\bf 62}, 085018 (2000). 

\bibitem{ad1} C. Adam, B. Muratori, and  C. Nash {\it Phys.Rev.D} {\bf 61}, 105018 (2000). 

\bibitem{ad2}  C. Adam, B. Muratori, and C. Nash, {\it Phys.Rev.D} {\bf 62}, 105027 (2000).

\bibitem{ay1} A. Ayala, J. Besprosvany, G. Pallares, and G. Piccinelli, {\it Phys.Rev. D} {\bf 64}, 123529 (2001).

\bibitem{ay2} A. Ayala and  J. Besprosvany {\it Nucl.Phys.B} {\bf 651},  211 (2003).

\bibitem{sikivie} P. Sikivie, {\it Phys. Rev. Lett.} {\bf 51}, 1415 (1983).

\bibitem{maiani} L. Maiani, R. Petronzio, and E. Zavattini, {\it Phys. Lett. B} {\bf 175}, 359 (1986). 

\bibitem{gasperini} M. Gasperini, {\it Phys. Rev. Lett.} {\bf 59}, 396 (1987). 

\bibitem{kim} J. Kim, {\it Phys. Rep.} {\bf 150}, 1 (1987); 

\bibitem{cheng} H.-Y. Cheng, {\it Phys. Rep.} {\bf 158}, 1 (1988).

\bibitem{biermann} L. Biermann, {\it Z. Naturf.} {\bf 5A}, 65 (1950).

\bibitem{LPA} M. Langer, J.-L. Puget, and N. Aghaim, {\it Phys. Rev. D}{\bf 67} 043505 (2003).

\bibitem{harrison1} E. Harrison, {\it Phys. Rev. Lett.} {\bf 18}, 1011 (1967).

\bibitem{harrison2}  E. Harrison, {\it Phys. Rev.} {\bf 167}, 1170 (1968).

\bibitem{harrison3} E. Harrison, {\it Mon. Not. R. Astr. Soc.} {\bf 147}, 279 (1970).

\bibitem{mishustin} I. Mishustin and A. Ruzmaikin, {\it Sov. Phys. JETP} {\bf 34}, 223 (1972).

\bibitem{vilenkin1} T. Vachaspati and A. Vilenkin, {\it Phys.Rev.Lett.} {\bf 67}, 1057 (1991).

\bibitem{shellard1} P. Avelino and P. Shellard, {\it Phys. Rev. D} {\bf 51}, 5946 (1995).

\bibitem{dm2} K. Dimopoulos, {\it Phys.Rev.D}{\bf 57}, 4629 (1998).  
 
\bibitem{dolr1} Z. Berezhiani and A. Dolgov, e-print Archive [astro-ph/0305595].

\bibitem{dolr2} A. Dolgov, e-print Archive [astro-ph/0306443].

\bibitem{hoganr} C. Hogan, e-print Archive [astro-ph/0005380].

\bibitem{gi1} H. Kodama and M. Sasaki, 
{\it Prog. Theor. Phys. Suppl.} {\bf 78}, 1  (1984).

\bibitem{gi2} V.F. Mukhanov, H.A. Feldman, and R. H. Brandenberger, 
{\it Phys. Rep.} {\bf 215}, 203 (1992).  

\bibitem{gi3} M. Giovannini,  {\it Class. Quant. Grav.} {\bf 20}, 5455 (2003).

\bibitem{mgst1} M. Gasperini, M. Giovannini, and G. Veneziano, {\it Phys. Rev. Lett.} {\bf 75}, 3796 (1995). 

\bibitem{FG}  F. Finelli  and A. Gruppuso, {\it Phys. Lett. B} {\bf 502}, 216 (2001).

\bibitem{gr1} L. Grishchuk, and  M. Solokhin, {\it Phys.Rev.D43}, 2566 (1991).

\bibitem{gr2} L. Grishchuk, {\it  Phys.Rev.D}, {\bf 53},  6784 (1996).

\bibitem{gr3}  L. Grishchuk,  and Yu. Sidorov,  {\it Phys.Rev.D} {\bf 42}, 3413 (1990). 

\bibitem{sqmg} M. Giovannini, {\it Phys.Rev.D} {\bf 61}, 087306 (2000). 

\bibitem{lou} R. Loudon, {\it The Quantum Theory of Light}, (Oxford University Press, 1991). 

\bibitem{mandel} L. Mandel and E. Wolf, {\it Optical Coherence and Quantum 
optics}, (Cambridge University Press, Cambridge, England, 1995).

\bibitem{car} P. Carruthers and C. C. Shih, {\it Int. J. Mod. Phys.A} {\bf 2}, 1447 (1987).

\bibitem{baym2} G. Baym, {\it Acta Phys.Polon.B} {\bf 29}, 1839 (1998).

\bibitem{lou2} R. Loudon, {\it Rep. Prog. Phys.} {\bf 43}, 913 (1980).

\bibitem{turner} M. S. Turner and L. M. Widrow, {\it Phys. Rev. D} {\bf 37}, 2734 (1988).

\bibitem{ratra} B. Ratra, {\it Astrophys. J. Lett.} {\bf 391}, L1 (1992).

\bibitem{dolgov} A. Dolgov, {\it Phys. Rev. D} {\bf 48}, 2499 (1993).

\bibitem{drummond} I. Drummond and S. Hathrell, {\it Phys.Rev. D} {\bf 22} 343 (1980).

\bibitem{carroll1}  S. Carroll, G. Field and R. Jackiw, {\it Phys. Rev. D} {\bf 41},  1231 (1990).

\bibitem{carroll2}  W. D. Garretson, G. Field and S. Carroll, {\it Phys. Rev. D} {\bf 46}, 5346 (1992).

\bibitem{carroll3} G. Field and S. Carroll {\it Phys.Rev.D}, {\bf 62}, 103008 (2000).

\bibitem{variation} M. Giovannini, {\it Phys. Rev. D} {\bf 64}, 061301 (2001).

\bibitem{bamba} K.~Bamba and J.~Yokoyama, e-print Archive [astro-ph/0310824].

\bibitem{gravphot} M. Gasperini, {\it Phys. Rev. D} {\bf 63}, 047301 (2001)

\bibitem{okun} L. Okun, {\it Sov. Phys. JETP} {\bf 56}, 502 (1982).

\bibitem{bertolami} O. Bertolami and D. Mota, {\it Phys. Lett. B} {\bf 455}, 96 (1999).

\bibitem{acdavis} A. C. Davis and K. Dimopoulos, {\it Phys. Rev. D} {\bf 55}, 7398 (1997). 

\bibitem{cal} E. Calzetta, A. Kandus and F. Mazzitelli, {\it Phys. Rev. D}, {\bf 57}, 7139 (1998).

\bibitem{cal1} A. Kandus,  E. Calzetta, F. Mazzitelli, and C. Wagner, {\it Phys.Lett. B} {\bf 472},  287 (2000).

\bibitem{davdim} 
K.~Dimopoulos, T.~Prokopec, O.~Tornkvist and A.~C.~Davis, {\it Phys. Rev. D} {\bf 65}, 063505 (2002).

\bibitem{mgs2} M. Giovannini, and M. Shaposhnikov, {\it Proc. of CAPP2000} (July 2000, Verbier Switzerland)
  eprint Archive [hep-ph/0011105]. 

\bibitem{mgint} M. Giovannini, {\it Phys. Rev. D} {\bf 62}, 123505 (2000).

\bibitem{star} U. Gunther, A. Starobinsky, and  A. Zhuk {\it  Phys.Rev.D}{\bf 69}, 044003 (2004).

\bibitem{ford} L. H. Ford, {\it Phys.Rev. D} {\bf 31}, 704 (1985).

\bibitem{berg} P. G. Bergmann, {\it Int. J. Theor. Phys.} {\bf 1}, 25 (1968).

\bibitem{bir} N. D. Birrel and P. C. W. Davies, {\em Quantum fields in curved 
space} (Cambridge University Press, Cambridge 1982).

\bibitem{gar} J. Garriga and E. Verdaguer, {\it Phys. Rev. D} {\bf 39}, 1072 (1991).

\bibitem{gio2} M. Gasperini and M. Giovannini, {\it Phys. Lett. B}{\bf 282}, 36 (1992);
{\it Phys.Rev.D} {\bf 47}, 1519 (1993).

\bibitem{lemoine} D. Lemoine and M. Lemoine, {\it Phys.Rev. D} {\bf 52}, 1955 (1995). 

\bibitem{mgst0}  M. Giovannini,  {\it Phys.Rev. D}{\bf 56},  631 (1997). 

\bibitem{infintdim1}  J.~M.~Cline, J.~Descheneau, M.~Giovannini and J.~Vinet,  JHEP {\bf 0306},  048 (2003)

\bibitem{infintdim2} M. Giovannini, {\it Phys.Rev.D}{\bf 66}, 044016 (2002 ).
 
\bibitem{infintdim3}M. Giovannini, J.-V. Le B\'e and S. Riederer, {\it Class.Quant.Grav.}{\bf 19}, 3357 (2002). 

\bibitem{baym3} G. Baym, D. Bodeker and L. McLerran, {\it Phys. Rev. D} {\bf 53}, 662 (1996).

\bibitem{kibblevil} T. Kibble and A. Vilenkin, {\it Phys. Rev. D} {\bf 52}, 679 (1995).

\bibitem{copeland1} P. Saffin and E. Copeland, {\it Phys. Rev. D} {\bf 56}, 1215 (1997).

\bibitem{copeland2} E. Copeland, P. Saffin and O. T\"ornkvist, {\it Phys. Rev. D} {\bf 61}, 105005 (2000).

\bibitem{ahonen2}  J. Ahonen and K. Enqvist, {\it Phys. Rev. D} {\bf 57}, 664 (1998).

\bibitem{torn2} O. T\"ornkvist, {\it Phys. Rev. D} {\bf 58}, 043501 (1998).

\bibitem{kocian} P. Kocian, e-print Archive [hep-ph/0006151].

\bibitem{vlasov2} V. Vlasov, V. Matveev,  A. Tavkhelidze, S. Khlebnikov, M. Shaposhnikov, {\it Sov. J. Part. Nucl.} {\bf 18}, 1 (1987).

\bibitem{vachaspati1} T. Vachaspati, {\it Phys. Lett. B} {\bf 265}, 258 (1991).

\bibitem{enqvistolesen} K. Enqvist and P. Olesen, {\it Phys. Lett. B} {\bf 329}, 195 (1994).

\bibitem{shap5} M. Joyce and M. Shaposhnikov, {\it Phys. Rev. Lett.} {\bf 79},  1193 (1997).

\bibitem{PT1} K. Kajantie, M. Laine, K. Rummukainen, and M. E. Shaposhnikov,
{\it Phys. Rev. Lett.} {\bf 77}, 2887 (1996).

\bibitem{PT2} K. Kajantie, M. Laine, K. Rummukainen, and M. E. Shaposhnikov,  {\it Nucl. Phys. B} {\bf 458}, 90 (1996).

\bibitem{spergel} J. Quashnock, A. Loeb and D. Spergel, {\it Astrophys. J. Lett.} {\bf 344}, L49 (1989).

\bibitem{cheng1} B. Cheng and A. Olinto, {\it Phys. Rev. D} {\bf 50}, 2421 (1994).

\bibitem{sigl2} G. Sigl, A. Olinto, and K. Jedamzik, {\it Phys. Rev. D} {\bf 55}, 4852 (1997).

\bibitem{bd2} D. Boyanovsky, M. Simionato, and  H. de Vega, {\it Phys.Rev.D} {\bf 67}, 023502 (2003).

\bibitem{bd3} D. Boyanovsky, H. de Vega, and M. Simionato, {\it Phys.Rev.D} {\bf 67}, 123505 (2003).

\bibitem{bd4} D. Boyanovsky, H. de Vega, and M. Simionato, e-Print Archive [astro-ph/0305131]. 

\bibitem{kainul} P. Elmfors, K. Enqvist, K. Kainulainen, {\it Phys.Lett.B} {\bf 440}, 269 (1998).

\bibitem{magnpt} K. Kajantie,  M. Laine,  J. Peisa,  K. Rummukainen, and  M. E. Shaposhnikov, 
{\it Nucl.Phys.B}{\bf 544}, 357 (1999).

\bibitem{volovik}  G. Volovik, {\it Phys.Rept.} {\bf 351}, 195 (2001)

\bibitem{bbn1} G. Greenstein, {\it Nature} {\bf 223}, 938 (1969).

\bibitem{bbn2} J. J. Matese and R. F. O' Connel, {\it Astrophys. J.} {\bf 160}, 451 (1970).

\bibitem{bbn3} B. Cheng, D. N. Schramm and J. Truran, {\it Phys. Rev. D} {\bf 
45}, 5006 (1994). 

\bibitem{bbn3a} B. Cheng, A. Olinto, D. N. Schramm and J. Truran,
{\it Phys. Rev. D} {\bf  54}, 4174 (1996).
 
\bibitem{bbn4} P. Kernan, G. Starkman and T. Vachaspati, {\it Phys. Rev. D}
{\bf 54}, 7207 (1996).

\bibitem{bbn4a}  P. Kernan, G. Starkman and T. Vachaspati, {\it Phys. Rev. D}
{\bf 56}, 3766 (1996).

\bibitem{bbn5} D. Grasso and H. Rubinstein, {\it Astropart. Phys.} {\bf 3}, 95 (1995).

\bibitem{bbn5a}  D. Grasso and H. Rubinstein, {\it Phys. Lett. B} {\bf  379}, 73 (1996).

\bibitem{iso1} K.  Thorne, {\it Astrophys. J.} {\bf 148}, 51 (1967).

\bibitem{iso2} S.  Hawking and R. J. Tayler, {\it Nature}{\bf 309}, 1278 (1966).

\bibitem{iso3} J. Barrow, {\it Mon. Not. R. Astron. Soc.}{\bf 175}, 359 (1976).

\bibitem{iso4} J. Barrow, {\it Phys.Rev.D} {\bf 55}, 7451 (1997) 

\bibitem{mam1} G. Steigman, {\it  Ann. Rev. Astron. Astrophys.} {\bf 14}, 339 (1976).

\bibitem{mam2} J. Rehm and K. Jedamzik, {\it Phys. Rev. Lett.} {\bf 81}, 3307 (1998).

\bibitem{mam3} H. Kurki-Suonio and E. Sihvola, {\it Phys. Rev. Lett.} {\bf 84}, 3756 (2000).

\bibitem{mam4} H. Kurki-Suonio and E. Sihvola, {\it Phys. Rev. D}{\bf 62}, 103508 (2000).

\bibitem{mam4a} E. Sihvola, {\it Phys. Rev. D} {\bf 63}, 103001 (2001).

\bibitem{mam4b} H. Kurki-Suonio, {\em BBN calculations}, astro-ph/0112182 .

\bibitem{mam5} J. Rehm and K. Jedamzik, {\it Phys. Rev. D}{\bf 63}, 043509 (2001).

\bibitem{mam6} M. Giovannini, H. Kurki-Suonio and E. Shivola, {\it Phys. Rev. D} {\bf 66},  043504 (2002).

\bibitem{zelm1} Ya. Zeldovich and I. Novikov, {\it The Structure and Evolution of the Universe}, (Chicago
University Press, Chicaggo, 1971), Vol.2.

\bibitem{zelm2} Ya. Zeldovich, {\it Sov. Phys. JETP}{\bf 21}, 656 (1965).

\bibitem{GDN} L. Grishchuk, A. Doroshkevich, and I. Novikov {\it Sov. Phys. JETP}{\bf 28}, 1210 (1969).

\bibitem{jed} K. Jedamzik, V. Katalinic, and A. V. Olinto,  {\it Phys. Rev. Lett.} {\bf 85} 700, (2000).

\bibitem{rg} J.Adams, U.Danielsson, D.Grasso, and H. Rubinstein, {\it Phys. Lett. B} {\bf 388}, 253 (1996). 

\bibitem{mgst2} M. Gasperini, M. Giovannini, and G. Veneziano, {\it Phys.Rev.D}, {\bf 52}, 6651 (1995).

\bibitem{durgas} R. Durrer, M. Gasperini, M. Sakellariadou and G. Veneziano, {\it Phys. Rev. D}{\bf 59}, 43511 (1999). 

\bibitem{ggvcmb} V. Bozza, et al. {\it Phys. Rev. D} {\bf 67}, 063514 (2003).

\bibitem{sw} R. K. Sachs and A. M. Wolfe, {\it Astrophys. J.} {\bf 147}, 73 (1967).

\bibitem{cobe} C. L. Bennet et al., Astrophys. J. Lett. {\bf 464}, L1 (1996).

\bibitem{cob2} M. Tegmark, Astrophys. J. Lett. {\bf 464}, L35 (1996).

\bibitem{boom} C. B. Netterfield et al., Astrophys. J. {\bf 571}, 604 (2002).

\bibitem{dasi} N. W. Halverson et al., Astrophys. J. {\bf 568}, 38 (2002).

\bibitem{maxima} A. T. Lee et al., Astrophys. J. {\bf 561}, L1 (2001).

\bibitem{archeops} A. Beno\^it et al.,  astro-ph/0210305.

\bibitem{WMAP1} C. Bennett {\it et al.} {\it Astrophys.J.Suppl.}{\bf 148}
1 (2003).

\bibitem{WMAP2} D. Spergel {\it et al.} {\it  Astrophys.J.Suppl.}
{\bf 148}, 175 (2003).

\bibitem{WMAP3} G. Hinshaw {\it et al.},  {\it 
Astrophys.J.Suppl.} {\bf 148}, 135 (2003).

\bibitem{koh} S. Koh, C. Lee, {\it Phys.Rev.D} {\bf 62}, 083509 (2000).

\bibitem{ferreirabarr} J. Barrow, P. Ferreira, and J. Silk, {\it Phys. Rev. Lett.} {\bf 78}, 3610 (1997).

\bibitem{mack} A. Mack, T. Kahniashvili, and A. Kosowsky, {\it Phys. Rev. D} {\bf 65}, 123004 (2002).

\bibitem{kan1} R. Durrer, T. Kahniashvili, and A. Yates, {\it Phys.Rev.D} {\bf 58} 123004, (1998).

\bibitem{kan0}  R. Durrer, T. Kahniashvili, {\it Helv.Phys.Acta} {\bf 71}, 445 (1998).

\bibitem{kan2} R. Durrer, P. Ferreira, and T. Kahniashvili, {\it Phys.Rev.D} {\bf 61}, 043001 (2000). 

\bibitem{caprdur} C. Caprini and  R. Durrer, {\it Phys.Rev.D} {\bf 65}, 023517 (2002).

\bibitem{subrapol} T. Seshadri and K. Subramanian, {\it Phys.Rev.Lett.} {\bf 87}, 101301 (2001).

\bibitem{subrabarrow2}  K. Subramanian, J. Barrow, {\it Mon. Not. Roy. Astron. Soc.} {\bf 335}, L57 (2002). 

\bibitem{subrabarrow1} K.  Subramanian, T. Seshadri, and  J. Barrow, {\it  Mon. Not. Roy. Astron. Soc.} {\bf 344}, L31 (2003).

\bibitem{pogosian} L. Pogosian, T. Vachaspati, S. Winitzki {\it  Phys.Rev.D} {\bf 65}, 083502 (2002).  

\bibitem{FR}  M. Giovannini, {\it Phys. Rev. D}{\bf 56},  3198 (1997). 

\bibitem{1a} M. Rees, {\it Astrophys. J.} {\bf 153}, L1 (1968).

\bibitem{2a} P. M. Lubin and G. F. Smoot, {\it Astrophys. J.} {\bf 245}, 1 (1981).

\bibitem{2ab} P. M. Lubin, P. Melese and G. F. Smoot, {\it Astrophys. J. Lett.} {\bf 273},
L51 (1983).

\bibitem{3a} M. Bersanelli et. al. [http://www.mrao.cam.ac.uk/projects/cpac/].

\bibitem{4a} E. Milaneschi and R. Fabbri, {\it Astron. Astrophys.} {\bf 151},
 7 (1985).

\bibitem{5a} D. Coulson, R. Crittenden and N. G. Turok, {\it Phys. Rev. D}
{\bf 52}, 5402 (1995).

\bibitem{5ab} D. Coulson, R. Crittenden and N. G. Turok, {\it Phys. Rev. Lett.} {\bf 73}, 2390 (1994).

\bibitem{6b} A. Kosowsky and A. Loeb, {\it Astrophys. J}{\bf 461}, 1 (1996).

\bibitem{6b2} D. Harari, J. Hayward and M. Zaldarriaga, {\it Phys. Rev. D} {\bf 55},  1841 (1997).

\bibitem{griru} D. Deryagin, D. Grigoriev, V. Rubakov and M. Sazhin, {\it Mod. Phys. Lett. A} {\bf 11}, 593 (1986). 

\bibitem{gw10} M. Giovannini, {\it Phys. Rev. D} {\bf 58}, 083504 (1998). 

\bibitem{gw1} M. Giovannini, {\it Phys. Rev. D} {\bf 60}, 123511 (1999).

\bibitem{gw1a} M. Giovannini,{\it  Class.Quant.Grav.} {\bf 16}, 2905 (1999). 

\bibitem{gw2} D. Babusci and M. Giovannini, {\it Int. J. Mod. Phys. D} {\bf 10}, 477 (2001).

\bibitem{gw3} D. Babusci and M. Giovannini, {\it Class. Quant. Grav.} {\bf 17}, 2621 (2000).

\bibitem{tsagas1} C. Tsagas and J. Barrow, {\it Class. Quantum Grav.} {\bf 14}, 2539 (1997).

\bibitem{tsagas2}  C. Tsagas and J. Barrow, {\it Class. Quantum Grav.} {\bf 15}, 3523 (1998).

\bibitem{papadop} D. Papadopoulos {\it Class.Quant.Grav.} {\bf 19}, 1 (2002).

\bibitem{anis1} M. Giovannini, {\it Phys. Rev. D} {\bf 59},  123518 (1999).

\bibitem{ryan} M. Ryan and L. Shepley, {\it Homogeneous Relativistic Cosmologies}, (Princeton University Press, Princeton 1978).

\bibitem{ks} R. Kantowski and P. K. Sachs. {\it J. Math. Phys.} {\bf 7}, 443 (1966).

\bibitem{bark1} J. Barrow and M. Dabrowski, {\it Phys. Rev. D}{\bf 55}, 630 (1997). 

\bibitem{bark2} J. Barrow and K. Kunze, {\it Phys. Rev. D}{\bf 55}, 623 (1997).

\bibitem{anis2} M. Giovannini, {\it Phys. Rev. D} {\bf 62}, 067301 (2000).

\bibitem{fordv} L. Ford, {\it Phys. Rev. D}{\bf 40}, 967 (1989).

\bibitem{lidseyv1} A. Burd and J. Lidsey, {\it Nucl. Phys. B}{\bf 351}, 679 (1991).

\bibitem{lidseyv2} J. Lidsey, {\it Nucl. Phys. B}{\bf 351}, 695 (1991).



\end{thebibliography}
\end{document}